\documentclass[aps,floatfix, twocolumn, superscriptaddress]{revtex4}
\usepackage[makeroom]{cancel}
\usepackage{bm}
\usepackage{amsmath,amssymb}
\usepackage{mathtools}
\mathtoolsset{showonlyrefs=true,showmanualtags=true}
\usepackage{amsfonts}
\usepackage{siunitx}
\usepackage{graphicx}
\usepackage{cases}

\newcommand{\be}{\begin{equation}}
\newcommand{\ee}{\end{equation}}

\newcommand{\dert}[1]{\frac{\partial #1}{\partial t}}
\newcommand{\tdert}[1]{\frac{d #1}{d t}}
\newcommand{\derf}[2]{\frac{\delta #1}{\delta #2}}
\newcommand{\mvec}{\bm m}
\newcommand{\hvec}{\bm h}
\newcommand{\rvec}{\bm r}

\newcommand{\heff}{\bm h_\mathrm{eff}}
\newcommand{\real}[1]{\Re\left\{#1\right\}}
\newcommand{\imag}[1]{\Im\left\{#1\right\}}
\usepackage{titlesec}
\titleformat{\paragraph}
{\normalfont\normalsize\bfseries}{\theparagraph}{1em}{}
\titlespacing*{\paragraph}
{0pt}{3.25ex plus 1ex minus .2ex}{1.5ex plus .2ex}

\renewcommand*{\theHsection}{\thesection}

\usepackage{adjustbox} 
\usepackage[dvipsnames]{xcolor}

\usepackage{xcolor} 
\usepackage{soul}   

\usepackage{hyperref}

\begin{document}

\title{Nonlinear interaction theory for parametrically-excited spin-wave modes in confined micromagnetic systems}

\author{Massimiliano d'Aquino}
\email{mdaquino@unina.it}
\affiliation{Department of Electrical Engineering and ICT, University of Naples Federico II, Naples, Italy}

\author{Salvatore Perna}
\affiliation{Department of Electrical Engineering and ICT, University of Naples Federico II, Naples, Italy}

\author{Hugo Merbouche}
\affiliation{SPEC, CEA, CNRS, Universit\'e Paris-Saclay, Gif-sur-Yvette, France}

\author{Grégoire de Loubens}
\affiliation{SPEC, CEA, CNRS, Universit\'e Paris-Saclay, Gif-sur-Yvette, France}

\begin{abstract}
We present a general theoretical approach for the quantitative description of parametric excitation of spin-wave modes in confined micromagnetic systems. This type of problem belongs to a broader class of nonlinear modal dynamics that arise across many areas of physics and engineering. The ferromagnetic sample is driven by parallel pumping with an external applied magnetic field having two tones at different frequencies, which are able to trigger parametric instability of two resonant modes. The two excited spin-wave modes interact in a strongly nonlinear fashion giving rise to quasiperiodicity, hysteresis and non-commutativity of steady-state oscillation regimes. To disentangle such a complex variety of dynamics, we develop a reduced-order model based on magnetization normal modes that is amenable of appropriate analytical treatment, leading to quantitative description of parametric instability thresholds, post-instability steady-state amplitude saturation and complete determination of phase diagrams for steady-state oscillation regimes. We have performed validation of the theory using numerical simulations. 
The phase diagrams allow to predict and explain all the features of the nonlinear interaction between the parametrically-excited spin-wave modes and can be directly compared with experimental results.
\end{abstract}
\maketitle

\section{Introduction}
The excitation of magnetization oscillations in ferromagnets is a central issue in spin dynamics for its applications relevant to microwave generation on the nanoscale in spintronic devices for magnetic storage\cite{Dieny2020opportunities}, reservoir, analog and neuromorphic computing\cite{bracher2018,nakane2018,hughes2019wave,Grollier2020,papp2021nanoscale,korber2023pattern}. Typical ways to trigger such oscillations in magnetic samples consist of applying appropriate time-varying external fields with suitable spectra that are able to excite natural magnetization modes via intrinsic ferromagnetic resonance\cite{Kittel1948} (FMR) at significant power level. 

Conventional FMR experiments are traditionally performed applying an external magnetic field having a microwave (ac) component perpendicular to a dc field that saturates the sample. At low power levels, resonance is exclusively induced by the component of the microwave magnetic field that aligns perpendicularly to the dc magnetization vector\cite{Kittel1948}. 

The characteristics of FMR undergo notable changes at elevated power levels, as elucidated by the pioneering works of Damon\cite{damon1953relaxation} and later investigations by Bloembergen and Wang\cite{Bloembergen1954}. At higher power intensities, a departure from the observations made at low power levels is evident. Commonly noted is the reduction in the FMR absorption peak's height with an escalation in power level. Moreover, an additional absorption peak can manifest at a dc field strength below that necessary for resonance. Harry Suhl's comprehensive explanation of these phenomena\cite{suhl1957theory} attributes these nonlinear effects to the destabilization of certain spin-wave modes when the amplitude of the uniform mode, propelled by the applied microwave field, surpasses a critical threshold.

Schl\"omann et al. and, independently, Morgenthaler have emphasized that the application of a microwave magnetic field with a sufficient amplitude parallel to the dc field induces nonlinear absorption due to the unstable growth of specific spin waves\cite{Schloemann1960,Morgenthaler1960}. This parallel pumping allows to excite spin-wave modes in a parametric resonance fashion\cite{landau_mechanics}, regardless of their spatial profiles provided that they have non-negligible ellipticity\cite{Gurevich2020}, and has been employed to generate them in several magnetic micro- and nanostructures such as extended  thin-films\cite{kurebayashi2011,sandweg2011,serga2012,hahn2013,lauer2016}, magnonic waveguides\cite{mohseni2020,heinz2022}, magnetic nanocontacts\cite{urazhdin2010}, magnetic tunnel junctions\cite{chen2017} and soft magnetic nanodots\cite{ulrichs2011,guo2014}. Recently, the parametric excitation of a large number of such spin-wave modes has been also demonstrated\cite{srivastava2023identification}, which paves the way to realize novel spin-wave based computing schemes\cite{bracher2018,nakane2018,hughes2019wave,papp2021nanoscale,korber2023pattern}. 

The theoretical interpretation of this phenomenon along with associated high-power effects, based on Landau-Lifshitz-Gilbert (LLG) equation\cite{landau1935dispersion}, has traditionally omitted a proper consideration of the boundary conditions at the sample surface. Instead, customary practice involved imposing periodic boundary conditions under which the sample's normal modes manifest as plane waves\cite{suhl1957theory}, allowing for a straightforward discussion of their stability when a "pump field" is applied either parallel or perpendicular to the dc field. While these results align well with the assumption that the wavelength of the considered modes is significantly smaller than the sample dimensions, the theory occasionally deviates from this premise. Notably, certain conditions predicted by the theory suggest an infinite wavelength for spin waves with the lowest instability threshold\cite{Schloemann1961}. Consequently, there has been an imperative need to formulate the instability criterion while considering the actual boundary conditions\cite{Schloemann1961,denton1961theoretical}. 

Much later, following the progress of fabrication of micro- and nano-scale samples, the theoretical treatment based on plane waves for the description of instability under parallel pumping in bulk bodies has been also extended to {uniformly-magnetized infinite magnetic thin-films\cite{kostylev1995parallel} and  ellipsoidal particles\cite{Krivosik2010hamiltonian}}.
However, these theoretical approaches are based on approximations related with the assumption of translational invariance along given direction(s) {or specific particle shape.}

In this paper, by using an approach based on magnetization normal modes\cite{daquino_novel_2009,perna2022computational}, we address the general theoretical formulation and consequent numerical calculation of magnetization dynamics driven by parallel pumping in arbitrarily shaped confined ferromagnets, achieved using ac external field with one or two tones at appropriate amplitude and frequencies such that parametric instability of one or two normal modes is triggered. 
The manifestations of such an instability occur in a very complex way due to the highly nonlinear nature of the mode coupling and dynamical interaction between selected modes.

{The instability driven by single-tone excitation has been theoretically studied for single-domain particles\cite{guo2014}, extended magnonic waveguides\cite{Verba2017waveguides}, spin-torque nanocontacts\cite{urazhdin2010} and perpendicular antiferromagnets\cite{Tomasello2022AFMVCMA}.} 
{Previous theoretical studies on the interactions between two spin-wave modes excited in nanostructures have been performed for nonlinear FMR\cite{Melkov2013nonlinear}, three-magnon scattering in vortex-state magnetic nanodots\cite{Slobodianiuk2019threemagnon,Verba2021threemagnon}, ultrathin ferromagnetic nanowires with antisymmetric exchange\cite{Verba2019DMI} and spin-torque oscillators\cite{Hamadeh2023multitone,Slobodianiuk2025multimode} by using a Hamiltonian formalism\cite{Lvov_book,Tyberkevych2020VHF}. }

The description of micromagnetic dynamics\cite{brown_micromagnetics_1963} using normal modes\cite{perna2022computational} has proven to be a powerful tool for the study of magnetization oscillations in ferromagnetic objects\cite{bruckner2019large,Perna2022AIP,Ngouagnia2025} by means of reduced-order models with few degrees of freedom, i.e. nonlinear coupled oscillators obtained projecting the LLG equation on the linear magnetization eigenmodes, that nevertheless have accuracy comparable with that of much more time-consuming full scale micromagnetic simulations\cite{abert2019micromagnetics}. 

We also note that the study of distributed (continuum) nonlinear dynamical systems through reduced‑order models created by performing appropriate projections onto a given set of basis functions is ubiquitous across mathematical physics, tracing back to the connection between the Korteweg-de Vries equation and the celebrated Fermi-Pasta-Ulam model\cite{fermi1955studies, zabusky1965interaction}, with examples that appear in areas as diverse as the analysis of vibrations in solid structures (e.g., using invariant‑manifold methods to reduce nonlinear elastic PDEs\cite{Kerschen2009,Touz2021}), the modeling of fluid flows (e.g., employing Proper-Orthogonal-Decomposition–Galerkin projection for Navier–Stokes equations\cite{Ballarin2016}), and the study of pattern formation and instabilities in thermo-acoustic\cite{Wildemans2023}, or combustion systems\cite{Doehner2022} (e.g., using coupled nonlinear oscillators like Van der Pol models to capture synchronization phenomena).
This places the problem of nonlinear spin-wave interactions under parametric excitation within the broader class of nonlinear modal dynamics studied across many fields of physics and engineering.

The main result of this paper is a quantitative recipe to determine the phase diagrams of steady-state oscillation regimes for parametrically-excited spin-wave mode pairs that mutually interact in a nonlinear fashion. Despite the multiplicity of regimes achievable changing the excitation conditions, the developed analytical theory shows that such diagrams depend on few parameters, namely appropriate self- and mutual nonlinear frequency shift coefficients and their combinations, which can be quantitatively determined from the knowledge of the eigenmodes profiles and influence the (co-)existence and stability of possible regimes. The most significant consequences of such complexity include a large number of manifestations of hysteresis and non-commutativity of oscillation regimes as function of external excitation conditions. {Remarkably, we point out that the diagrams produced according to the developed theory have been directly and successfully compared with recent experiments which probe the steady-state regime of parametrically-excited spin-wave mode pairs in a magnetic microdisk using two-tone spectroscopy\cite{soares_knet_flagship_paper_2025}.}

The paper is structured as follows. 
In section \ref{sec:LLG dynamics}, we introduce the relevant dynamical equations in micromagnetics along with the Normal Modes Model (NMM). In section \ref{sec:self interactions}, we derive the analytical expressions for parametric instability thresholds associated with the normal mode matching the frequency selection rule imposed by an external field at given frequency. 
Then, by extending the analysis to the post-instability weakly nonlinear oscillation regime, we develop an analytical theory for the steady-state amplitude of the parametrically-excited normal mode. Subsequently, in Section \ref{sec:mutual modes interaction}, we consider the more complex (and interesting) case of interactions between pairs of parametrically-excited normal modes via two-tone external signal. In this respect, 
we develop an analytical theory for the steady-state post-instability mode amplitudes in the coupled regime, showing that either indirect mode suppression or enhancement is possible. Remarkably, we demonstrate that such pairwise interaction is non-reciprocal (i.e. the influence of the former mode on the latter is not equal to the vice versa) and may also lead to hysteresis, multi-stability, quasiperiodicity and non-commutative oscillation dynamics depending on history of the external excitation (e.g. the temporal sequence used for activation/deactivation of each external signal tone). 
Finally, validation of the theory is performed in section \ref{sec:numerical results} by using numerical micromagnetic simulations for a nanoscale thin-disk-shaped sample. The paper also includes some more technical derivations and additional numerical results in the appendices and supplementary material.

\section{Magnetization dynamics under parallel pumping in micromagnetic systems}\label{sec:LLG dynamics}

Magnetization dynamics of a ferromagnetic body is described by the LLG equation written in dimensionless form\cite{BMS2009}:
\begin{equation}\label{eq:LLG}
    \dert{\mvec} = -\mvec \times \heff[\mvec]+\alpha\mvec\times\dert{\mvec}  \quad,
\end{equation}
where  $\mvec(\rvec,t)$ is the magnetization unit-vector (normalized by the saturation magnetization $M_s$) at each location $\rvec\in \Omega$ ($\Omega$ is the region occupied by the magnetic body), time is measured in units of $(\gamma M_s)^{-1}$ (corresponding to 5.7 ps for $\gamma=2.21\times 10^5 A^{-1}s^{-1} m$ and $\mu_0 M_s=1$T), $\alpha$ is the Gilbert (dimensionless and positive, typically in the order $\sim 10^{-4}\div 10^{-2}$) damping parameter.

The effective field $\heff(\rvec,t)$ is given by\cite{brown_micromagnetics_1963}:
\begin{equation}\label{eq:heff}
\heff=-\derf{g}{\mvec}  \,\,,
\end{equation}
which takes into account interactions (exchange, anisotropy, magnetostatics, Zeeman) among magnetic moments and is expressed as the variational derivative of the free energy functional (the dimensionless energy is measured in units of $\mu_0 M_s^2 V$, with $V$ being the volume of region $\Omega$)
\begin{equation}\label{eq:free energy}
    g(\mvec,\hvec_a)=\frac{1}{V} \int_\Omega \frac{l_\mathrm{ex}^2}{2} (\nabla\mvec)^2 + f_\mathrm{an} - \frac{1}{2}\hvec_m\cdot \mvec - \hvec_a\cdot \mvec \,dV \,,
\end{equation}
where $A_\mathrm{ex}$ and $l_\mathrm{ex}=\sqrt{(2A_\mathrm{ex})/(\mu_0 M_s^2)}$ are the exchange stiffness constant and length, respectively, $f_\mathrm{an}$ is the anisotropy energy density, $\hvec_m$ is the magnetostatic (demagnetizing) field and $\hvec_a(\rvec, t)$ the external applied field. 

When the anisotropy is of uniaxial type, such that $f_\mathrm{an}=\kappa_\mathrm{an} [1-(\bm m\cdot \bm e_\mathrm{an})^2]$ with $\kappa_\mathrm{an}$ and $\bm e_\mathrm{an}$ being the uniaxial anisotropy constant and unit-vector, respectively,  the effective field can be expressed by the sum of a linear operator $\mathcal{C}$ acting on magnetization vector field plus the applied field:
\begin{equation}
    \bm h_\mathrm{eff}(\bm r, t)= -\mathcal{C} \bm m + \bm h_a \,,
\end{equation}
where $\mathcal{C}=-l_\mathrm{ex}^2\nabla^2+\mathcal{N}-\kappa_\mathrm{an} e_\mathrm{an}\otimes e_\mathrm{an}$ and $\mathcal{N}$ is the (symmetric-positive definite) demagnetizing operator such that:
\begin{equation} \label{eq:demagnetizing field}
    \bm h_m(\bm r) = \frac{1}{4\pi}\nabla\nabla\cdot \int_\Omega \frac{\bm m(\bm r')}{|\bm r - \bm r'|} \, dV = -\mathcal{N}\bm m\,.
\end{equation}
Equation \eqref{eq:LLG} is usually complemented with the natural boundary conditions $\partial\bm m/\partial\bm n=0$ at the body surface $\partial\Omega$, which is typical when no surface anisotropy is considered\cite{rado1959}.
It can be shown that the operator $\mathcal{C}$ with the aforementioned boundary conditions is self-adjoint in the appropriate subspace of square-integrable vector fields\cite{brown_micromagnetics_1963}.

The LLG dynamics fulfills the preservation of magnetization amplitude at each spatial location:
\begin{equation}\label{eq:micromagnetic constraint}
    |\mvec(\rvec)|^2=1 \quad,\quad\forall\rvec\in\Omega \,,\,\forall t\geq t_0
\end{equation}
as it can be seen by dot multiplying both sides of eq.\eqref{eq:LLG} by $\mvec$, provided that it is initially satisfied (for $t=t_0)$.

We consider a stable micromagnetic equilibrium configuration $\mvec_0(\rvec)$ under a dc applied field such that it satisfies Brown's equations\cite{brown_micromagnetics_1963}
while also being a local minimizer of the free energy functional \eqref{eq:free energy}.

We now decompose magnetization $\mvec(\rvec,t)$ as 
\begin{equation}
    \mvec(\rvec,t)=\mvec_0(\rvec)+\delta\mvec(\rvec,t) \quad.
\end{equation}
The micromagnetic constraint \eqref{eq:micromagnetic constraint} implies that 
\begin{equation}
    |\mvec_0(\rvec)+\delta\mvec(\rvec,t)|^2=1 \Rightarrow \mvec_0\cdot\delta\mvec=-\frac{1}{2}|\delta\mvec|^2\quad,
\end{equation}
meaning that magnetization deviation $\delta\mvec$ can be decomposed into a transverse plus longitudinal component (with respect to the equilibrium magnetization $\mvec_0)$ in the following form
\begin{equation}\label{eq:magnetization deviation}
    \delta\mvec=\delta\mvec_\perp +\delta m_0 \mvec_0= \delta\mvec_\perp -\frac{1}{2}|\delta\mvec|^2 \mvec_0 \quad.
\end{equation}

\subsection{Linear magnetization dynamics and magnetization eigenmodes}
The above equation states that, when an additional small-amplitude ac field $\delta\bm h_a(t)$ is applied and the deviation is small enough such that $|\delta\mvec|\ll 1$, at first-order in $|\delta\mvec|$ the dynamics occurs transverse to the equilibrium, namely $\delta\mvec\approx\delta\mvec_\perp$. 

In the latter situation of small oscillations, the LLG equation \eqref{eq:LLG} can be linearized as follows:
\begin{equation}\label{eq:linearized LLG}
    \dert{\delta\mvec}=\mvec_0\times \mathcal{A}_0\delta\mvec  -(\mvec_0+\delta\mvec)\times\delta\hvec_a +\alpha\mvec_0\times\dert{\delta\mvec}  \quad,
\end{equation}
where the operator $\mathcal{A}_0$ is associated with the Hessian\cite{daquino_novel_2009} of the free energy functional \eqref{eq:free energy} and has the expression:
\begin{equation}\label{eq:A0 operator}
    \mathcal{A}_0=\mathcal{C}+h_0\mathcal{I} \quad,
\end{equation}
with $h_0(\rvec)=\mvec_0\cdot\heff[\mvec_0]$ being the projection of the effective field at equilibrium onto equilibrium magnetization and $\mathcal{I}$ the identity operator.

In the conservative and unforced (i.e. zero damping $\alpha=0$ and zero ac field $\delta\hvec_a(t)=0$) situation, the linearized dynamics can be expressed in the form of a generalized eigenvalue problem\cite{daquino_novel_2009}:
\begin{equation}\label{eq:generlized eigenproblem}
    \mathcal{A}_{0\perp}\bm\varphi_h =\omega_h\mathcal{B}_0\bm\varphi_h \quad,
\end{equation}
where $\mathcal{A}_{0\perp}=\mathcal{P}_\perp\mathcal{A}_0$ denotes projection of $\mathcal{A}_0$ on the plane point-wise transverse to $\mvec_0$ ($\mathcal{P_\perp}=\mathcal{I}-\mvec_0\otimes\mvec_0$),  $\mathcal{B}_0=-j\Lambda[\mvec_0(\rvec)]$, and $\Lambda$ denotes the operator notation for cross-product such that $\Lambda[\bm v]\bm w=\bm v\times\bm w$.

The couple $(\omega_h,\bm\varphi_h(\rvec))$ represent the $h$-th eigenpair that is solution of the problem \eqref{eq:generlized eigenproblem} (composed of eigenfrequency and eigenmode, respectively). It has been shown\cite{daquino_novel_2009} that the spectrum of the eigenproblem is discrete and the eigenfrequencies are all real due to the self-adjointness of the operators $\mathcal{A}_{0\perp},\mathcal{B}_{0}$ and positive definiteness of $\mathcal{A}_{0\perp}$ ($\mvec_0$ is a stable equilibrium).

As a further consequence of the real nature of the operator $\mathcal{A}_0$ and the definition of $\mathcal{B}_0$, it happens that, for each eigenpair $(\omega_h,\bm\varphi_h)$, there exists another eigenpair given by $(-\omega_h,\bm\varphi_h^*)$ (the notation $^*$ denotes complex conjugate). It has  been also shown that the absolute value of the spectrum is bounded from below, namely $|\omega_h|\geq\omega_1, h=1,2,\ldots$ where $\omega_1>0$ denotes the frequency of the lowest mode.  Moreover, the eigenfunctions satisfy the weighted orthonormality condition\cite{daquino_novel_2009}:
\begin{equation}\label{eq:orthonormality}
    (\bm \varphi_h,\bm\varphi_k)_{\mathcal{A}_{0\perp}}=(\bm\varphi_h,\mathcal{A}_{0\perp}\bm \varphi_k)=\frac{1}{V}\int_\Omega \bm\varphi_h^*\cdot\mathcal{A}_{0\perp}\bm \varphi_k\,dV=\delta_{hk} \,,
\end{equation}
where the compact notation $(\cdot,\cdot)$ denotes the usual (complex) inner product in $\mathbb{L}^2(\Omega)$ and $\delta_{hk}$ is Kronecker's symbol.

Magnetization deviation $\delta\mvec_\perp$ can be represented by using normal modes (eigenmodes) $\bm \varphi_h$:
\begin{equation}\label{eq:magnetization expansion}
    \delta\mvec_\perp(\rvec,t)=\sum_h a_h(t)\bm\varphi_h(\rvec) \quad,
\end{equation}
where the complex amplitudes $a_h(t)$ must guarantee that the magnetization deviation $\delta\mvec_\perp$ is a real field. Thus, denoting $\omega_{-h}=-\omega_h\quad,\quad \bm\varphi_{-h}=\bm\varphi_h^*$
one can rewrite eq.\eqref{eq:magnetization expansion} as
\begin{equation}\label{eq:magnetization expansion 2}
    \delta\mvec_\perp(\rvec,t)=
    \sum_{h\geq 1} a_h(t)\bm\varphi_h(\rvec)+ a_{-h}(t)\bm\varphi_h^*(\rvec) \quad,
\end{equation}
which implies that necessarily must be $a_{-h}(t)=a_h^*(t)$.

We recall that, in the absence of ac forcing but in the presence of nonzero (small) damping $\alpha$, the linear LLG dynamics can be described by an appropriate perturbation of the eigenproblem \eqref{eq:generlized eigenproblem}, which yields the following perturbed eigenfrequencies\cite{daquino_novel_2009}:
\begin{equation}\label{eq:perturbed eigenfrequency}
    \omega_h^{(\alpha)} =\omega_h+j\alpha\omega_h^2||\bm\varphi_h||^2 \quad,
\end{equation}
where notation $||\cdot||$ means the usual norm in $\mathbb{L}^2(\Omega)$.

\subsection{Weakly-nonlinear magnetization dynamics and Normal Modes Model}

We analyze now the conservative small-amplitude magnetization dynamics obtained by setting zero damping (i.e. $\alpha=0$) in eq.\eqref{eq:LLG} 
and follow the approach described in Ref.\cite{perna2022computational} and termed Normal Modes Model (NMM) where, by performing series expansion of the longitudinal magnetization deviation $\delta m_0(\rvec,t)$ within the so-called 'parabolic approximation'
\begin{equation}
    \delta m_0(\rvec,t)=\sqrt{1-\delta\mvec_\perp^2} \approx 1-\frac{1}{2}\delta\mvec_\perp^2 \quad,
\end{equation}
and projecting magnetization dynamics on the $h$-th eigenfunction $\bm \varphi_h$, the dynamical equation for the complex amplitude $a_h(t)$ (here reported as truncated at third-order terms) is obtained:
\begin{equation}\label{eq:NMM}
    \dot{a_h}=j\omega_h\left(b_h+\sum_i b_{hi} a_i+\sum_{i,j} c_{hij}a_ia_j + \sum_{i,j,k} \frac{d_{hijk}}{2} a_i a_j a_k\right) ,
\end{equation}
where 'dot' notation indicates time-derivative and coefficients $b,c,d$ are given by\cite{perna2022computational}:
\begin{align}
    b_h&=-(\bm\varphi_h,\delta\hvec_a) \,,\, \nonumber \\ 
    b_{hi} &= \delta_{hi}+\alpha b_{hi}^\alpha + b_{hi}^f\,,\, b_{hi}^\alpha = 
    j\,\omega_i\,\left(\bm\varphi_h, \bm\varphi_i\right)\,,\, \nonumber \\b_{hi}^f&= -\left(\bm\varphi_h,\bm m_0\cdot\bm \delta\bm h_a\,\bm\varphi_i\right)  \label{eq:NMM coef b}\\
 c_{hij} &= -\frac{1}{2}\left(\bm\varphi_h,\mathcal{C}\psi_{ij}\bm m_0\right) -\left(\bm\varphi_h,\bm m_0\cdot\mathcal{C}\bm\varphi_j\bm\varphi_i\right)\,,\label{eq:NMM coef c} \\
 d_{hijk} &=-\left(\bm\varphi_h, \psi_{jk}\,\mathcal{C}\bm\varphi_i\right) + \left(\bm\varphi_h,\bm m_0\cdot\mathcal{C}\psi_{jk}\bm m_0\,\bm \varphi_i \right) \,, \label{eq:NMM coef d}
\end{align}
where $\psi_{ij} = \bm\varphi_i\cdot\bm\varphi_j$. In eq.\eqref{eq:NMM}, the third and fourth terms resemble those arising from three- and four-magnon scattering interactions in classical spin waves theory\cite{Lvov_book} but the weighting coefficients are determined by the spatial profiles of the eigenmodes.

It is worthwhile considering the linear unforced conservative dynamics arising from the NMM for each amplitude $a_h$:
\begin{equation}
\dot{a_h}=j\omega_h a_h \quad,   \label{eq:linear conservative mode dynamics} 
\end{equation}
which obviously leads to a solution $a_h=a_h(0)e^{j(\omega_h t+\phi_{h0})}$ that describes harmonic oscillations at frequency $\omega_h$. According to the expansion \eqref{eq:magnetization expansion 2}, the term $a_h\bm\varphi_h+a_h^*\bm\varphi_h^*$ in the above situation describes a real vector field 'rotating' with angular frequency $\omega_h$ in the plane transverse to $\mvec_0$ at each spatial location $\rvec$. Such a rotation can be equivalently described for any arbitrary choice of the initial phase $\phi_{h0}$. 
We also observe that an orthonormal eigenfunction $\bm \varphi_h$ scaled with a factor $e^{j\phi}$ is still an orthonormal eigenfunction of the problem \eqref{eq:generlized eigenproblem}. For reasons that will be more clear in the sequel, we will assume hereafter that orthonormal eigenfunctions $\bm \varphi_h$ are also such that:
\begin{equation}\label{eq:gauge phi0}   \imag{(\bm\varphi_h,\bm\varphi_h^*)}=\imag{(\bm\varphi_h,\bm\varphi_{-h})}=0 \quad.
\end{equation}
This is equivalent to making a special choice of $\phi_{h0}$ that will be instrumental in the subsequent derivations. We point out that the gauge property \eqref{eq:gauge phi0} independently holds for each normal mode $\bm\varphi_h$.  

\section{Excitation of single modes driven by parallel pumping}\label{sec:self interactions}

We now analyze magnetization oscillations driven by parallel pumping using a single-tone time-harmonic signal. We follow an incremental approach starting from the analysis of the linear dynamics and subsequently introducing nonlinear effects.

\subsection{Linear dynamics}

Consistently with the parallel pumping situation, we assume that the equilibrium and, consequently, the ac field are mostly aligned with the $x$-axis, namely 
\begin{equation}\label{eq:parallel pumping assumptions}
\mvec_0(\rvec)\approx\bm e_x \,,\, \delta\hvec_a(t)=\delta h_a(t)\mvec_0\approx \delta h_a(t)\bm e_x \,.
\end{equation}

The result of the above assumptions is the following  equation that describes conservative linear dynamics for modes' amplitude $a_h(t)$ driven by parallel pumping:
\begin{equation}\label{eq:linear conservative parametric mode dynamics}
    \dot{a_h}=j\omega_h \sum_i [\delta_{hi} - \delta h_a(t) (\bm\varphi_h,\bm\varphi_i)]a_i \quad. 
\end{equation}

When a sufficiently small damping $0<\alpha\ll 1$ is considered in the dynamics,  one can formally replace $\omega_h$ in eq.\eqref{eq:linear conservative mode dynamics}  with the perturbed frequency $\omega_h^{(\alpha)}$ (see eq.\eqref{eq:perturbed eigenfrequency}), which leads to the following modification of eq.\eqref{eq:linear conservative parametric mode dynamics}:
\begin{equation}\label{eq:linear parametric mode dynamics}
   \dot{a_h}=j\omega_h \sum_i [\delta_{hi} - \delta h_a(t) (\bm\varphi_h,\bm\varphi_i)]a_i - \lambda_h a_h\quad,    
\end{equation}
where the damping parameter $\lambda_h=\alpha\omega_h^2||\bm\varphi_h||^2$.

Let us now consider the equations \eqref{eq:linear parametric mode dynamics} for the dynamics of mode $h$ and its conjugate $-h$ where we retain only 'self-interaction' terms with indices $i=h,-h$, which yields:
\begin{align}
    \dot{a_h}&\!=\!+j \omega_h \!\!\left[(1\!-\!\delta h_a(t)||\bm\varphi_h||^2) a_h \!-\! \delta h_a(t) (\bm \varphi_h,\bm \varphi_{h}^*) a_h^*\right]  \!-\! \lambda_h a_h\,, \label{eq:linear parametric ah}\\
\dot{a_h^*}&\!=\!-j \omega_h \![(1\!-\!\delta h_a(t) ||\bm\varphi_h||^2) {a_h^*} \!-\!\delta h_a(t)  (\bm \varphi_h,\bm \varphi_h^*)^*a_{h} ]  \!-\! \lambda_h a_h^*\,. \label{eq:linear parametric ah*}
\end{align}

It can be shown that eqs.\eqref{eq:linear parametric ah},\eqref{eq:linear parametric ah*} admit dynamics with diverging amplitude $|a_h|$ only due to the second term. In fact, by multiplying eq.\eqref{eq:linear parametric ah} by $a_h^*$ and eq.\eqref{eq:linear parametric ah*} by $a_h$ and summing both sides, one has:
\begin{align}
    \tdert{|a_h|^2} &\!= \!-j\omega_h\{(\bm\varphi_h,\bm\varphi_h^*){a_h^*}^2 \!\!-\! [(\bm\varphi_h,\bm\varphi_h^*){a_h^*}^2]^*\}\delta h_a(t) \!-\! 2\lambda_h |a_h|^2\nonumber \\
    &= 2\omega_h \imag{(\bm\varphi_h,\bm\varphi_h^*){a_h^*}^2}\delta h_a(t) - 2\lambda_h |a_h|^2\quad, \label{eq:ah square dynamics}   
\end{align}
which implies that, under periodic forcing $\delta h_a(t)$, the average value of $|a_h|^2$ may increase (starting from a tiny nonzero value) if the right-hand side has positive average over the forcing period. 

In this respect, considering a forcing field of the type
\begin{equation}
    \delta h_a(t)=\delta h \sin(\omega_\mathrm{rf} t) \quad,\quad \omega_\mathrm{rf}=2\omega_h+\epsilon_h \approx 2\omega_h \,,
\end{equation}
where $\epsilon_h=\omega_\mathrm{rf}-2\omega_h$ denotes a small frequency detuning (i.e. such that $|\epsilon_h|\ll 2\omega_h$), we search for solutions $a_h$ oscillating at half the forcing frequency\cite{landau_mechanics} $\omega_\mathrm{rf}/2=\omega_h+\epsilon_h/2$:
\begin{equation}\label{eq:ah ansatz}
    a_h(t)=A(t)e^{j[(\omega_h+\frac{\epsilon_h}{2})t+\Phi]}.
\end{equation}
By plugging eq.\eqref{eq:ah ansatz} into eq.\eqref{eq:ah square dynamics} and remembering that $(\bm\varphi_h,\bm\varphi_h^*)$ can be chosen real according to the gauge property \eqref{eq:gauge phi0}, one ends up with
\begin{align}
       2A\dot{A}&= -2\omega_h (\bm\varphi_h,\bm\varphi_h^*) A^2 \delta h \cdot \\
       &\cdot \sin((2\omega_h +\epsilon_h)t +2\Phi) \sin(\omega_\mathrm{rf} t) - 2\lambda_h A^2\,.\label{eq:ah square dynamics 2}    
\end{align}
By assuming nonzero amplitude $A\neq 0$, dividing both sides by $A$, and retaining only 'resonant' terms such that $2\sin((2\omega_h +\epsilon_h)t +2\Phi) \sin(\omega_\mathrm{rf} t)\approx \cos(2\Phi)$, one can integrate both sides of eq.\eqref{eq:ah square dynamics 2} over the forcing period and see that the right-hand side is positive (that is, the amplitude $A$ grows over time) provided that:
\begin{equation}
 -\frac{\omega_h (\bm\varphi_h,\bm\varphi_h^*) \delta h \cos(2\Phi)}{2} - \lambda_h >0 \quad,
\end{equation}
which means that the field amplitude must overcome the critical value:
\begin{equation}
      \delta h_\mathrm{thres,h}=\frac{2\lambda_h}{\omega_h|(\bm\varphi_h,\bm\varphi_h^*)|} =\frac{2\alpha\omega_h||\bm\varphi_h||^2}{|(\bm\varphi_h,\bm\varphi_h^*)|}\quad.  \label{eq:threshold field parametric resonance} 
\end{equation}

The latter expression is amenable to physical interpretation if one makes the macrospin assumption (i.e. spatially uniform magnetization).
In fact, by solving the eigenproblem \eqref{eq:generlized eigenproblem} in such a condition where applied field and equilibrium magnetization are directed along $\bm e_x$, one obtains the eigenfrequencies $\omega_{1,2}=\pm\sqrt{\omega_{0y}\omega_{0z}}$ and eigenmodes $\bm\varphi_{1}=(i/\sqrt{2\omega_{0y}})\bm e_y + (1/\sqrt{2\omega_{0z}})\bm e_z \,,\, \bm\varphi_2=\bm\varphi_1^*$, where $\omega_{0y}=D_y-D_x+h_{ax}$ and $\omega_{0z}=D_z-D_x+h_{ax}$ and $D_x\leq D_y <D_z$ are the effective demagnetizing factors taking into account shape and crystalline anisotropy.
In this respect, the ellipticity of magnetization motion is given by:
\begin{equation}\label{eq:macrospin ellipticity}
    e=\sqrt{\frac{\omega_{0y}}{\omega_{0z}}} \quad,
\end{equation}
which varies between $0$ (linear) and $1$ (circular) polarizations. Now we point out that:
\begin{equation}\label{eq:macrospin L2 phi norm}
    ||\bm\varphi_{h}||^2=\frac{1}{2\omega_{0y}}+\frac{1}{2\omega_{0z}} \quad,\quad h=1,2\,,
\end{equation}
and, consequently, one can express the ellipticity of both macrospin normal modes as follows:
\begin{equation}\label{eq:ellipticity as norms}
    e_h^2=\frac{||\bm\varphi_h||^2-|(\bm\varphi_h,\bm\varphi_h^*)|}{||\bm\varphi_h||^2 + |(\bm\varphi_h,\bm\varphi_h^*)|} \quad.
\end{equation}
It is apparent that the above definition guarantees $e_h\leq 1$. Now, recalling the threshold for parametric resonance expressed by eq.\eqref{eq:threshold field parametric resonance}, one can reinterpret it in terms of the ellipticity (we only report the expression for $\omega_h>0$):
\begin{equation}\label{eq:threshold vs ellipticity}
    \delta h_\mathrm{thres,h}=\frac{2\alpha\omega_h||\bm\varphi_h||^2}{|(\bm\varphi_h,\bm\varphi_h^*)|}= 2\alpha\omega_h\frac{1+e_h^2}{1-e_h^2} \quad.
\end{equation}
Equation \eqref{eq:threshold vs ellipticity} immediately tells that, in order to have parametric resonance, one needs asymmetric bodies such that $\omega_{0y}\neq\omega_{0z}$, which rules out circularly polarized precessional motion that would occur, for instance, if $D_y=D_z$ (e.g. rotationally-symmetric (spheroidal) magnetic body around the $x$-axis).
The relationship between eigenmodes norms and ellipticity offers the possibility to recover the classical expressions for parametric thresholds reported in literature (see, for instance, ref.\cite{Gurevich2020}).

\begin{figure*}[t]
    (a)\hspace{5cm}(b)\hspace{3cm}(c)\hspace{5cm}(d)\\
    \centering
    \includegraphics[width=18cm]{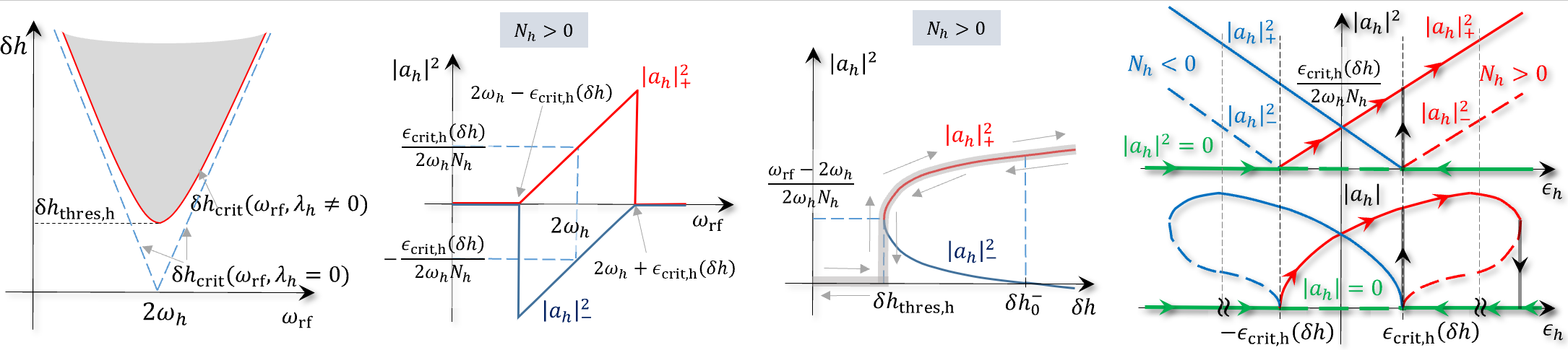}
    \caption{(a) Detuning cone for parametric excitation of mode $h$ via parallel pumping (see eq.\eqref{eq:detuning cone}). The solid red (dashed blue) line refers to eq.\eqref{eq:critical detuning cone} when the damping $\lambda_h>0$ (resp. $\lambda_h=0$). (b) Steady-state amplitude $|a_h|^2$ as function of ac field frequency $\omega_\mathrm{rf}$  for PWM excitation above threshold $\delta h>\delta h_\mathrm{thres,h}$. (c) Steady-state amplitude $|a_h|^2$ as function of ac field amplitude $\delta h$. In both (b),(c) it is assumed that the NFS $N_h>0$. The solid red (resp. blue) line refers to $|a_h|^2_+$ (resp. $|a_h|^2_-$) in eq.\eqref{eq:saturation amplitude mode h}. (d) Steady-state amplitude $|a_h|^2$ as function of ac field frequency detuning $\epsilon_h$ under CW parametric excitation above threshold $\delta h>\delta h_\mathrm{thres,h}$. Solid (dashed) lines refer to stable (unstable) P-modes. Red (blue) lines refer to $N_h>0$ ($N_h<0$), green lines refer to P-mode $|a_h|^2=0$. Arrows elucidate the hysteresis in the case $N_h>0$ and the irreversible downward jump that may occur for large positive detuning (notice the axis breaks in the lower panel) not described by the theory.}
    \label{fig:parametric resonance cone panels}
\end{figure*}

Now we look for the excitation conditions in terms of amplitude and frequency of the external field suitable to trigger the parametric instability.
By expressing eqs.\eqref{eq:linear parametric ah}-\eqref{eq:linear parametric ah*} in polar coordinates $A(t),\Phi(t)$
(see the appendix \ref{sec:derivation of detuning cone}) 
and averaging the equations with respect to the fast time scale with period $T=2\pi/\omega_\mathrm{rf}$, one obtains:
\begin{align}
    \dot{A}&=\left[-\frac{\omega_h (\bm\varphi_h,\bm\varphi_h^*) \delta h}{2} \cos(2\Phi)-\lambda_h\right] A \,, \label{eq:averaged linear parametric A} \\
    \dot{\Phi}+\frac{\epsilon_h}{2}&= \frac{\omega_h (\bm\varphi_h,\bm\varphi_h^*) \delta h}{2} \sin(2\Phi) \,. \label{eq:averaged linear parametric Phi}
\end{align}
where, with little abuse of notation, we have denoted averaged amplitude and phase with the same letters $A$ and $\Phi$.
Now we search for the stationary periodic solutions (termed P-modes\cite{BMS2009}) of eqs.\eqref{eq:averaged linear parametric A}-\eqref{eq:averaged linear parametric Phi} such that $\dot{A}=0,\dot{\Phi}=0$. It is worth noting that stationary solutions of eqs.\eqref{eq:averaged linear parametric A}-\eqref{eq:averaged linear parametric Phi} correspond to periodic oscillations (P-modes) at half-forcing frequency $\omega_\mathrm{rf}/2$. Thus, one can derive the following condition for parametric excitation of the mode $h$,
termed the detuning cone (Arnold's tongue) $\delta h>\delta h_\mathrm{crit,h}$, in the control plane $(\omega_\mathrm{rf},\delta h)$:
\begin{equation}
    \delta h > \sqrt{\frac{(\omega_\mathrm{rf}-2\omega_h)^2}{\omega_h^2|(\bm\varphi_h,\bm\varphi_h^*)|^2}+\delta h_\mathrm{thres,h}^2}  \label{eq:detuning cone}
\end{equation}
that is reported in fig.\ref{fig:parametric resonance cone panels}(a). For what follows, it is also useful to express the critical curve in terms of detuning:
\begin{equation}
        \epsilon_\mathrm{crit,h}^2 = (\omega_\mathrm{rf,crit,h}-2\omega_h)^2= \omega_h^2|(\bm\varphi_h,\bm\varphi_h^*)|^2(\delta h^2-\delta h_\mathrm{thres,h}^2)  \,. \label{eq:critical detuning cone}
\end{equation}
which leads to the following parametric excitation condition:
\begin{equation}
    |\epsilon_h|=|\omega_\mathrm{rf}-2\omega_h|<\epsilon_\mathrm{crit,h}(\delta h) \quad. \label{eq:detuning cone eps}
\end{equation}

It is apparent that, for zero damping $\alpha=\lambda_h=0$, the threshold $\delta h_\mathrm{thres}$ also vanishes and the critical curve described by eq.\eqref{eq:critical detuning cone} degenerates into a couple of straight lines $\delta h_\mathrm{crit,h}(\omega_\mathrm{rf},\lambda_h=0)$ (detuning cone) departing from the point $(2\omega_h,0)$. The effect of nonzero damping $\lambda_h>0$ implies that there is a minimum field threshold required for parametric resonance (linear in $\alpha$) and the critical curve $\delta h_\mathrm{crit,h}(\omega_\mathrm{rf},\lambda_h>0)$ asymptotically approaches the detuning cone for large frequency mismatch with respect to $2\omega_h$.

\subsection{Nonlinear effects. Steady-state regime and amplitude saturation in parallel pumping} \label{sec:nonlinear self interaction}

Here we extend the above theory to derive the maximum steady-state amplitude of the mode parametrically-excited above the threshold. To this end, we assume that the natural frequency has the following dependence on the amplitude $|a_h|$:
\begin{equation}   \label{eq:frequency vs amplitude NFS}
\omega_h'(|a_h|)=\omega_h(1+N_h|a_h|^2)\quad, 
\end{equation}
where $\omega_h>0$ is the positive eigenfrequency of normal mode $h$. Incidentally, it is worth stressing that throughout the paper we will term all the mathematical quantities $a_h,|a_h|,|a_h|^2$ without distinction as 'mode amplitudes' but, strictly speaking, $|a_h|^2$ physically represents the mode power (or, mode intensity), which is very important for the comparison with experiments. The assumption \eqref{eq:frequency vs amplitude NFS} can be justified by looking at the NMM equation \eqref{eq:NMM} and performing the following steps:
\begin{enumerate}
    \item We consider only self-interactions, namely the mode indices within the summations $i,j,k$ can only assume values $h,-h$; in the specific case of $d_{hijk}$, we only consider the sequence $\{i,j,k\}=\{h,-h,h\}$ and all its circular permutations.
    \item We neglect terms $c_{hij}a_i a_j$ since we search for solutions oscillating at frequency $\sim\omega_h$. Conversely, the aforementioned terms either produce oscillations with double frequency $\sim2\omega_h$ (when $i=j=h$) that are off-resonance or almost stationary terms (when $i=\pm h,j=\mp h$).
\end{enumerate}
The result of these assumptions is that the only nonlinear term to include in the mode amplitude dynamical equations is:
\begin{align}
         j\omega_h&\real{\frac{d_{h,h,-h,h} + d_{h,h,h,-h}+ d_{h,-h,h,h}}{2}}|a_h|^2 a_h \\
         &= j\omega_h N_h |a_h|^2 a_h \quad, \label{eq:NFS mode h}
\end{align}
which defines the $h$-th mode's nonlinear frequency shift (NFS) $N_h=\real{\tilde{N}_h}$. Due to the symmetry of the tensor $d_{hijk}$  with respect to the exchange of the indices $j,k$ (see eq.\eqref{eq:NMM coef d}), the NFS coefficient \eqref{eq:NFS mode h} is reduced to:
\begin{equation}
    \tilde{N}_h= d_{h,h,-h,h} + \frac{d_{h,-h,h,h}}{2} \quad.   \label{eq:NFS mode h 2}
\end{equation}

It can be proved that the complex NFS $\tilde{N}_h$ defined in eq.\eqref{eq:NFS mode h 2} is a purely real number (the proof is reported in section \ref{sec:appendix NFS real} in the supplementary material).

For the specific case of the macrospin assumption discussed in the previous section, defining $D_{xz}=D_x-D_z$ and $D_{xy}=D_x-D_y$, a compact expression for the NFS is:
\begin{equation}
    N_h = \frac{3D_{xz}\omega_{0y}^2+3D_{xy}\omega_{0z}^2+(D_{xz}+D_{xy})\omega_{0y}\omega_{0z}}{8\omega_{0y}^2\omega_{0z}^2} \,. \label{eq:macrospin NFS}    
\end{equation}

\par
Now we proceed to derive equations for the nonlinear dynamics of mode $h$.
The term reported in eq.\eqref{eq:NFS mode h} augments the linear equations \eqref{eq:linear parametric ah}-\eqref{eq:linear parametric ah*} in the following way:
\begin{align}
    \dot{a_h}&=j \omega_h [(1+N_h|a_h|^2-\delta h_a(t) ||\bm\varphi_h||^2)a_h  \nonumber \\
    &-\delta h_a(t)(\bm \varphi_h,\bm \varphi_{h}^*) a_h^*)]  - \lambda_h a_h\,,\, \text{   and c.c.}\label{eq:nonlinear parametric ah} 
\end{align} 
The NFS term does not affect the threshold for the parametric excitation of the mode, as one can clearly see in the framework of the threshold derivation carried out in the previous section. Nevertheless, it plays an important role in the achievement of a steady-state regime after the mode amplitude growth has been triggered by an ac field exceeding the threshold $\delta h_\mathrm{thres,h}$. In fact, when the mode amplitude grows, due to the NFS in eq.\eqref{eq:frequency vs amplitude NFS}, the oscillation frequency varies with the changing amplitude $|a_h|$, increasing or decreasing with respect to the eigenfrequency $\omega_h$ depending on the sign $\pm$ of the coefficient $N_h$. The frequency change ends when the amplitude $|a_h|$ reaches a steady-state constant value, which happens when the amplitude-dependent detuning matches the critical curve for the parametric resonance\cite{guo2014}, namely:
\begin{align}
    &(\omega_\mathrm{rf}-2\omega_h'(|a_h|))^2
    =(\epsilon_h-2\omega_h N_h|a_h|^2)^2= \nonumber \\ 
    &=\epsilon_\mathrm{crit,h}^2=  
    \omega_h^2|(\bm\varphi_h,\bm\varphi_h^*)|^2(\delta h^2-\delta h_\mathrm{thres,h}^2) \,. \label{eq:nonlinear detuning cone}
\end{align}
Equation \eqref{eq:nonlinear detuning cone} imposes that the amplitude-dependent detuning must match the critical detuning at the given field $\delta h$ (the other parameters are fixed, notice that $\epsilon_h=\omega_\mathrm{rf}-2\omega_h$ denotes the detuning at negligible amplitude $|a_h|^2\approx 0$). This condition translates into an equation for the steady-state amplitude $|a_h|$ that will describe the nonlinear phenomenon of amplitude saturation after the growth due to parametric excitation above the threshold.
  
The solutions $|a_h|^2$ of eq.\eqref{eq:nonlinear detuning cone} 
can be written as:
\begin{align}
    |a_h|_\pm^2&=\frac{\omega_\mathrm{rf}\!-\!2\omega_h \!\pm\! \text{sign}(N_h)\omega_h|(\bm\varphi_h,\bm\varphi_h^*)|\sqrt{\delta h^2\!-\!\delta h_\mathrm{thres,h}^2}}{2\omega_h N_h}= \nonumber \\ &=\frac{\epsilon_h + s^\pm_h\epsilon_\mathrm{crit,h}(\delta h)}{2\omega_h N_h}\,. \label{eq:saturation amplitude mode h pm}
\end{align}
where $s^\pm_h=\pm\text{sign}(N_h)$ (the compact notation will be useful in the sequel). Now we discuss the outcome of eq.\eqref{eq:saturation amplitude mode h pm} concerning steady-state regimes that always start from a tiny nonzero initial amplitude. This resembles what can be done experimentally\cite{soares_knet_flagship_paper_2025} by applying pulse-width-modulated (PWM) excitation field to separately probe the steady-state regime at given amplitude and frequency, that occurs when switching the ac excitation off for enough time to let the system relax before applying the excitation field again with modified amplitude and/or frequency (hereafter this type of excitation will be referred to as PWM). We also remark that situations where amplitude and frequency of the external field are continuously changed (hereafter referred to as continuous-wave (CW)) may lead to different steady-states, as we will see in the sequel.

Under these assumptions, the qualitative behavior of $|a_h|^2$ as function of $\omega_\mathrm{rf}$ is reported in fig.\ref{fig:parametric resonance cone panels}(b). This graph can be interpreted with the help of fig.\ref{fig:parametric resonance cone panels}(a), imagining to follow a horizontal line enclosed within the shaded region at a given field $\delta h>\delta h_\mathrm{thres,h}$. It is apparent that, due to the positive definiteness of $|a_h|^2$, only one of the solutions $|a_h|^2_\pm$ makes sense and depends on the sign of the NFS $N_h$. When the frequency detuning $\epsilon_h=\omega_\mathrm{rf}-2\omega_h$ goes outside the interval $[-\epsilon_\mathrm{crit,h}(\delta h),\epsilon_\mathrm{crit,h}(\delta h)]$ the steady-state output amplitude vanishes since the parametric excitation condition \eqref{eq:detuning cone eps} is not matched anymore (we recall that each PWM excitation starts from almost zero initial mode amplitude). In fig.\ref{fig:parametric resonance cone panels}(b), it is assumed $N_h>0$, but when $N_h<0$ eq.\eqref{eq:saturation amplitude mode h pm} immediately suggests that the  $|a_h|^2_\pm$ (red and blue lines in fig.\ref{fig:parametric resonance cone panels}(b)) undergo a mirror-reflection around the abscissa axis.  

Thus, one can conclude that, under PWM excitation, the only admissible (non-negative) solution between $|a_h|^2_\pm$ according to Eq.\eqref{eq:saturation amplitude mode h pm} can be expressed as
\begin{equation}\label{eq:saturation amplitude mode h}
    |a_h|_+^2
    = \frac{\epsilon_h +s^+_h \epsilon_\mathrm{crit,h}(\delta h)}{2\omega_h N_h}\,,
\end{equation}
{which describes the typical sawtooth curve\cite{guo2014,urazhdin2010} for power versus driving frequency (with positive/negative slope depending on positive/negative sign of NFS $N_h$).}

It is worth mentioning that, in this steady-state regime, the parametrically-excited mode $h$ produces a time-domain periodic signal that oscillates with amplitude $|a_h|_+$ (given by \eqref{eq:saturation amplitude mode h}) and frequency $\omega_\mathrm{rf}/2$ that is now detuned with respect to the amplitude-dependent natural frequency \eqref{eq:frequency vs amplitude NFS} by $\omega_\mathrm{rf}/2-\omega_h'(|a_h|^2_+)$. Thus, parametric resonance occurs through synchronization of the forced response of the system at frequency $\omega_\mathrm{rf}/2$ (half frequency of the external signal) slightly detuned with respect to the (amplitude-dependent) natural mode frequency $\omega_h'(|a_h|_+^2)$.
We also remark that, for constant field amplitude $\delta h$, the maximum amplitude obtainable under PWM excitation by changing the applied rf frequency is given by $|a_h|^2_\mathrm{max}=\frac{\epsilon_\mathrm{crit,h}(\delta h)}{\omega_h|N_h|}$ (see fig.\ref{fig:parametric resonance cone panels}(b)).

It is also interesting to look at the behavior of $|a_h|^2_\pm$ as function of field amplitude $\delta h$ at a given external ac frequency $\omega_\mathrm{rf}$. This is reported in fig.\ref{fig:parametric resonance cone panels}(c). As in the previous fig.\ref{fig:parametric resonance cone panels}(b), we assume $N_h>0$ and positive detuning $\omega_\mathrm{rf}-2\omega_h>0$. 
When the amplitude $\delta h$ exceeds the threshold $\delta h_\mathrm{thres,h}$, two branches $|a_h|_\pm^2$ exist simultaneously (solid red and blue curves) 
but, as mentioned before, the system will follow $|a_h|^2_+$ (see the shaded thicker gray line in fig.\ref{fig:parametric resonance cone panels}(c)) even when changing the field amplitude back and forth. Thus, the only jump occurs when the field amplitude passes through the threshold value $\delta h_\mathrm{thres,h}$.
Analogous considerations can be made for the remaining combination of signs of NFS and detuning.

The existence and stability of steady-state solutions \eqref{eq:saturation amplitude mode h pm} under generic excitation (not limited to PWM) can be studied by introducing polar coordinates and using the same averaging technique as before (see Appendix \ref{sec:stability single P modes} for details).

The P-modes amplitudes $|a_h|^2$ are then given by eq.\eqref{eq:saturation amplitude mode h pm} (considering both signs $\pm)$ when the parameters are such that $|a_h|^2\geq 0$. 
We report here for convenience the existence conditions
\begin{align}
    &\, s_h^+(\epsilon_h+s_h^\pm \epsilon_\mathrm{crit,h})\geq 0 \quad, \label{eq:nonzero Pmodes single tone}
\end{align}
as well as stability conditions
\begin{equation}
    s_h^\pm (\epsilon_h+s_h^\pm\epsilon_\mathrm{crit,h})\geq 0 \quad,\quad s_h^\pm=\pm\text{sign}(N_h)\,, \label{eq:stability condition nonzero P-mode}
\end{equation}
which have been compactly expressed as function of the sign of the NFS $N_h$.
These will be instrumental later to study interactions between pairs of parametrically-excited modes (notice that stability condition \eqref{eq:stability condition nonzero P-mode} differs from the existence condition \eqref{eq:nonzero Pmodes single tone} in the leading coefficients $s_h^\pm$ and $s_h^+$, respectively).

A first outcome of the stability analysis (see Appendix \ref{sec:stability single P modes}) is that the only stable P-mode is always $|a_h|_+^2$ regardless of the sign of the NFS, as depicted in fig.\ref{fig:parametric resonance cone panels}(d).

A second outcome is that the steady-state oscillation can be either nonzero and synchronized (phase-locked) with half frequency $\omega_\mathrm{rf}/2$ of the pumping signal or identically zero. In the sequel, we will see that this condition can be violated in the case of parametric excitation of a spin-wave modes' pair, giving rise to the onset of nonzero unsynchronized (quasiperiodic) regimes, termed Q-modes\cite{BMS2009}. 

One can observe that hysteresis is possible. For instance, when $N_h>0$, if one starts from negative detuning and increases $\epsilon_h$, the solid green and red branch is followed (see arrows in the upper panel of fig.\ref{fig:parametric resonance cone panels}(d)). Conversely, starting from positive large enough $\epsilon_h$ and decreasing it, one follows the solid green branch descending down to $\epsilon_h=\epsilon_\mathrm{crit,h}$ when an upward jump on the solid red branch occurs. Analogous reasoning can be done when $N_h<0$ following blue and green branches.

As final remark, we observe that, despite the theory seems to predict stability of nonzero P-mode $|a_h|_+^2$ for arbitrary large detuning (in absolute value), one has to recall that the theory relies on the assumption of small detuning\cite{landau_mechanics}, which means that for large $|\epsilon_h|$, the P-mode $|a_h|_+^2$ will lose stability and will fall to $|a_h|^2=0$ (solid green branch). This is also illustrated in the lower panel of fig.\ref{fig:parametric resonance cone panels}(d) after the axis breaks denoting large $|\epsilon_h|$. 

In addition, to reconcile fig.\ref{fig:parametric resonance cone panels}(b) with fig.\ref{fig:parametric resonance cone panels}(d), we observe that the former refers to steady-state amplitude vs frequency when the initial amplitude is tiny but nonzero $|a_h|\approx 0$ (PWM excitation), while the latter refers to stability of steady-state amplitude under continuous change of frequency (i.e. as it would happen in a continuous wave (CW) experiment). Thus, PWM excitation does not produce hysteresis while CW excitation does. Accordingly, figure \ref{fig:parametric resonance cone panels}(b) does not exhibit the hysteresis depicted in fig.\ref{fig:parametric resonance cone panels}(d).

\section{Mutual interactions between parametrically-excited modes}\label{sec:mutual modes interaction}

We now consider interactions between pairs of modes with positive indices $h\neq n$, with amplitudes denoted as $a_h$ and $a_n$, that are excited by a two-tone signal:
\begin{equation}
    \delta h_a(t)=\delta h_1\sin(\omega_\mathrm{rf1}t) + \delta h_2\sin(\omega_\mathrm{rf2}t) \quad, \label{eq:two tones signal}
\end{equation}
where the frequencies are close to twice the respective eigenfrequencies $\omega_\mathrm{rf1}\sim2\omega_h, \omega_\mathrm{rf2}\sim 2\omega_n$ and the initial phases (not important in the discussion) are chosen equal to zero for the sake of simplicity. 
The NMM equations \eqref{eq:NMM} for the amplitude dynamics of modes $h,n$ are:
\begin{align} 
    \dot{a_h}&=j\omega_h\left(b_h+\sum_i b_{hi} a_i+\sum_{i,j} c_{hij}a_ia_j + \sum_{i,j,k} \frac{d_{hijk}}{2} a_i a_j a_k\right) \,, \label{eq:ah NMM} \\ 
    \dot{a_n}&=j\omega_n\left(b_n+\sum_i b_{ni} a_i+\sum_{i,j} c_{nij}a_ia_j + \sum_{i,j,k} \frac{d_{nijk}}{2} a_i a_j a_k\right) \,. \label{eq:an NMM}
\end{align}
Now we make the following assumptions:
\begin{enumerate}
    \item we consider self-interactions, namely the mode indices within the summation triples $\{i,j,k\}$ can only assume values $h,-h$ or $n,-n$; this is similar to what has been outlined in the previous section \ref{sec:self interactions}.
    \item We consider mutual interactions such that the mode indices within the summation triples $\{i,j,k\}$ can only assume values $\{h,-h, n\}, \{-h,h, n\}$ or $\{n,-n, h\}, \{-n,n, h\}$ and circular permutations of the latter sequences (we observe that, contrary to the self-interaction case, here $h\neq n$ implies considering also sequences with inverted indices, leading to six terms instead of three).
    \item We neglect terms $c_{hij}a_i a_j$ since we search for solutions oscillating at frequency $\sim\omega_h$ and $\sim\omega_n$. Conversely, the aforementioned terms either produce oscillations that are off-resonance or almost stationary terms (also similar to what has been outlined in section \ref{sec:nonlinear self interaction}).
\end{enumerate}
Under these assumptions, by following a line of reasoning similar to what has been described in section \ref{sec:nonlinear self interaction}, we define the following modes interaction matrix hereafter referred to as (complex) mutual NFS matrix:
\begin{equation}\label{eq:complex NMFS matrix}
    \tilde{N}_{hn}=d_{h,n,h,-n} + d_{h,h,n,-n}+ d_{h,-n,h,n} \quad.
\end{equation}
where the expression of $d_{hijk}$ is given by eq.\eqref{eq:NMM coef d} and  the symmetry of the tensor $d_{hijk}$ with respect to the exchange of indices $j,k$, has been used.

By observing equations \eqref{eq:ah NMM}-\eqref{eq:an NMM}, one can infer that the real part of the coefficient $\tilde{N}_{hn}$ would act as a pure mutual nonlinear  frequency shift in that it produces a purely imaginary term proportional to $j\omega_h |a_n|^2a_h$ in the right-hand side, while the imaginary part would play the role of a nonlinear damping-like term since a purely real term proportional to $-|a_n|^2a_h$ appears in the right-hand side. More specifically, one can be easily convinced that the coefficients
\begin{equation}
    N_{hn}=\real{\tilde{N}_{hn}} \quad, \quad \lambda_{hn}=\omega_h\imag{\tilde{N}_{hn}} \quad,
\end{equation}
respectively represent the NFS and the nonlinear damping of the mode $h$ controlled by the amplitude $|a_n|^2$ of mode $n$. Similar argument holds for the coefficients with exchanged indices $N_{nh}$ and $\lambda_{nh}$. 

It can be shown (see sections \ref{sec:appendix mutual NFS real}-\ref{sec:appendix symmetry NFS real} in supplementary material) that, in general, the mutual NFS matrix is real and symmetric, namely:
\begin{equation}
    \lambda_{hn}=\lambda_{nh}=0 \quad,\quad N_{hn}= N_{nh}\quad. \label{eq:reality and symmetry Nhn}
\end{equation}

\begin{figure}[t]
    \centering
    \includegraphics[width=6cm]{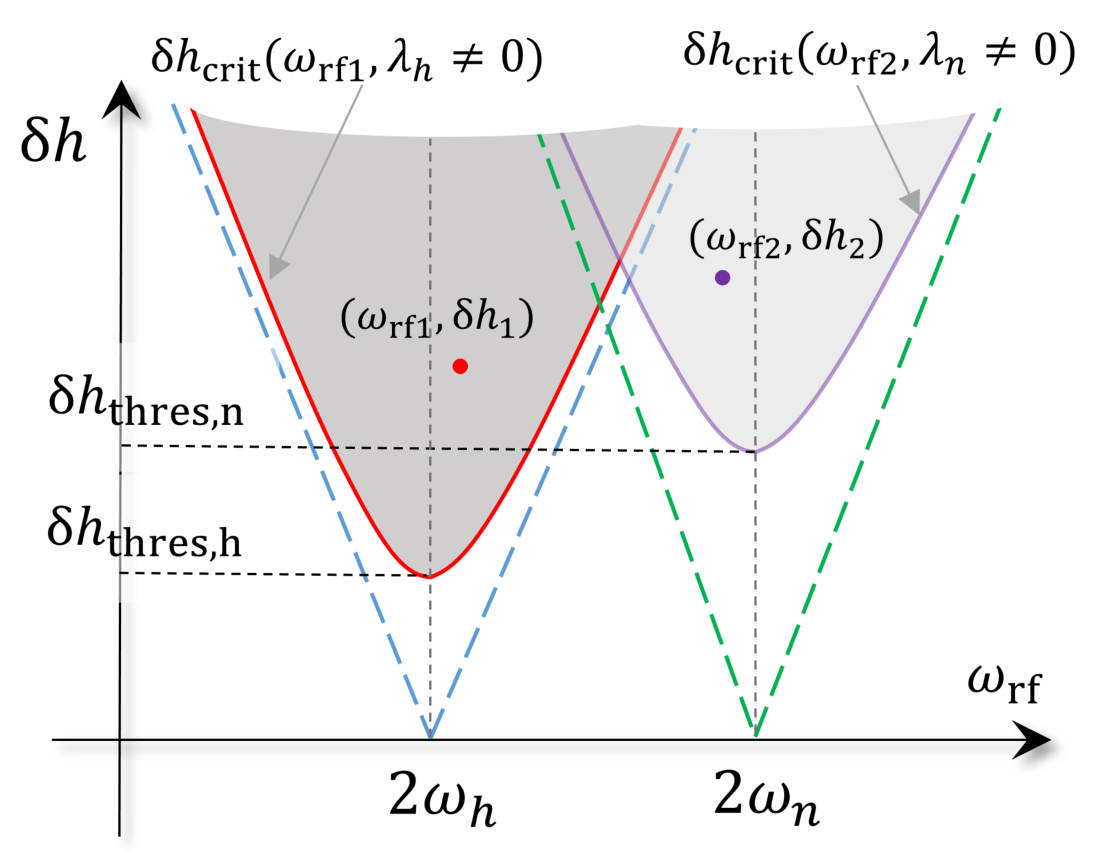}
    \caption{Graphical representation of conditions for parametric excitation of modes $h$ and $n$ via parallel pumping with two-tone signal. The shaded regions depict pairs of excitation parameters $(\omega_\mathrm{rf1},\delta h_1)$ and $(\omega_\mathrm{rf2},\delta h_2)$ that trigger the amplitude growth of mode $h$ and $n$ from nonzero initial values according to eq.\eqref{eq:detuning cone}. The solid red (dashed blue) and solid purple (dashed green) lines refer to eq.\eqref{eq:critical detuning cone} when the damping $\lambda_h, \lambda_n>0$ (resp. $\lambda_h=\lambda_n=0$). The minimum thresholds $\delta h_\mathrm{thres,h}$ and $\delta h_\mathrm{thres,n}$ (see eq.\eqref{eq:threshold field parametric resonance}) are also reported.}
    \label{fig:parametric resonance cone interacting}
\end{figure}

Note that, even if the mutual NFS coefficients coincide, the interaction
between modes $h$ and $n$ will be termed \emph{non-reciprocal} if the effect on the dynamics of mode $h$ produced by a nonzero amplitude of mode $n$ is not the same as that on dynamics of mode $n$ produced by a nonzero amplitude of mode $h$.

Now we rewrite the equations \eqref{eq:ah NMM}-\eqref{eq:an NMM} and their complex conjugates as follows, hereafter referred to as \emph{two-modes model}:
\begin{align}
    \dot{a_h}&=j \omega_h [1+N_{hh}|a_h|^2+N_{hn}|a_n|^2 -\delta h_a(t) ||\bm\varphi_h||^2]a_h + \\
    &-j \omega_h \delta h_a(t) (\bm \varphi_h,\bm \varphi_{h}^*) a_h^*  -\lambda_h a_h \quad, \text{ and c.c.}\label{eq:nonlinear coupled parametric ah} \\
    \dot{a_n}&=j \omega_n [1+N_{nn}|a_n|^2+N_{nh}|a_h|^2-\delta h_a(t) ||\bm\varphi_n||^2]{a_n} + \\ 
    &-j \omega_n \delta h_a(t) (\bm \varphi_n,\bm \varphi_{n}^*) a_n^* -\lambda_n a_n \quad, \text{ and c.c.}  \label{eq:nonlinear coupled parametric an}
\end{align}

Then, let us consider the two detuning cones defined by eq.\eqref{eq:detuning cone} for each mode $h$ and $n$ (see fig.\ref{fig:parametric resonance cone panels}(b)). 
If we assume that the excitation parameters $(\omega_\mathrm{rf1},\delta h_1)$ and $(\omega_\mathrm{rf2},\delta h_2)$ are chosen such that they do not fall into overlapping regions between the respective detuning cones (Arnold's tongues defined by \eqref{eq:detuning cone} independently for each mode), that is when excitation amplitudes are not too large as it is graphically illustrated in fig.\ref{fig:parametric resonance cone interacting},  we can reasonably state that solutions of eqs. \eqref{eq:nonlinear coupled parametric ah}-\eqref{eq:nonlinear coupled parametric an} will not differ significantly from those of the same equations where the two tones act separately in each one, namely $\delta h_a(t)\approx\delta h_1\sin(\omega_\mathrm{rf1} t)$ in eq.\eqref{eq:nonlinear coupled parametric ah} (and in its c.c.) and $\delta h_a(t)\approx\delta h_2\sin(\omega_\mathrm{rf2} t)$ in eq.\eqref{eq:nonlinear coupled parametric an} (and in its c.c.).

This means that each of eqs.\eqref{eq:nonlinear coupled parametric ah}-\eqref{eq:nonlinear coupled parametric an} is formally identical to eq.\eqref{eq:nonlinear parametric ah}, provided that the following primed quantities $\omega_h'(|a_h|,|a_n|)$, $\omega_n'(|a_n|,|a_h|)$ 
 are defined:
\begin{align}   
\omega_h'(|a_h|,|a_n|)&=\omega_h(1+N_{hh}|a_h|^2+N_{hn}|a_n|^2)\,, \label{eq:scaled omega mode h}\\ \omega_n'(|a_n|,|a_h|)&=\omega_n(1+N_{nn}|a_n|^2+N_{nh}|a_h|^2)  \,. \label{eq:scaled omega mode n}  
\end{align}
By using the same line of reasoning as in section \ref{sec:nonlinear self interaction} to determine the amplitude saturation after parametric excitation above the threshold, which led to eq.\eqref{eq:nonlinear detuning cone}, one can see that the steady-state amplitudes $|a_h|^2,|a_n|^2$ are given by the following coupled nonlinear equations obtained imposing that the amplitude-dependent detuning associated separately to each mode equals the corresponding critical detuning, namely:
\begin{align}
    (\omega_\mathrm{rf1}&-2\omega_h'(|a_h|,|a_n|))^2= \nonumber \\
    &={\epsilon}_\mathrm{crit,h}^2=  \omega_h^2|(\bm\varphi_h,\bm\varphi_h^*)|^2(\delta h_1^2-\delta h_\mathrm{thres,h}^2) \,, \label{eq:nonlinear detuning cone primed mode h} \\
    (\omega_\mathrm{rf2}&-2\omega_n'(|a_n|,|a_h|))^2 \nonumber \\
    &={\epsilon}_\mathrm{crit,n}^2=  \omega_n^2|(\bm\varphi_n,\bm\varphi_n^*)|^2(\delta h_2^2-\delta h_\mathrm{thres,n}^2) \,. \label{eq:nonlinear detuning cone primed mode n}   
\end{align}
We remark that the above equations depend on the steady-state amplitudes $|a_h|^2,|a_n|^2$ in the left-hand side (through the 'primed' $\omega_h',\omega_n'$ according to eqs.\eqref{eq:scaled omega mode h},\eqref{eq:scaled omega mode n}). 

It is worth mentioning that, in the steady-state regime defined above, each of the interacting parametrically-excited modes $h$ (resp. $n$) produces a time-domain periodic signal that oscillates with amplitude $|a_h|$ (resp. $|a_n|$) satisfying \eqref{eq:nonlinear detuning cone primed mode h}-\eqref{eq:nonlinear detuning cone primed mode n} and frequency $\omega_\mathrm{rf1}/2$ (resp. $\omega_\mathrm{rf2}/2$) that is slightly detuned with respect to the amplitude-dependent natural frequency \eqref{eq:scaled omega mode h} by $\omega_\mathrm{rf1}/2-\omega_h'(|a_h|,|a_n|)$ (resp. amplitude-dependent natural frequency \eqref{eq:scaled omega mode n} by $\omega_\mathrm{rf2}/2-\omega_n'(|a_h|,|a_n|)$). 

Equations \eqref{eq:nonlinear detuning cone primed mode h}-\eqref{eq:nonlinear detuning cone primed mode n} hold if the following existence constraints are satisfied:
\begin{align}
    |a_h|^2, |a_n|^2 &\geq 0  \quad, \label{eq:constraint real amplitudes}  
\end{align}
which are imposed by the fact that the mode amplitudes $|a_h|,|a_n|$ must be real quantities.

Using an appropriate averaging technique on the two-modes model \eqref{eq:nonlinear coupled parametric ah}-\eqref{eq:nonlinear coupled parametric an}, it can be shown that the amplitudes $|a_h|^2,|a_n|^2$ satisfying eqs.\eqref{eq:nonlinear detuning cone primed mode h}-\eqref{eq:nonlinear detuning cone primed mode n} are stationary solutions of the following dynamical system with state variables $A_h,\Phi_h,A_n,\Phi_n$:
\begin{align}
    \dot{A_h}&=\left[-\frac{\omega_h (\bm \varphi_h,\bm \varphi_h^*) \delta h_1}{2} \cos(2\Phi_h)-\lambda_h\right] A_h \,, \label{eq:averaged two nonlinear parametric Ah} \\
    \dot{\Phi}_h\!+\!\frac{\epsilon_h}{2}&\!=\!\omega_h ( N_{hn} A_n^2 + N_{hh} A_h^2)+ \frac{\omega_h (\bm \varphi_h,\bm \varphi_h^*) \delta h_1}{2} \sin(2\Phi_h) \,, \label{eq:averaged two nonlinear parametric Phi_h} \\
    \dot{A_n}&=\left[-\frac{\omega_n (\bm \varphi_n,\bm \varphi_n^*) \delta h_2}{2} \cos(2\Phi_n)-\lambda_n\right] A_n \,, \label{eq:averaged two nonlinear parametric An} \\
    \dot{\Phi}_n\!+\!\frac{\epsilon_n}{2}&\!=\!\omega_n ( N_{nh} A_h^2 + N_{nn} A_n^2)+ \frac{\omega_n (\bm \varphi_n,\bm \varphi_n^*) \delta h_2}{2} \sin(2\Phi_n) \,, \label{eq:averaged two nonlinear parametric Phi_n}
\end{align}
hereafter referred to as \emph{averaged two-modes model}. In equations \eqref{eq:averaged two nonlinear parametric Ah}-\eqref{eq:averaged two nonlinear parametric Phi_n}, $A_h,A_n$ are the amplitudes of modes $h,n$ averaged on the fast time scales (the forcing periods $2\pi/\omega_\mathrm{rf1},2\pi/\omega_\mathrm{rf2}$, respectively),  $\Phi_h,\Phi_n$ are the phases and $\epsilon_h=\omega_\mathrm{rf1}-2\omega_h, \epsilon_n=\omega_\mathrm{rf2}-2\omega_n$ are the detunings.

\subsection{Existence and stability of steady-state regimes under two-tone excitation}
\label{sec:existence and stability two-tone}

The interaction between two parametrically-excited modes is governed by only three nonlinear coefficients: the self-NFS coefficients $N_{hh},N_{nn}$ and the mutual NFS coefficient $N_{hn}=N_{nh}$. Depending on their signs and magnitudes, the interaction may suppress or enhance one mode with respect to the uncoupled regime, produce coexistence of multiple steady-state solutions, hysteresis, and quasiperiodic oscillations. The following analysis aims at determining the possible steady-state regimes arising from mode interactions and the excitation conditions under which they can be observed.

\begin{figure}[t]
    \centering
    \includegraphics[width=8cm]{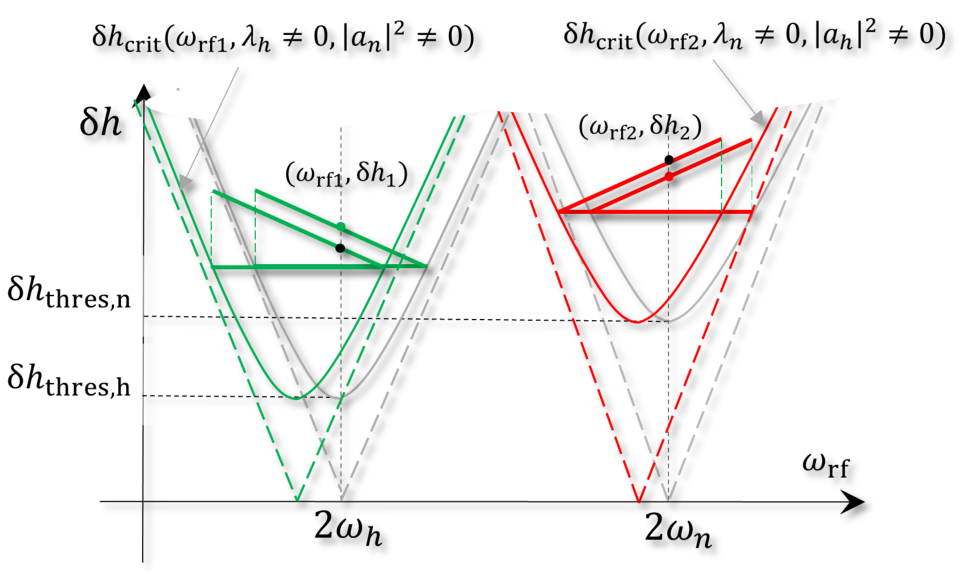}
    \caption{Graphical representation of interaction for parametrically-excited modes $h$ and $n$ via parallel pumping with two-tone signal. The shaded lines (dashed/solid meaning absence/presence of damping) represent the detuning cones for the modes $h,n$ in the uncoupled regime (i.e. the case of single tone excitation). Notice that mode $h$ (resp. $n$) has been assumed to have negative (resp. positive) NFS for illustrative purpose. Green and red lines represent 'interacting' detuning cones shifted by mutual NFS $N_{hn}=N_{nh}$ (assumed negative here). The left-shifting of detuning cones evidences suppression of mode $h$ and enhancement of mode $n$ (black dots compared to green/red dots).}
    \label{fig:parametric resonance cone interacting type1}
\end{figure}

We introduce the amplitude-dependent detunings (with respect to the mutually shifted natural frequencies) as 
\begin{align}
\epsilon_h'&=\epsilon_h\!-\!2\omega_h N_{hn}|a_n|^2=\omega_\mathrm{rf1}\!-\!2\omega_h-2\omega_h N_{hn}|a_n|^2 \,, \label{eq:eps h prime}\\ 
\epsilon_n'&=\epsilon_n\!-\!2\omega_n N_{nh}|a_h|^2=\omega_\mathrm{rf2}\!-\!2\omega_n-2\omega_n N_{nh}|a_h|^2 \,. \label{eq:eps n prime}
\end{align}
By following the same line of reasoning as in the self-interaction case (section \ref{sec:nonlinear self interaction}), one can consider the aforementioned detunings $\epsilon'_h$ and $\epsilon'_n$ in the place of $\epsilon_h,\epsilon_n$ in eq.\eqref{eq:nonlinear detuning cone}, which leads to the following expressions for the steady-state amplitudes $|a_h|^2,|a_n|^2$ that resemble eq.\eqref{eq:saturation amplitude mode h pm}:
\begin{align}
        |a_h|^2&=
        \frac{\epsilon'_h + s^\pm_h \epsilon_\mathrm{crit,h}}{2\omega_h N_{hh}} =|a_h|^2_\mathrm{\pm,unc}(\omega_\mathrm{rf1},\delta h_1) -\frac{N_{hn}}{N_{hh}}|a_n|^2 \label{eq:saturation amplitude mode h type 1} \,, \\
        |a_n|^2&=
        \frac{\epsilon'_n + s^\pm_n \epsilon_\mathrm{crit,n}}{2\omega_n N_{nn}} =|a_n|^2_\mathrm{\pm,unc}(\omega_\mathrm{rf2},\delta h_2) -\frac{N_{nh}}{N_{nn}}|a_h|^2 \label{eq:saturation amplitude mode n type 1} \,,
\end{align}
where we have considered all possible signs $s^\pm_h=\pm \text{sign}(N_{hh}) , s^\pm_n=\pm \text{sign}(N_{nn})$ in front of the critical detunings.
The latter equations state that, in the coupled regime, the steady-state amplitudes of modes $h$ and $n$ may be enhanced or attenuated in a linear fashion depending on the signs of the self- and mutual NFS coefficients, compared to the corresponding 'uncoupled' ones 
\begin{align}
    |a_h|^2_\mathrm{\pm,unc}(\omega_\mathrm{rf1},\delta h_1) &= \frac{\epsilon_h + s^\pm_h \epsilon_\mathrm{crit,h}}{2\omega_h N_{hh}} \quad,\label{eq:uncoupled amp h} \\
    |a_n|^2_\mathrm{\pm,unc}(\omega_\mathrm{rf2},\delta h_2) &= \frac{\epsilon_n + s^\pm_n \epsilon_\mathrm{crit,n}}{2\omega_n N_{nn}} \quad, \label{eq:uncoupled amp n}
\end{align}
denoted with the subscript 'unc', computed according to eq.\eqref{eq:saturation amplitude mode h pm}. One can see fig.\ref{fig:parametric resonance cone interacting type1} for a graphical illustration.

The solution of the linear system \eqref{eq:saturation amplitude mode h type 1}-\eqref{eq:saturation amplitude mode n type 1} is:
\begin{align}
    &|a_h|^2=|a_h|^2_\mathrm{\pm\pm,coup}(\omega_\mathrm{rf1},\delta h_1,\omega_\mathrm{rf2},\delta h_2)= \nonumber \\
    &=L_{hn}\left[|a_h|^2_\mathrm{\pm,unc}(\omega_\mathrm{rf1},\delta h_1)-\frac{N_{hn}}{N_{hh}}|a_n|^2_\mathrm{\pm,unc}(\omega_\mathrm{rf2},\delta h_2)\right] \,, \label{eq:saturation amplitude mode h type 1 linear}\\
    &|a_n|^2=|a_n|^2_\mathrm{\pm\pm,coup}(\omega_\mathrm{rf1},\delta h_1,\omega_\mathrm{rf2},\delta h_2)= \nonumber \\
    &=L_{nh}\left[|a_n|^2_\mathrm{\pm,unc}(\omega_\mathrm{rf2},\delta h_2)-\frac{N_{nh}}{N_{nn}}|a_h|^2_\mathrm{\pm,unc}(\omega_\mathrm{rf1},\delta h_1)\right] \label{eq:saturation amplitude mode n type 1 linear} \,,
\end{align}
where the 'coupled' steady-state amplitudes are expressed as linear combination of the 'uncoupled' ones, the leading coefficient $L_{hn}=L_{nh}=\frac{N_{hh}N_{nn}}{N_{hh}N_{nn}-N_{hn}^2}$ and the right-hand sides are functions of the control parameters $(\omega_\mathrm{rf1},\delta h_1,\omega_\mathrm{rf2},\delta h_2)$ and (self-, mutual) NFS. The first (resp. second) $\pm$ subscript in $|a_h|^2_\mathrm{\pm\pm,coup}, |a_n|^2_\mathrm{\pm\pm,coup}$  refers to the subscript in the uncoupled amplitude $|a_h|^2_\mathrm{\pm,unc}$ (resp. uncoupled amplitude $|a_n|^2_\mathrm{\pm,unc}$) at the right-hand side.

It is important to stress that, despite eqs.\eqref{eq:saturation amplitude mode h type 1}-\eqref{eq:saturation amplitude mode n type 1} appear as linear equations for $|a_h|^2,|a_n|^2$, they are indeed nonlinear in that the amplitudes fulfill the aforementioned equations only if they are nonnegative, namely if the existence constraints \eqref{eq:constraint real amplitudes} hold. 
These amplitude-dependent conditions define regions that partition the plane $(|a_h|^2,|a_n|^2)$ into sets of different admissible solutions and make the whole system of equations \eqref{eq:saturation amplitude mode h type 1}-\eqref{eq:saturation amplitude mode n type 1} with constraints \eqref{eq:constraint real amplitudes} nonlinear, which produces multiplicity of solutions/regimes for given excitation parameters $(\omega_\mathrm{rf1},\delta h_1,\omega_\mathrm{rf2},\delta h_2)$.
As a consequence of the above reasoning, it is important discussing the ranges of validity of the linear relations \eqref{eq:saturation amplitude mode h type 1}-\eqref{eq:saturation amplitude mode n type 1} and, thus, that of the related 'coupled' solutions \eqref{eq:saturation amplitude mode h type 1 linear}-\eqref{eq:saturation amplitude mode n type 1 linear}. 

First, we observe that Eqs.\eqref{eq:saturation amplitude mode h type 1 linear}-\eqref{eq:saturation amplitude mode n type 1 linear} are backward compatible with the solutions where only one mode amplitude is nonzero, which were discussed in the case of single-mode excitation (see section \ref{sec:nonlinear self interaction}).
Then, under two-tone excitation, the steady-state regimes can correspond to either of the following solutions that we term with the shorthand notation $z,u_h,u_n,c$ meaning 'zero', 'uncoupled mode $h$', 'uncoupled mode $n$', 'coupled modes', respectively:
\begin{align}
      z:\,  |a_h|^2&=0 \quad,\quad |a_n|^2=0 \quad. \label{eq:type 1 mode h off mode n off} \\
    u_h:\, |a_h|^2&=|a_h|^2_\mathrm{+,unc} \,,\, |a_n|^2=0 \,, \label{eq:type 1 mode h on mode n off} \\
    u_n:\, |a_n|^2&=|a_n|^2_\mathrm{+,unc} \,,\, |a_h|^2=0 \,, \label{eq:type 1 mode h off mode n on} \\
    c=c_{\pm\pm}:\, |a_h|^2&=|a_h|^2_\mathrm{\pm\pm,coup} \,,\, |a_n|^2=|a_n|^2_\mathrm{\pm\pm,coup} \,.  \label{eq:type 1 coupled modes}
\end{align}
where $|a_h|^2_\mathrm{+,unc}, |a_n|^2_\mathrm{+,unc}$ are defined by eqs.\eqref{eq:uncoupled amp h}-\eqref{eq:uncoupled amp n} and $|a_h|^2_\mathrm{\pm\pm,coup},|a_n|^2_\mathrm{\pm\pm,coup}$ by eqs.\eqref{eq:saturation amplitude mode h type 1 linear}-\eqref{eq:saturation amplitude mode n type 1 linear}.
We have not included 'uncoupled' solutions $|a_h|^2=|a_h|^2_\mathrm{-,unc}, |a_n|^2=0$ and $|a_n|^2=|a_n|^2_\mathrm{-,unc}, |a_h|^2=0$ in the above list because single mode analysis (see section \ref{sec:nonlinear self interaction}) already established that they are always unstable in their region of existence.  

\begin{figure}[t]
    \centering
 \includegraphics[width=8cm]{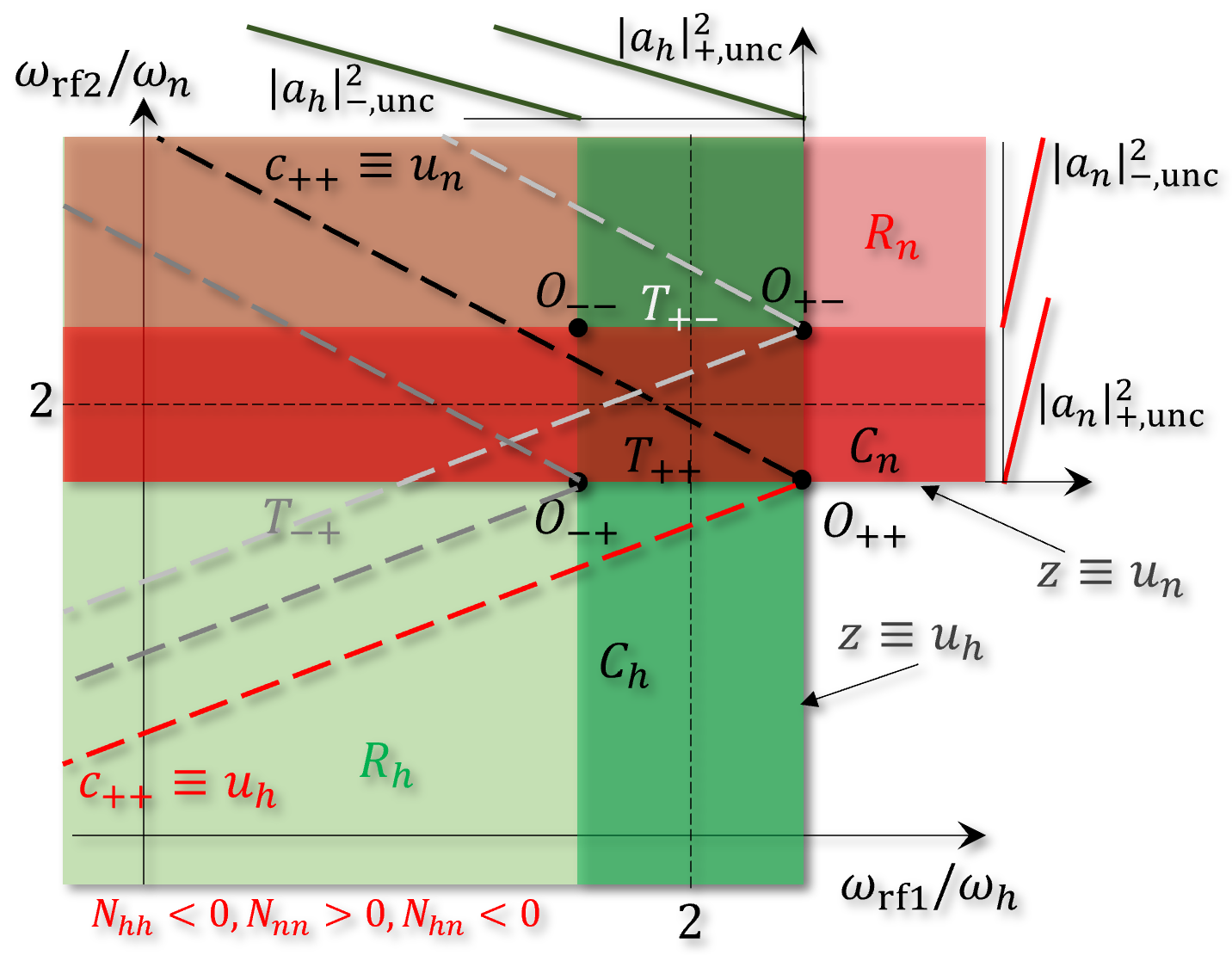}
    \caption{Phase diagram of steady-state regimes (see eqs.\eqref{eq:type 1 mode h off mode n off}-\eqref{eq:type 1 coupled modes}) in the control plane $(\omega_\mathrm{rf1},\omega_\mathrm{rf2})$ for parametrically-excited modes $h$ and $n$ via parallel pumping with two-tone signal at fixed above-threshold amplitudes $\delta h_1,\delta h_2$. It is assumed that $N_{hh}<0,N_{nn}>0,N_{hn},N_{nh}<0, N_{hh}N_{nn}/(N_{hh}N_{nn}-N_{hn}^2)>0$. Different colored regions are depicted corresponding to the existence conditions \eqref{eq:constraint real amplitudes}    
    for uncoupled regimes $u_h$ (green), $u_n$ (red). Coupled $c$ regimes exist in regions $T_{\pm\pm}$ enclosed between pairs of dashed lines $c_{\pm\pm}\equiv u_h$ and $c_{\pm\pm}\equiv u_n$ (notice that regions $T_{--}$ is not reported for the sake of visualization).  Black dashed line corresponds to $c_{++}\equiv u_n$ 
        while red dashed line refers to $c_{++}\equiv u_h$. 
    }
    \label{fig:Pmodes existence regions}
\end{figure}

Which of the aforementioned steady-state solutions will be actually reached will depend on the history of excitation parameters $(\omega_\mathrm{rf1},\delta h_1,\omega_\mathrm{rf2},\delta h_2)$ determined, for instance, by the sequential application of the single tones. 
In this respect, the mode amplitudes dependence may be very complex for arbitrary history of the excitation parameters in the four-dimensional space $(\omega_\mathrm{rf1},\delta h_1,\omega_\mathrm{rf2},\delta h_2)$. Usual experimental situations either refer to changing tone frequencies of the signal while keeping the amplitude fixed or vice-versa. 
In the sequel, we will consider the former situation where $(\delta h_1,\delta h_2)$ are fixed and $(\omega_\mathrm{rf1},\omega_\mathrm{rf2})$ are changed as it is the case of experiments described in ref.\cite{soares_knet_flagship_paper_2025}. 

In that situation, steady-state regimes described by eqs.\eqref{eq:type 1 mode h off mode n off}-\eqref{eq:type 1 coupled modes} exist in regions (possibly overlapping meaning co-existence) of the forcing plane $(\omega_\mathrm{rf1},\omega_\mathrm{rf2})$. The detailed analytical characterization of existence regions is reported in appendix \ref{sec:existence conditions}. For illustrative purpose, these regions are reported in fig.\ref{fig:Pmodes existence regions}. One can see that, under CW excitation, uncoupled modes $u_h$ and $u_n$ exist in rectangular regions $R_h$ (green) and $R_n$ (red) while, for PWM excitation, they exist in $C_h,C_n$ forming a 'cross-shaped' region with center in $(2\omega_h,2\omega_n)$. Coupled modes $c_{\pm\pm}$ exist within triangular regions denoted as $T_{\pm\pm}$ having vertices in points $O_{\pm\pm}$. In general, existence regions are separated by straight lines that imply coincidence, also termed 'collision', between different types of regimes. Such a coincidence is denoted with the notation $\equiv$ (see for instance lines $z\equiv u_h, z\equiv u_n, c_{++}\equiv u_h, c_{++}\equiv u_n$ in fig.\ref{fig:Pmodes existence regions}). 
We stress that the lines $c_{\pm\pm}\equiv u_h$ are mutually parallel, and the same holds for the lines $c_{\pm\pm}\equiv u_n$, since their slopes only depend on the ratios between self- and mutual NFS coefficients.
More technical details on critical lines are reported in appendix \ref{sec:stability conditions}.

\begin{figure*}[t]
    \centering
    \includegraphics[width=9cm]{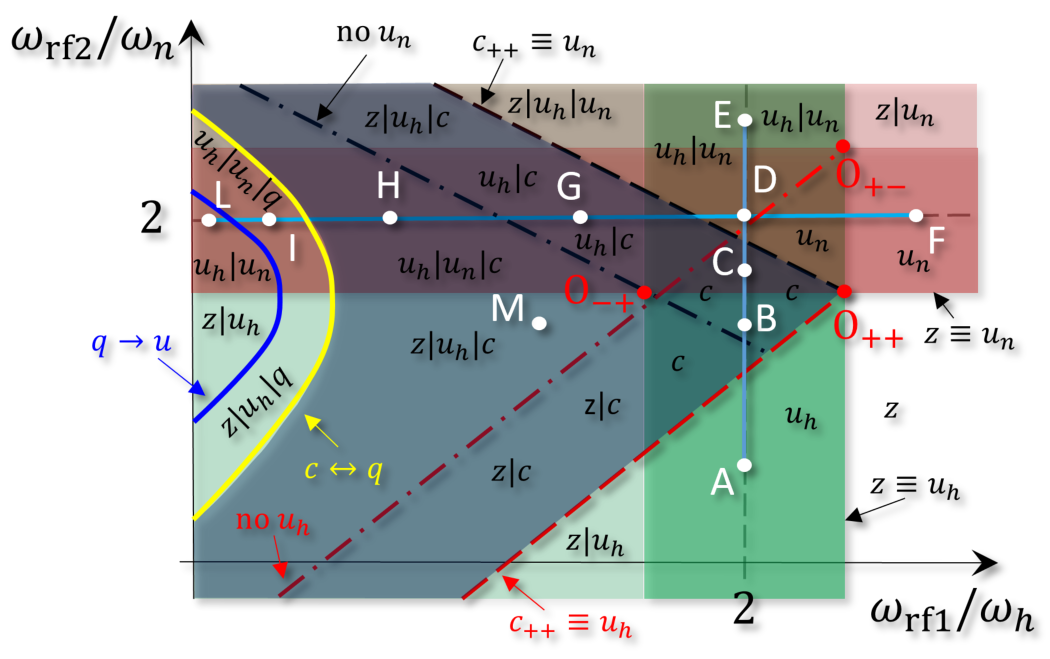} \\
    $N_{hh}<0,N_{nn}>0, N_{hn}=N_{nh}<0, N_{hh}N_{nn}/(N_{hh}N_{nn}-N_{hn}^2)>0$ \\
    \includegraphics[width=7.5cm]{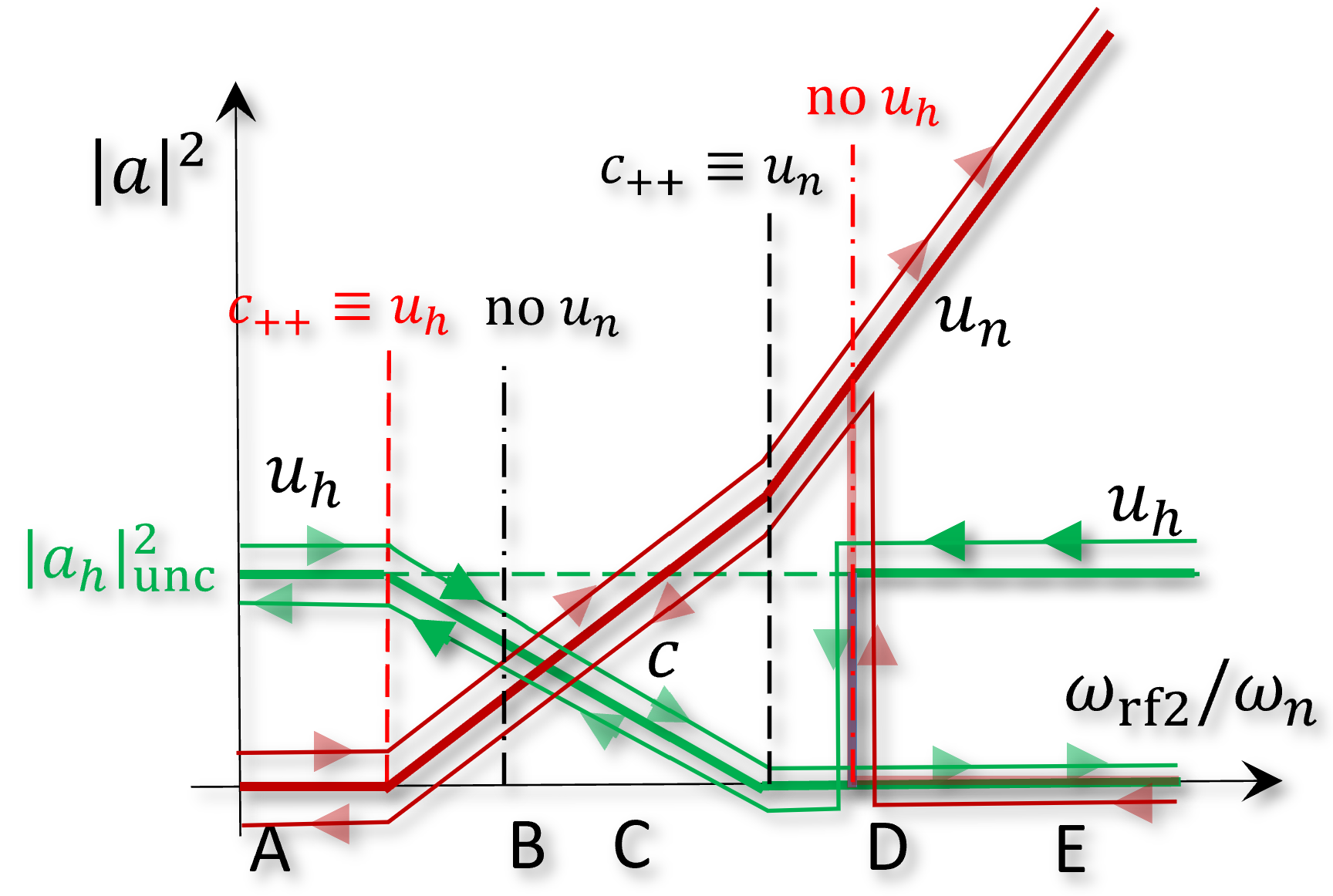}\includegraphics[width=8.5cm]{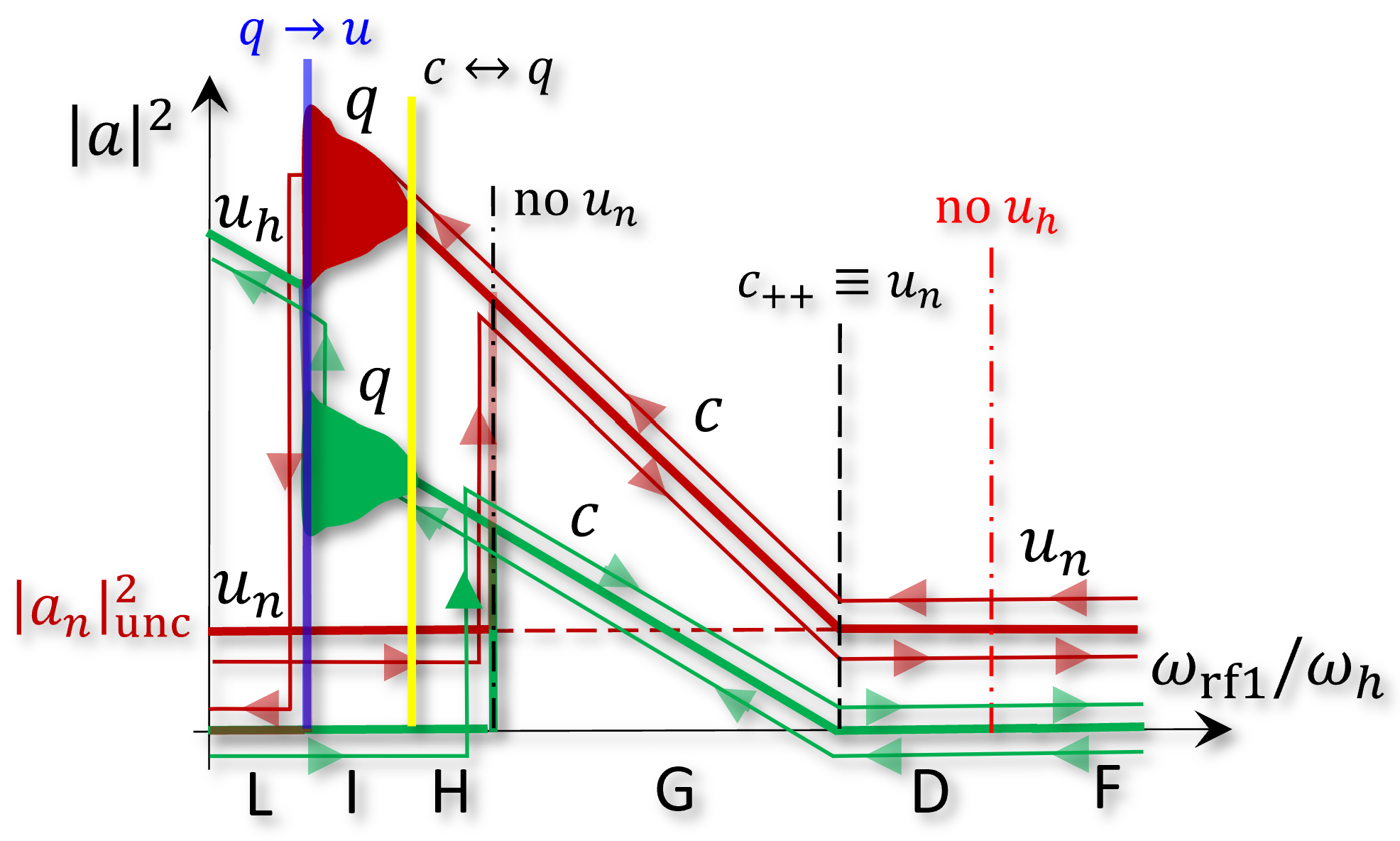}
    \caption{Phase diagram of interaction in $(\omega_\mathrm{rf1},\omega_\mathrm{rf2})$ plane for parametrically-excited modes $h$ and $n$ via parallel pumping with two-tone signal at fixed above-threshold amplitudes $\delta h_1,\delta h_2$.  
    Top panel depicts different regions, colored when associated with stable regimes, corresponding to the existence conditions \eqref{eq:constraint real amplitudes}
    for uncoupled $u_h$ (green), $u_n$ (red) and coupled $c$ (blue) regimes (see eqs.\eqref{eq:type 1 mode h off mode n off}-\eqref{eq:type 1 coupled modes}).  Black dashed line corresponds to the condition $c_{++}\equiv u_n$ 
        while red dashed line refers to $c_{++}\equiv u_h$. 
    Black dash-dotted line refers to the condition $c_{-+}\equiv u_n$ (denoted as 'no $u_n$'), while red dash-dotted line refers to the condition $c_{+-}\equiv u_h$ (denoted as 'no $u_h$'). Solid yellow (Hopf bifurcation) line $c\leftrightarrow q$ refer to the boundary of region $T_{++}$ due to the onset of Q-modes, solid blue (homoclinic bifurcation) line $q\rightarrow u$ denotes the boundary of the region where Q-modes can exist. White dots labeled with letters A-L indicate some of the possible combinations of coexisting steady-state regimes. Coexistence of several regimes in a given region of the control plane is denoted using the compact notation '$|$'.    
    Light-blue solid lines refer to possible experiments keeping one tone at fixed frequency (e.g. twice its natural resonant frequency) while sweeping the other tone's frequency in a continuous wave (CW) fashion. Lower panels report steady-state amplitudes $|a_h|^2$ (green) and $|a_n|^2$ (red) corresponding to 'vertical'  and 'horizontal' (light-blue) slow variation of excitation frequencies. Both situations exhibit hysteretic behavior (see thinner lines with arrowheads referring to forward or backward frequency sweeps). The vertical forward (resp. backward) sequence of steady states is $u_h \rightarrow c \rightarrow u_n$ (resp. $u_h \rightarrow u_n \rightarrow c \rightarrow u_h$). The horizontal forward (resp. backward) sequence of steady states is $u_n\rightarrow c\rightarrow u_n$ (resp. $u_n\rightarrow c \rightarrow q \rightarrow u_h$).}
    \label{fig:type1 phase diagram and hysteresis}
\end{figure*}

\begin{table*}[t]
    \centering {\footnotesize \textbf{Algorithm to construct the phase diagram of mode pairs interaction in control plane $(\omega_\mathrm{rf1},\omega_\mathrm{rf2})$}
    \begin{tabular}{|c|c|c|c|c|c|c|c|}
    \hline
       $N_{hh}$  & $N_{nn}$ & $N_{hn}=N_{nh}$ & $c_{\pm\pm}\equiv u_n$ slope sign($\frac{N_{hn}}{N_{nn}}$) & $c_{\pm\pm}\equiv u_h$ slope sign($\frac{N_{nh}}{N_{hh}}$) & $O_{++}$ & $O_{+-}$ & $O_{-+}$ \\
       \hline
        $-$     &   $+$    &   $-,+$       &      $-,+$       &    $+,-$                     & LR  & UR & LL \\
        $-$     &   $-$    &   $-,+$       &      $+,-$       &    $+,-$                     & UR  & LR & UL \\
        $+$     &   $-$    &   $+,-$       &      $-,+$       &    $+,-$                     & UL  & LL & UR \\
        $+$     &   $+$    &   $-,+$       &      $-,+$       &    $-,+$                     & LL  & UL & LR \\
    \hline
    \end{tabular}

    \begin{enumerate}
        \item Draw the 'cross' region $C_h\cup C_n$ using eqs.\eqref{eq:Ch},\eqref{eq:Cn} included in existence regions $R_h,R_n$ (see eqs. \eqref{eq:Rh},\eqref{eq:Rn}) of regimes $u_h,u_n$. \label{item:E1}
        \item Locate points $O_{++},O_{+-},O_{-+}$ (see the above table and eq.\eqref{eq:c triangles vertices}) that are vertices of the existence regions $T_{++},T_{+-},T_{-+}$ of regimes $c_{++},c_{+-},c_{-+}$.
        \item Determine slopes $\omega_h N_{hn}/(\omega_n N_{nn}), \omega_n N_{nh}/(\omega_h N_{hh})$ (see the above table) of the four (parallel two by two) lines $c_{\pm\pm}\equiv u_n, c_{\pm\pm}\equiv u_h$ (eqs.\eqref{eq:line ah_coup_eq_zero},\eqref{eq:line ah_coup_eq_zero bound},\eqref{eq:line an_coup_eq_zero},\eqref{eq:line an_coup_eq_zero bound}) intersecting vertices $O_{\pm\pm}$.
         
        \item 
        Determine existence regions $T_{\pm\pm}$ (see eqs. \eqref{eq:T}). 
        Practically, such triangular regions always extend from the vertex towards the unbounded portion of $R_h\cup R_n$ (green and red) region.
        \label{item:E4}
    
        \item The zero P-mode $z$, given by \eqref{eq:type 1 mode h off mode n off}, is unstable in the interior of the 'cross' $C_h\cup C_n$ (dark green and dark red) region; \label{item:S1}
    \item the coupled P-modes $c_{++}$, given by \eqref{eq:type 1 coupled modes}, are stable in $T_{++}$ close to the lines $c_{++}\equiv u_h$ and $c_{++}\equiv u_n$ if  $\text{sign}(N_{hh})\text{sign}(N_{nn}) (N_{hh}N_{nn}-N_{hn}^2) >0$; \label{item:S2}
    \item the coupled P-modes $c_{+-}$, given by \eqref{eq:type 1 coupled modes}, are stable in $T_{+-}$ close to the line $c_{+-}\equiv u_h$ if $\text{sign}(N_{hh})\text{sign}(N_{nn}) (N_{hh}N_{nn}-N_{hn}^2)< 0$; \label{item:S3}
    \item the coupled P-modes $c_{-+}$, given by \eqref{eq:type 1 coupled modes}, are stable in $T_{-+}$ close to the line $c_{-+}\equiv u_n$ if $\text{sign}(N_{hh})\text{sign}(N_{nn}) (N_{hh}N_{nn}-N_{hn}^2)< 0$; \label{item:S4}
    \item the coupled P-modes $c_{--}$, given by \eqref{eq:type 1 coupled modes}, are unstable in $T_{--}$; \label{item:S5}
    \item the uncoupled P-mode $u_h$, given by \eqref{eq:type 1 mode h on mode n off}, is unstable in the region enclosed within the red ($c_{++}\equiv u_h$ and dash-dotted $c_{+-}\equiv u_h$ '$\text{no } u_h$') lines; otherwise, it is stable in the remaining part of (green) region $R_h$; \label{item:S6}
    \item the uncoupled P-mode $u_n$, given by \eqref{eq:type 1 mode h off mode n on}, is unstable in the region enclosed within the black (dashed $c_{++}\equiv u_n$ and dash-dotted $c_{-+}\equiv u_n$ '$\text{no } u_n$') lines; otherwise, it is stable in the remaining part of (red) region $R_n$; \label{item:S7}
    \item Q-modes can exist either in region $T_{++}$ (when $c_{++}$ P-modes are stable) sufficiently far from the vertex $O_{++}$ between lines $c\leftrightarrow q$ and $q\rightarrow u$ or in regions $T_{+-},T_{-+}$ (when $c_{+-},c_{-+}$ are stable) close to  the vertices $O_{+-},O_{-+}$ and between lines $c\leftrightarrow q$ and $q\rightarrow u$, respectively. \label{item:S8}
    
    \end{enumerate}
    \caption{Summary of qualitative features and algorithm to construct  the phase diagram in the $(\omega_\mathrm{rf1},\omega_\mathrm{rf2})$ control plane (see the examples in figs.\ref{fig:type1 phase diagram and hysteresis}, \ref{fig:type1 phase diagram and hysteresis h1n3}) when the amplitudes of the tones $\delta h_1,\delta h_2$ are both above the respective thresholds.  
    Notation XY with X=L (Lower),U (Upper) and Y=L (Left),R (Right) denotes the corresponding corner of the rectangular central region \(C_h\cap C_n\) (brown in figs.\ref{fig:type1 phase diagram and hysteresis},\ref{fig:type1 phase diagram and hysteresis h1n3}).  The rules \ref{item:E1}-\ref{item:E4} refer to existence whereas the rules \ref{item:S1}-\ref{item:S8} refer to stability of interacting steady-state regimes.} \label{tab:type1 interactions}}
    
\end{table*}

The above discussion based on physical arguments concerning the coexistence and collision of regimes can be made rigorous by performing stability and bifurcation analysis\cite{Perko_book} of periodic solutions, termed P-modes, for the two-modes model \eqref{eq:nonlinear coupled parametric ah}-\eqref{eq:nonlinear coupled parametric an} that can be carried out mostly with analytical techniques using the averaged two-modes model \eqref{eq:averaged two nonlinear parametric Ah}-\eqref{eq:averaged two nonlinear parametric Phi_n}. This analysis of more technical character is also reported in appendix \ref{sec:stability multi P modes} for interested readers. 

Besides P-modes solutions, the dynamical system \eqref{eq:nonlinear coupled parametric ah}-\eqref{eq:nonlinear coupled parametric an} may admit quasiperiodic solutions (limit cycles) termed Q-modes that, for each mode, combine two frequencies.
The onset of Q-modes carries the physical meaning of the loss of synchronization between the system response and the driving frequencies (half those of the external signal tones, namely $\omega_\mathrm{rf1}/2$ and $\omega_\mathrm{rf2}/2$), which produces nonzero unsynchronized steady-state oscillation regimes, being allowed by the higher dimensionality of the (four-dimensional) dynamical state-space compared to the situation of single mode excitation where this is forbidden (see section \ref{sec:nonlinear self interaction}). For this reason, Q-modes can only originate from P-mode coupled regimes $c_{\pm\pm}$ that eventually become unstable for appropriate combinations of tone frequencies.
Specifically, for mode $h$ the Q-mode combines the applied tone frequency $\omega_\mathrm{rf1}/2$ with a (typically lower) frequency intrinsic of the system and related with the small spiraling motion around the unstable P-mode, giving rise to low-frequency modulation. Analogous reasoning holds for mode $n$ that combines the applied tone frequency $\omega_\mathrm{rf2}/2$ with a (typically lower) frequency. 

\begin{figure*}[t]
    \centering
     \includegraphics[width=0.5\textwidth]{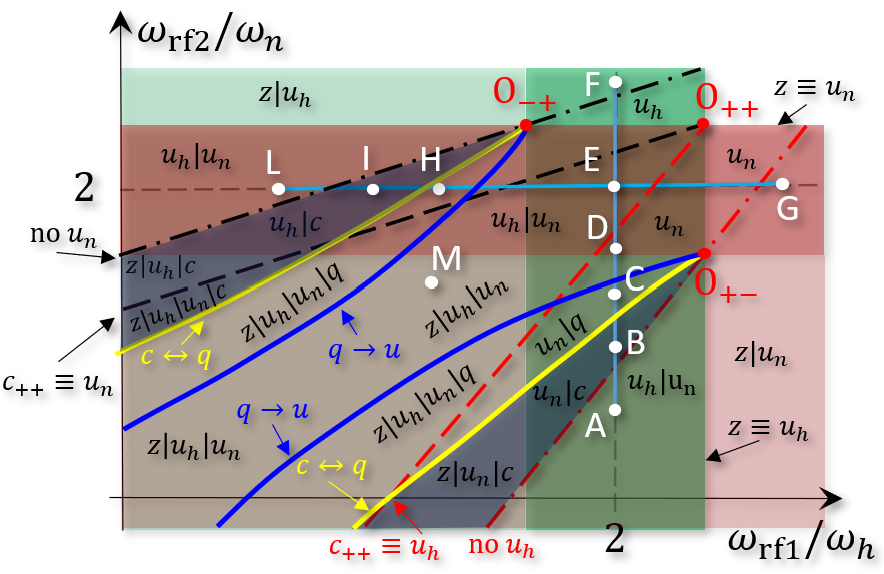} 
     \\ 
     $N_{hh},N_{nn}<0,N_{hn}=N_{nh}<0,N_{hh}N_{nn}/(N_{hh}N_{nn}-N_{hn}^2)<0$ \\
     \includegraphics[width=7.5cm]{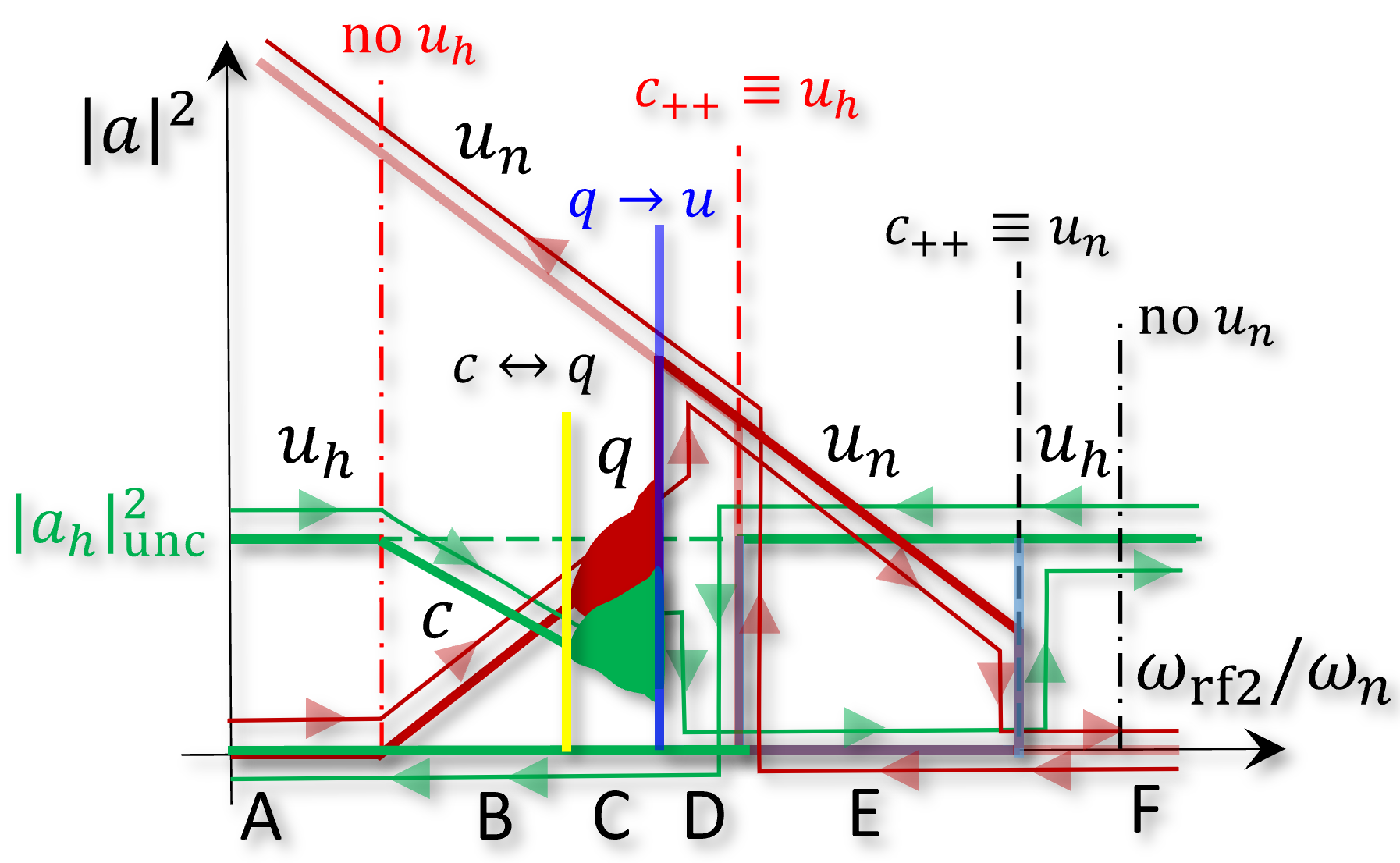}\includegraphics[width=8.5cm]{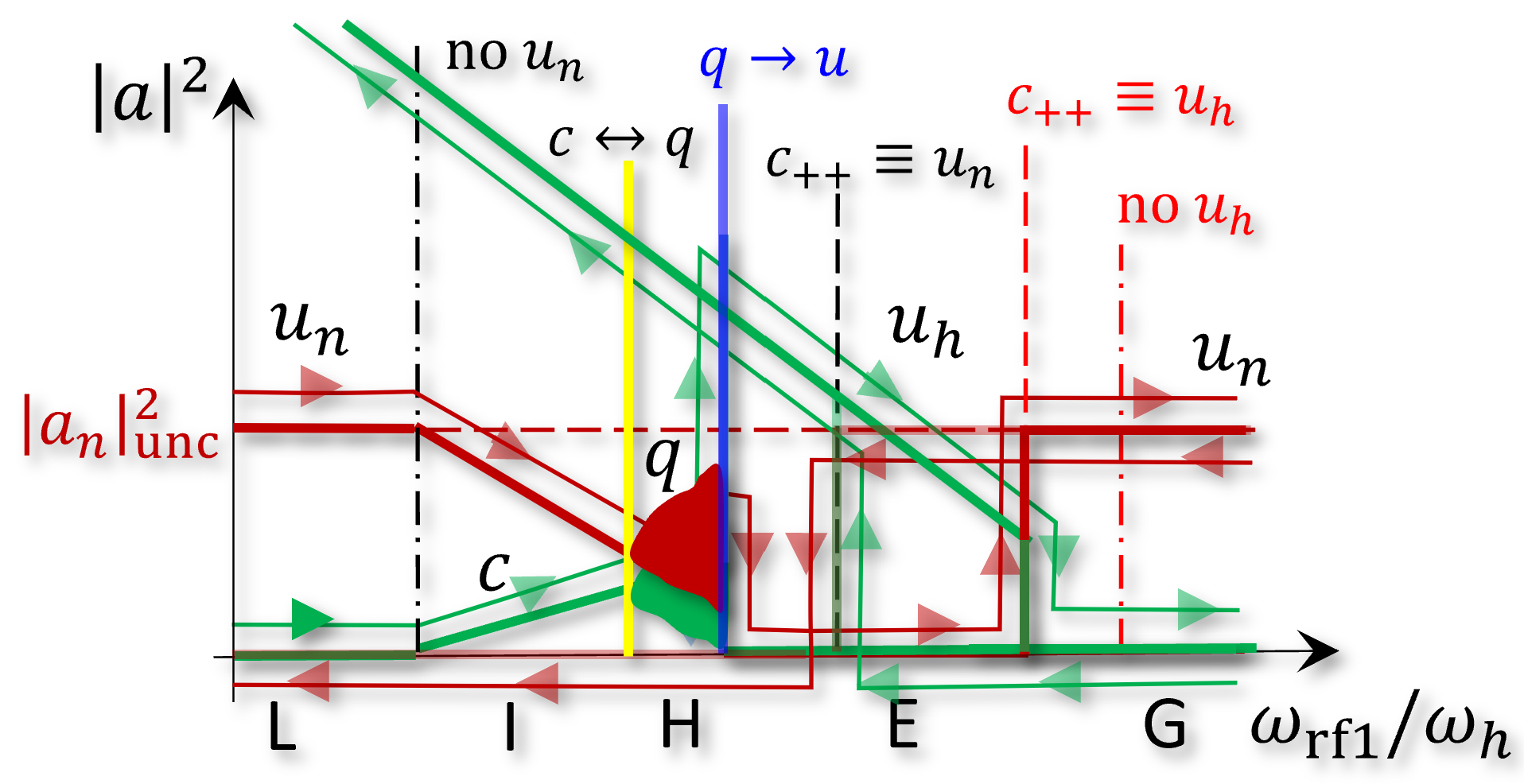}
    \caption{Phase diagram of interaction in $(\omega_\mathrm{rf1},\omega_\mathrm{rf2})$ plane for parametrically-excited modes $h$ and $n$ via parallel pumping with two-tone signal at fixed above-threshold amplitudes $\delta h_1,\delta h_2$. 
    The different regions, colored when associated with stable regimes, correspond to the existence conditions \eqref{eq:constraint real amplitudes}
     for uncoupled $u_h$ (green), $u_n$ (red) and coupled $c$ (blue) regimes (see eqs.\eqref{eq:type 1 mode h off mode n off}-\eqref{eq:type 1 coupled modes}).  Black dashed line corresponds to the condition $c_{++}\equiv u_n$ 
    while red dashed line refers to $c_{++}\equiv u_h$. 
    Black dash-dotted line refers to the condition $c_{-+}\equiv u_n$ (denoted as 'no $u_n$'), while red dash-dotted line refers to the condition $c_{+-}\equiv u_h$ (denoted as 'no $u_h$'). Solid yellow (Hopf bifurcation) line $c\leftrightarrow q$ refer to the boundary of regions $T_{+-},T_{-+}$ due to the onset of Q-modes, solid blue (homoclinic bifurcation) line $q\rightarrow u$ denotes the boundary of the region where Q-modes can exist. White dots labeled with letters A-L indicate some of the possible combinations of coexisting steady-state regimes. Coexistence of several regimes in a given region of the control plane is denoted using the compact notation '$|$'.    
    Light-blue solid lines refer to possible experiments keeping one tone at fixed frequency (e.g. twice its natural resonant frequency) while sweeping the other tone's frequency in a continuous wave (CW) fashion. 
    Lower panels report steady-state amplitudes $|a_h|^2$ (green) and $|a_n|^2$ (red) corresponding to 'vertical' and 'horizontal' (light-blue) slow variation of excitation frequencies. Both situations exhibit hysteretic behavior (see thinner lines with arrowheads referring to forward or backward frequency sweep). The vertical forward (resp. backward) sequence of steady states is $u_h \rightarrow c \rightarrow q \rightarrow u_n \rightarrow u_h$ (resp. $u_h \rightarrow u_n$). The horizontal forward (resp. backward) sequence of steady states is $u_n\rightarrow c\rightarrow q \rightarrow u_h \rightarrow u_n$ (resp. $u_n\rightarrow u_h$).
    }
    \label{fig:type1 phase diagram and hysteresis h1n3}
\end{figure*}

The general outcome of the stability analysis (see appendix \ref{sec:stability multi P modes}) defines the regions of stability for each possible P/Q-mode regime and leads to the complete determination of steady-state regimes phase diagrams. It turns out that such phase diagrams only depend on the NFS coefficients $N_{hh},N_{nn},N_{hn}$ and their combinations, leading to two qualitatively different scenarios where: 
\begin{itemize}
\item  $\text{sign}(N_{hh})\text{sign}(N_{nn}) (N_{hh}N_{nn}-N_{hn}^2) >0$ (or, equivalently, $N_{hh}N_{nn}/(N_{hh}N_{nn}-N_{hn}^2)>0$), depicted in fig.\ref{fig:type1 phase diagram and hysteresis}; 
\item $\text{sign}(N_{hh})\text{sign}(N_{nn}) (N_{hh}N_{nn}-N_{hn}^2) <0$ (or, equivalently $N_{hh}N_{nn}/(N_{hh}N_{nn}-N_{hn}^2)<0$), depicted in fig.\ref{fig:type1 phase diagram and hysteresis h1n3}. 
\end{itemize}
Remarkably, we stress that the remaining situations associated with sign changes of the parameters $N_{hh},N_{nn},N_{hn}$ are related to those of figs.\ref{fig:type1 phase diagram and hysteresis} and \ref{fig:type1 phase diagram and hysteresis h1n3} by using simple symmetry arguments (reflections in the $(\omega_\mathrm{rf1},\omega_\mathrm{rf2})$ control plane). 

In all situations, in order to have an observable P-mode regime of a given type (among those listed in \eqref{eq:type 1 mode h off mode n off}-\eqref{eq:type 1 coupled modes}), the associated stability region must have nonempty intersection with the corresponding existence region defined before and depicted in fig.\ref{fig:Pmodes existence regions} (details in appendix \ref{sec:existence conditions}). Stability regions in figs.\ref{fig:type1 phase diagram and hysteresis} and \ref{fig:type1 phase diagram and hysteresis h1n3} are colored green for uncoupled $u_h$, red for $u_n$ and blue for coupled $c$ regimes. Moreover, Q-modes are stable in regions of the control plane bounded by lines $c\leftrightarrow q$ (solid yellow in figs.\ref{fig:type1 phase diagram and hysteresis} and \ref{fig:type1 phase diagram and hysteresis h1n3}) corresponding to Hopf bifurcation and lines $q\rightarrow u$ associated with homoclinic bifurcation\cite{Perko_book} (solid blue in figs.\ref{fig:type1 phase diagram and hysteresis} and \ref{fig:type1 phase diagram and hysteresis h1n3}). The location of these regions depends on the values and signs of the NFS coefficients. In section \ref{sec:numerical results}, we will discuss quantitative examples of mode pairs interactions involving both P- and Q-modes. 

As a consequence of existence and stability analyses, the algorithm to construct the phase diagram for any choice of the parameters is reported in table \ref{tab:type1 interactions}, where the rules \ref{item:E1}-\ref{item:E4} refer to existence and the rules \ref{item:S1}-\ref{item:S8} refer to stability of interacting steady-state regimes (rules \ref{item:S1}-\ref{item:S5} for zero and coupled P-modes, rules \ref{item:S6}-\ref{item:S7} for uncoupled P-modes, and rule \ref{item:S8} for Q-modes). It is worth stressing that, regardless of the scenario, rules \ref{item:S6}-\ref{item:S7} state that $u_h$ is unstable in the region enclosed by the lines $c_{++}\equiv u_h$ and $c_{+-}\equiv u_h$ (red dashed and dash-dotted lines in figs.\ref{fig:type1 phase diagram and hysteresis} and \ref{fig:type1 phase diagram and hysteresis h1n3}), whereas $u_n$ is unstable in the region enclosed by the lines $c_{++}\equiv u_n$ and $c_{-+}\equiv u_n$ (black dashed and dash-dotted lines in figs.\ref{fig:type1 phase diagram and hysteresis} and \ref{fig:type1 phase diagram and hysteresis h1n3}). The obtained phase diagram can be used to compare and/or interpret the experimental results, as it is done in ref.\cite{soares_knet_flagship_paper_2025}.

For illustrative purpose, one can see in the top panel of fig.\ref{fig:type1 phase diagram and hysteresis} that 
there are several regions where coexistence of multiple stable solutions occurs. Such coexistence in a given region of the control plane is denoted using the compact notation '$|$'. For instance, a region labeled as '$u_h|u_n$' indicates that both 'uncoupled' modes may be excited at steady-state regime but only one at the time. 

It is worth noting that, when excitation conditions are chosen such that coupled solutions are stable (blue regions, e.g. points C,B,G,H,M in fig.\ref{fig:type1 phase diagram and hysteresis}), the solution produces regimes that are different (or even not reachable) by the uncoupled single modes. In this respect, with the above assumptions on NFSs, mode $h$ can be attenuated and mode $n$ enhanced compared to the uncoupled amplitudes $|a_h|^2_\mathrm{+,unc},|a_n|^2_\mathrm{+,unc}$ at the same frequencies, respectively. 

One can also see that coupled P-modes of type $c_{++}$ exist and are stable in $T_{++}$, while  $c_{+-},c_{-+}$ P-modes exist in regions $T_{+-},T_{-+}$ but are unstable (according to rules \ref{item:S2}-\ref{item:S4} in table \ref{tab:type1 interactions}), while Q-modes $q$ exist and are stable in the region enclosed between (yellow) Hopf bifurcation line $c\leftrightarrow q$  and (blue) homoclinic bifurcation line $q\rightarrow u$ (according to rule \ref{item:S8} in table \ref{tab:type1 interactions}).

In the case of fig.\ref{fig:type1 phase diagram and hysteresis h1n3}, $c_{++}$ P-modes exist but are unstable in $T_{++}$, while regions $T_{+-},T_{-+}$ where $c_{+-},c_{-+}$ P-modes exist and are stable, are bounded by Hopf bifurcation lines $c\leftrightarrow q$ (yellow) determining the transitions from coupled regimes to Q-modes that in turn disappear after crossing homoclinic bifurcation lines $q\rightarrow u$ (blue). We remark that, in this situation, two pairs of Hopf and homoclinic bifurcation lines originate from the same points $O_{+-}$ and $O_{-+}$, giving rise to Bogdanov-Takens bifurcations\cite{Kuznetsov2023} (see appendix \ref{sec:stability multi P modes} for details).

It is important remarking again that the coupled regime region overlaps with the others determining a complex picture of coexistence with other regimes, which gives rise to hysteresis and non-commutativity.
To elucidate such circumstances, we will first consider the situation of continuous wave (CW) excitation (section \ref{sec:CW excitation}) where frequencies are slowly enough changed in a continuous fashion without switching the field tones off. Then, we will address the situation of PWM excitation (section \ref{sec:PWM excitation}) where the two tones are switched off letting the system relax towards (almost) zero mode amplitudes before changing the frequencies, as it is the case of recent experiments reported in ref.\cite{soares_knet_flagship_paper_2025}.

\subsection{Continuous wave excitation}\label{sec:CW excitation}

In order to predict the possible outcome of an experiment performed by keeping one frequency fixed while sweeping the other in a continuous-wave (CW) fashion, we have illustrated the sequence of steady-state solutions $|a_h|^2,|a_n|^2$ corresponding to 'vertical'  or 'horizontal' (light-blue lines) frequency sweeps in the top panel of fig.\ref{fig:type1 phase diagram and hysteresis}. We have analyzed both directions $A\rightarrow E,E\rightarrow A$ and $L\rightarrow F,F\rightarrow L$ which refer to forward and backward frequency sweeps, respectively. 

The former 'vertical' sweep starts from point $A$ in the green region where only the regime $u_h$ exists. Then, by increasing $\omega_\mathrm{rf2}$, crossing of the red dashed line (continuity condition $c_{++}\equiv u_h$) occurs that corresponds to entering the coupled regime $c_{++}$ at $|a_n|^2_\mathrm{coup}=0$ and $|a_h|^2=|a_h|^2_\mathrm{unc}$ (see eq. \eqref{eq:saturation amplitude mode n type 1 linear}), as one can see in lower left panel of fig.\ref{fig:type1 phase diagram and hysteresis} (solid green and red lines). The rightwards/leftwards arrows in the same panel denote branches corresponding to forward/backward external frequency sweeps. 

After entering such coupled regime region, further increasing $\omega_\mathrm{rf2}$ (blue region, points B,C) produces attenuation of mode $h$ and enhancement of mode $n$ until the mode $h$ vanishes when intersecting the dashed black line (continuity condition $c_{++}\equiv u_n$), which means $|a_h|^2_\mathrm{coup}=0$ and $|a_n|^2=|a_n|^2_\mathrm{unc}$ (see eq.\eqref{eq:saturation amplitude mode h type 1 linear}). We observe that crossing the (dash-dotted black) line $c_{-+}\equiv u_n \Leftrightarrow \text{ 'no $u_n$'}$ produces no effect since the system is already in the coupled regime $c_{++}$. 

Then, by continuing to increase $\omega_\mathrm{rf2}$, one enters the brown region where mode $u_n$ is still allowed (point D in region of coexistence of uncoupled $|a_h|^2,|a_n|^2$). We observe that crossing the (red dash-dotted) line $c_{+-}\equiv u_h \Leftrightarrow \text{ 'no $u_h$'}$ has no effect since the system is already in the uncoupled regime $u_n$. By further increasing the frequency $\omega_\mathrm{rf2}$, the uncoupled amplitude of mode $n$ increases while mode $h$ is inhibited (notice that there is a change of slope due to different coefficients multiplying $\omega_\mathrm{rf2}$ in the expressions of $|a_n|^2_\mathrm{coup}$ and $|a_n|^2_\mathrm{unc}$, eqs.\eqref{eq:saturation amplitude mode n type 1} and \eqref{eq:saturation amplitude mode n type 1 linear}). The linear increase of amplitude also continues when one enters the upper green region (transition $D\rightarrow E$) since this overlaps with the light red region, where linearly increasing amplitude of uncoupled regime is allowed outside the critical detuning cone $\omega_\mathrm{rf2}>2\omega_n+\epsilon_\mathrm{crit,n}$ (we recall that $N_{nn}>0$ has been assumed). 
 
Now we suppose to start the experiment from large enough frequency $\omega_\mathrm{rf2}$ such that only mode $h$ is excited at uncoupled amplitude $|a_h|^2_\mathrm{unc}$ while mode $n$ has zero amplitude (that is $u_h$ regime), and we slowly decrease the frequency down to point $E$. 
The backward frequency sweep from $E$ towards $A$ reveals a different amplitude branch due to the fact that through the green region only mode $h$ is allowed at uncoupled amplitude $|a_h|^2_\mathrm{unc}$. Then, after crossing the (red dash-dotted) line $c_{+-}\equiv u_h \Leftrightarrow \text{ 'no $u_h$'}$ between $D$ and $C$, there is an irreversible downward jump of $|a_h|^2$ to zero, while $|a_n|^2$ jumps upwards to the uncoupled solution (regime $u_n$) that is followed until the dashed black line $c_{++}\equiv u_n$ is crossed. In this situation, the uncoupled solution $u_n$ will coincide with the coupled solution $c_{++}$ at the maximum value of the amplitude. 
The rest of the downward branch is followed with a change of slope associated with coupled solutions $c_{++}$ until crossing the dashed red line $c_{++}\equiv u_h$ restores the uncoupled regime $u_h$.
It is important to observe that crossing the (dash-dotted black) line $c_{-+}\equiv u_n \Leftrightarrow \text{ 'no $u_n$'}$ below point $B$ does not alter the coupled regime $c_{++}$ since it is allowed in the contiguous blue region. 

As far the 'horizontal' (light blue) line cut in fig.\ref{fig:type1 phase diagram and hysteresis} is concerned, one can be easily convinced that the amplitude response versus frequency $\omega_\mathrm{rf1}$ is that reported in the lower-right panel of the same figure, by using a similar reasoning as above. There is a difference compared to the 'vertical' line cut. In fact, in the forward frequency sweep coming from large enough negative detuning, the system is in the $u_n$ regime. Then, after crossing the (dash-dotted black) line $c_{-+}\equiv u_n \Leftrightarrow \text{ 'no $u_n$'}$, the $u_n$ regime disappears and two possible regimes are possible, namely $u_h|c$. In this situation, to assess which of the two is actually reached, one has to necessarily consider the (transient) dynamical process that brings the system in the final steady-state, since no continuity condition can be exploited. Figure \ref{fig:type1 phase diagram and hysteresis} depicts the case where the system irreversibly jumps in the coupled regime $c_{++}$. 
Then, by increasing the frequency, the coupled regime persists until the continuity condition $c_{++}\equiv u_n$ (dashed black line) is crossed, meaning $|a_h|^2=0, |a_n|^2=|a_n|^2_\mathrm{unc}$. Further increasing of $\omega_\mathrm{rf1}$ does not alter the regime. 

In the alternative situation (not shown in the figure), after crossing the line $c_{-+}\equiv u_n \Leftrightarrow \text{ 'no $u_n$'}$, the system jumps in the uncoupled regime $u_h$ where it persists until the $u_h$ regime disappears for sufficiently large frequency $\omega_\mathrm{rf1}>2\omega_h+\epsilon_\mathrm{crit,h}$ (outside green region $R_h$).  

Conversely, in the backward sweep experiment, the system enters the coupled regime (transitions $F\rightarrow  D\rightarrow G$) by crossing the (black dashed) line $c_{++}\equiv u_n$, namely through the continuity condition $|a_h|^2_\mathrm{coup}=0,|a_n|^2=|a_n|^2_\mathrm{unc}$ (see eq. \eqref{eq:saturation amplitude mode h type 1 linear}), which inhibits the uncoupled regime $u_h$ despite the coexistence $u_h|c$. We observe that crossing the (dash-dotted red) line $c_{+-}\equiv u_h \Leftrightarrow \text{ 'no $u_h$'}$ does not produce any effect since the system is already in the uncoupled regime $u_n$. By further decreasing frequency $\omega_\mathrm{rf1}$, the (yellow Hopf bifurcation) line $c\leftrightarrow q$ is crossed and a Q-mode denoted as $q$ arises from the coupled regime $c_{++}$ become unstable ($H\rightarrow I$). By continuing to decrease the frequency ($I\rightarrow L$), the Q-mode disappears after crossing the critical (blue homoclinic bifurcation) line $q\rightarrow u$, ending in one of the stable uncoupled regimes $u_h,u_n$ ($u_h$ in the specific case of fig.\ref{fig:type1 phase diagram and hysteresis}).

Analogous lines of reasoning can be followed to analyze the other cases corresponding to different combinations of the values and signs of the NFS coefficients $N_{hh},N_{nn},N_{hn}=N_{nh}$, combining the qualitative structures of the phase portraits depicted in figs.\ref{fig:type1 phase diagram and hysteresis} and \ref{fig:type1 phase diagram and hysteresis h1n3} with simple symmetry arguments (reflections in the $(\omega_\mathrm{rf1},\omega_\mathrm{rf2})$ control plane) as mentioned in the previous section.

\subsection{Pulse-width-modulated excitation}\label{sec:PWM excitation}

\begin{figure*}[t]
    \centering
    \includegraphics[width=9cm]{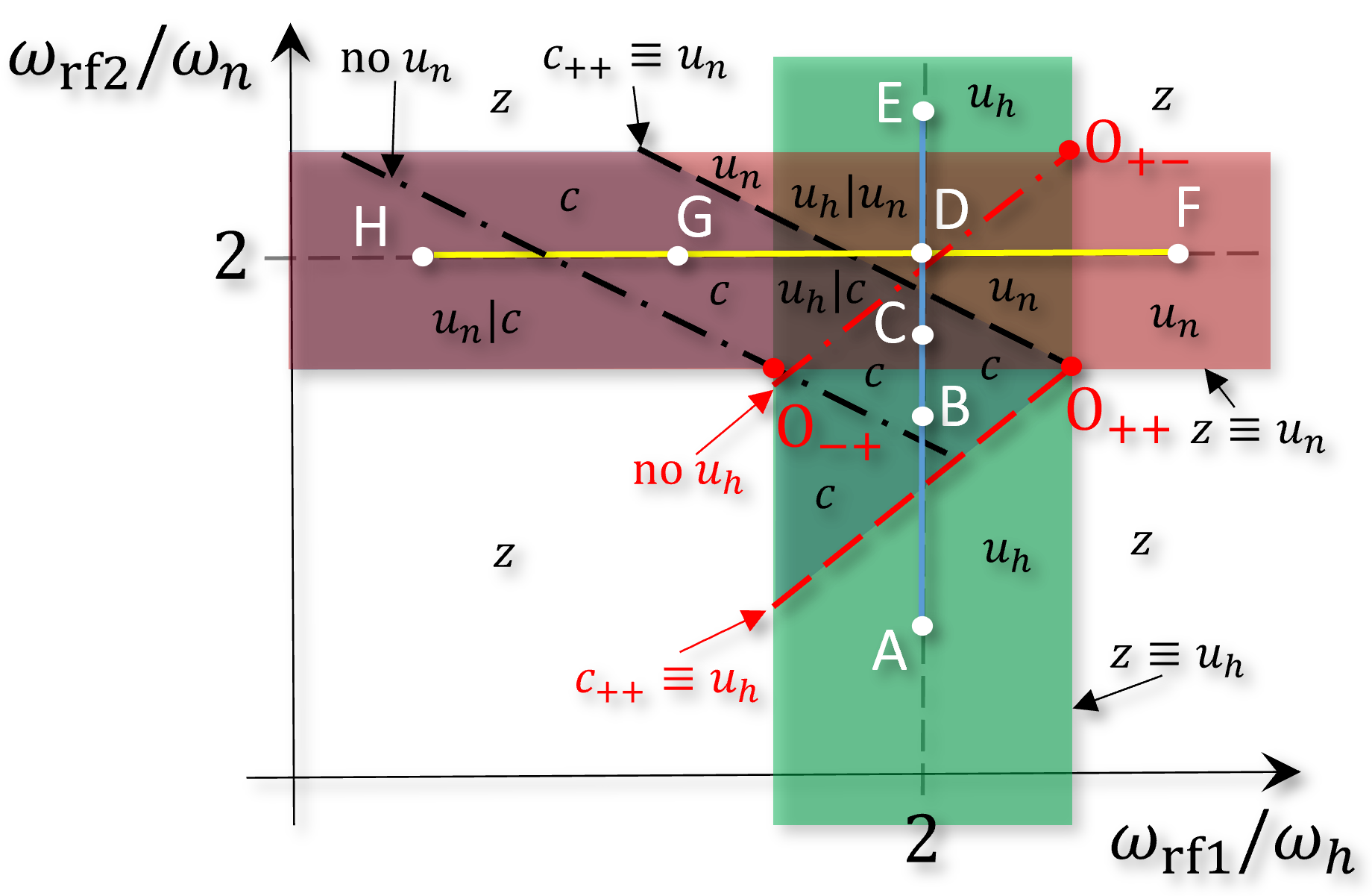} \\ $N_{hh}<0,N_{nn}>0, N_{hn}=N_{nh}<0, N_{hh}N_{nn}/(N_{hh}N_{nn}-N_{hn}^2)>0$ \\ \includegraphics[width=8cm]{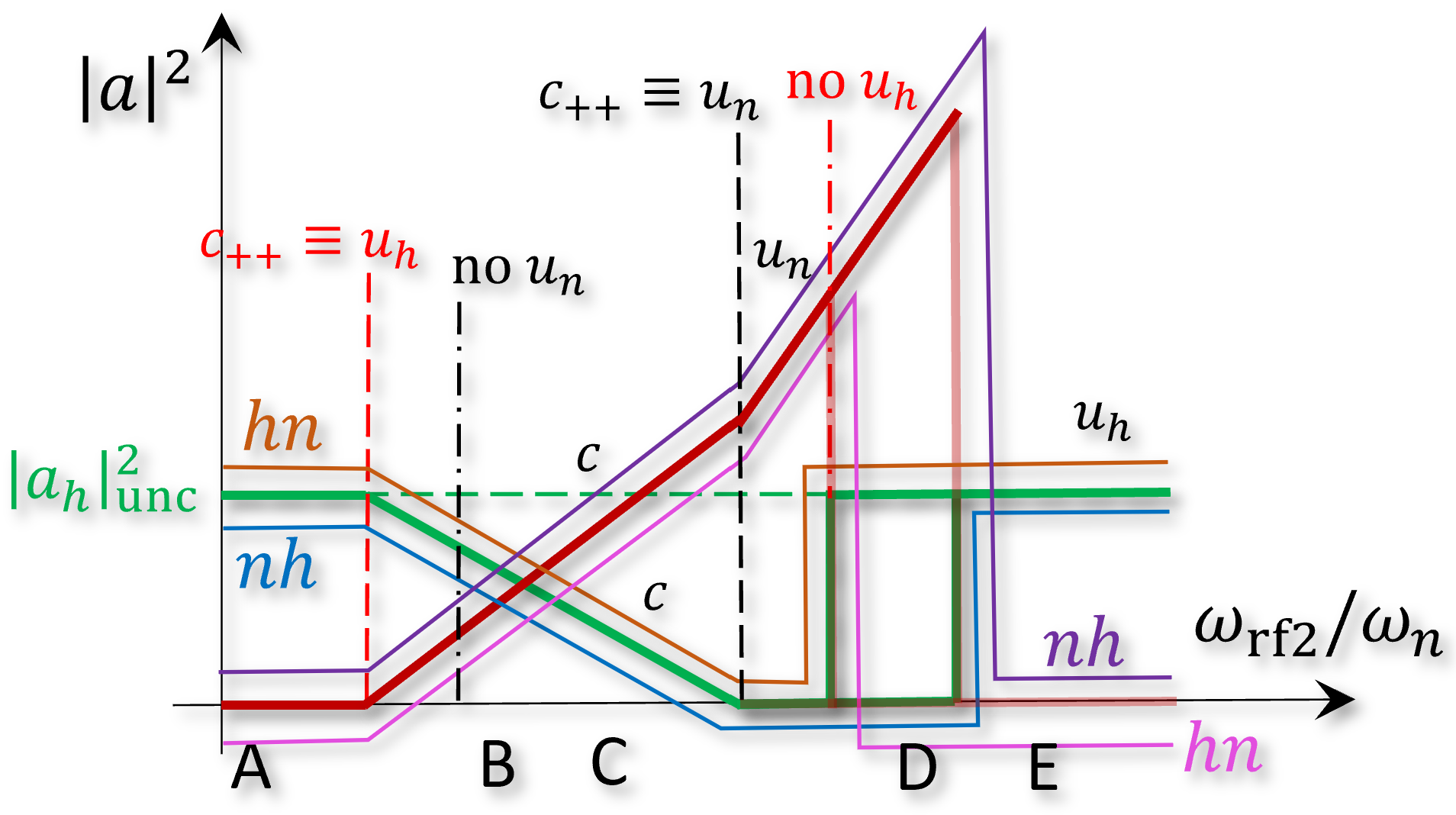}\includegraphics[width=8cm]{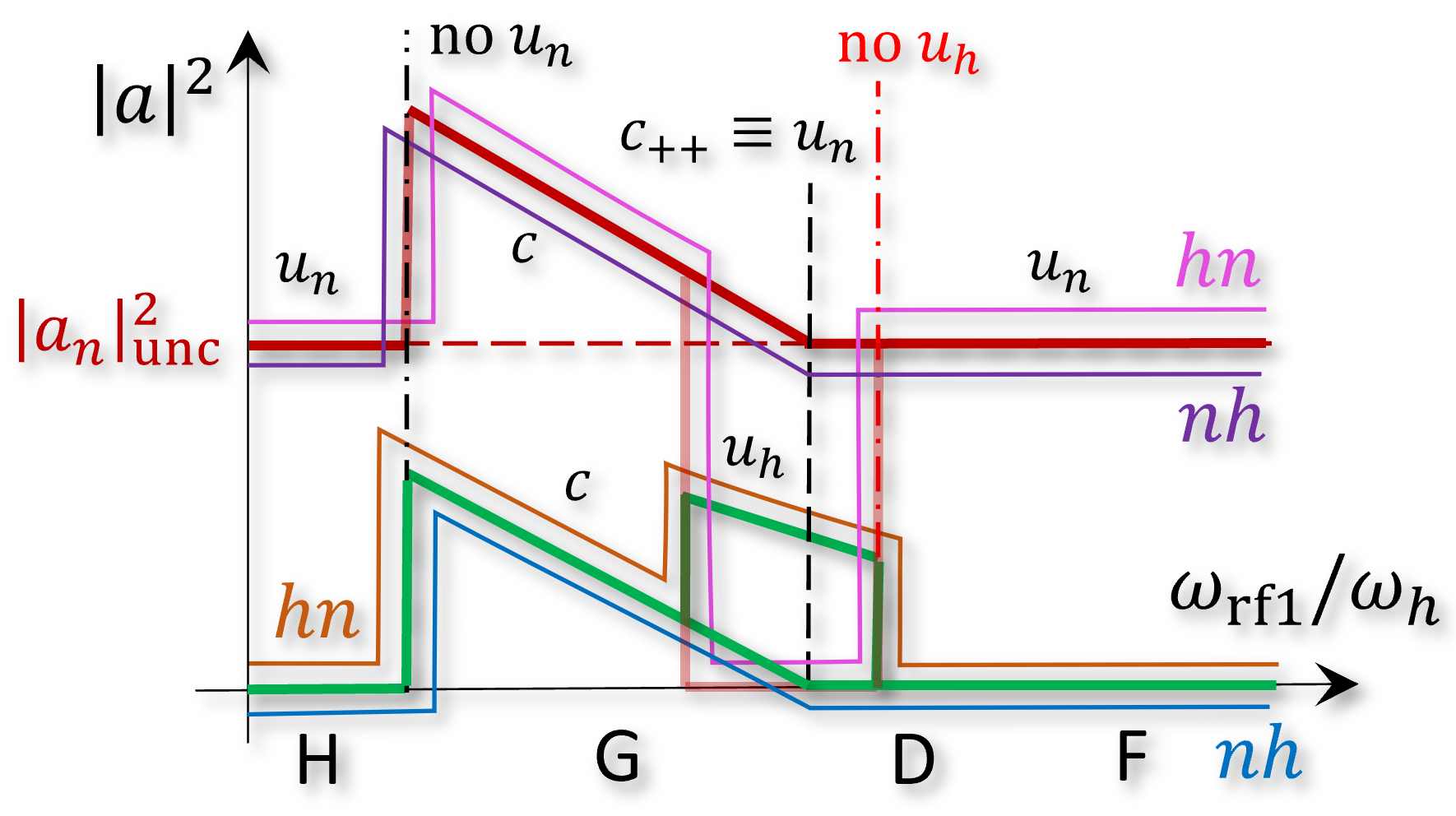}
    \caption{Phase diagram of interaction in $(\omega_\mathrm{rf1},\omega_\mathrm{rf2})$ plane for parametrically-excited modes $h$ and $n$ via parallel pumping with two-tone PWM signal at fixed above-threshold amplitudes $\delta h_1,\delta h_2$.  
    Top panel depicts different regions, colored when associated with stable regimes, corresponding to the existence conditions \eqref{eq:constraint real amplitudes}
     for uncoupled $u_h$ (green), $u_n$ (red) and coupled $c$ (blue) regimes (see eqs.\eqref{eq:type 1 mode h off mode n off}-\eqref{eq:type 1 coupled modes}).  Black dashed line corresponds to the condition $c_{++}\equiv u_n$ while red dashed line refers to $c_{++}\equiv u_h$. 
    Black dash-dotted line refers to the condition $c_{-+}\equiv u_n$ (denoted as 'no $u_n$'), while red dash-dotted line refers to the condition $c_{+-}\equiv u_h$ (denoted as 'no $u_h$'). White dots labeled with letters A-H indicate some of the possible combinations of coexisting steady-state regimes. Coexistence of several regimes in a given region of the control plane is denoted using the compact notation '$|$'.    
    Light-blue and yellow solid lines refer to possible experiments keeping one tone at fixed frequency (e.g. twice its natural resonant frequency) while changing the other tone's frequency in a PWM fashion. Lower panels report steady-state amplitudes $|a_h|^2$ (green) and $|a_n|^2$ (red) corresponding to (light-blue) 'vertical'  and (yellow) 'horizontal'  variation of excitation frequencies. Both situations exhibit non-commutative behavior depending on the sequence of tones $hn$ or $nh$. Orange and purple (resp. blue and violet) lines refer to $|a_h|^2,|a_n|^2$ when the tone sequence is $hn$ (resp. $nh$).}
    \label{fig:type1 phase diagram and hysteresis PWM}
\end{figure*}

We now suppose that the two tones are switched on sequentially always starting from tiny nonzero amplitudes $|a_h|^2,|a_n|^2$ of the modes, as it is performed in experiments described in ref.\cite{soares_knet_flagship_paper_2025}. It is assumed that the delay between the first and the second tone is long enough to let the system reach the steady state. First, we point out that one cannot observe nonzero steady-state amplitude outside the 'cross' region $C_h\cup C_n$ since the zero solution $z$ is stable outside the aforementioned region, as it is apparent in fig.\ref{fig:type1 phase diagram and hysteresis}. Moreover, the uncoupled regime $u_h$ (resp. $u_n$) is stable only in the green region $C_h$ (resp. red region $C_n$) under PWM excitation. Thus, the phase diagram of fig.\ref{fig:type1 phase diagram and hysteresis} reduces to that reported in fig.\ref{fig:type1 phase diagram and hysteresis PWM}.

As a consequence of the coexistence of different regimes in each subregion of the 'cross' phase diagram, depending on the sequence of tone excitation ($hn$ or $nh$) one will observe a different steady-state. In order to illustrate possible outcomes of experiments with PWM excitation, one can refer to (light blue) 'vertical'  and (yellow) 'horizontal' lines in fig.\ref{fig:type1 phase diagram and hysteresis PWM} where only one frequency is changed while the other is kept constant. 

We remark that, as happens for the single-mode PWM excitation (see section \ref{sec:nonlinear self interaction}), the steady-state response does not depend on the direction (forward or backward) of the frequency sweep (i.e. there is no hysteresis). Rather, the different response is ascribed to the sequence of tone excitation ($hn$ or $nh$), which we term non-commutativity. 

The situation corresponding to the 'vertical' line is described in the lower left panel of fig.\ref{fig:type1 phase diagram and hysteresis PWM}. Let us first consider the sequence $hn$. If one starts from point $A$ and increases the frequency $\omega_\mathrm{rf2}$, the system will stay in the uncoupled regime $u_h$ (orange line for $|a_h|^2$, purple line for $|a_n|^2)$ till the red dashed line $c_{++}\equiv u_h$ is crossed. After crossing this line, the system will go into the coupled regime $c_{++}$ until the critical (dashed black) line $c_{++}\equiv u_n$ is crossed. In that situation, the coupled regime $c_{++}$ disappears and the system ends up in the uncoupled regime $u_n$ because $u_h$ is forbidden between red lines according to the rule \ref{item:S6} in table \ref{tab:type1 interactions}. 
Further increase of frequency \(\omega_\mathrm{rf2}\) will produce a linear increase of the amplitude \(|a_n|^2\) (we recall that we assumed \(N_{nn}>0\)) with a different slope. This situation persists until the critical (red dash-dotted) line \(c_{+-}\equiv u_h\Leftrightarrow \text{`no }u_h\text{'}\) is crossed. Beyond this line, the uncoupled regime \(u_h\) becomes admissible again; since mode \(h\) is excited before mode \(n\) in the \(hn\) sequence, the system goes into the uncoupled regime \(u_h\). Further increase of \(\omega_\mathrm{rf2}\) results in a constant steady-state amplitude of the uncoupled mode \(h\).

Now we consider the sequence $nh$. Again, starting from $A$ and increasing $\omega_\mathrm{rf2}$, the system persists in the uncoupled regime $u_h$ (blue line for $|a_h|^2$, violet line for $|a_n|^2$) which is the only stable one. After crossing the line $c_{++}\equiv u_h$, the system enters the region of the coupled regime $c_{++}$ until the critical line $c_{++}\equiv u_n$ is crossed. In that situation, the amplitude $|a_h|^2=0$ and $|a_n|^2=|a_n|^2_\mathrm{unc}$. By increasing the frequency $\omega_\mathrm{rf2}$, the mode $h$ is inhibited while the (first excited) mode $n$ increases its amplitude being in the regime $u_n$. The limit situation for this increase is represented by the boundary of the brown region in the phase diagram (transition $D\rightarrow E$) beyond which the uncoupled regime $u_n$ is no longer stable. 

In the following, we analyze the situation corresponding to the (yellow) 'horizontal' line and sequence $hn$ that is described in the lower right panel of fig.\ref{fig:type1 phase diagram and hysteresis PWM}. We suppose to start from point $H$ where only uncoupled regime $u_n$ (orange line for $|a_h|^2$, purple line for $|a_n|^2$) is stable. Then, by increasing the frequency $\omega_\mathrm{rf1}$, the $u_n$ regime persists until the critical (black dash-dotted) line '$\text{no } u_n$' is crossed giving rise to the coupled regime $c$ (negative slopes arise from eqs.\eqref{eq:saturation amplitude mode h type 1 linear}-\eqref{eq:saturation amplitude mode n type 1 linear} and assumptions on NFS coefficients). After that, by increasing $\omega_\mathrm{rf1}$ in such a way to enter the (green) region $C_h$ that stabilizes $u_h$, subsequent excitation of mode $n$ has no effect and the system ends up in the uncoupled regime $u_h$. By increasing the frequency, mode $h$ exhibits a linear decrease in $|a_h|^2$ (remember that we assumed $N_{hh}<0$) until the (red dash-dotted) critical line '$\text{no } u_h$' is crossed. By increasing $\omega_\mathrm{rf1}$ beyond such critical value, mode $h$ disappears and mode $n$ comes up towards the uncoupled  regime $u_n$. No further change occurs as a consequence of a frequency increase.

Now we consider the inverted sequence of tones $nh$. Always starting from point $H$, after crossing the critical line '$\text{no } u_n$', the coupled regime $c_{++}$ is excited (mode $h$ is turned on when mode $n$ is already on). Then, both mode amplitudes will undergo linear decrease until the critical (dashed black) line $c_{++}\equiv u_n$ is crossed, which means that regimes $c_{++}$ and $u_n$ coincide. By further increasing $\omega_\mathrm{rf1}$, the mode $n$ persists in the uncoupled regime. 

Analogous lines of reasoning can be followed to analyze the other cases corresponding to different combinations of the signs of the NFS coefficients $N_{hh},N_{nn},N_{hn}=N_{nh}$. 

{We stress that the phase diagrams described in this section (such as that depicted in the top panel of Fig.\ref{fig:type1 phase diagram and hysteresis PWM}) have been successfully used to explain the recent experiments performed on a 1$\mu$m YIG disk where pairs of spin-wave modes are parametrically excited using two-tone spectroscopy\cite{soares_knet_flagship_paper_2025} (see also section \ref{sec:num PWM}).}

\section{Numerical results}\label{sec:numerical results}

\subsection{Methods}

In order to test and validate the developed theory, we perform a numerical study of a YIG disk with the following material parameters: saturation magnetization $M_s=140.7\,$kA/m, exchange stiffness $A_\mathrm{ex}=3.7\,$pJ/m, absolute value of the gyromagnetic ratio $\gamma=2.2329\times 10^5\,m A^{-1}s^{-1}$. With these parameters the dimensionless time is measured in units $(\gamma M_s)^{-1}\approx 32\,$ps. We assume a damping parameter $\alpha=0.01$ which is quite large for YIG samples only in order to make time-domain simulations feasible (a small damping would imply longer time to simulate). 
We consider a 100 nm YIG disk, with dimensions $100\times 100\times 5\,$ nm$^3$. 
A static external field with amplitude $\mu_0 h_{ax} M_s=30$ mT is applied along the $x$ axis.

In the sequel, we will model such YIG disks first as macrospin, then by means of full micromagnetic simulations. The results of macrospin simulations are reported in section \ref{sec:macrospin simulations} of the supplementary material showing excellent agreement with the theory.

\begin{figure}[t]
    \centering
    \includegraphics[width=8cm]{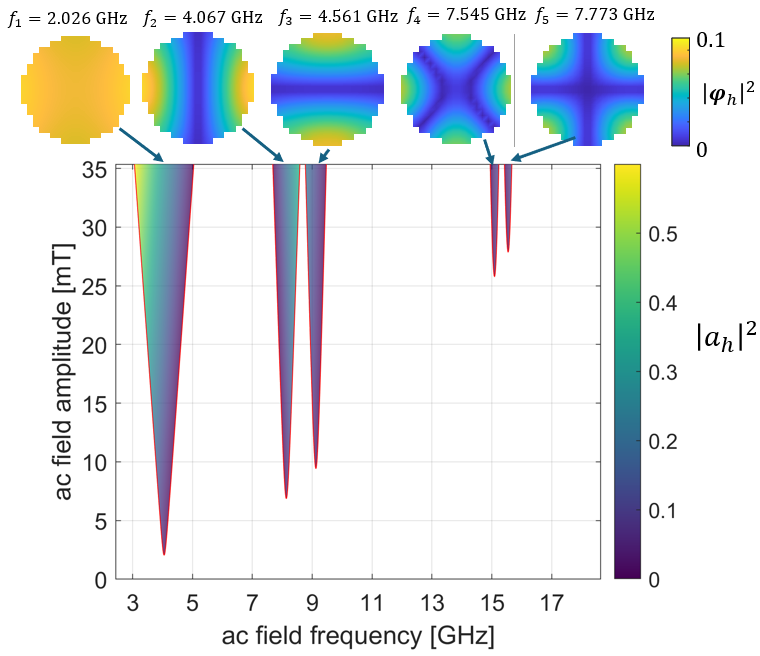} 
    \caption{Parametric resonance phase diagram for 100 nm YIG disk. The first 5 modes have been computed and their spatial amplitude profiles appear on the top along with their frequencies $f_h$ (the color code denotes local oscillation amplitude normalized to its maximum value). The main panel reports the detuning cones (Arnold's tongues) for parametric excitation of the 5 modes in the control plane $(\omega_\mathrm{rf},\delta h)$. The color code here denotes the saturation of steady-state amplitude $|a_h|^2$ according to eq.\eqref{eq:saturation amplitude mode h} and the delimiting curves (red solid line) are computed according to eq.\eqref{eq:detuning cone}.}
    \label{fig:YIG disk 100 phase diagram}
\end{figure}

Here we consider 
the YIG disk with diameter 100 nm and thickness 5 nm, discretized into $20\times 20\times 1$ computational cells of size \(5\times5\times5\,\mathrm{nm}^3\). 
The eigenmodes are calculated by using the large-scale formulation described in ref.\cite{dAquino_JAP_2023} implemented in the numerical code MaGICo\cite{MaGICo}. In figure \ref{fig:YIG disk 100 phase diagram}, the phase diagram for parametric resonance of the first five modes is reported. The five eigenfrequencies are in the range between 2.02 and 7.77 GHz and the spatial profiles of normal magnetization oscillation modes are also visible in the top section of fig.\ref{fig:YIG disk 100 phase diagram}. For instance, one immediately sees that the fundamental mode $h=1$ has a frequency $f_1=2.0263\,$GHz that is very close to the value $2.0384\,$GHz computed within the macrospin approximation (see supplementary section \ref{sec:macrospin simulations}).

\subsection{Excitation of single modes driven by parametric pumping}

In the main panel of fig.\ref{fig:YIG disk 100 phase diagram}, one can see the detuning cones centered around each eigenfrequency, whose boundaries are given by eq.\eqref{eq:detuning cone}. 
The color scale depicting the interior region of each cone is associated with the steady-state mode power $|a_h|^2$ computed according to the analytical theory (see eq.\eqref{eq:saturation amplitude mode h}). From the observation of the filling pattern of each cone, one can infer the sign of the NFS coefficient $N_h$ associated with each mode $h=1,\ldots,5$.  

The NFS coefficients have been computed according to eq.\eqref{eq:NFS mode h} from the knowledge of the spatial profiles of the eigenmodes $\bm \varphi_1,\ldots\bm\varphi_5$. The values of eigenfrequencies, self-NFS coefficients and threshold field amplitudes computed according to eq.\eqref{eq:threshold field parametric resonance} are reported in table \ref{tab:YIG disk 100nm} within the supplementary material. 
One can see that the fundamental (Kittel) mode has negative NFS $N_1=-0.8159$, which is in good agreement with the value $-0.9256$ obtained in the macrospin computation. The higher order modes have NFS with both positive and negative sign, which reflects in different filling patterns of the regions enclosed within the detuning cones in the phase diagram reported in fig.\ref{fig:YIG disk 100 phase diagram}. 

\begin{figure*}[t]
    \centering
    \includegraphics[width=0.37\textwidth]{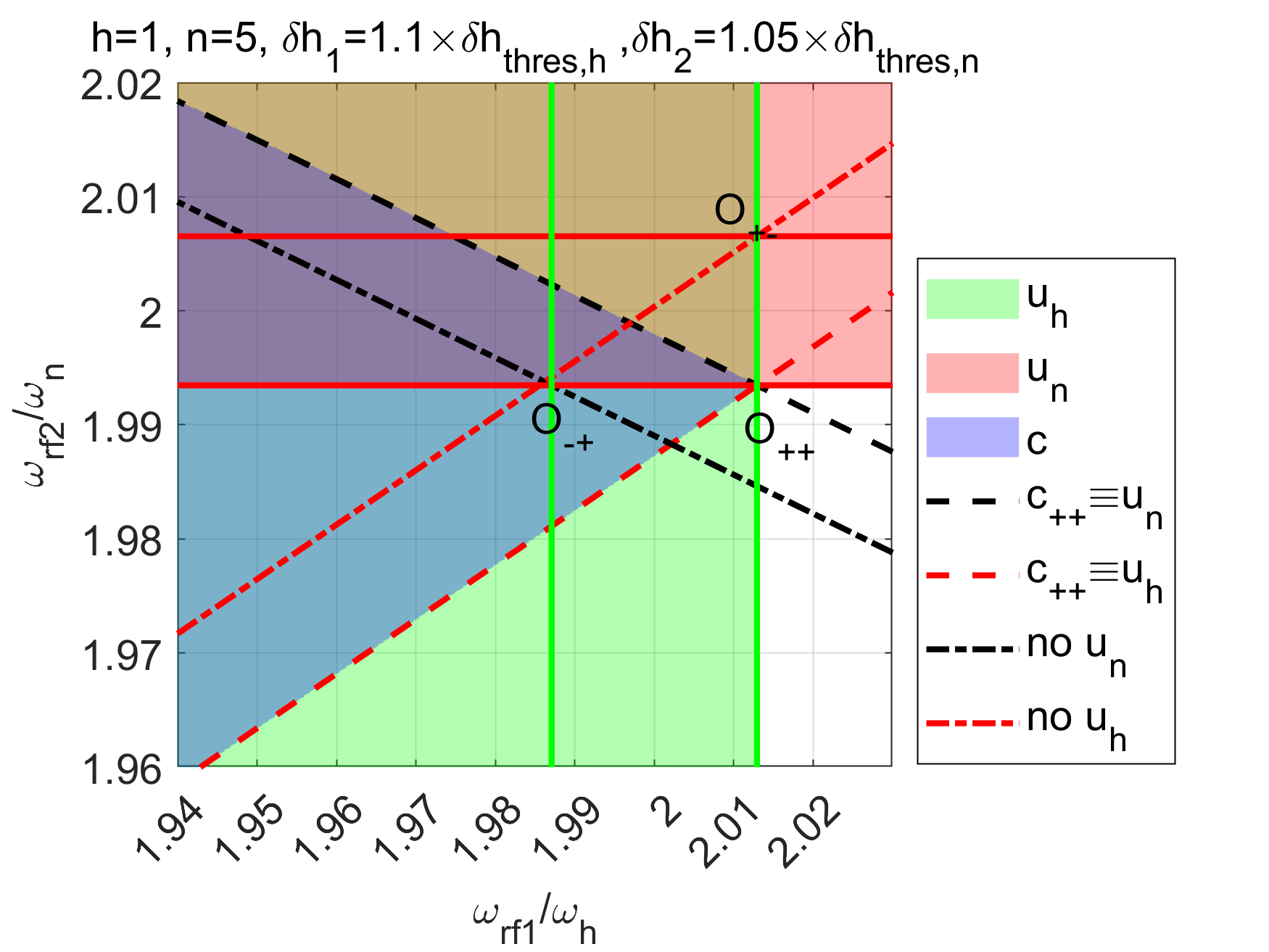} \includegraphics[width=0.31\textwidth]{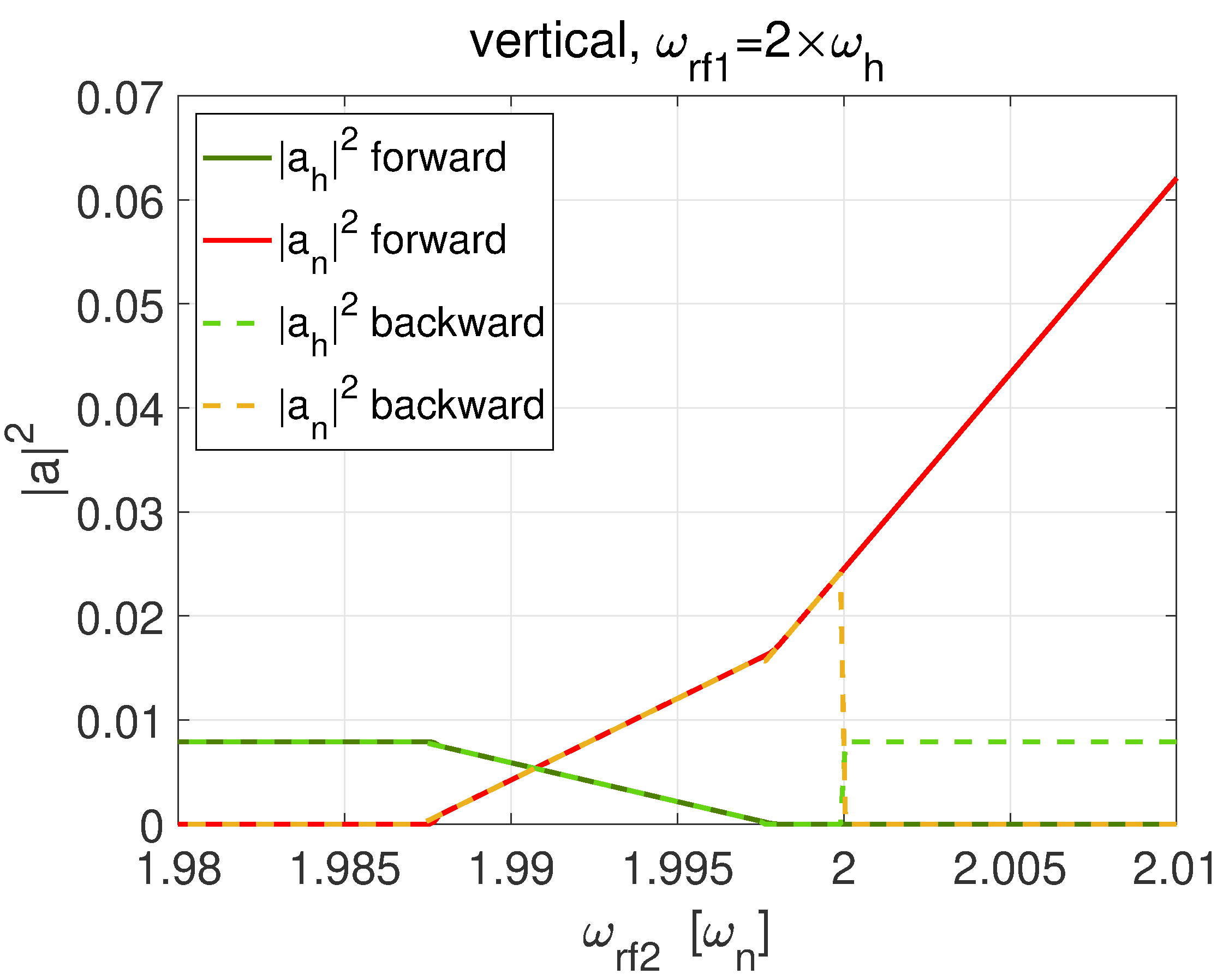}\includegraphics[width=0.31\textwidth]{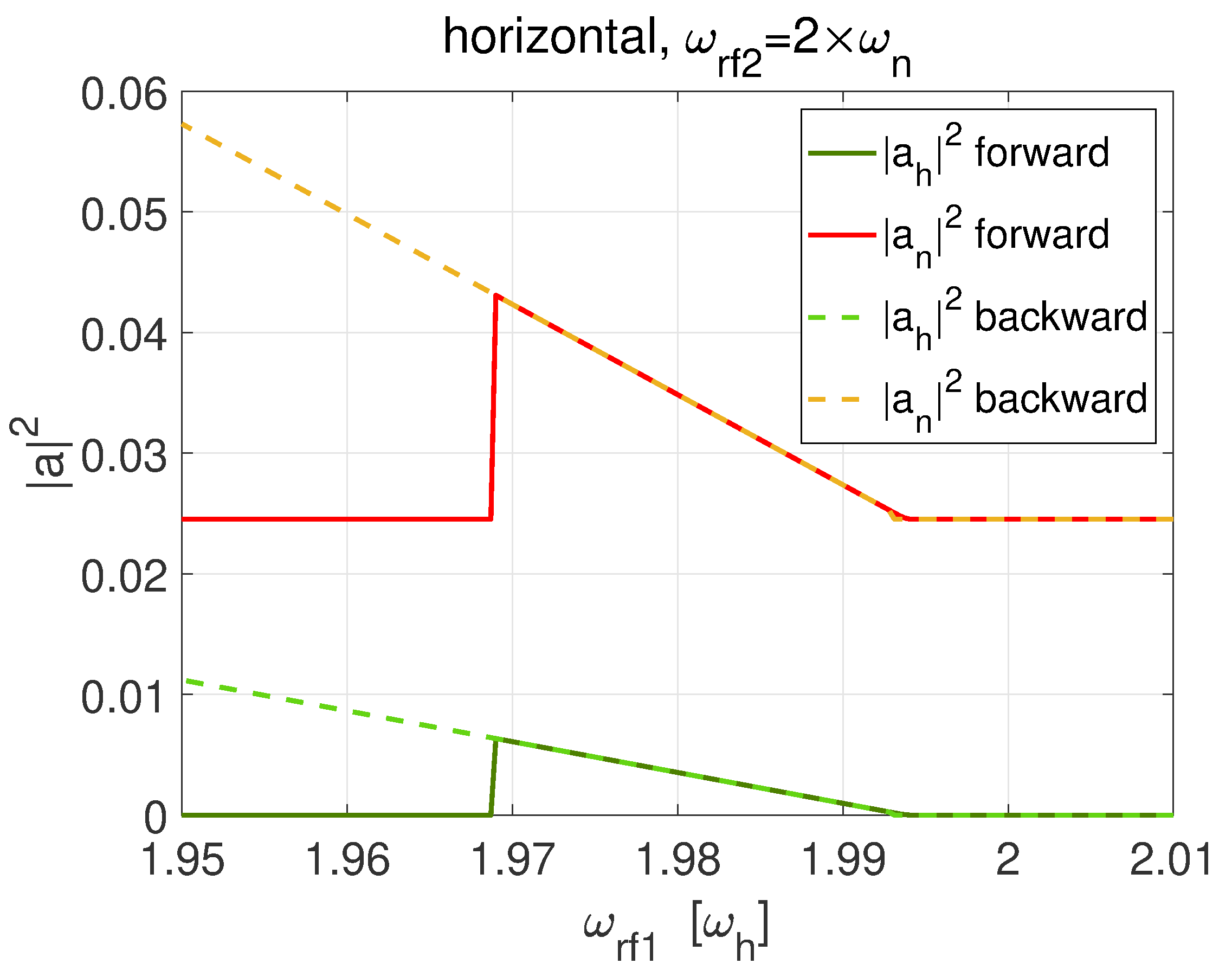}
    \caption{Phase diagram of interaction in $(\omega_\mathrm{rf1},\omega_\mathrm{rf2})$ plane for parametrically-excited modes $h=1$ and $n=5$ via parallel pumping with two-tone signal at fixed above-threshold amplitudes $\delta h_1=1.1\times\delta h_\mathrm{thres,h},\delta h_2=1.05\times\delta h_\mathrm{thres,n}$. NFS coefficients are $N_{hh}=-0.8159, N_{hn}=-0.3897=N_{nh}, N_{nn}=0.1332$. Left panel depicts different regions corresponding to the conditions \eqref{eq:constraint real amplitudes} 
    for which modes $u_h,u_n,c_{++}$ exist (green, red, blue). The boundaries of green and red regions are marked with green and red solid lines, respectively.  Dashed black (resp. red) line corresponds to the condition $c_{++}\equiv u_n$ (resp. $c_{++}\equiv u_h$).  
    Middle and right panels report steady-state amplitudes $|a_h|^2,|a_n|^2$ as function of frequency corresponding to 'vertical' ('horizontal') slow variation of (CW) excitation frequency $\omega_\mathrm{rf2}$ (resp. $\omega_\mathrm{rf1}$) while $\omega_\mathrm{rf1}=2\omega_h$ (resp. $\omega_\mathrm{rf2}=2\omega_n$), obtained by numerical integration of the averaged two-modes model \eqref{eq:averaged two nonlinear parametric Ah}-\eqref{eq:averaged two nonlinear parametric Phi_n}. Both processes exhibit hysteretic behavior (see solid/dashed lines referring to forward/backward frequency sweep).}
    \label{fig:type1 phase diagram and hysteresis modes 1 and 5}
\end{figure*}

Time-domain micromagnetic simulations\cite{dAquino_JCP_2005} of single mode excitation have been performed and the results are reported in supplementary section \ref{sec:micromagnetic simulation single tone} showing very good agreement with the analytical theory both concerning thresholds for parametric instability and steady-state amplitude saturation. 

\subsection{Parametric excitation of mode pairs}

We now analyze the parametric excitation of modes by using a two-tone signal of the type \eqref{eq:two tones signal}.
In this respect, we first use the developed analytical theory and then compare the predictions with numerical simulations of the full NMM \eqref{eq:NMM}, the two-modes model \eqref{eq:nonlinear coupled parametric ah}-\eqref{eq:nonlinear coupled parametric an} and the averaged two-modes model \eqref{eq:averaged two nonlinear parametric Ah}-\eqref{eq:averaged two nonlinear parametric Phi_n}. 

To this end, we have computed the mutual NFS interaction matrix given by eq. \eqref{eq:complex NMFS matrix} that is real and symmetric due to \eqref{eq:reality and symmetry Nhn}:
\begin{align}
&N_{hn}= 
\small\left(
\begin{array}{rrrrr}
   -0.8159  &  0.0558  & -1.2972 &  -0.2726 &  -0.3897\\
    0.0558  &  0.2853  & -0.4796 &   0.1301 &  -0.0895\\
   -1.2972  & -0.4796  & -0.3390 &  -0.3563 &  -0.2488\\
   -0.2726  &  0.1301  & -0.3563 &   0.1582 &  -0.4179\\
   -0.3897  & -0.0895  & -0.2488 &  -0.4179 &   0.1332
\end{array} \right) 
\label{eq:NMFS YIG100 correct} \,.\\
\end{align}
We remark that the diagonal terms of matrix $N_{hn}$ coincide with the NFS coefficients $N_h$ defined by eq.\eqref{eq:NFS mode h} and reported in supplementary material (table \ref{tab:YIG disk 100nm}).

We consider possible experiments 
similar to those described in ref.\cite{soares_knet_flagship_paper_2025}, 
where both modes are excited by two tones with above-threshold amplitudes $\delta h_1,\delta h_2$. The frequency of one tone is kept constant (e.g. at twice the natural frequency of the associated mode) while the other tone's frequency is swept forward or backward in CW fashion. The forward/backward frequency sweeping is instrumental to check hysteresis of the steady-state amplitude regimes. We also perform numerical simulations to assess non-commutativity upon the temporal sequence in PWM experiments. For both CW and PWM excitations, we analyze two qualitatively different situations determined by appropriate choice of the indices $h,n$ corresponding to the phase diagrams in figs.\ref{fig:type1 phase diagram and hysteresis} and \ref{fig:type1 phase diagram and hysteresis h1n3}.

\begin{figure*}[t]
    \centering
    \hspace{0.1\textwidth} NMM, sequence $15$ \hspace{0.1\textwidth} two-modes, sequence $15$  \\
    \includegraphics[width=0.3\textwidth]{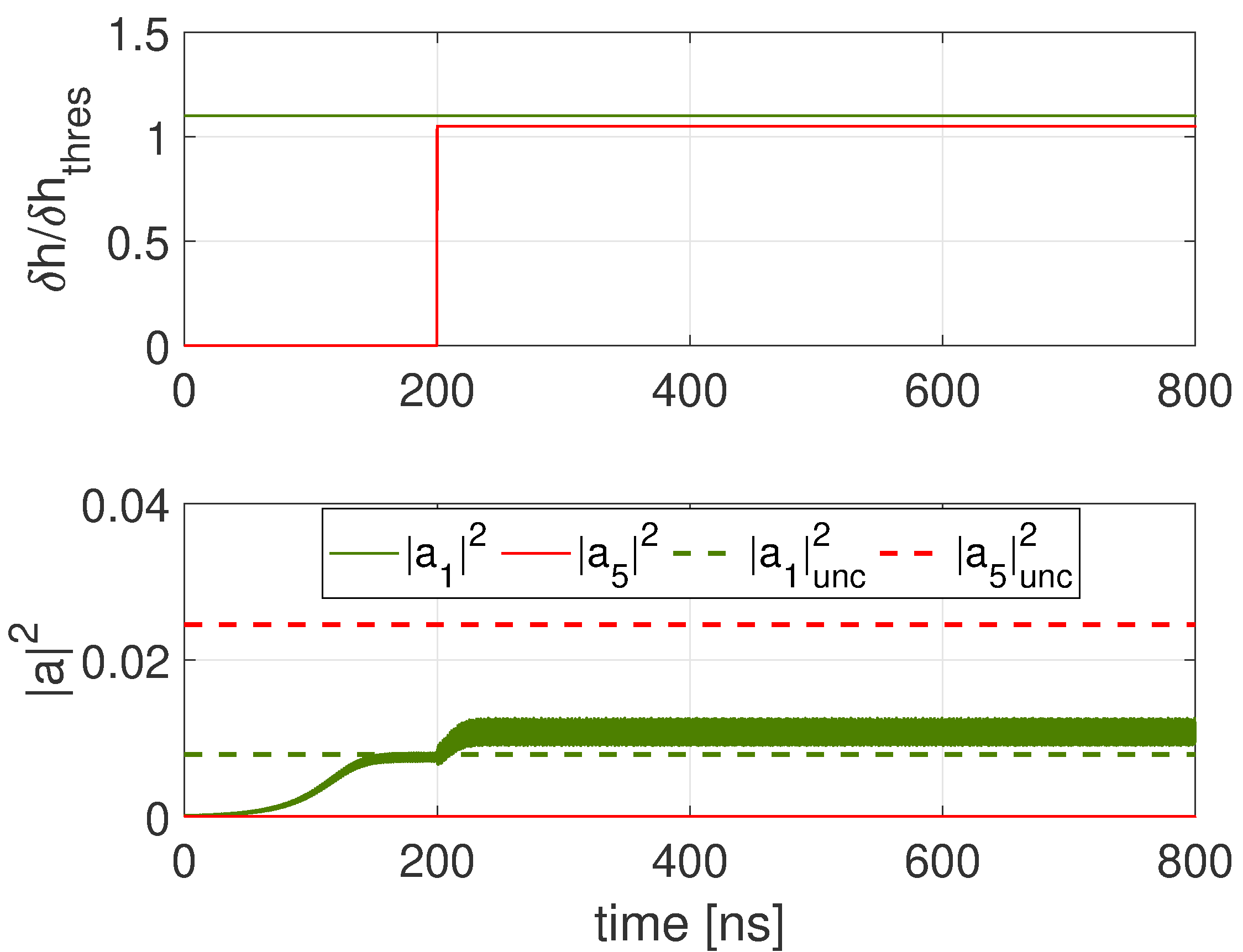}
    \includegraphics[width=0.3\textwidth]{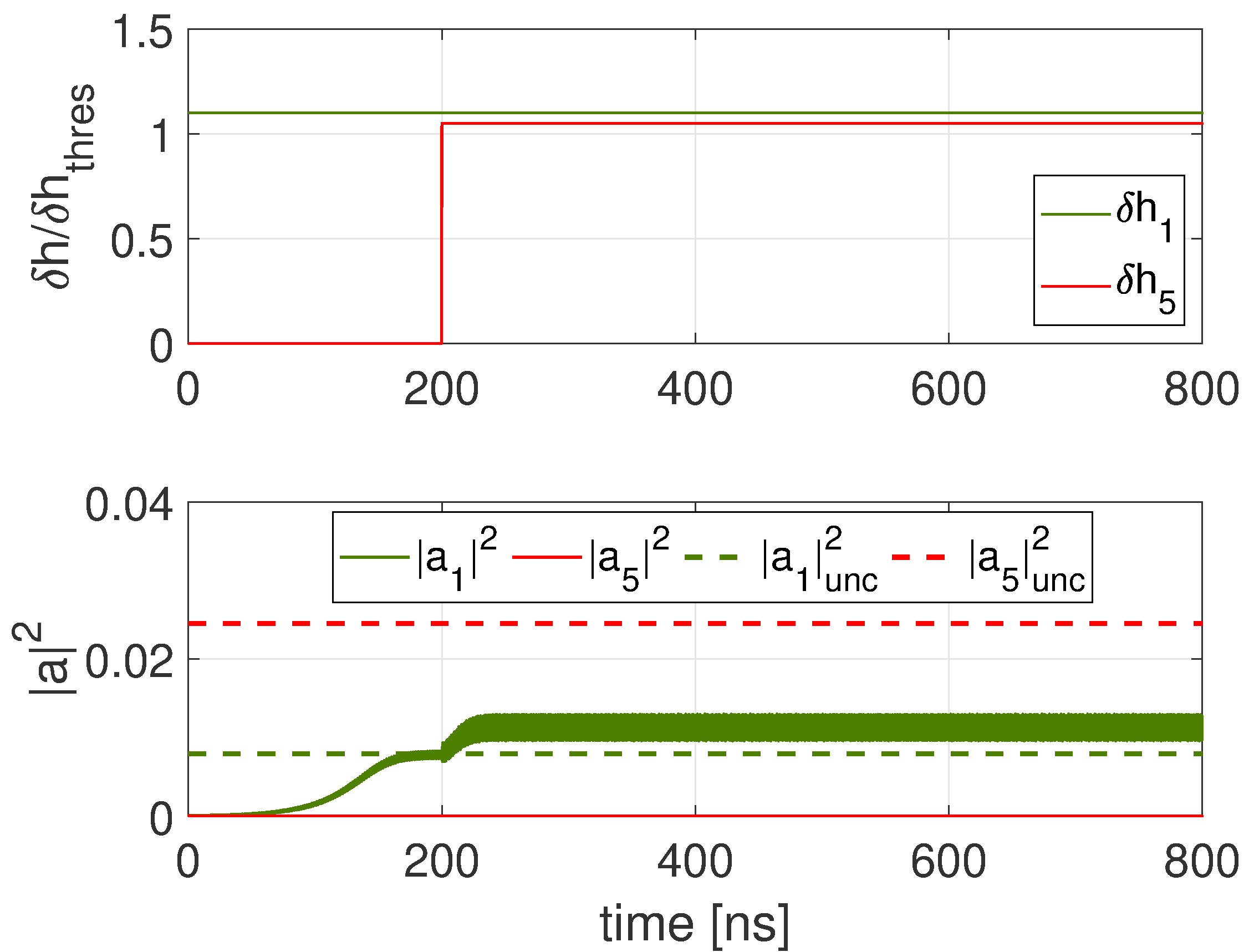} \\
    \hspace{0.1\textwidth} NMM, sequence $51$ \hspace{0.1\textwidth} two-modes, sequence $51$  \\
    \includegraphics[width=0.3\textwidth]{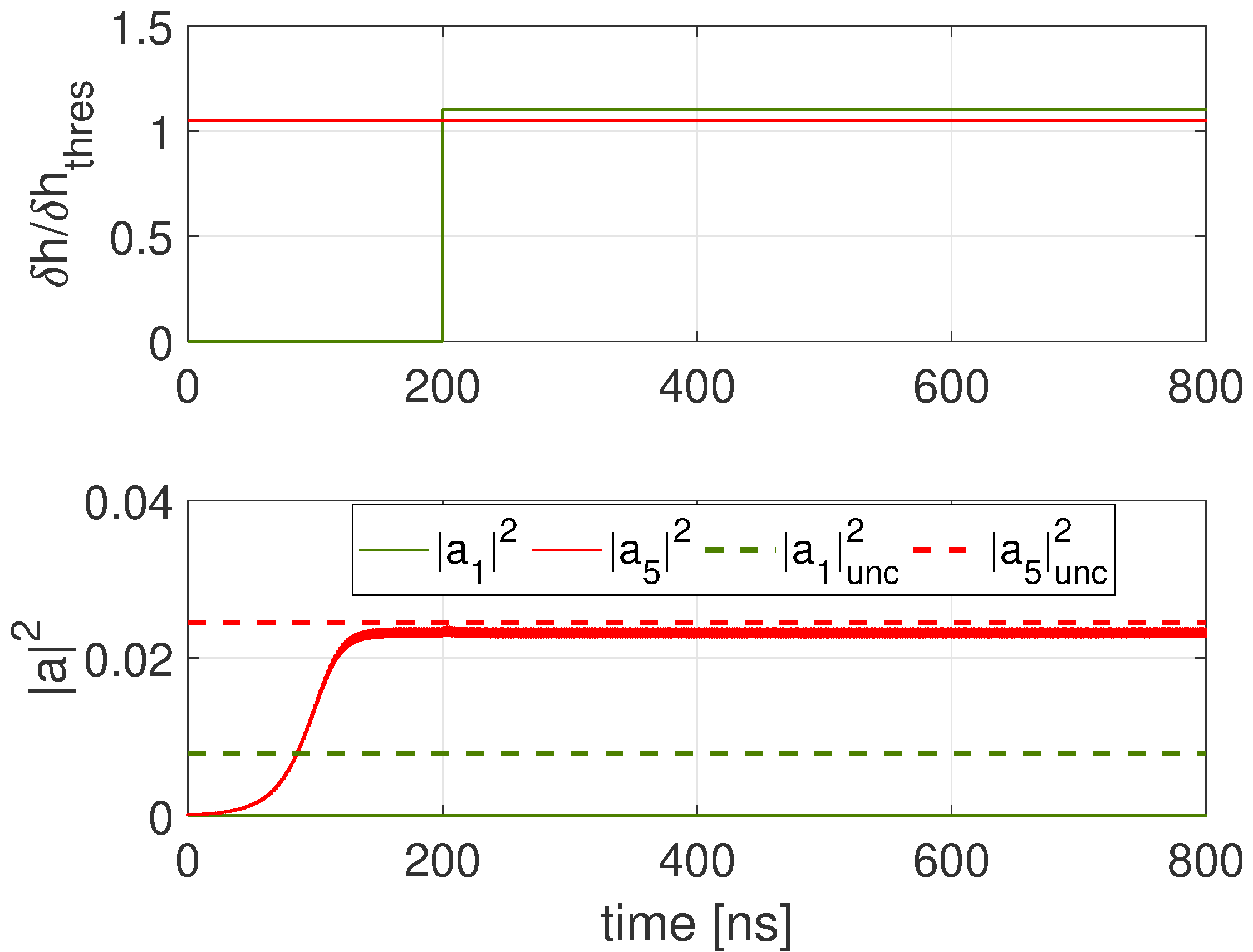}
    \includegraphics[width=0.3\textwidth]{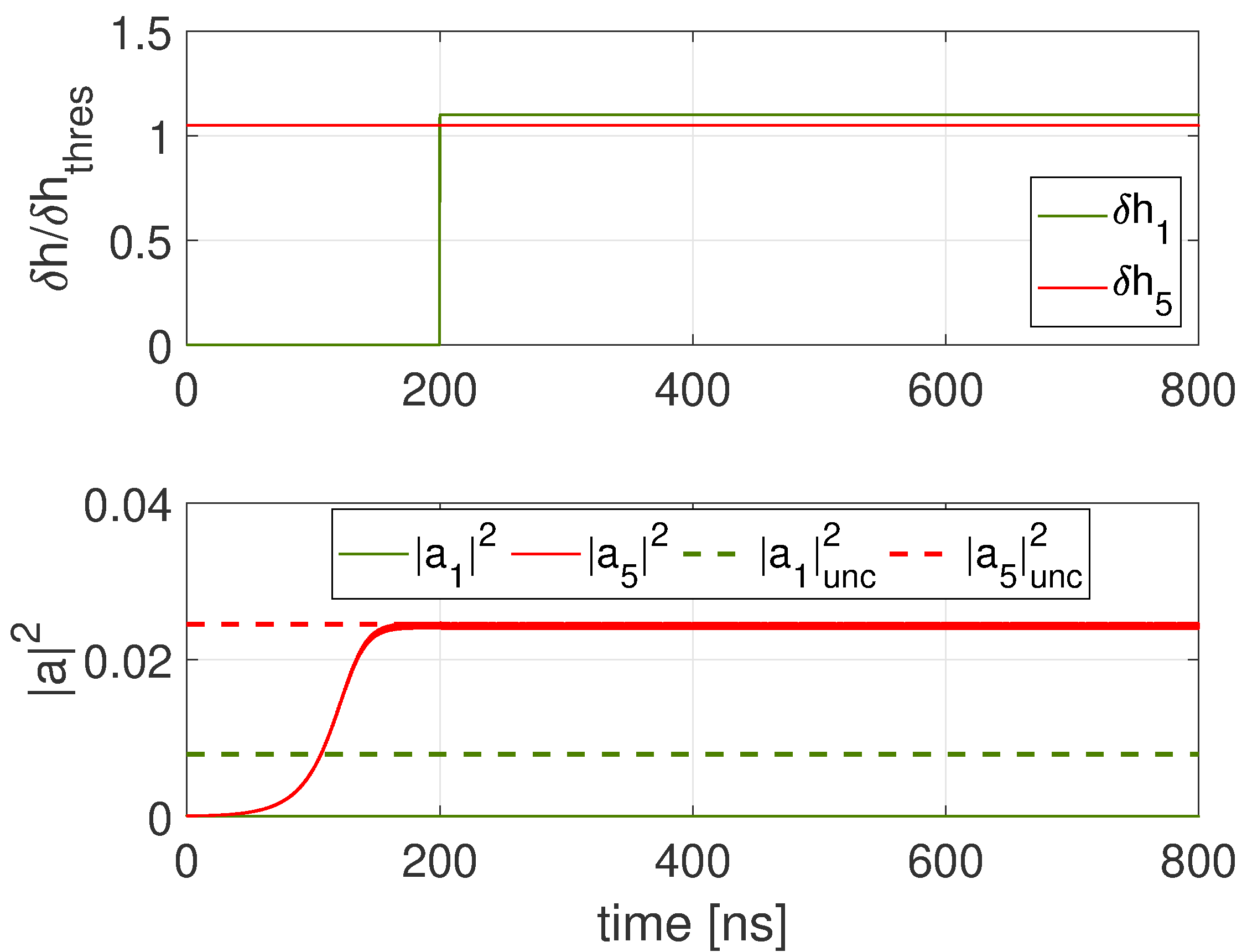}    
    \caption{Non-commutativity of interaction between parametrically-excited modes $h=1$ and $n=5$.  Left and right panels report comparisons between numerical simulation of full NMM \eqref{eq:NMM} with 10 modes, numerical simulation of the two-modes model \eqref{eq:nonlinear coupled parametric ah}-\eqref{eq:nonlinear coupled parametric an} and analytical formulas for steady-state amplitudes of uncoupled modes \eqref{eq:saturation amplitude mode h}. The ac field frequencies are $\omega_\mathrm{rf1}=2\omega_h, \omega_\mathrm{rf2}=2\omega_n$ while the field amplitudes are $\delta h_1=1.1\times\delta h_\mathrm{thres,1}, \delta h_2=1.05\times\delta h_\mathrm{thres,5}$. Upper (lower) panels refers to mode 1 excited before (after) mode 5.}
    \label{fig:YIG_disk_100_mode1_1p1_mode5_1p1}
\end{figure*}

In this respect, we first choose the modes $h=1$ and $n=5$ that have $N_{hh}=-0.8159, N_{hn}=-0.3897=N_{nh}, N_{nn}=0.1332$ (first scenario); results concerning the second scenario are reported in the supplementary material. The first step is computing the phase diagram of the interactions (see fig.\ref{fig:type1 phase diagram and hysteresis}), that is reported in the left panel of fig.\ref{fig:type1 phase diagram and hysteresis modes 1 and 5}. 
We can infer from the theory that the (dashed black) line $c_{++}\equiv u_n$ corresponding to the condition $|a_h|^2_\mathrm{coup}=0,|a_n|^2=|a_n|^2_\mathrm{unc}$ (see eq. \eqref{eq:saturation amplitude mode h type 1 linear}) and the (dashed red) line $c_{++}\equiv u_h$ referring to $|a_n|^2_\mathrm{coup}=0,|a_h|^2=|a_h|^2_\mathrm{unc}$ (see eq. \eqref{eq:saturation amplitude mode n type 1 linear}) have  slopes with opposite signs (see also table \ref{tab:type1 interactions}) given by $N_{hn}/N_{nn}<0$ and $N_{hn}/N_{hh}>0$, respectively. Finally, the critical lines intersect at the lower right corner $O_{++}$ of the brown region as established by eqs.\eqref{eq:c triangles vertices}.

The coefficient $L_{hn}=N_{hh}N_{nn}/(N_{hh}N_{nn}-N_{hn}^2)$ is positive, meaning that triangular regions $T_{\pm\pm}$ are located to the left of the lines $c_{\pm\pm}\equiv u_n$ and above the lines $c_{\pm\pm}\equiv u_h$ that are grouped in pairs with the same slopes $\omega_h N_{hn}/(\omega_n N_{nn})<0$ and $\omega_nN_{nh}/(\omega_h N_{hh})>0$, respectively. In the figure, for the sake of simplicity, we have only reported the lines $c_{++}\equiv u_n$, $c_{++}\equiv u_h$, $c_{-+}\equiv u_n$ (also labelled as 'no $u_n$'), $c_{+-}\equiv u_h$ ('no $u_h$').

Moreover, the stability condition $\text{sign}(N_{hh})\text{sign}(N_{nn})(N_{hh}N_{nn}-N_{hn}^2)>0$ means that $c_{+-},c_{-+}$ regimes are unstable whereas only $c_{++}$ P-modes are stable close to $O_{++}$ and critical (dashed red and black) lines $c_{++}\equiv u_h, c_{++}\equiv u_n$.

We first consider excitation of the two modes at low power above the threshold. 

\subsubsection{Hysteresis}

The middle and right panels in fig.\ref{fig:type1 phase diagram and hysteresis modes 1 and 5} 
refer to 'vertical' and 'horizontal' variations of excitation frequencies in a forward (solid lines) and backward (dashed lines) fashion. One can clearly see that the numerical integration of the averaged two-modes model \eqref{eq:averaged two nonlinear parametric Ah}-\eqref{eq:averaged two nonlinear parametric Phi_n} reproduces the qualitative behaviors outlined in the previous section and depicted in fig.\ref{fig:type1 phase diagram and hysteresis}. Specifically, the steady-state behavior of modes $h=1$ and $n=5$ triggered by 'vertical' excitation is hysteretic. The hysteresis is due to the different sequence of steady states, $u_h\rightarrow c \rightarrow u_n$ for forward and $u_h\rightarrow u_n \rightarrow c\rightarrow u_h$ for backward sweep, respectively. The different sequences are determined by the coexistence of regimes $u_h,u_n$ in the brown region combined with the reversed crossing order of dashed black $c_{++}\equiv u_n$ and dash-dotted red 'no $u_h$' lines  in forward and backward sweeps (see section \ref{sec:CW excitation}).
Moreover, the steady-state behavior corresponding to 'horizontal' excitation is also hysteretic, where hysteresis occurs due to the coexistence between $c_{++}$ and $u_n$ modes for small enough $\omega_\mathrm{rf1}$.

\subsubsection{Non-commutativity}

In order to assess the non-commutativity of steady-state regimes due to the different temporal sequence for the activation of excitation tones, we have performed time-domain simulation of the full NMM equation \eqref{eq:NMM} with 10 modes (modes $\bm\varphi_h, h=1\ldots,5$ and their conjugates $\bm\varphi_h^*$)
and compared the result with the solution of the reduced-order two-modes model \eqref{eq:nonlinear coupled parametric ah}-\eqref{eq:nonlinear coupled parametric an} and the analytical theory developed in section \ref{sec:mutual modes interaction}. The excitation parameters are the same as those used in fig.\ref{fig:type1 phase diagram and hysteresis modes 1 and 5}: $\omega_\mathrm{rf1}=2\omega_h,\delta h_1=1.1\times\delta h_\mathrm{thres,1}, \omega_\mathrm{rf2}=2\omega_n, \delta h_2=1.05\times\delta h_\mathrm{thres,5}$. The location of the point $(2\omega_h,2\omega_n)$ in the coexistence (brown) region of the phase diagram (left 
panel in fig.\ref{fig:type1 phase diagram and hysteresis modes 1 and 5}) suggests that, starting from the zero P-mode $z$, only one uncoupled mode, either $u_h$ or $u_n$ can be present at the time, depending on which mode is excited before the other (i.e. the first excited wins). This is in perfect agreement with results of time-domain simulations reported in fig.\ref{fig:YIG_disk_100_mode1_1p1_mode5_1p1}, which demonstrate that in such non-commutative regime, the steady-state amplitudes remain close to their 'uncoupled' levels.

\begin{figure*}[t]
    \centering
    \includegraphics[width=0.37\textwidth]{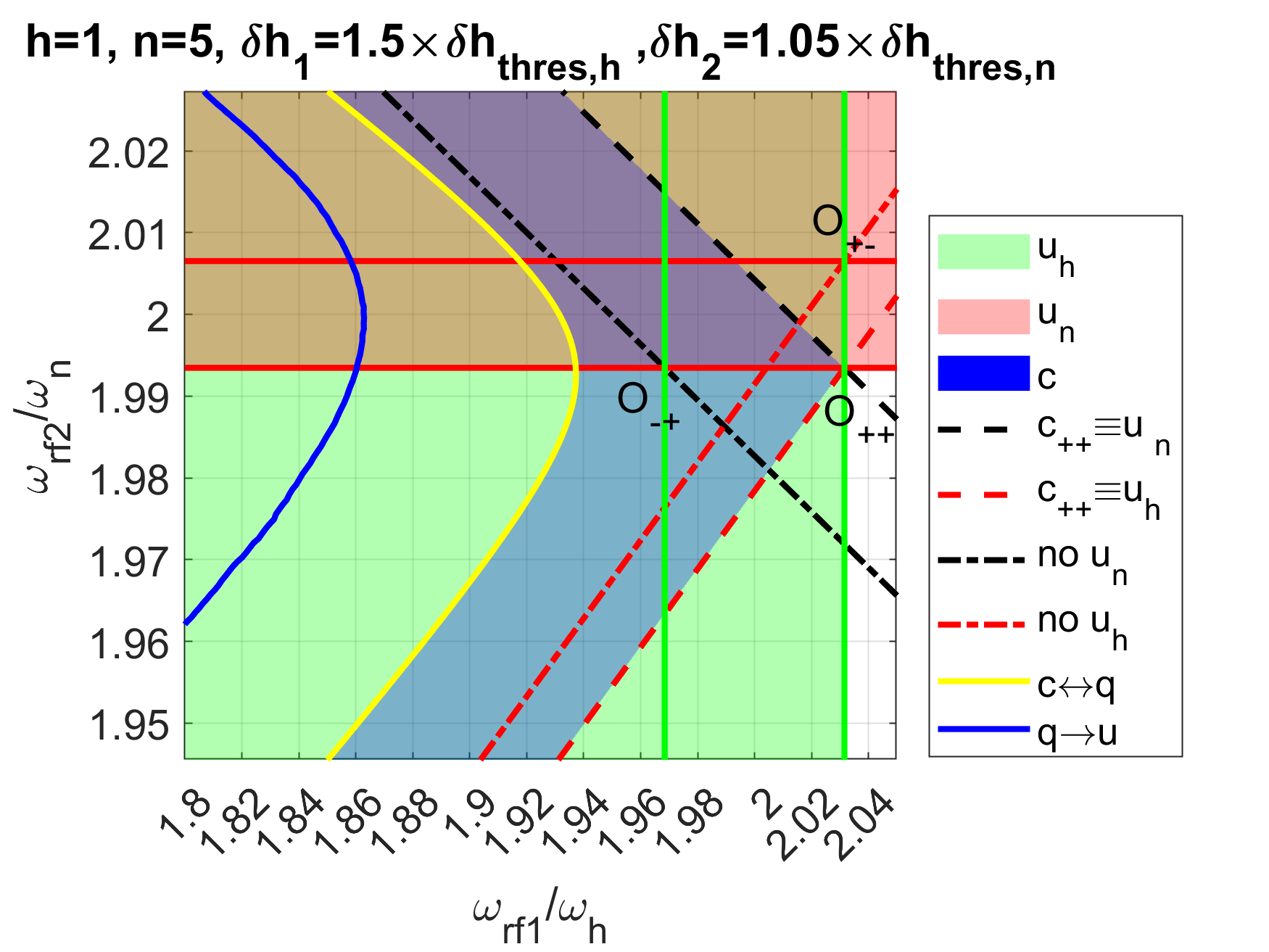} \includegraphics[width=0.3\textwidth]{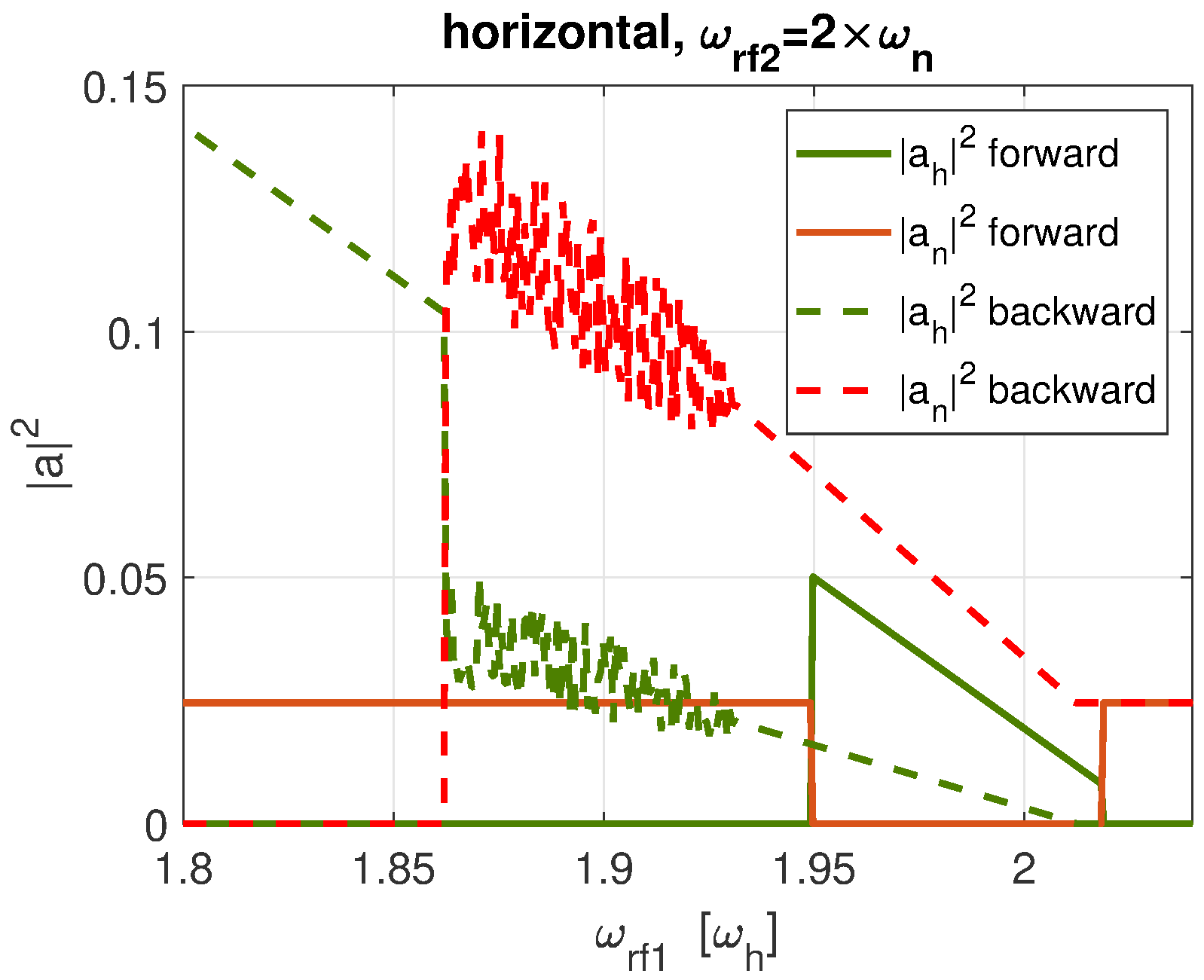} \\ \vspace{0.5cm}
    \includegraphics[width=0.34\textwidth]{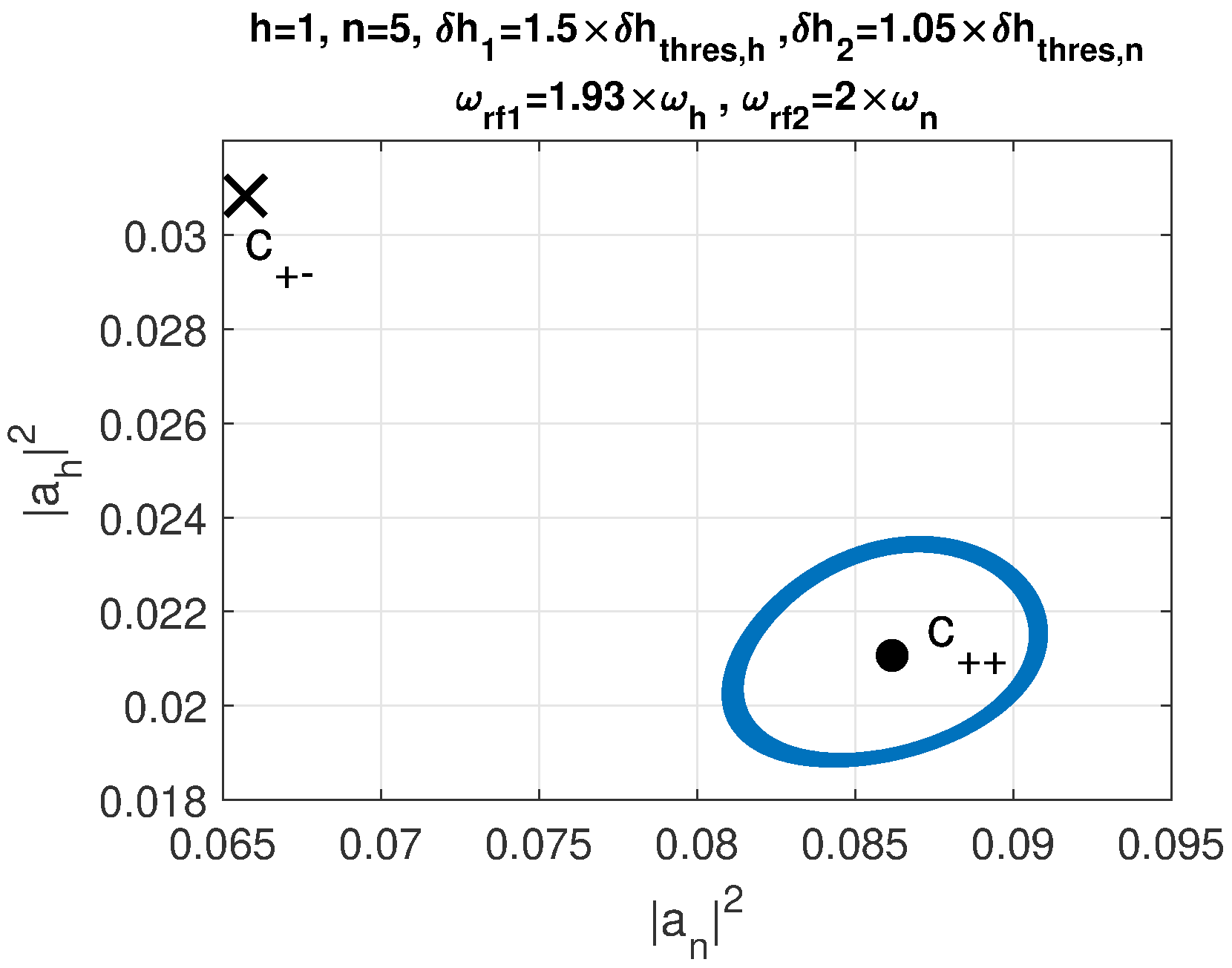} \includegraphics[width=0.34\textwidth]{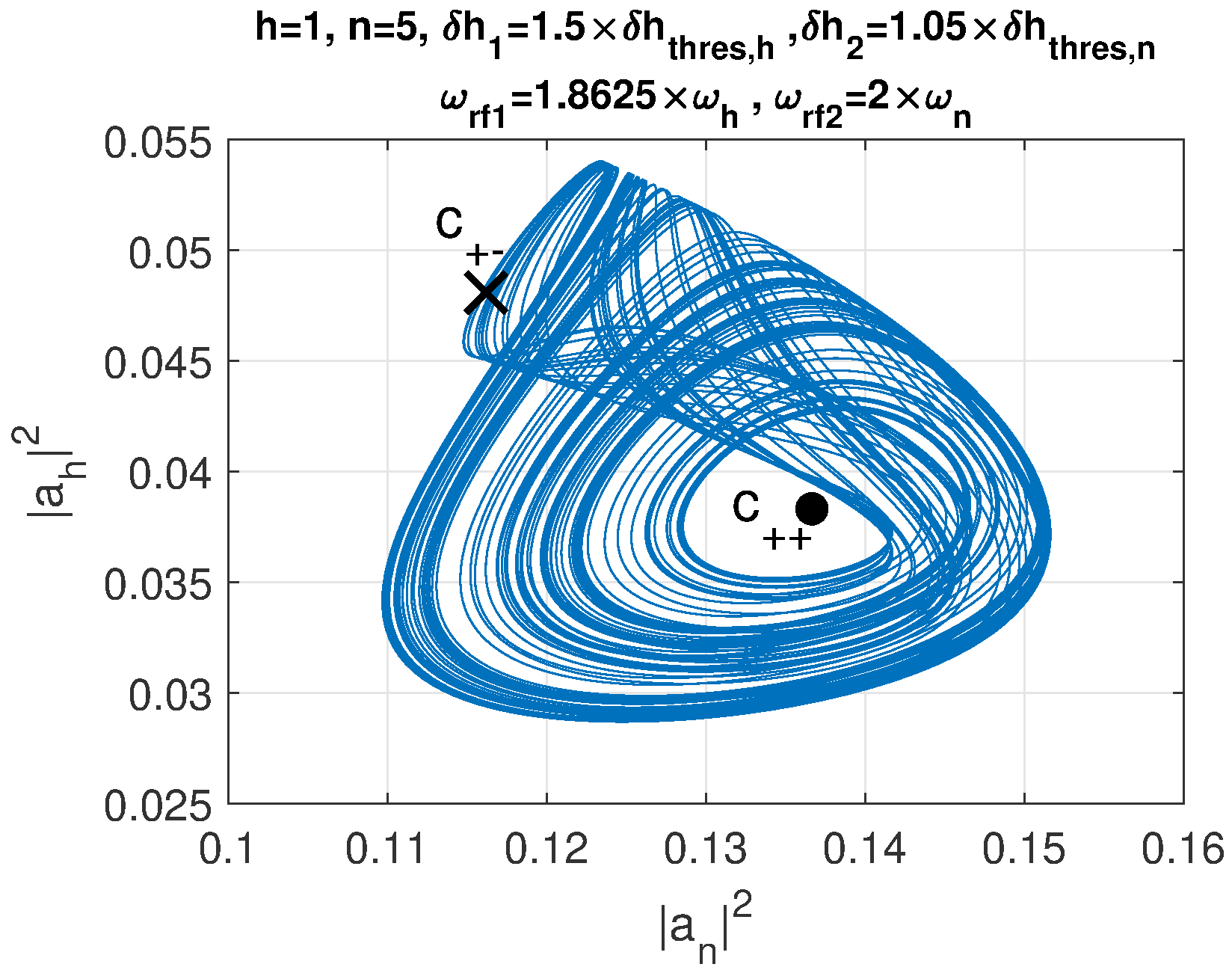}
    \caption{Phase diagram of interaction in $(\omega_\mathrm{rf1},\omega_\mathrm{rf2})$ plane for parametrically-excited modes $h=1$ and $n=5$ via parallel pumping with two-tone signal at fixed above-threshold amplitudes $\delta h_1=1.5\times\delta h_\mathrm{thres,h},\delta h_2=1.05\times\delta h_\mathrm{thres,n}$. NFS coefficients are $N_{hh}=-0.8159, N_{hn}=-0.3897=N_{nh}, N_{nn}=0.1332$. Upper left panel depicts different regions (green, red, blue) corresponding to the conditions \eqref{eq:constraint real amplitudes} 
    for which modes $u_h, u_n, c_{++}$ exist and are stable (the boundaries of the 'cross' region $C_h\cup C_n$ are marked with green and red solid lines, respectively). Stable Q-modes exist in the region enclosed between yellow $c\leftrightarrow q$ and blue $q\rightarrow u$ lines associated with Hopf and homoclinic bifurcation, respectively. Upper right panel reports steady-state response to forward/backward horizontal excitation at frequency $2\omega_n$ obtained by numerical integration of the averaged two-modes model \eqref{eq:averaged two nonlinear parametric Ah}-\eqref{eq:averaged two nonlinear parametric Phi_n}. Lower left and right panels report projection of the phase portrait of the averaged two-modes model in the plane $(|a_n|^2,|a_h|^2)$ for excitation conditions close to Hopf and homoclinic bifurcations, respectively, such that a Q-mode (solid blue) exist.}
    \label{fig:type1 phase diagram and hysteresis modes 1 and 5 large power}
\end{figure*}

Next we have compared the theory and time-domain simulations more specifically in the coupled regime. To this end, we have slightly increased the amplitude of the first tone to $\delta h_1=1.5\times\delta h_\mathrm{thres,1}$ while keeping the second tone at $\delta h_2=1.05\times\delta h_\mathrm{thres,5}$ with same frequencies $\omega_\mathrm{rf1}=2\omega_h, \omega_\mathrm{rf2}=2\omega_n$. The results are reported in fig.\ref{fig:type1 phase diagram and hysteresis modes 1 and 5 large power}. First, we observe that the upper-left panel of fig.\ref{fig:type1 phase diagram and hysteresis modes 1 and 5 large power} shows that the excitation $(2\omega_h,2\omega_n)$ lies in the coupled (blue) region of the phase diagram. This means that both coupled and uncoupled P-modes are stable (coexisting) and the actual dynamics will depend on the history of the excitation, leading to non-commutativity that also involves coupled regimes. This prediction is confirmed by time-domain simulations reported in figs.\ref{fig:YIG_disk_100_mode5_1p05_mode1_1p5}-\ref{fig:YIG_disk_100_mode5_1p05_mode1_1p5_coupled_non_comm} in the supplementary material.

\begin{figure*}[t]
    \centering
     CW horizontal sweeps\\ 
    \includegraphics[width=0.32\linewidth] {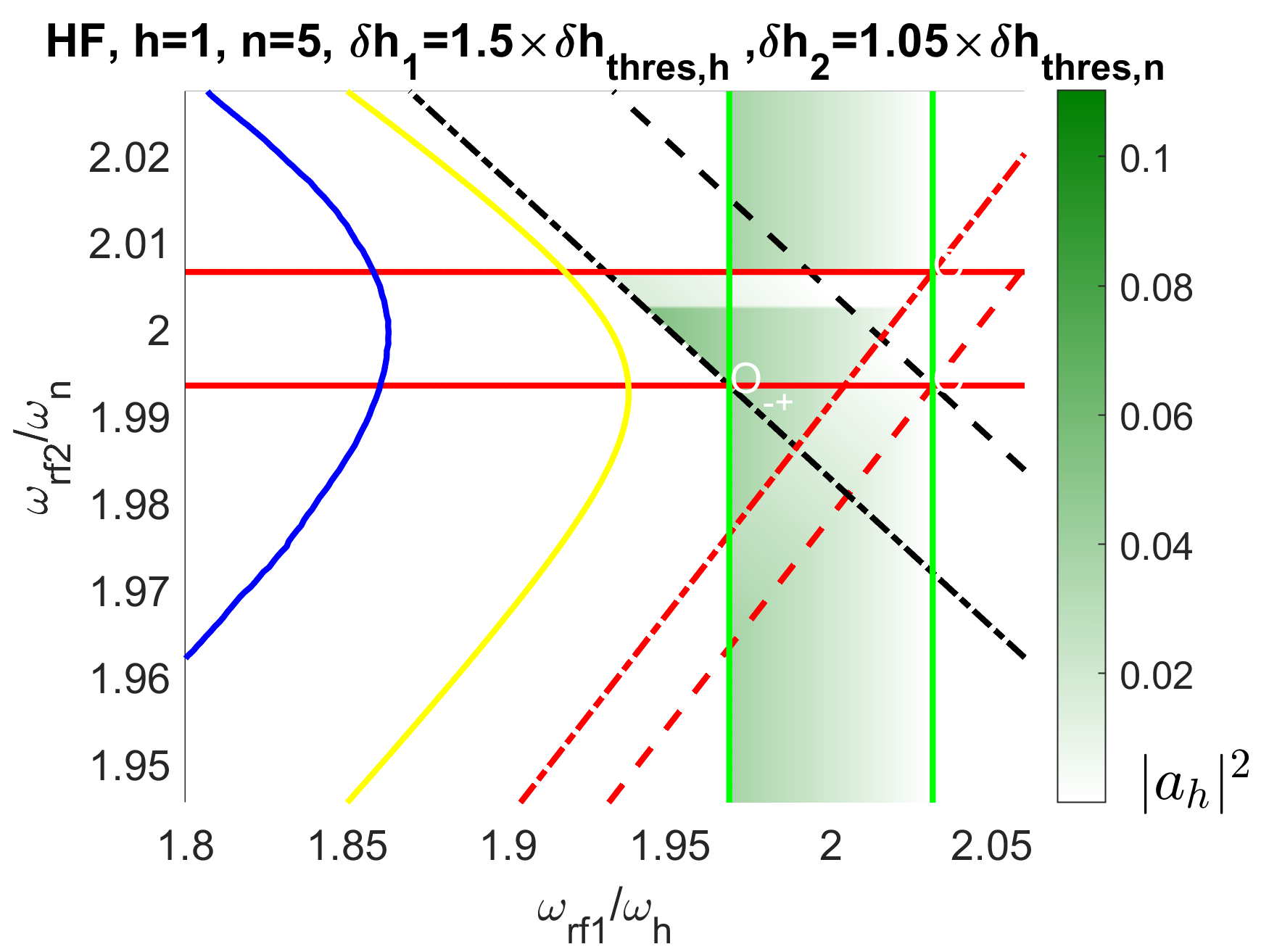}     \includegraphics[width=0.32\linewidth]{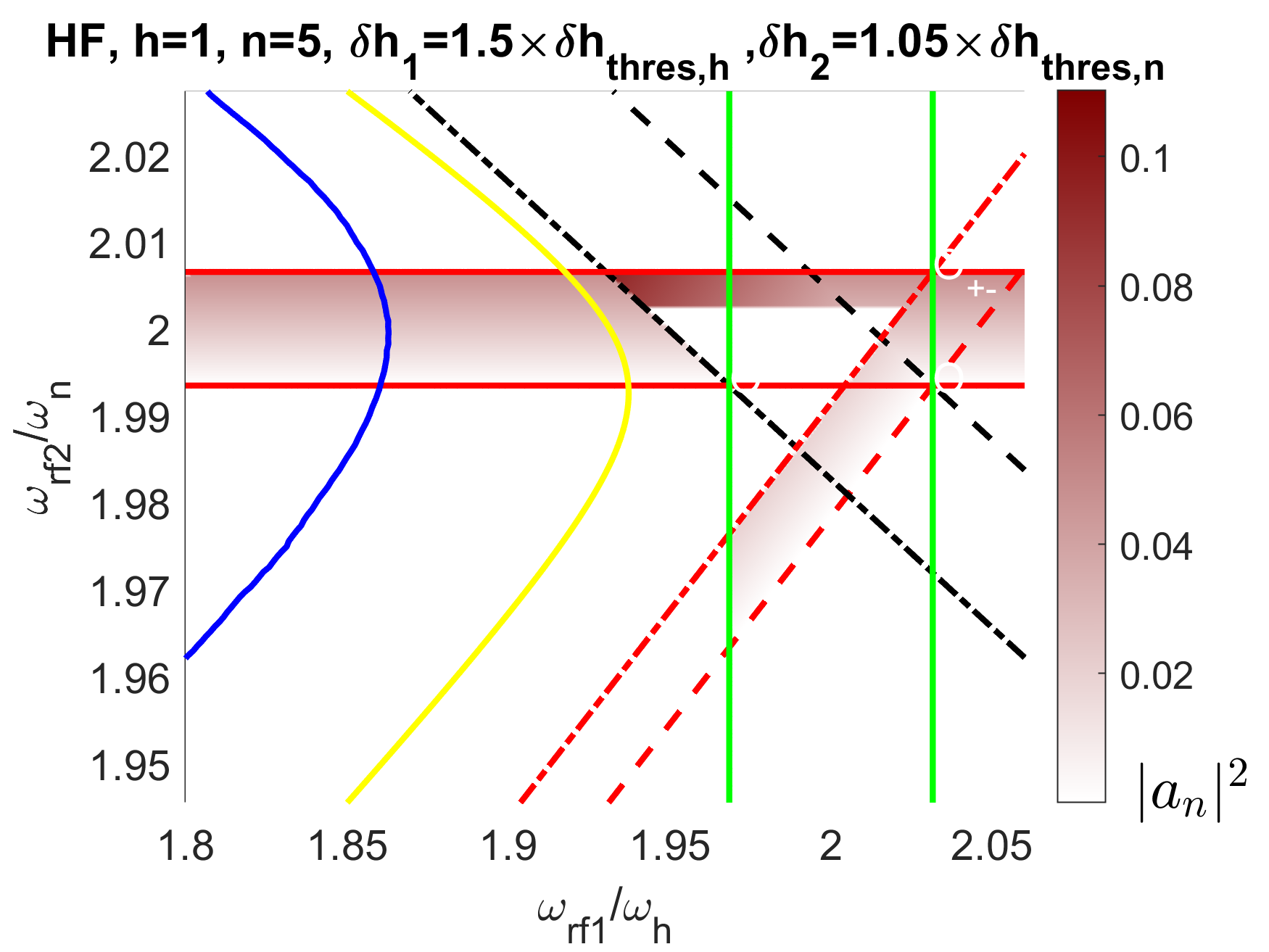}    \includegraphics[width=0.32\linewidth]{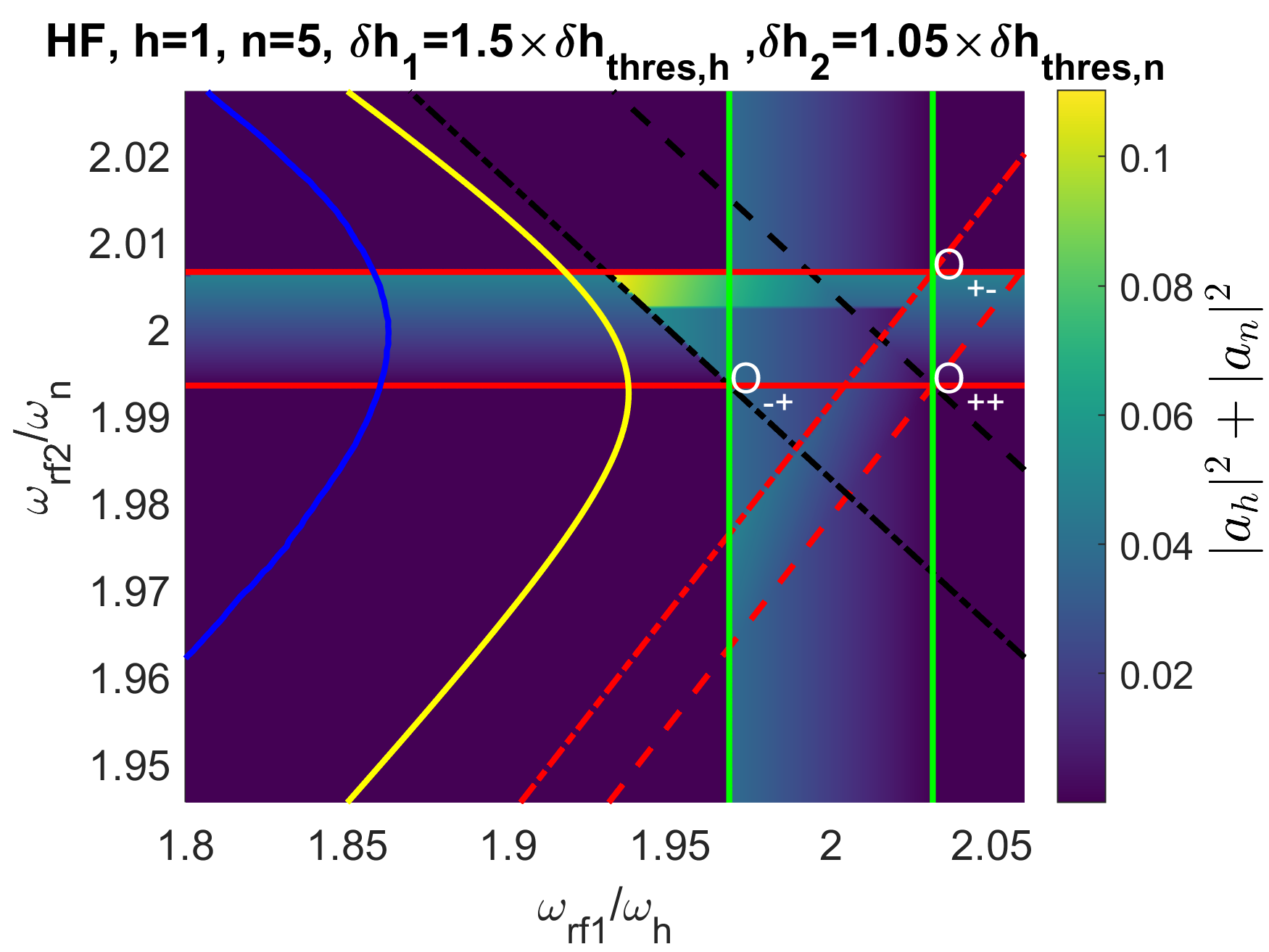} \\
    \includegraphics[width=0.32\linewidth]{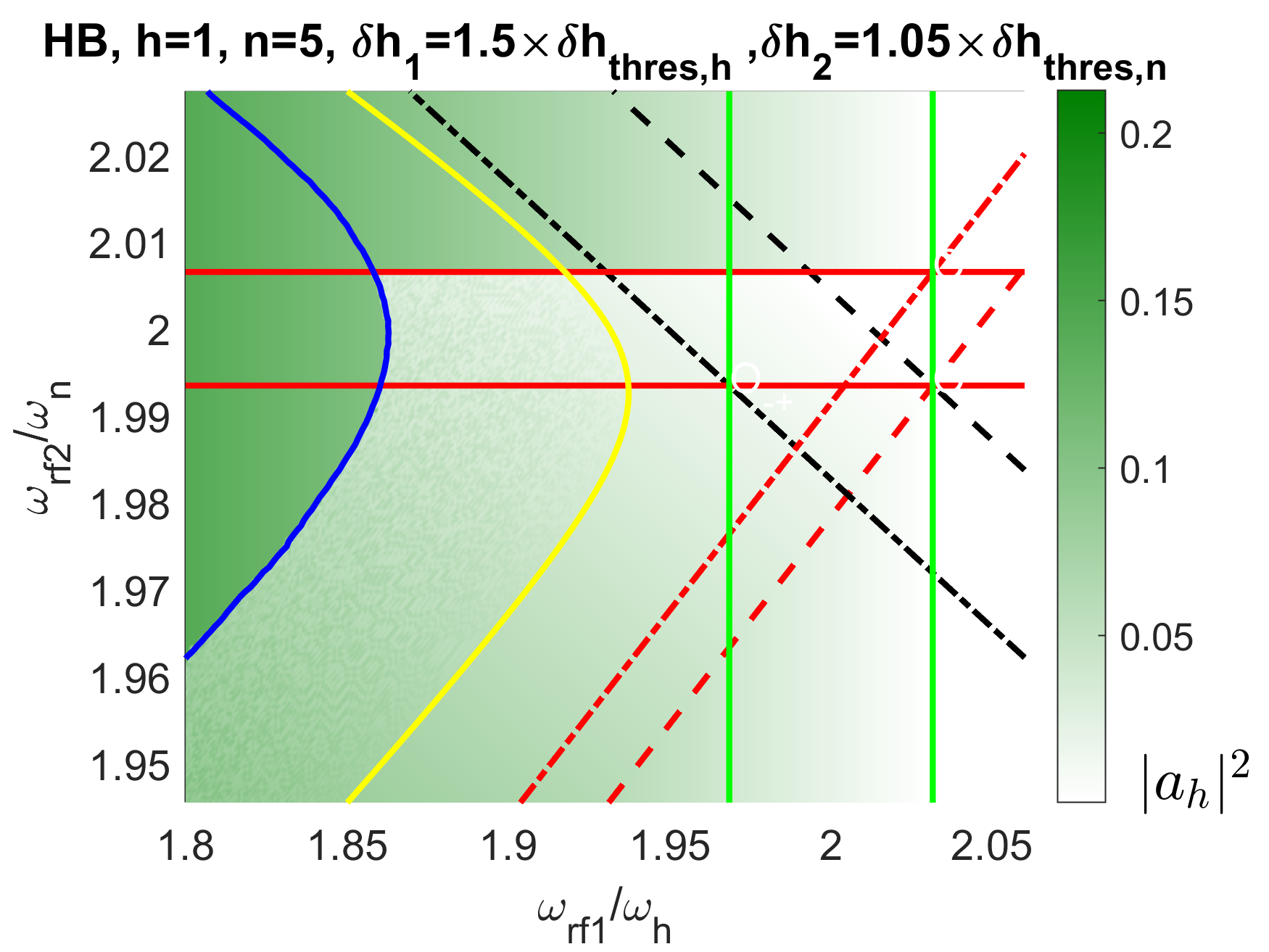}     \includegraphics[width=0.32\linewidth]{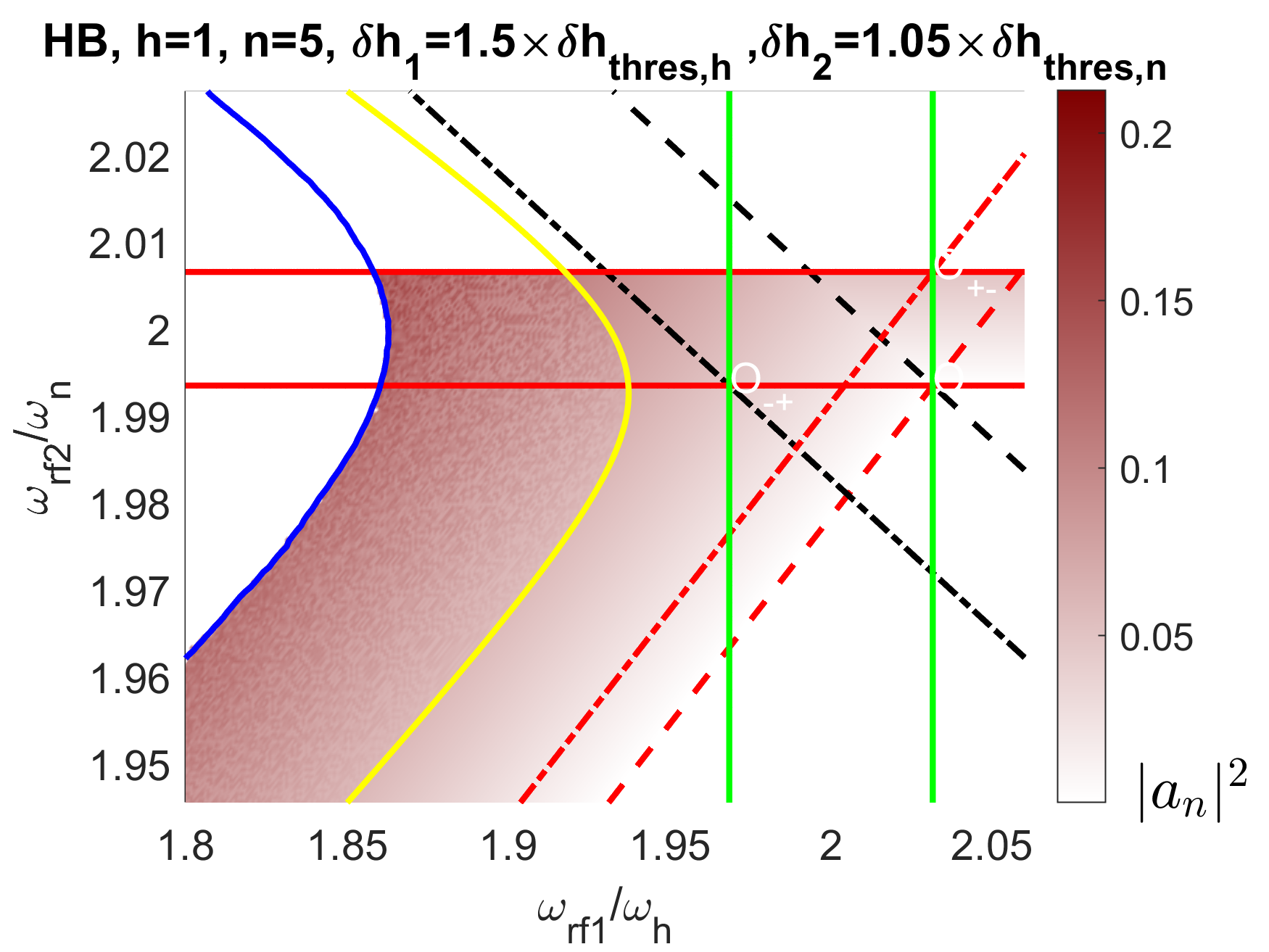}    \includegraphics[width=0.32\linewidth]{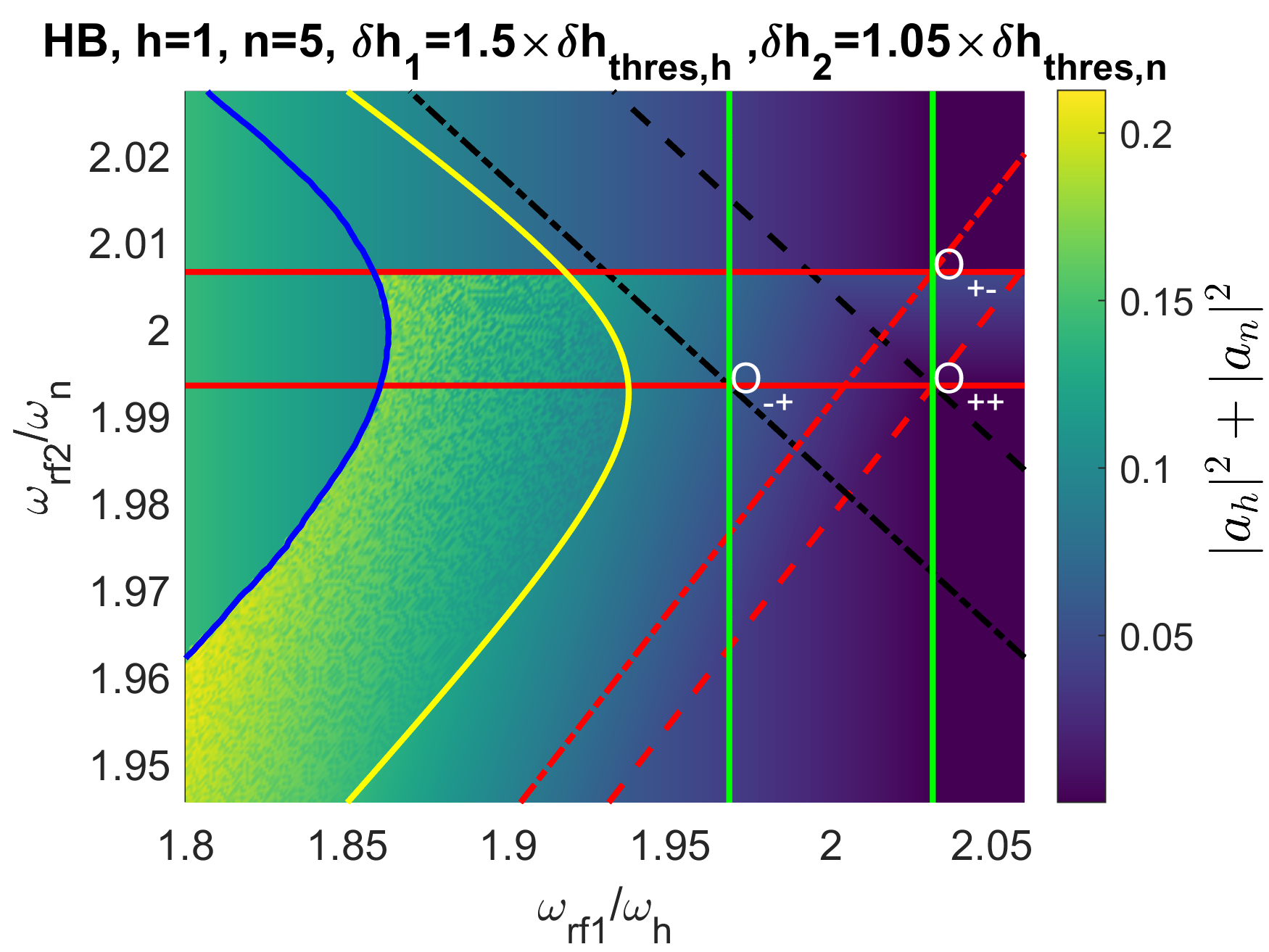} \\

    \caption{Continuous wave (CW) dynamics for parametrically-excited modes $h=1$ and $n=5$ via parallel pumping with two-tone signal at fixed above-threshold amplitudes $\delta h_1=1.5\times\delta h_\mathrm{thres,h},\delta h_2=1.05\times\delta h_\mathrm{thres,n}$. Each point in the colormap is obtained integrating the averaged two-modes model \eqref{eq:averaged two nonlinear parametric Ah}-\eqref{eq:averaged two nonlinear parametric Phi_n} for long enough time to approach the steady-state. The panels report the amplitude diagrams associated to steady-state for: (left) mode $h$, amplitude  $|a_h|^2$; (middle) mode $n$, amplitude $|a_n|^2$; (right) summed amplitudes $|a_h|^2+|a_n|^2$. Notation HF/HB in the panel title bar means Horizontal Forward/Backward frequency sweep, respectively.  Black dashed line corresponds to the condition $c_{++}\equiv u_n$ while red dashed line refers to $c_{++}\equiv u_h$. The  boundaries of regions $C_h$ and $C_n$ are marked with green and red solid lines, respectively.
    Black dash-dotted line refers to the condition $c_{-+}\equiv u_n$ ('no $u_n$'), while red dash-dotted line refers to the condition $c_{+-}\equiv u_h$ ('no $u_h$'). Solid yellow (Hopf bifurcation) line $c\leftrightarrow q$ refer to the boundary of region $T_{++}$ due to the onset of Q-modes, solid blue (homoclinic bifurcation) line $q\rightarrow u$ denotes the boundary of the region where Q-modes can exist.}
    \label{fig:CW array h1 n5 horizontal}
\end{figure*}

\subsubsection{Quasiperiodicity}

Next we discuss further interesting dynamics that appears under 'horizontal' excitations where the frequency $\omega_\mathrm{rf1}$ is moderately smaller than the natural frequency  $2\omega_h$. One can clearly see in the top-left panel of fig.\ref{fig:type1 phase diagram and hysteresis modes 1 and 5 large power} the presence of two additional lines (yellow $c\leftrightarrow q$ and  blue $q\rightarrow u$), which correspond to the occurrence of Hopf and homoclinic (saddle-connection) bifurcations, respectively, as discussed in section \ref{sec:existence and stability two-tone} (and also in appendix \ref{sec:stability multi P modes}). 
It is apparent that, in the region enclosed between yellow $c\leftrightarrow q$ and dark blue $q\rightarrow u$ lines, a limit cycle (Q-mode) exist. We remark that a limit cycle for the averaged two-modes dynamical eqs.\eqref{eq:averaged two nonlinear parametric Ah}-\eqref{eq:averaged two nonlinear parametric Phi_n} corresponds to a quasiperiodic time-domain signal for the two-modes dynamics \eqref{eq:nonlinear coupled parametric ah}-\eqref{eq:nonlinear coupled parametric an} that, for each mode, combines two frequencies. Specifically, for mode $h$ the Q-mode combines the applied tone frequency $\omega_\mathrm{rf1}/2$ with the (lower) frequency of the limit cycle, giving rise to low-frequency modulation. Analogous reasoning holds for mode $n$ that combines the applied tone frequency $\omega_\mathrm{rf2}/2$ with the (lower) frequency of the limit cycle. Figure \ref{fig:type1 phase diagram and hysteresis modes 1 and 5 large power} reports (top-right panel) the amplitude response related to horizontal excitation sweep at constant $\omega_\mathrm{rf2}=2\omega_n$. The forward frequency sweep (see solid green and red lines) exhibits suppression of regime $u_n$ in favor of $u_h$ when the (black dash-dotted) line 'no $u_n$' is crossed at $\omega_\mathrm{rf1}\approx 1.95\omega_h$. We notice that this is qualitatively different from what has been depicted in the lower right panel of fig.\ref{fig:type1 phase diagram and hysteresis}, but still consistent with the coexistence of regimes $c$ and $u_h$. Simulations performed at slightly larger $\omega_\mathrm{rf2}=2.005\omega_n$ (not reported for brevity) reproduce the behavior reported in fig.\ref{fig:type1 phase diagram and hysteresis}. 

As far as the backward frequency sweep (see dashed green and red lines) is concerned, one can clearly see that, in the range $(1.85\omega_h, 1.95\omega_h)$ it reveals a Q-mode arising from a coupled regime of type $c_{++}$. 
The phase portrait of the P-mode dynamics projected in the amplitude plane $(|a_n|^2,|a_h|^2)$ shows (bottom-left panel of fig.\ref{fig:type1 phase diagram and hysteresis modes 1 and 5 large power}) the Q-mode at the onset, i.e. immediately after the supercritical Hopf bifurcation when the excitation crosses the (yellow) Hopf bifurcation line.  Computation of the eigenvalues of the Jacobian matrix \eqref{eq: 4x4 nonzero P modes Jacobian} of the averaged two-modes model \eqref{eq:averaged two nonlinear parametric Ah}-\eqref{eq:averaged two nonlinear parametric Phi_n} yields the frequency of the limit cycle $\approx 0.0151$ (corresponding to about 75.5 MHz) that produces low-frequency modulation in the resulting time-domain signal for the two-modes model \eqref{eq:nonlinear coupled parametric ah}-\eqref{eq:nonlinear coupled parametric an}. By further decreasing the frequency, the Q-mode enlarges until the homoclinic bifurcation (blue) line is crossed. At that point, the limit cycle intersects the invariant manifold pertaining to the saddle P-mode $c_{+-}$ (bottom right panel of fig.\ref{fig:type1 phase diagram and hysteresis modes 1 and 5 large power}) and annihilates stabilizing one of the uncoupled modes $u_h,u_n$. In the specific case, the system relaxes towards the $u_n$ regime.

\begin{figure*}
    \centering
     PWM sequence $hn$\\ 
    \includegraphics[width=0.32\linewidth] {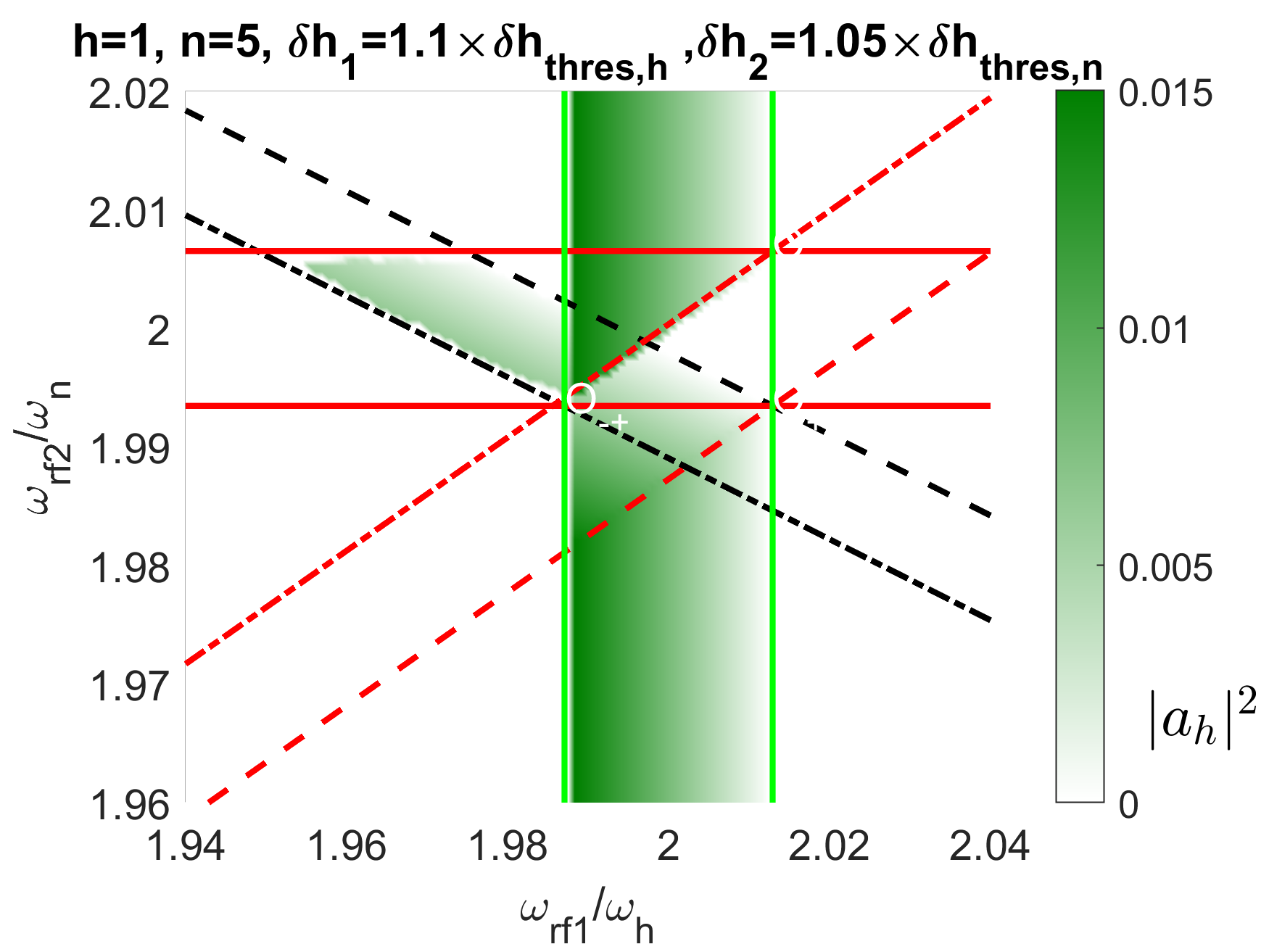}     \includegraphics[width=0.32\linewidth]{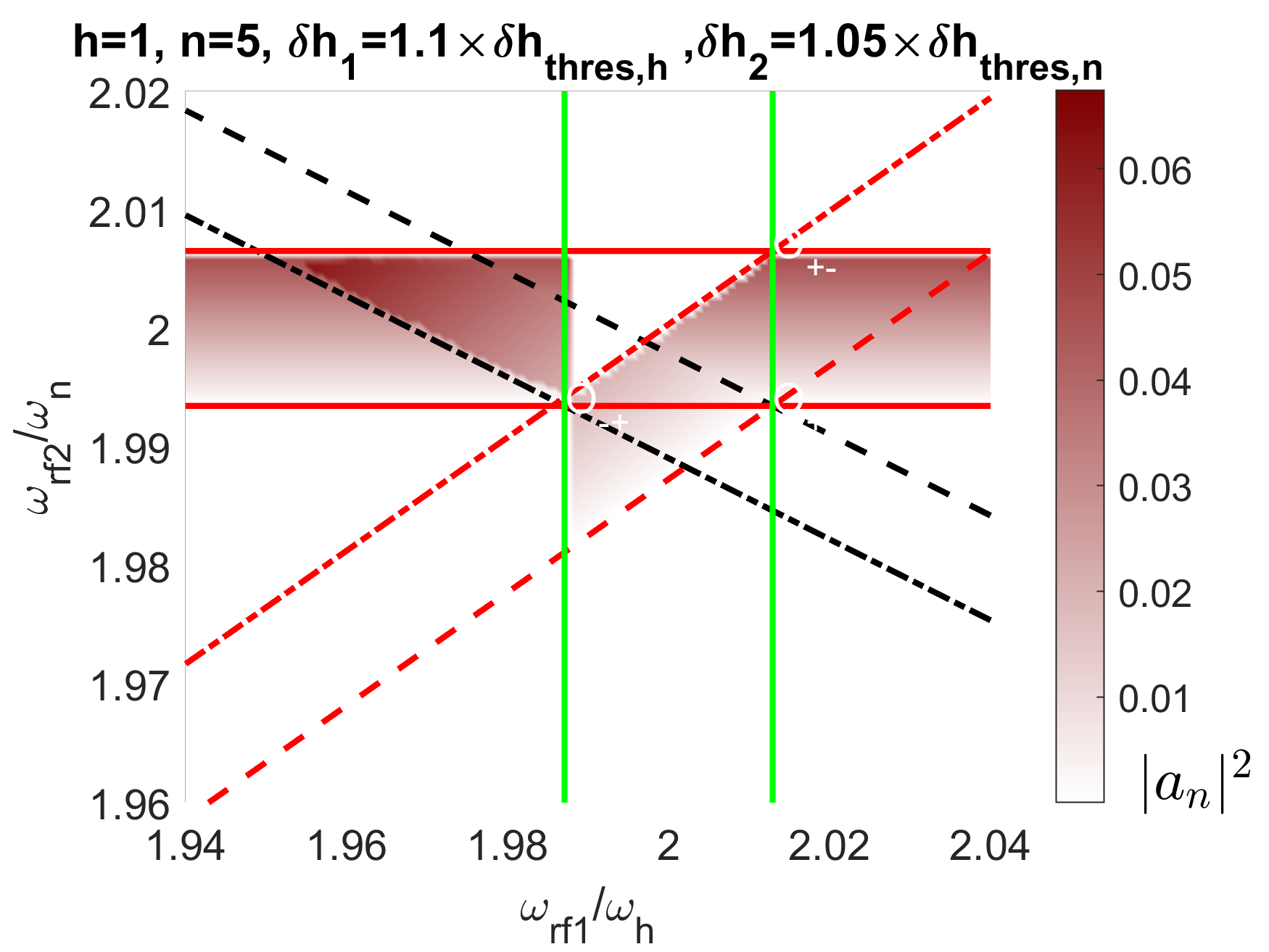}    \includegraphics[width=0.32\linewidth]{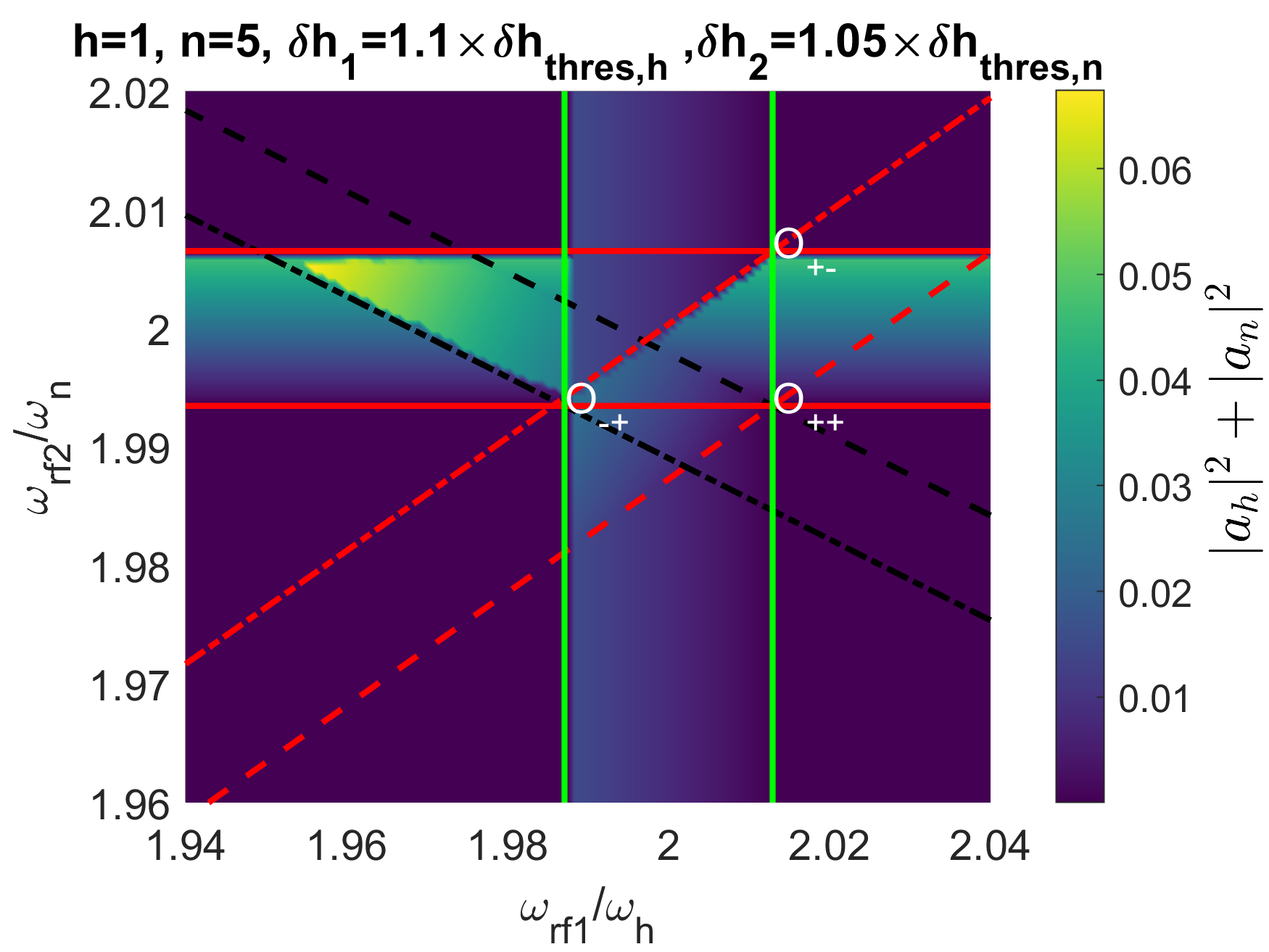} \\
   PWM sequence $nh$\\ 
    \includegraphics[width=0.32\linewidth] {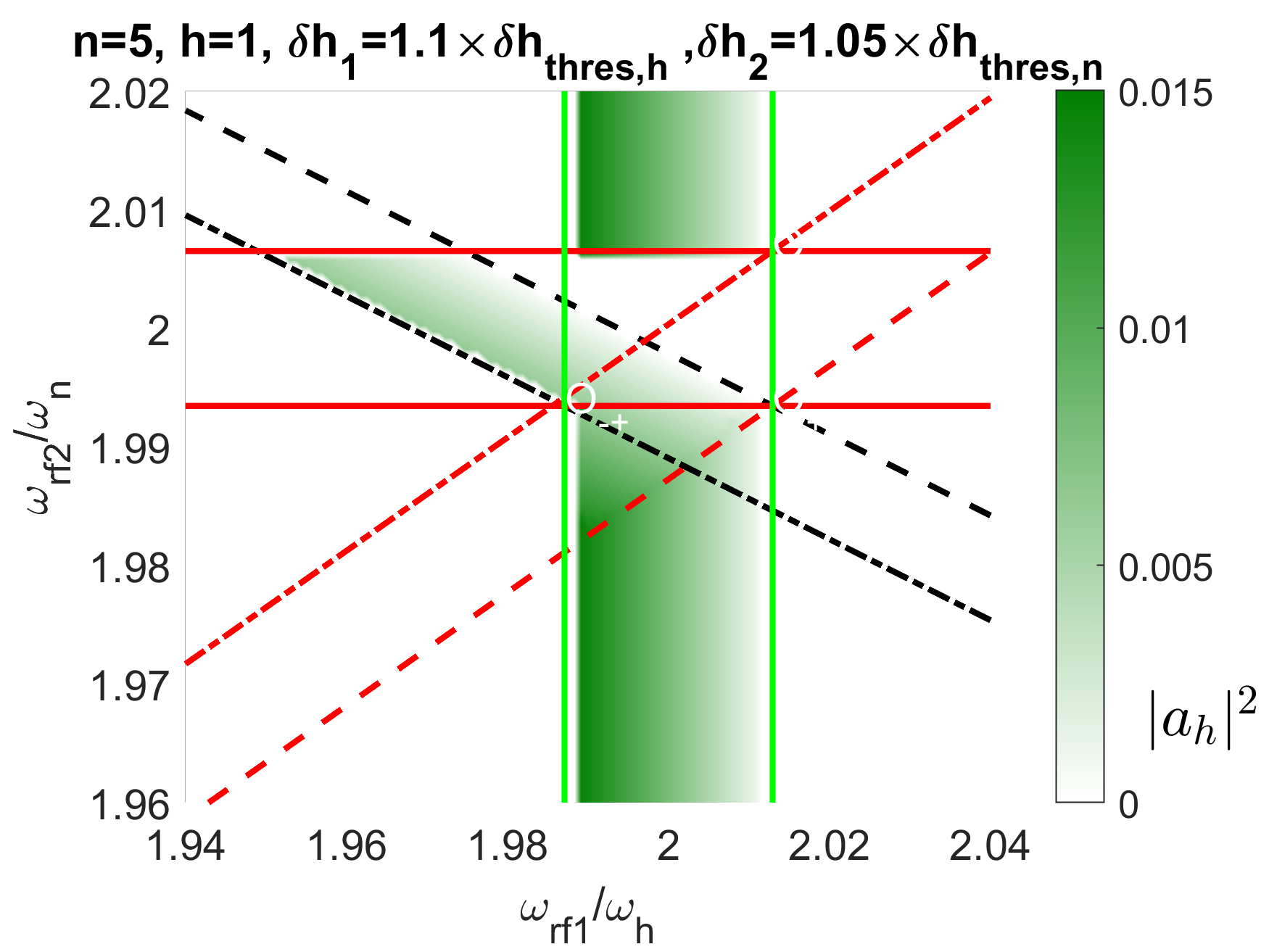}     \includegraphics[width=0.32\linewidth]{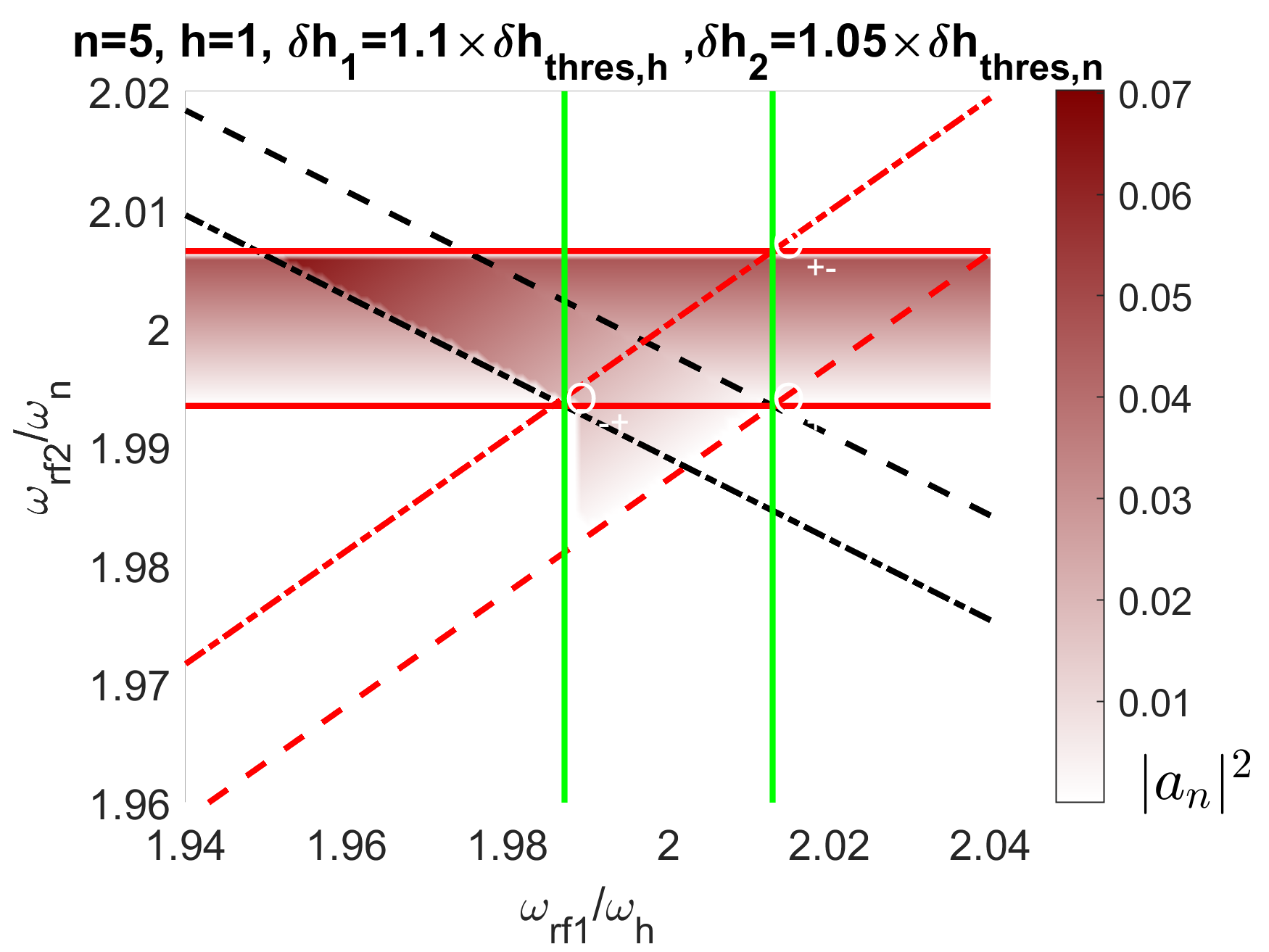}    \includegraphics[width=0.32\linewidth]{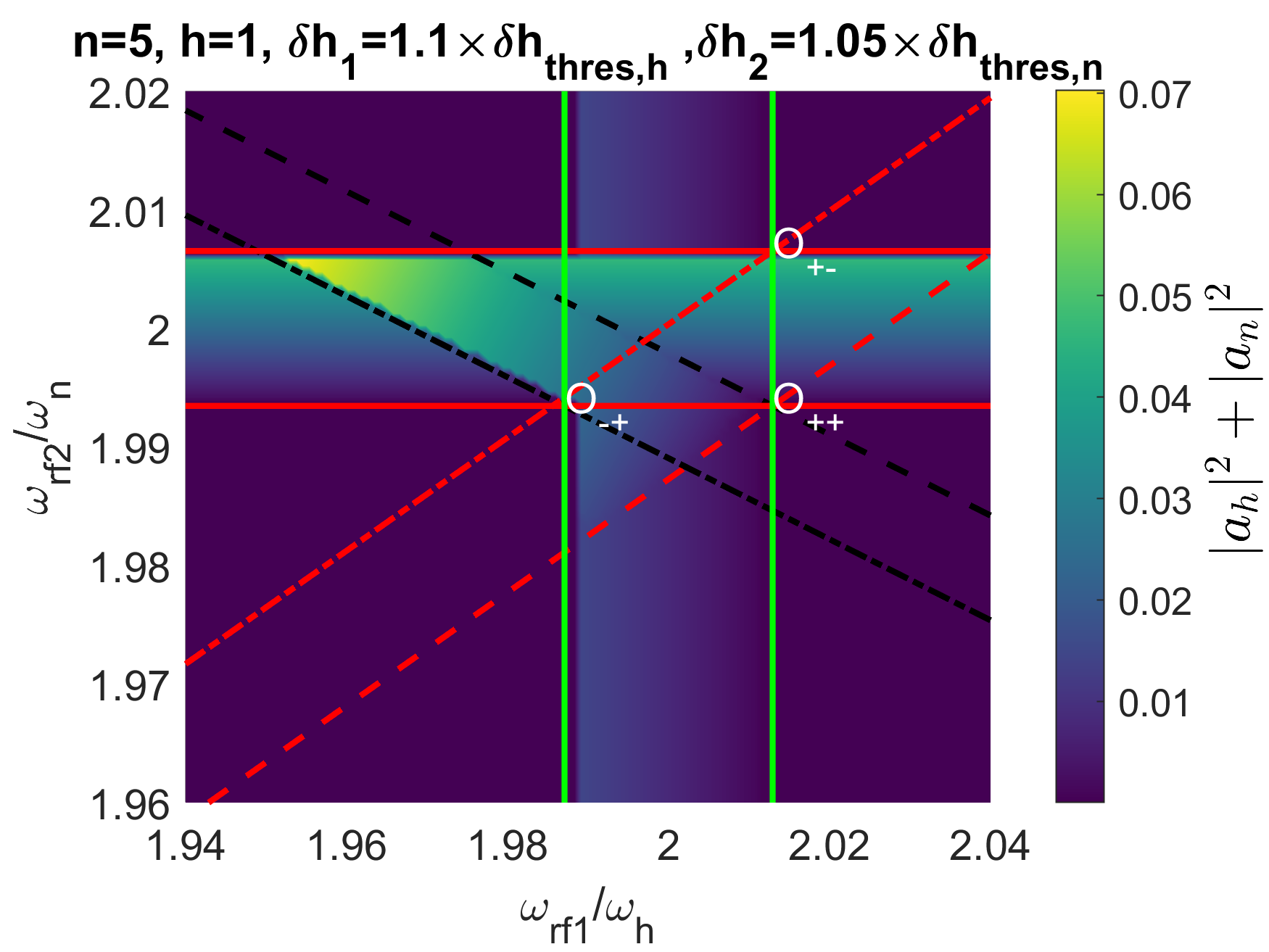} 

    \caption{Pulse-width Modulated (PWM) dynamics for parametrically-excited modes $h=1$ and $n=5$ via parallel pumping with two-tone signal at fixed above-threshold amplitudes $\delta h_1=1.1\times\delta h_\mathrm{thres,h},\delta h_2=1.05\times\delta h_\mathrm{thres,n}$. Each point in the colormap is obtained integrating the averaged two-modes model \eqref{eq:averaged two nonlinear parametric Ah}-\eqref{eq:averaged two nonlinear parametric Phi_n} for long enough time to approach the steady-state. The panels report the amplitude diagrams associated to steady-state for: (left) mode $h$, amplitude  $|a_h|^2$; (middle) mode $n$, amplitude $|a_n|^2$; (right) summed amplitudes $|a_h|^2+|a_n|^2$. Top row: $hn$ PWM excitation tone sequence; bottom row: $nh$ PWM excitation tone sequence. Black dashed line corresponds to the condition $c_{++}\equiv u_n$ while red dashed line refers to $c_{++}\equiv u_h$. The  boundaries of regions $C_h$ and $C_n$ are marked with green and red solid lines, respectively.
    Black dash-dotted line refers to the condition $c_{-+}\equiv u_n$ ('no $u_n$'), while red dash-dotted line refers to the condition $c_{+-}\equiv u_h$ ('no $u_h$'). The phase diagram in fig.\ref{fig:type1 phase diagram and hysteresis modes 1 and 5} is instrumental for the interpretation of the amplitude patterns.}
    \label{fig:PWM array h1 n5}
\end{figure*}

\subsubsection{Numerical calculation of phase diagrams in CW excitation}

Finally, we have performed numerical simulations of forward/backward horizontal and vertical frequency sweeps by time-integrating the averaged two-modes model \eqref{eq:averaged two nonlinear parametric Ah}-\eqref{eq:averaged two nonlinear parametric Phi_n}. The integration has been carried out for a time long enough to let the system go to the steady state. The final amplitudes $|a_h|^2,|a_n|^2$ have been used to compute colored amplitude maps (i.e. physically corresponding to modes' power maps) in the control plane $(\omega_\mathrm{rf1},\omega_\mathrm{rf2})$ that can be directly compared with experiments. The results are reported in figure \ref{fig:CW array h1 n5 horizontal} and can be interpreted with the help of the phase diagram discussed in figures \ref{fig:type1 phase diagram and hysteresis} and \ref{fig:type1 phase diagram and hysteresis modes 1 and 5}. In this respect, greenish (resp. reddish) color maps in the left (resp. middle) panels of fig.\ref{fig:CW array h1 n5 horizontal} refer to $|a_h|^2$ (resp. $|a_n|^2$), whereas the color map in the right panel refers to the summed amplitudes $|a_h|^2+|a_n|^2$.

The forward (resp. backward) horizontal (denoted as HF/HB in figure panels) sweep has been performed by exciting both modes at constant $\omega_\mathrm{rf2}$ starting from sufficiently small (resp. large) frequency $\omega_\mathrm{rf1}$ and then linearly increasing (resp. decreasing) it in a CW fashion.  One can clearly see that horizontal forward frequency sweep does not involve Q-modes owing to the fact that coupled regimes $c_{++}$ are only excited at the right of the (black dash-dotted) line $c_{-+}\equiv u_n$ ('no $u_n$') that bounds the region where $u_n$ is not allowed. We recall that the (dashed black) line $c_{++}\equiv u_n$ indicates continuity between coupled $c_{++}$ and uncoupled $u_n$ regimes, giving rise to smooth transitions in the amplitude map. We also point out that, depending on the constant value of $\omega_\mathrm{rf2}$, different regimes are excited when the (black dash-dotted) line $c_{-+}\equiv u_n$ ('no $u_n$') is crossed. This occurs due to the coexistence between $c_{++}$ and $u_h$ P-modes in that region of the phase diagram (see fig.\ref{fig:type1 phase diagram and hysteresis}), given that $u_n$ is unstable within the black lines according to the rule \ref{item:S7} reported in table \ref{tab:type1 interactions}. As mentioned in section \ref{sec:CW excitation}, in this situation, to assess which of the two regimes is actually reached, one has to necessarily consider the transient dynamical process that brings the system in the final steady-state, since no continuity condition can be exploited. 

The result is rather different when the horizontal backward sweep is considered. One can clearly recognize the onset and the subsequent annihilation of a Q-mode in correspondence of (yellow, Hopf bifurcation) line $c\leftrightarrow q$ and (blue, homoclinic bifuraction) $q\rightarrow u$ line. The Q-mode originates from coupled $c_{++}$ P-modes excited within the triangular region $T_{++}$ with vertex $O_{++}$. At the left of the (blue) homoclinic bifurcation line $q\rightarrow u$ only the P-mode $u_h$ is stable. We observe that the dashed red $c_{++}\equiv u_h$ (resp. dashed black $c_{++}\equiv u_n$) line implies the continuity between coupled $c_{++}$ and uncoupled $u_h$ (resp. $u_n$) P-modes, giving rise to smooth transitions in the amplitude map. 

The results concerning vertical frequency sweeps are reported in section \ref{sec:further numerical results h1n5} of the supplementary material and can be discussed by similar reasoning as that made for the horizontal sweeps.

\subsubsection{Numerical calculation of phase diagrams in PWM excitation}\label{sec:num PWM}

We have performed also numerical simulations of mode pairs dynamics driven by PWM excitation. The system has been excited with two sequences, denoted as $hn$ and $nh$, which refer to the order of activation of each tone of the driving signal. The results are reported in figure \ref{fig:PWM array h1 n5} and can be interpreted with the help of the diagrams reported in figures \ref{fig:type1 phase diagram and hysteresis PWM} and \ref{fig:type1 phase diagram and hysteresis modes 1 and 5}. 

We conclude this analysis observing that the specific color patterns in the amplitude diagrams can be completely explained using the theory developed in section \ref{sec:mutual modes interaction}. 
In order to explore the range of possibilities offered by the choice of different parameters, we have also analyzed the case where the NFS coefficients $N_{hh},N_{nn},N_{hn}=N_{nh}$ lead to the second scenario. The results are reported in section \ref{sec:further numerical results h1n3} of the supplementary material. 

{Finally, as mentioned before, we point out that the developed theory has been effectively applied to quantitatively explain recent experiments on the parametric excitation of spin-wave modes in a 1$\mu$m YIG disk using two-tone PWM spectroscopy\cite{soares_knet_flagship_paper_2025}. Specifically, the steady-state amplitude diagrams such as those depicted in Fig.\ref{fig:PWM array h1 n5} have been successfully compared with the measured maps of total steady-state spin-wave intensity.} 

{Remarkably, several other interesting behaviors predicted by the theory, such as hysteresis and quasiperiodicity under continuous-wave (CW) excitation, remain still unexplored and pave the way to future experiments probing highly nonlinear dynamical phenomena.}

\section{Conclusions}

In this work, we have developed a comprehensive analytical framework to describe the nonlinear dynamics of parametrically-excited spin-wave modes in confined ferromagnetic systems under parallel pumping. Starting from the micromagnetic dynamical equations, we employed a Normal Modes Model (NMM) to derive explicit expressions for parametric instability thresholds and for the post-instability steady-state amplitudes of the excited modes. These results were obtained without relying on restrictive approximations, thereby ensuring wide applicability to various system geometries and excitation conditions.

Our analysis was first focused on the single-mode excitation regime, providing a quantitative understanding of amplitude saturation following the onset of parametric instability. We then extended the theory to the case of two-tone excitation ({both situations of continuous-wave (CW) and pulse width modulated (PWM) excitations have been considered as in realistic experiments}), where two resonant modes are simultaneously driven into instability. In this coupled-mode regime, we derived analytical expressions that capture the rich and complex behavior emerging from nonlinear mode interactions. The theory predicts non-reciprocal coupling effects, where the influence of one mode on another is not symmetric, as well as the emergence of hysteresis, multi-stability, quasiperiodic dynamics, and non-commutative responses depending on the temporal protocol of external signal activation. Despite such a large variety of dynamical behaviors, we have demonstrated that they are completely governed by the self- and mutual NFS coefficients (three parameters) that can be computed from the knowledge of the modes spatial profiles.

The analytical results were validated through numerical micromagnetic simulations performed on a nanoscale YIG disk, demonstrating excellent agreement with the predicted instability thresholds, amplitude dynamics, and phase diagrams. These diagrams effectively map the variety of steady-state oscillation regimes and provide a powerful tool for interpreting and designing experiments involving parametrically driven magnetization dynamics. 

{Indeed, the developed theory was successfully used for quantitative explanation of recent experiments on the parametric excitation of spin-wave modes in a 1$\mu$m YIG disk via two-tone PWM spectroscopy. Future experimental investigations could be designed to explore the whole variety of predicted dynamical behaviors such as, for instance, hysteresis in mode coupling and low-frequency modulation produced by quasiperiodicity under CW excitation.}

Overall, the theoretical approach introduced here not only offers a predictive framework for understanding nonlinear spin-wave interactions in confined systems but also lays the groundwork for future studies aimed at exploiting such dynamics in magnonic devices, signal processing applications, and energy-efficient spintronic technologies.

\begin{acknowledgements}
This work is supported by the Horizon2020 Research
Framework Programme of the European Commission
under Grant No. 899646 (k-NET).
\end{acknowledgements}


\appendix
\section{Derivation of the detuning cone}\label{sec:derivation of detuning cone}

We rewrite eq.\eqref{eq:linear parametric ah} in a more compact notation:
\begin{equation}
    \dot{a}=j\omega_0(1+f h \sin(\omega t))a+j\omega_0 g h \sin(\omega t)a^* -\lambda a \quad, \label{eq:linear parametric a 0}
\end{equation}
where $a= a_h,\omega_0= \omega_h,f=||\bm\varphi_h||^2,h=-\delta h,\omega=\omega_\mathrm{rf}, g=(\bm\varphi_h,\bm\varphi_h^*),\lambda=\lambda_h$.
Let us introduce the polar coordinates $A$ and $\Phi$ such that:
\begin{equation}
    a(t)=A(t)e^{j\Phi(t)}\quad\Rightarrow\quad \dot{a}=\dot{A}e^{j\Phi}+j\dot{\Phi}Ae^{j\Phi} \quad.
\end{equation}
By substituting in eq.\eqref{eq:linear parametric a 0}, one has:
\begin{equation}
    \dot{A}e^{j\Phi}+j\dot{\Phi}Ae^{j\Phi}=j\omega_0 F Ae^{j\Phi}+j\omega_0 G Ae^{-j\Phi} -\lambda Ae^{j\Phi}
\end{equation}
where $F(t)=(1+f h \sin(\omega t)), G(t)=g h \sin(\omega t)$. Now let us divide by $e^{j\Phi}$ and decompose the result into real and imaginary part. The result is:
\begin{align}
    \dot{A}&=\omega_0 G A\sin{2\Phi}-\lambda A \quad, \label{eq:linear parametric A}\\
    A\dot{\Phi}&=\omega_0 F A+\omega_0 G A\cos(2\Phi) \quad. \label{eq:linear parametric Phi}
\end{align}
Assuming nonzero amplitude $A\neq 0$ and using expressions for $F(t),G(t)$, one has:
\begin{align}
    \dot{A}&=\omega_0 g h \sin(\omega t) A\sin{2\Phi}-\lambda A \quad, \label{eq:linear parametric A 2}\\
    \dot{\Phi}&=\omega_0 (1+f h \sin(\omega t))+ \omega_0 g h \sin(\omega t) \cos(2\Phi) \quad. \label{eq:linear parametric Phi 2}
\end{align}
We now assume external forcing at double frequency $\omega\sim2\omega_0$:
\begin{equation}
    \omega=2\omega_0+\epsilon \quad.
\end{equation}
Equation \eqref{eq:linear parametric A 2}-\eqref{eq:linear parametric Phi 2} become:
\begin{align}
    \dot{A}&=\left[\omega_0 g h \sin((2\omega_0+\epsilon)t) \sin{2\Phi}-\lambda\right] A \quad, \label{eq:linear parametric A 3}\\
    \dot{\Phi}&=\omega_0 (1+f h \sin((2\omega_0+\epsilon)t))+ \omega_0 g h \sin(\omega t) \cos(2\Phi) \quad. \label{eq:linear parametric Phi 3}
\end{align}
We search for solutions oscillating at half the forcing frequency $\omega/2=\omega_0+\epsilon/2$:
\begin{equation}
    \Phi(t)=\left(\omega_0+\frac{\epsilon}{2}\right)t+\Phi_0 \quad, \label{eq:ansatz resonant Phi}
\end{equation}
which implies that: 
\begin{align}   &\sin((2\omega_0+\epsilon)t)\sin((2\omega_0+\epsilon)t+2\Phi_0)=
\\&=\frac{1}{2}\left[\cos(2\Phi_0)-\cos((4\omega_0+2\epsilon)t+2\Phi_0)\right] \quad, \label{eq:Werner 1}\\
&\sin((2\omega_0+\epsilon)t)\cos((2\omega_0+\epsilon)t+2\Phi_0)= \\
&=\frac{1}{2}\left[-\sin(2\Phi_0) + \sin((4\omega_0+2\epsilon)t+2\Phi_0)\right] \quad.\label{eq:Werner 2} 
\end{align}
By using eqs.\eqref{eq:Werner 1}-\eqref{eq:Werner 2} in eq.\eqref{eq:linear parametric A 3}-\eqref{eq:linear parametric Phi 3}, one ends up with:    
\begin{align}
    \dot{A}&\!=\!\bigg[\frac{\omega_0 g h}{2} \cos(2\Phi_0)-\lambda\bigg] A - \frac{\omega_0 g h}{2}\cos((4\omega_0 \!+\! 2\epsilon)t \!+\!2\Phi_0) A , \label{eq:linear parametric A 4}\\
    \omega_0&+\frac{\epsilon}{2}+\dot{\Phi}_0=\omega_0 + \omega_0 f h \sin((2\omega_0+\epsilon)t) \\ &- \frac{\omega_0 g h}{2} \sin(2\Phi_0) + \frac{\omega_0 g h}{2}\sin((4\omega_0+2\epsilon)t+2\Phi_0) . \label{eq:linear parametric Phi 4}
\end{align}
By averaging the latter equations with respect to the fast time scale with period $T=2\pi/\omega$, one obtains:
\begin{align}
    \dot{A}&=\left[\frac{\omega_0 g h}{2} \cos(2\Phi_0)-\lambda\right] A \quad, \label{eq:averaged linear parametric A appendix} 
    \\
    \dot{\Phi}_0+\frac{\epsilon}{2}&=- \frac{\omega_0 g h}{2} \sin(2\Phi_0) \quad. \label{eq:averaged linear parametric Phi appendix}
\end{align}
where, with little abuse of notation, we have denoted averaged amplitude and phase with the same letters $A$ and $\Phi$.

Now we search for the stationary solutions (P-modes) of eqs.\eqref{eq:averaged linear parametric A appendix}-\eqref{eq:averaged linear parametric Phi appendix}:
\begin{equation} 
    \frac{\omega_0 g h}{2} \cos(2\Phi_0)-\lambda =0, \quad,\quad \frac{\epsilon}{2}=- \frac{\omega_0 g h}{2} \sin(2\Phi_0) \quad. \label{eq:linear parametric Pmodes}
\end{equation}
The dependence on the phase $\Phi_0$ can be eliminated expressing 
\begin{equation}
    \cos(2\Phi_0)=\frac{2\lambda}{\omega_0 g h}, \quad,\quad \sin(2\Phi_0)=\frac{-\epsilon}{\omega_0 g h} \quad.
\end{equation}
By squaring and summing the latter equations, one ends up with 
\begin{equation}
    1=\left(\frac{2\lambda}{\omega_0 g h}\right)^2+ \left(\frac{\epsilon}{\omega_0 g h}\right)^2 \Rightarrow (\omega_0g h)^2=4\lambda^2+\epsilon^2 \quad,
\end{equation}
which gives the expression of the critical detuning cone:
\begin{equation}
    \frac{(\omega_0 g h)^2-\epsilon^2}{4}=\lambda^2 \quad, \label{eq:critical detuning cone proof}
\end{equation}
that in the original variables yields eq.
\eqref{eq:critical detuning cone}.

It is also possible evaluating the exponential growth rate for the amplitude of the P-mode due to the instability. To this end, we observe that eq.\eqref{eq:averaged linear parametric Phi appendix} can be solved for $\Phi_0(t)$, yielding the solution:
\begin{equation}
    \sin(2\Phi_0)=-\dfrac{\epsilon}{\omega_0 g h} \quad,    
\end{equation}
The above solution requires $|\epsilon|\leq |\omega_0 g h|$, namely that the excitation lies inside the detuning cone. Thus, one has:
\begin{equation}
    \cos(2\Phi_0)=\pm \sqrt{1-\frac{\epsilon^2}{(\omega_0 g h)^2}} \quad.
\end{equation}
By substituting it in eq.\eqref{eq:averaged linear parametric A appendix}, one ends up with:
\begin{align}
    \dot{A}&=\left[\pm \frac{\omega_0 g h}{2} \sqrt{1-\frac{\epsilon^2}{(\omega_0 g h)^2}}-\lambda\right]A =\\
    &=\left[\pm \sqrt{\frac{(\omega_0 g h)^2-\epsilon^2}{4}}-\lambda\right]A \quad.
\end{align}
By choosing the right sign $\pm$ depending on the sign of $g$, one can infer that the amplitude growth rate is:
\begin{equation}
    \zeta=\sqrt{\frac{(\omega_0 g h)^2-\epsilon^2}{4}}-\lambda \quad\Rightarrow\quad A(t)=A(0)e^{\zeta t} \quad.
\end{equation}
We remark that, in critical conditions, eq.\eqref{eq:critical detuning cone proof} states that 
\begin{equation}
    \frac{(\omega_0 g h_\mathrm{crit})^2-\epsilon_\mathrm{crit}^2}{4}=\lambda^2 \quad\Rightarrow\quad \zeta=0\quad.
\end{equation}

\section{Steady-state P-mode regimes under single-tone excitation}\label{sec:stability single P modes}

The steady-state solutions \eqref{eq:saturation amplitude mode h pm} can be also derived by introducing polar coordinates and using the same averaging technique as before, leading to:
\begin{align}
    \dot{A}&=\left[-\frac{\omega_h (\bm\varphi_h,\bm\varphi_h^*) \delta h}{2} \cos(2\Phi)-\lambda_h\right] A \quad, \label{eq:averaged nonlinear parametric A} \\
    \dot{\Phi}+\frac{\epsilon_h}{2}&=\omega_h N_h A^2 + \frac{\omega_h (\bm\varphi_h,\bm\varphi_h^*) \delta h}{2} \sin(2\Phi) \quad. \label{eq:averaged nonlinear parametric Phi}
\end{align}

The P-modes amplitudes $A_0^2$ are then given by eq.\eqref{eq:saturation amplitude mode h pm} (considering both signs $\pm)$ when the parameters are such that the squared amplitudes $A_0^2$ are non-negative. 
We report them here for convenience along with the existence conditions:
\begin{align}
    &A_{0\pm}^2=\frac{\epsilon_h+s_h^\pm \epsilon_\mathrm{crit,h}}{2\omega_h N_h}\geq 0 \,,\, s_h^\pm=\pm \text{sign}(N_h)\, \Rightarrow \nonumber \\
    &\, s_h^+(\epsilon_h+s_h^\pm \epsilon_\mathrm{crit,h})\geq 0 \!\Rightarrow \! \small\left\{\begin{array}{c}  
         A^2_{0+}\geq 0\text{ if } N_h>0\,,\, \epsilon_h+ \epsilon_\mathrm{crit,h}\geq 0   \\
         A^2_{0+}\geq 0\text{ if } N_h<0\,,\, \epsilon_h- \epsilon_\mathrm{crit,h}\leq 0   \\ 
         A^2_{0-}\geq 0\text{ if } N_h>0\,,\, \epsilon_h- \epsilon_\mathrm{crit,h}\geq 0   \\
         A^2_{0-}\geq 0\text{ if } N_h<0\,,\, \epsilon_h+ \epsilon_\mathrm{crit,h}\leq 0            
    \end{array} \right. . \label{eq:nonzero Pmodes single tone0}
\end{align}

It is also possible studying the stability of P-modes solutions \eqref{eq:nonzero Pmodes single tone0}. To this end, eqs.\eqref{eq:averaged nonlinear parametric A}-\eqref{eq:averaged nonlinear parametric Phi} can be linearized around stationary P-modes solutions $A_0,\Phi_0$:
\begin{equation}
    \left(\begin{array}{c}
         \dot{\Delta A}   \\
         \dot{\Delta\Phi}
    \end{array}\right) =  J(A_0,\Phi_0)\cdot  \left(\begin{array}{c}
         {\Delta A}   \\
         {\Delta\Phi} 
    \end{array}\right) \label{eq:linearized P modes} \quad.
\end{equation}
where $J(A_0,\Phi_0)$ is the Jacobian matrix:
\begin{equation}
    J(A_0,\Phi_0)=\left[\begin{array}{cc}
       -\frac{\omega_h g_h \delta h}{2}\cos(2\Phi_0)-\lambda_h   &  \omega_h g_h \delta h \sin(2\Phi_0) A_0\\
       2\omega_h N_h A_0  & \omega_h g_h \delta h \cos(2\Phi_0) 
    \end{array}\right],   
\end{equation}
where $g_h=(\bm\varphi_h,\bm \varphi^*_h)$.

We remark that the P-mode $A_0=0$ is unstable (stable) when $(\epsilon_h,\delta h)$ lies inside (outside) the detuning cone \eqref{eq:nonlinear detuning cone}. Conversely, when $A_0\neq 0$, 
one can show that the eigenvalues of $J(A_0,\Phi_0)$ are
\begin{equation}
    \zeta_{1,2}=-\lambda_h\pm\sqrt{\lambda_h^2-(\epsilon_h+s_h^\pm \epsilon_\mathrm{crit,h})s_h^\pm\epsilon_\mathrm{crit,h}} \quad.
\end{equation}
This means that the P-mode $A_{0\pm}^2$ is stable (node or focus, both eigenvalues have non-positive real part) when
\begin{equation}
    (\epsilon_h+s_h^\pm \epsilon_\mathrm{crit,h})s_h^\pm\epsilon_\mathrm{crit,h}\geq 0 \quad. \label{eq:nonzero Pmodes single tone stability}
\end{equation}
Interestingly, we notice that, being $\lambda_h>0$, only bifurcations\cite{Perko_book} of saddle-node type can occur (Hopf bifurcations would require the trace of the matrix $J$ to vanish), which from the physical viewpoint means that the steady-state oscillation can be either nonzero and synchronized (phase-locked) with half frequency $\omega_\mathrm{rf}/2$ of the pumping signal or identically zero. In the sequel, we will see that this condition can be violated in the case of parametric excitation of a spin-wave modes' pair, giving rise to the onset of nonzero unsynchronized (quasiperiodic) regimes, termed Q-modes\cite{BMS2009}. 

Coming back to discussing the single mode stability we observe that, for instance, when $N_h>0$, $A_{0+}$ exists when $\epsilon_h+\epsilon_\mathrm{crit,h}\geq 0$ (see eqs.\eqref{eq:nonzero Pmodes single tone0}) and is always stable according to eq.\eqref{eq:nonzero Pmodes single tone stability}. Conversely, $A_{0-}$ exists when $\epsilon_h-\epsilon_\mathrm{crit,h}\geq 0$ but is always unstable.  Analogous reasoning holds when $N_h<0$. 
Finally, the P-mode $A_{0\pm}^2$ becomes a saddle when
\begin{equation}
    s_h^\pm\epsilon_\mathrm{crit,h}(\epsilon_h+s_h^\pm\epsilon_\mathrm{crit,h})=0 \quad. \label{eq:saddle node single mode}
\end{equation}
Thus, summarizing the outcome of the above study of P-modes, one has:
\begin{equation}
        A_0^2=\left\{\begin{array}{lcl}
          0 & \rightarrow &
    \left\{\begin{array}{l}
              \text{stable if   }|\epsilon_h|> \epsilon_\mathrm{crit,h}\geq 0\,, \\
              \text{unstable if }|\epsilon_h|<\epsilon_\mathrm{crit,h}\geq 0\,, 
         \end{array} \right. \\
         A_{0+}^2
             & \rightarrow & \left\{\begin{array}{l}
              \text{stable if   }\epsilon_h+\epsilon_\mathrm{crit,h}\geq 0\,,\,N_h>0  \\
              \text{stable if }\epsilon_h-\epsilon_\mathrm{crit,h}\leq 0\,,\,N_h<0 
         \end{array} \right.\\
         A_{0-}^2
         & \rightarrow & \left\{\begin{array}{l}
              \text{unstable if   }\epsilon_h-\epsilon_\mathrm{crit,h}> 0\,,\,N_h>0 \\
              \text{unstable if }\epsilon_h+ \epsilon_\mathrm{crit,h}< 0\,,\,N_h<0 
         \end{array} \right.
    \end{array}\right. , \label{eq:stability single mode summary}
\end{equation} 
which is depicted in fig.\ref{fig:parametric resonance cone panels}(d).

\section{Steady-state P-mode regimes and stability under two-tone excitations}\label{sec:existence and stability multi P modes}

\subsection{Existence conditions for interacting steady-state regime pairs}\label{sec:existence conditions}

Here we derive the sets of control parameters $(\omega_\mathrm{rf1},\omega_\mathrm{rf2})$ such that the existence conditions \eqref{eq:constraint real amplitudes} are fulfilled. 
For illustrative purpose, we assume that mode $h$ (resp. $n$) has negative (positive) NFS, namely $N_{hh}<0,N_{nn}>0$ (meaning $s^+_h=-1,s^+_n=1$) and mutual NFSs $N_{hn}=N_{nh}<0$ as it was done in fig.\ref{fig:parametric resonance cone interacting type1}. 

 The existence conditions are functions of the mode amplitudes $|a_h|^2,|a_n|^2$. Thus, in order to eliminate such dependence, we plug the expressions of possible steady-state regimes $z,u_h,u_n,c$ according to eqs.\eqref{eq:type 1 mode h off mode n off}-\eqref{eq:type 1 coupled modes}, which are in turn function of $\omega_\mathrm{rf1},\omega_\mathrm{rf2}$, into the existence conditions \eqref{eq:constraint real amplitudes}. The results are then represented as colored regions in the $(\omega_\mathrm{rf1},\omega_\mathrm{rf2})$ plane (stability will be discussed in sec.\ref{sec:stability conditions}), as one can see in fig.\ref{fig:Pmodes existence regions}. 

In this respect, the plane is partitioned into regions where:
\begin{itemize}
\item mode $h$ exists in the uncoupled regime $u_h$ (green rectangle $R_h$), given by eq.\eqref{eq:nonzero Pmodes single tone}
\begin{equation}
    R_h: s_h^+(\epsilon_h+ s^+_h \epsilon_\mathrm{crit,h})\geq 0 \quad, \label{eq:Rh}
\end{equation}
that is bounded on one side by the vertical line
\begin{equation}
    z\equiv u_h: \epsilon_h+ s^+_h \epsilon_\mathrm{crit,h}=0 \Rightarrow \omega_\mathrm{rf1}=2\omega_h-s^+_h\epsilon_\mathrm{crit,h} \,. \label{eq:z eq uh line}
\end{equation}
For the case of fig.\ref{fig:Pmodes existence regions}, this gives
\begin{align}
    N_{hh}<0\,,\,s^+_h=-1 \quad \Rightarrow \quad \epsilon_h+ s^+_h \epsilon_\mathrm{crit,h}\leq 0 \quad \Rightarrow \nonumber \\
    \Rightarrow \quad \epsilon_h \leq -s^+_h \epsilon_\mathrm{crit,h} \quad \Rightarrow \quad\omega_\mathrm{rf1}\leq 2\omega_h+\epsilon_\mathrm{crit,h} \quad.
\end{align}

It is also worth defining the bounded subregion $C_h\subset R_h$ (dark green rectangle) corresponding to nonzero uncoupled regimes $u_h$ obtainable under PWM excitation. These are given by excitations that make the 'zero' steady-state unstable (see eq.\eqref{eq:stability single mode summary} in section \ref{sec:nonlinear self interaction}):
\begin{equation}
    C_h: |\epsilon_h|\leq \epsilon_\mathrm{crit,h} \quad. \label{eq:Ch}
\end{equation}

\item Mode $n$ exists in the uncoupled regime $u_n$ (red rectangle $R_n$), given by eq.\eqref{eq:nonzero Pmodes single tone}
\begin{equation}
    R_n: s_n^+(\epsilon_n+ s^+_n \epsilon_\mathrm{crit,n})\geq 0 \quad, \label{eq:Rn}
\end{equation}
that is bounded on one side by the horizontal line
\begin{equation}
    z\equiv u_n: \epsilon_n+ s^+_n \epsilon_\mathrm{crit,n}=0 \Rightarrow \omega_\mathrm{rf2}=2\omega_n-s^+_n\epsilon_\mathrm{crit,n} \,. \label{eq:z eq un line}
\end{equation}
For the case of fig.\ref{fig:Pmodes existence regions}, this gives
\begin{align}
        N_{nn}>0\,,\,s^+_n=+1 \quad \Rightarrow \quad \epsilon_n+ s^+_n \epsilon_\mathrm{crit,n}\geq 0 \quad \Rightarrow \nonumber \\
        \Rightarrow \quad \epsilon_n \geq -s^+_n \epsilon_\mathrm{crit,n} \quad \Rightarrow \quad
    \omega_\mathrm{rf2}\geq 2\omega_n-\epsilon_\mathrm{crit,n} \quad.
\end{align}

As before, we define the bounded subregion $C_n\subset R_n$ (dark red rectangle) corresponding to nonzero uncoupled regimes $u_n$ obtainable under PWM excitation. These are given by:
\begin{equation}
    C_n: |\epsilon_n|\leq \epsilon_\mathrm{crit,n} \quad. \label{eq:Cn}
\end{equation}

\item The two rectangles (dark green) $C_h$ and (dark red) $C_n$ forming the 'cross' region $C_h\cup C_n$ intersect in a rectangular (brown) region $C_h\cap C_n$ whose four vertices $O_{\pm\pm}$ are very important to understand the phase diagram concerning coupled regimes.

Their coordinates in $(\omega_\mathrm{rf1},\omega_\mathrm{rf2})$ plane are given by the conditions $\epsilon_h+s^\pm_h\epsilon_\mathrm{crit,h}=0 \Rightarrow  |a_h|^2_\mathrm{\pm,unc}=0$ and $\epsilon_n+s^\pm_n\epsilon_\mathrm{crit,n}=0 \Rightarrow  |a_n|^2_\mathrm{\pm,unc}=0$:
\begin{equation}
    O_{\pm\pm}: \left\{ \begin{array}{c}
        \omega_\mathrm{rf1}=2\omega_h \mp \text{sign}(N_{hh})\epsilon_\mathrm{crit,h}(\delta h_1) \\
        \omega_\mathrm{rf2}=2\omega_n \mp \text{sign}(N_{nn})\epsilon_\mathrm{crit,n}(\delta h_2)
    \end{array}\right. \,, \label{eq:c triangles vertices}
\end{equation}

Under the present assumptions $N_{hh}<0,N_{nn}>0$, eqs.\eqref{eq:c triangles vertices} imply that point $O_{++}$ is located at the lower right corner of the brown region in fig.\ref{fig:Pmodes existence regions}. Analogous reasoning yields that $O_{+-},O_{-+}$ and $O_{--}$ are located in the upper-right, lower-left and upper-left corners, respectively.

\item Modes $h,n$ exist in coupled regimes $c_{\pm\pm}$ (four unbounded triangles $T_{\pm\pm}$ delimited by dashed lines in fig.\ref{fig:Pmodes existence regions}), given by eqs.\eqref{eq:saturation amplitude mode h type 1 linear}-\eqref{eq:saturation amplitude mode n type 1 linear} where the expressions of 'uncoupled' amplitudes defined by eqs.\eqref{eq:uncoupled amp h}-\eqref{eq:uncoupled amp n} are used and all of the four combinations of $s^\pm_h, s^\pm_n$ signs are considered, plugged into the existence conditions \eqref{eq:constraint real amplitudes}:  
\begin{equation}
T_{\pm\pm}:\left\{\begin{array}{l}
      L_{hn}\left[\frac{\epsilon_h + s^\pm_h \epsilon_\mathrm{crit,h}}{2\omega_h N_{hh}}-\frac{N_{hn}}{N_{hh}}\frac{\epsilon_n + s^\pm_n \epsilon_\mathrm{crit,n}}{2\omega_n N_{nn}}\right] \geq 0  \\
      L_{hn}\left[\frac{\epsilon_n + s^\pm_n \epsilon_\mathrm{crit,n}}{2\omega_n N_{nn}}-\frac{N_{nh}}{N_{nn}}\frac{\epsilon_h + s^\pm_h \epsilon_\mathrm{crit,h}}{2\omega_h N_{hh}}\right] \geq 0 \,.
\end{array} \right. \label{eq:T}
\end{equation} 
We observe that the vertices of $T_{\pm\pm}$ are the points $O_{\pm\pm}$, as defined before by eqs.\eqref{eq:c triangles vertices} that  make the terms in square brackets vanish simultaneously in the latter inequalities. 
We remark that such inequalities are linear with respect to $\omega_\mathrm{rf1},\omega_\mathrm{rf2}$ owing to the obvious relationship with the detuning, and define triangular regions $T_{\pm\pm}$ in the $(\omega_\mathrm{rf1},\omega_\mathrm{rf2})$ plane, delimited by two straight lines defined imposing that one of the terms in square brackets vanishes. 
In this respect, the former relation defines a region in the plane $(\omega_\mathrm{rf1},\omega_\mathrm{rf2})$ delimited by a straight line, denoted as
\begin{align}
    c_{\pm\pm}\equiv u_n : &\text{ line passing through $O_{\pm\pm}$ } \nonumber \\
    &\text{with slope $\omega_h N_{hn}/(\omega_n N_{nn})$} \quad, \label{eq:def c eq un}
\end{align}
for reasons explained in the subsequent section, whereas the second defines a region delimited by a straight line, denoted as 
\begin{align}
    c_{\pm\pm}\equiv u_h : &\text{ line passing through $O_{\pm\pm}$ }\nonumber \\
    &\text{with slope $\omega_n N_{nh}/(\omega_h N_{hh})$} \quad. \label{eq:def c eq uh}
\end{align}
We stress that these two lines intersect at $O_{\pm\pm}$. 
For instance, one can see in fig.\ref{fig:Pmodes existence regions} the lines $c_{++}\equiv u_n$ and $c_{++}\equiv u_h$ depicted as dashed black and red lines, respectively.
We also observe that the slopes of all the lines $c_{\pm\pm}\equiv u_n, c_{\pm\pm}\equiv u_h$ only depend on the signs of the self- and mutual NFS coefficients, so they are grouped in pairs of parallel straight lines. 

In more practical terms, one can show that triangular regions $T_{\pm\pm}$ always extend from their vertex towards the unbounded portion of $R_h\cup R_n$ regardless of the sign of the leading coefficient $L_{hn}$.

For the situation of fig.\ref{fig:Pmodes existence regions}, the first slope is negative owing to $N_{hn}/N_{nn}<0$ and the second is positive due to $N_{nh}/N_{hh}>0$. 
This implies that coupled modes $c_{++}$ exist in $T_{++}$ to the left of line $c_{++}\equiv u_n$ (dashed black in fig.\ref{fig:Pmodes existence regions}) and above line $c_{++}\equiv u_h$ (dashed red in fig.\ref{fig:Pmodes existence regions}). 
Analogously, the coupled modes $c_{+-}$ exist in region $T_{+-}$ to the left of line $c_{+-}\equiv u_n$ and above line $c_{+-}\equiv u_h$. Similar reasoning holds for regimes of type $c_{-+}, c_{--}$.
\end{itemize}

Finally, we point out that the light green (resp. light red) subregions belonging to $R_h$ (resp. $R_n$) in fig.\ref{fig:Pmodes existence regions} correspond to large negative detuning $\omega_\mathrm{rf1}-2\omega_h < -\epsilon_\mathrm{crit,h}$ (resp. large positive detuning $\omega_\mathrm{rf2}-2\omega_n > \epsilon_\mathrm{crit,n}$) where uncoupled steady-state amplitudes may linearly increase as function of decreasing (resp. increasing) rf frequency if arriving from inside the corresponding dark region in a CW fashion, as it is visible in the amplitude diagrams reported in the upper and right parts of fig.\ref{fig:Pmodes existence regions} (similar to what is depicted in fig.\ref{fig:parametric resonance cone panels}(d) for single-mode parametric excitation in section \ref{sec:self interactions}). 

\subsection{Discussion on the stability of interacting steady-state regime pairs} \label{sec:stability conditions}

Once the existence conditions for multiple steady-state regimes have been established, the problem of assessing the stability for each regime arises. Here we provide some qualitative criteria that allow to complete the construction of the phase diagram in the control plane $(\omega_\mathrm{rf1},\omega_\mathrm{rf2})$ reported in fig.\ref{fig:Pmodes existence regions}. More specifically, we anticipate that this will lead to the phase diagram depicted in the top panel of fig.\ref{fig:type1 phase diagram and hysteresis}, where stable $c$ regimes are associated with regions colored blue in the $(\omega_\mathrm{rf1},\omega_\mathrm{rf2})$ plane.

First of all, we observe that the 'zero' regime is unstable in the 'cross' region $C_h\cup C_n$ defined in the previous section (see eqs.\eqref{eq:Ch},\eqref{eq:Cn}). Then, we proceed to analyze nonzero  regimes $u_h,u_n,c_{\pm\pm}$.

To this end, it is worth making a general observation on the lines that originate from the four vertices $O_{\pm\pm}$ and enclose the regions $T_{\pm\pm}$ of coupled regimes $c_{\pm\pm}$ (e.g. blue region $T_{++}$ of  $c_{++}$ regimes in fig.\ref{fig:type1 phase diagram and hysteresis}). We have denoted them with the notations $c_{\pm\pm}\equiv u_h$ and $c_{\pm\pm}\equiv u_n$, defined as in eqs.\eqref{eq:def c eq un},\eqref{eq:def c eq uh}, that have the meaning of continuity (collision) between coupled $c_{\pm\pm}$ and uncoupled $u_h,u_n$ regimes. They are essential for understanding the stability of coupled regimes as we will see in the sequel.  

We start considering the vertex $O_{++}$.
The first line also defined before by eq.\eqref{eq:def c eq un} (dashed black line in fig.\ref{fig:type1 phase diagram and hysteresis}) corresponds to the condition labeled as '$c_{++}\equiv u_n$' meaning 
\begin{equation}
   c_{++}\equiv u_n:\, |a_h|^2_\mathrm{++,coup}=0 \,, \, |a_n|^2_\mathrm{++,coup}=|a_n|_\mathrm{+,unc} \quad, \label{eq:def c++ eq un line}  
\end{equation}
 given by eq. \eqref{eq:saturation amplitude mode h type 1 linear}. Analogously, the second line also defined before by eq.\eqref{eq:def c eq uh} (dashed red line in fig.\ref{fig:type1 phase diagram and hysteresis}) refers to the condition '$c_{++}\equiv u_h$' meaning 
 \begin{equation}
   c_{++}\equiv u_h:\, |a_n|^2_\mathrm{++,coup}=0 \,, \, |a_h|_\mathrm{++,coup}^2=|a_h|^2_\mathrm{+,unc}  \,, \label{eq:c++ eq uh line}
 \end{equation}
given by eq. \eqref{eq:saturation amplitude mode n type 1 linear}. By expressing them with respect to the applied frequencies $\omega_\mathrm{rf1},\omega_\mathrm{rf2}$, one has the analytical expressions:
\begin{align}
    c_{++}\equiv u_n:
    \frac{\epsilon_h + s^+_h \epsilon_\mathrm{crit,h}}{2\omega_h N_{hh}}-\left(\frac{N_{hn}}{N_{hh}}\right)\frac{\epsilon_n + s^+_n \epsilon_\mathrm{crit,n}}{2\omega_n N_{nn}}=0
    \,, \label{eq:line ah_coup_eq_zero}\\
    c_{++}\equiv u_h:
    \frac{\epsilon_n + s^+_n \epsilon_\mathrm{crit,n}}{2\omega_n N_{nn}}-\left(\frac{N_{nh}}{N_{nn}}\right)\frac{\epsilon_h + s^+_h \epsilon_\mathrm{crit,h}}{2\omega_h N_{hh}}=0
    \label{eq:line an_coup_eq_zero} \,.
\end{align}

As anticipated, these lines are extremely important since they represent continuity (collision) conditions for the steady-state history owing to the fact that, for frequencies $\omega_\mathrm{rf1},\omega_\mathrm{rf2}$ satisfying \eqref{eq:line ah_coup_eq_zero},\eqref{eq:line an_coup_eq_zero}, the coupled regime $c_{++}$ coincides with one of the uncoupled solutions $u_h,u_n$ (see eqs.\eqref{eq:saturation amplitude mode h type 1 linear}-\eqref{eq:saturation amplitude mode n type 1 linear}). 
As mentioned in the previous section, their slopes are controlled by the self- and mutual NFS of the selected modes' pair. In fact, looking at coefficients multiplying $\omega_\mathrm{rf1},\omega_\mathrm{rf2}$ in eq.\eqref{eq:line ah_coup_eq_zero}
one can see that the slope is $\omega_h N_{hn}/(\omega_n N_{nn})$. Conversely, observing coefficients multiplying $\omega_\mathrm{rf2},\omega_\mathrm{rf1}$ in eq.\eqref{eq:line an_coup_eq_zero} 
one infers that the slope is $\omega_n N_{nh}/(\omega_h N_{hh})$. Under the above assumptions on the signs of NFS coefficients, this yields negative (positive) slope for the black (red) line. 

In this respect, we introduce the following additional critical lines (black and red dash-dotted lines in fig.\ref{fig:type1 phase diagram and hysteresis}, respectively) that correspond to collisions between coupled $c_{-+}$ (resp. $c_{+-}$) and uncoupled $u_n$ (resp. $u_h$) regimes, respectively (see eqs.\eqref{eq:saturation amplitude mode h type 1 linear}-\eqref{eq:saturation amplitude mode n type 1 linear}):
\begin{align}    
   c_{-+}\equiv u_n:\, |a_h|^2_\mathrm{-+,coup}=0 \,, \, |a_n|^2_\mathrm{-+,coup}=|a_n|_\mathrm{+,unc} \,, \label{eq:def c-+ eq un line}  \\
   c_{+-}\equiv u_h:\, |a_n|^2_\mathrm{+-,coup}=0 \,, \, |a_h|_\mathrm{+-,coup}^2=|a_h|^2_\mathrm{+,unc}  \,. \label{eq:c+- eq uh line} 
\end{align}
that in the variables $\omega_\mathrm{rf1},\omega_\mathrm{rf2}$ read as:
\begin{align}
    c_{-+}\equiv u_n:
    \frac{\epsilon_h + s^-_h \epsilon_\mathrm{crit,h}}{2\omega_h N_{hh}}-\left(\frac{N_{hn}}{N_{hh}}\right)\frac{\epsilon_n + s^+_n \epsilon_\mathrm{crit,n}}{2\omega_n N_{nn}}=0
    \,, \label{eq:line ah_coup_eq_zero bound}\\
    c_{+-}\equiv u_h:
    \frac{\epsilon_n + s^+_n \epsilon_\mathrm{crit,n}}{2\omega_n N_{nn}}-\left(\frac{N_{nh}}{N_{nn}}\right)\frac{\epsilon_h + s^-_h \epsilon_\mathrm{crit,h}}{2\omega_h N_{hh}}=0
    \label{eq:line an_coup_eq_zero bound} \,.
\end{align}
We observe that each of these lines is one of the boundaries of the triangular region $T_{-+}$ and $T_{+-}$ originating at the vertices $O_{-+}$ and $O_{+-}$, respectively.
The above lines play an additional role in that they represent the boundary of the regions where the 'uncoupled' regime $u_n$ expressed by \eqref{eq:type 1 mode h off mode n on} (resp. $u_h$ expressed by \eqref{eq:type 1 mode h on mode n off}) is annihilated by the coupled regime $c_{-+}$ (resp. $c_{+-}$), as one can infer from eqs.\eqref{eq:saturation amplitude mode h type 1 linear}-\eqref{eq:saturation amplitude mode n type 1 linear}. For this reason, they are also denoted as 'no $u_n$' and 'no $u_h$', respectively. 

Interestingly, we observe that critical lines $c_{+-}\equiv u_n$ and $c_{-+}\equiv u_h$ are associated with the remaining boundaries of the related triangular regions $T_{-+}$ and $T_{+-}$ of vertices $O_{-+}$ and $O_{+-}$, respectively. At the same time, they refer to collisions between uncoupled solution $|a_n|^2_\mathrm{-,unc}$ (resp. $|a_h|^2_\mathrm{-,unc}$) and coupled regime $c_{+-}$ (resp. $c_{-+}$) that would never give rise to an observable steady-state due to the unconditional instability of $|a_h|^2_\mathrm{-,unc}, |a_n|^2_\mathrm{-,unc}$ (see section \ref{sec:nonlinear self interaction}, eq.\eqref{eq:stability single mode summary}). Therefore, the lines $c_{+-}\equiv u_n$ and $c_{-+}\equiv u_h$ do not affect the dynamics of the interacting modes.

We also recall the existence of two additional critical lines, introduced in the previous section, having the meaning of continuity (collision) conditions between the 'zero' regime $z$ and the uncoupled regimes $u_h$ and $u_n$, respectively, that are labeled as '$z\equiv u_h$' (vertical) and '$z\equiv u_n$' (horizontal) in fig.\ref{fig:type1 phase diagram and hysteresis} and defined by eqs.\eqref{eq:z eq uh line},\eqref{eq:z eq un line}.

These general remarks are useful to figure out the qualitative partitioning of the phase diagram that can be instrumental to the interpretation of possible experiments. 

\subsection{P-mode stability under two-tone excitations}\label{sec:stability multi P modes}

The phase diagrams of figures \ref{fig:type1 phase diagram and hysteresis} and \ref{fig:type1 phase diagram and hysteresis PWM} can be obtained by studying the stability of periodic solutions (termed P-modes) of normal modes dynamics. This study can be carried out analytically by using the averaging technique to extend the theory developed for the single-mode dynamics in section \ref{sec:nonlinear self interaction}.

In fact, under the assumption made in section \ref{sec:mutual modes interaction}, the  
P-mode solutions $(A_h,A_n)$ of eqs.\eqref{eq:averaged two nonlinear parametric Ah}-\eqref{eq:averaged two nonlinear parametric Phi_n} are given by:
\begin{align}
    (A_h^2,A_n^2)&=(0,0)\quad, \label{eq:Pmode zero}\\
    (A_h^2,A_n^2)&=(A^2_\mathrm{h\pm,unc},0) \quad,  \label{eq:Pmode unc h}\\
    (A_h^2,A_n^2)&=(0,A^2_\mathrm{n\pm,unc}) \quad, \label{eq:Pmode unc n} \\
    (A_h^2,A_n^2)&=L_{hn}\bigg(A^2_\mathrm{h\pm,unc}-\frac{N_{hn}}{N_{hh}}A^2_\mathrm{n\pm,unc} \,,\, A^2_\mathrm{n\pm,unc}- \nonumber \\
    &\frac{N_{nh}}{N_{nn}}A^2_\mathrm{h\pm,unc} \bigg) \quad,  \label{eq:Pmode coup}
\end{align}
which, apart from the $\pm$ that takes into account all possibilities, coincide with zero solution $z$ given by \eqref{eq:type 1 mode h off mode n off}, uncoupled solutions $u_h,u_n$ given by \eqref{eq:type 1 mode h on mode n off}-\eqref{eq:type 1 mode h off mode n on}, and coupled solutions $c_{\pm\pm}$ given by \eqref{eq:saturation amplitude mode h type 1 linear}-\eqref{eq:saturation amplitude mode n type 1 linear} described in section \ref{sec:mutual modes interaction}.
 
 The stability analysis for the above P-modes is governed by the linearized equation:
\begin{equation}
    \left(\begin{array}{c}
         \dot{\Delta A_h}  \\
         \dot{\Delta \Phi_h}\\         
         \dot{\Delta A_n}  \\
         \dot{\Delta \Phi_n}
    \end{array} \right) = J(A_h,\Phi_h,A_n,\Phi_n) \cdot \left(\begin{array}{c}
         {\Delta A_h}  \\
         {\Delta \Phi_h}\\
         {\Delta A_n}  \\
         {\Delta \Phi_n}
    \end{array} \right)
\end{equation}
where the Jacobian matrix $J$ is in the $2\times 2$ block form:
\begin{equation}
    J=\left[\begin{array}{cc}
       J_{hh}  &   J_{hn} \\
       J_{nh}  &  J_{nn}
    \end{array}   \right] \quad, \label{eq:block 2x2 P modes Jacobian}
\end{equation}
where blocks $J_{pq}$ with $p,q=n,h$ are given by expressing the P-modes phases (stationary solutions of eqs.\eqref{eq:averaged two nonlinear parametric Ah}-\eqref{eq:averaged two nonlinear parametric Phi_n}) as function of the P-modes amplitudes leading to
\begin{align}
    J_{pp}&=\left[\begin{array}{cc}
       \sqrt{\frac{(\omega_p g_p h_p)^2-{\epsilon''}_p^2}{4}}-\lambda_p  &   -(2\omega_p N_{pp}A_p^2-\epsilon'_p) A_p\\
       2\omega_p N_{pp} A_p  &  -2 \sqrt{\frac{(\omega_p g_p h_p)^2-{\epsilon''}_p^2}{4}}
    \end{array}   \right] ,\label{eq:block P modes Jacobian} \\
    \quad J_{pq}&=\left[ \begin{array}{cc}
       0  & 0 \\
       2\omega_p N_{pq} A_q  & 0
    \end{array}\right] \,, 
\end{align}
where $\epsilon''_p=(\epsilon'_p-2\omega_pN_{pp}A_p^2)$ is the full detuning including self- and mutual-NFS and $g_p=(\bm\varphi_p,\bm \varphi_p^*), h_p=\delta h_1,\delta h_2$.
For nonzero P-modes 
\begin{align}
        J_{pp}&=\left[\begin{array}{cc}
       0  &   -(2\omega_p N_{pp}A_{p\pm}^2-\epsilon'_p) A_{p\pm}\\
       2\omega_p N_{pp} A_{p\pm}  &  -2 \lambda_p
    \end{array}   \right] ,
    \label{eq:block P modes Jacobian 1} 
    \\   
    J_{pp}&=\left[\begin{array}{cc}
       0  &   -s^\pm_p\epsilon_\mathrm{crit,p} A_{p\pm}\\
       (\epsilon'_p + s^\pm_p \epsilon_\mathrm{crit,p})/A_{p\pm}  &  -2 \lambda_p
    \end{array}   \right] , 
    \label{eq:block P modes Jacobian 2} 
\end{align}

In the sequel, the scalar entries of matrix $J$ will be denoted as $j_{ik}$, with $i,k=1,\ldots,4$.

Now we start analyzing bifurcations of co-dimension one\cite{Perko_book} that only involve one parameter of the dynamical system (e.g. the frequency of a single tone). 

First, we immediately observe that the stability of the zero P-mode $z$ retains the same conditions as in the single mode case, namely it is unstable inside the 'cross' region $C_h\cup C_n$ (dark green and dark red in fig. \ref{fig:type1 phase diagram and hysteresis}).

For nonzero P-modes $(A_h,A_n)\neq (0,0)$, the stability is controlled by the Jacobian matrix \eqref{eq:block 2x2 P modes Jacobian} that can be rewritten using the simplification \eqref{eq:block P modes Jacobian 1}:
\begin{widetext}
\begin{equation}
     J=\left[\begin{array}{cccc}
       0  &   -(2\omega_h N_{hh}A_h^2-\epsilon'_h) A_h & 0 & 0 \\
       2\omega_h N_{hh} A_h  &  -2 \lambda_h & 2\omega_h N_{hn}A_n & 0 \\
       0 & 0 & 0 & -(2\omega_n N_{nn}A_n^2-\epsilon'_n) A_n \\
       2\omega_n N_{nh} A_h & 0 & 2\omega_n N_{nn} A_n & -2\lambda_n
    \end{array}   \right] \quad. \label{eq: 4x4 nonzero P modes Jacobian}
\end{equation}   
\end{widetext}
We observe that saddle-node type bifurcations imply that the following condition is verified\cite{Perko_book,kuznetsov2012,bosi2019local}:
\begin{equation}
    \det(J(A_h,A_n))=0 \quad. \label{eq:saddle node condition}
\end{equation}
Thus, considering the uncoupled $u_h,u_n$ P-modes \eqref{eq:Pmode unc h}-\eqref{eq:Pmode unc n}, one can easily check that one off-diagonal $2\times 2$ block vanishes in the Jacobian matrix $J$ (due to vanishing $A_q$ in the expression of $J_{pq}$ in eq.\eqref{eq:block P modes Jacobian}). This means that the eigenvalues of $J$ are given by both the eigenvalues of single blocks $J_{hh}$ and $J_{nn}$.  Moreover, the condition \eqref{eq:saddle node condition} for saddle-node type bifurcations is given by the simpler one involving only $2\times 2$ diagonal blocks:
\begin{equation}
    \det(J)=\det(J_{hh})\det(J_{nn}) = 0 \quad. \label{eq:saddle node condition block diagonal}
\end{equation}
This allows one to obtain stability conditions by combining the analysis already performed for the single-mode stability (see eqs.\eqref{eq:stability single mode summary}, \eqref{eq:stability condition nonzero P-mode} in section \ref{sec:nonlinear self interaction}) for modes $h$ and $n$ as they were independent. A first important consequence of the above result is that the involved uncoupled P-modes $u_h,u_n$ cannot undergo Hopf bifurcations, as happens for the single-mode excitation (see section \ref{sec:nonlinear self interaction}). Thus, in such situation, the condition \eqref{eq:saddle node condition block diagonal} is expressed as a couple of equations (see eq.\eqref{eq:saddle node single mode}):
\begin{align}
    s_h^\pm (\epsilon'_h+s_h^\pm\epsilon_\mathrm{crit,h})=0 \,,\, & s_h^\pm=\pm\text{sign}(N_{hh})\,, \label{eq:stability condition h}   \\
    s_n^\pm (\epsilon'_n+s_n^\pm\epsilon_\mathrm{crit,n})=0 \,,\, & s_n^\pm=\pm\text{sign}(N_{nn})\,. \label{eq:stability condition n} 
\end{align}
We remark that in the latter equations, the amplitude-dependent detunings $\epsilon'_h(\omega_\mathrm{rf1},A_n^2),\epsilon'_n(\omega_\mathrm{rf2},A_h^2)$ appear. Due to the linearity of the equations with respect to squared amplitudes $A_h^2,A_n^2$, one can eliminate them by substitution and end up with expressions that only depend on control parameters $(\omega_\mathrm{rf1},\delta h_1,\omega_\mathrm{rf2},\delta h_2)$ (we recall that $\epsilon_\mathrm{crit,h},\epsilon_\mathrm{crit,n}$ depend on the field amplitudes $\delta h_1,\delta h_2$, respectively). 

In general, these equations will describe surfaces in the four-dimensional space $(\omega_\mathrm{rf1},\delta h_1,\omega_\mathrm{rf2},\delta h_2)$.
However, in the particular case of excitation with two fixed parameters (e.g. tone amplitudes, frequencies, or other combinations), the eqs.\eqref{eq:stability condition h}-\eqref{eq:stability condition n} determine critical (bifurcation) curves in the control plane separating different types of P-mode solutions.

In the case of constant amplitudes $\delta h_1,\delta h_2$, one can easily see that these curves are all straight lines in the plane $(\omega_\mathrm{rf1},\omega_\mathrm{rf2})$, owing to the obvious linear dependence of detuning $\epsilon'_h, \epsilon'_n$ on the external field frequencies $\omega_\mathrm{rf1},\omega_\mathrm{rf2}$, respectively.

In particular, considering the situation depicted in fig.\ref{fig:type1 phase diagram and hysteresis}, eqs.\eqref{eq:stability condition h}-\eqref{eq:stability condition n} yield:
\begin{itemize}
    \item the boundaries of the 'cross' $C_h\cup C_n$ (dark green and dark red regions $C_h,C_n$ in fig.\ref{fig:type1 phase diagram and hysteresis}) when computed on the zero P-mode $z$ given by \eqref{eq:Pmode zero};
    \item the red (dashed $c_{++}\equiv u_h$ and dash-dotted $c_{+-}\equiv u_h$ '$\text{no } u_h$') lines when computed on the uncoupled P-mode $u_h$ given by \eqref{eq:Pmode unc h};
    \item the black (dashed $c_{++}\equiv u_n$ and dash-dotted $c_{-+}\equiv u_n$ '$\text{no } u_n$') lines when computed on the uncoupled P-mode $u_n$ given by \eqref{eq:Pmode unc n}. 
\end{itemize}

Now we discuss the stability of uncoupled P-modes $u_h,u_n$ in the presence of both signal tones turned on. This can be done considering that, being one of the two modes amplitudes equal to zero, the Jacobian matrix \eqref{eq:block 2x2 P modes Jacobian},\eqref{eq: 4x4 nonzero P modes Jacobian} has one off-diagonal block that vanishes. This means that the eigenvalues $\zeta$ are given by those of the single diagonal blocks $J_{hh},J_{nn}$. 

For instance, searching for the instability conditions for P-mode $u_h$ under nonzero excitation of mode $n$ implies imposing that $A^2_h=A^2_\mathrm{h+,unc}\neq 0, A_n^2=0$ so that the eigenvalues of $J$ are:
\begin{align}
    \zeta_h=-\lambda_h\pm\sqrt{\lambda_h^2-s_h^+\epsilon_\mathrm{crit,h}(\epsilon_h+s_h^+\epsilon_\mathrm{crit,h}) } \,, \\ 
    \zeta_n=-\lambda_n\pm\sqrt{\lambda_n^2-s_n^\pm\epsilon_\mathrm{crit,n}(\epsilon'_n+s_n^\pm\epsilon_\mathrm{crit,n}) } \,. \label{eq:eig J un}
\end{align}
This means that the uncoupled mode $u_h$ is unstable when either of the following conditions (see eq.\eqref{eq:stability condition nonzero P-mode}) holds
\begin{align}
        s_h^+ (\epsilon_h+s_h^+\epsilon_\mathrm{crit,h})< 0 \quad, \label{eq:uh unstable condition 1} \\
        s_n^\pm (\epsilon'_n+s_n^\pm\epsilon_\mathrm{crit,n})\geq 0 \quad.  \label{eq:uh unstable condition 2}
\end{align}
The former condition corresponds to points in the control plane outside the green region $R_h$, while the latter corresponds to $|\epsilon'_n|\leq\epsilon_\mathrm{crit,n}$ namely
\begin{equation}
    s_n^\pm (\epsilon_n+s_n^\pm\epsilon_\mathrm{crit,n}-2\omega_n N_{nh}A^2_\mathrm{h+,unc})\geq 0 \quad.
\end{equation}
By remembering that $A^2_\mathrm{h+,unc}=(\epsilon_h+s^+_h\epsilon_\mathrm{crit,h})/(2\omega_h N_{hh})$, one can show that the two inequalities arising from the choice of $s^\pm_n$ correspond to points belonging to the region enclosed between red (dashed $c_{++}\equiv u_h$ and dash-dotted $c_{+-}\equiv u_h$) lines. 
Analogously, exchanging the roles of the regimes $u_h$ and $u_n$, one has that the uncoupled P-mode $u_n$ is unstable when 
\begin{align}
        s_n^+ (\epsilon_n+s_n^+\epsilon_\mathrm{crit,n})< 0 \quad,   \label{eq:un unstable condition 1}\\
        s_h^\pm (\epsilon'_h+s_h^\pm\epsilon_\mathrm{crit,h})\geq 0 \quad,
     \label{eq:un unstable condition 2}
\end{align}
namely for excitation frequencies that lie outside the red region $R_n$ or within the region enclosed between black (dashed $c_{++}\equiv u_n$ and dash-dotted $c_{-+}\equiv u_n$) lines.

We now analyze the stability of coupled P-modes $c_{\pm\pm}$. In general, to assess that a given P-mode is stable, one has to check that all the eigenvalues of the Jacobian matrix $J$ have negative real part. Being $J$ a $4\times 4$ matrix, closed-form expressions for the eigenvalues would be unpractical although feasible in principle. For this reason, in the following we will derive some simpler criteria that help to infer the stability of coupled regimes without performing the eigenvalue calculation.

We start from $c_{++}$, observing that it is enough considering what happens around the critical point $O_{++}$ where coincidence between several P-modes occurs, namely $z\equiv u_h\equiv u_n\equiv c_{++}$, as one can see from the intersection of the corresponding critical lines (see fig.\ref{fig:type1 phase diagram and hysteresis} for example). By considering P-modes associated with excitation conditions sufficiently close to $O_{++}$ and lying on the dashed (black or red) lines $c_{++}\equiv u_n, c_{++}\equiv u_h$, one can immediately conclude that only saddle-node bifurcations are possible because one of the mode amplitudes is zero (as before, the matrix $J$ becomes block-diagonal and single-mode analysis holds independently for each mode). 

To this end, we use the expressions \eqref{eq:Pmode coup} for coupled P-modes $c_{++}$ in the $4\times 4$ Jacobian matrix $J$ using simplification \eqref{eq:block P modes Jacobian 2}:
\begin{widetext}
\begin{equation}
     J=\left[\begin{array}{cccc}
       0  &   -s^\pm_h\epsilon_\mathrm{crit,h} A_h & 0 & 0 \\
       (\epsilon'_h + s^\pm_h \epsilon_\mathrm{crit,h})/A_h  &  -2 \lambda_h & 2\omega_h N_{hn}A_n & 0 \\
       0 & 0 & 0 & -s^\pm_n\epsilon_\mathrm{crit,n} A_n \\
       2\omega_n N_{nh} A_h & 0 & (\epsilon'_n + s^\pm_n \epsilon_\mathrm{crit,n})/A_n & -2\lambda_n
    \end{array}   \right] \quad. \label{eq: 4x4 P modes Jacobian coupled}
\end{equation}   
\end{widetext}

The condition \eqref{eq:saddle node condition} for saddle-node type bifurcations can be imposed by computing the determinant performing Laplace development around the only nonzero coefficient in the first row and recursively doing the same around the coefficient of indices $(2,3)$ of the $3\times 3$ cofactor. The calculation again yields the expression of the dashed black $c_{++}\equiv u_n$ and dashed red $c_{++}\equiv u_h$ lines.
In this respect, one could have inferred such result in an alternative way by considering the above saddle-node type bifurcation (precisely, it is a transcritical bifurcation\cite{Perko_book}) as 'collision' between the coupled P-mode $c_{++}$ given by \eqref{eq:Pmode coup} with either uncoupled P-mode $u_h$ given \eqref{eq:Pmode unc h} or $u_n$ given by \eqref{eq:Pmode unc n} as originally done for the derivation of eqs.\eqref{eq:line ah_coup_eq_zero},\eqref{eq:line an_coup_eq_zero}.

The stability of given $c_{++}$ solutions can be quantitatively studied by eigenvalue analysis for the Jacobian $J$ expressed by eq.\eqref{eq: 4x4 P modes Jacobian coupled} as mentioned before. Nevertheless, some general guidelines can be obtained exploiting the continuity condition between uncoupled  $u_h,u_n$ and coupled regimes, namely considering P-modes of $c_{++}$ type lying sufficiently close to the critical lines $c_{++}\equiv u_h$ or $c_{++}\equiv u_n$ that are associated with saddle-node bifurcations implying $\det(J)=0$ according to eq.\eqref{eq:saddle node condition} (at least one eigenvalue $\zeta_i, i=1,\ldots,4$ is zero). 

To this end, we recall the Routh-Hurwitz stability criterion\cite{Perko_book}. In this respect, we express the fourth-order characteristic polynomial $P(\zeta)$ of $J$ as
\begin{align}
    P(\zeta)&=(\zeta-\zeta_1)(\zeta-\zeta_2)(\zeta-\zeta_3)(\zeta-\zeta_4)= \\
    =&\zeta^4-S_1\zeta^3+S_2\zeta^2-S_3\zeta+S_4 \quad, \label{eq:characteristic polynomial J coup}
\end{align}
where $S_1,S_2,S_3,S_4$ defined as
\begin{align}
    S_1&=\zeta_1+\zeta_2+\zeta_3+\zeta_4= \text{tr}(J)=-(-j_{22}-j_{44}) \quad,\\    S_2&=\zeta_1\zeta_2+\zeta_1\zeta_3+\zeta_1\zeta_4+\zeta_2\zeta_3+\zeta_2\zeta_4+\zeta_3\zeta_4 = \\
    &=j_{22}j_{44}-j_{12}j_{21}-j_{34}j_{43}\quad,\\
    S_3&=\zeta_1\zeta_2\zeta_3+\zeta_1\zeta_2\zeta_4+\zeta_1\zeta_3\zeta_4+\zeta_2\zeta_3\zeta_4 = \\
    &=-(j_{12}j_{21}j_{44}+j_{22}j_{34}j_{43})\quad, \\
    S_4&=\zeta_1\zeta_2\zeta_3\zeta_4=\det(J) = j_{12}j_{21}j_{34}j_{43}-j_{12}j_{34} j_{23} j_{41}\quad,
\end{align}
are the sums of the principal minors of order one, two, three and four of the matrix $J$, while $\zeta_1,\ldots,\zeta_4$ are the eigenvalues of $J$.

Among various technical conditions, the Routh-Hurwitz criterion states that a necessary condition for the stability of P-mode $c_{++}$ is that all the coefficients of the polynomial are positive. Then, if one analyzes P-modes close to saddle-node critical lines $c_{++}\equiv u_h$ or $c_{++}\equiv u_n$, the instability will be mainly controlled by the zero-order term of the characteristic polynomial $S_4=\det(J)$. For this reason, we compute the determinant of matrix \eqref{eq: 4x4 P modes Jacobian coupled} using the Schur formula for block-matrices $\det(J)=S_4=\det(J_{11})\det(J_{22}-J_{21}J_{11}^{-1} J_{12})$ and impose the negativeness:
\begin{equation}
    S_4=j_{12}j_{21}j_{34}j_{43}-j_{12}j_{34} j_{23} j_{41} = j_{12}j_{34}(j_{21}j_{43}-j_{23}j_{41}) < 0 \,.
\end{equation}
By using the expression \eqref{eq: 4x4 P modes Jacobian coupled} of the Jacobian $J$, one immediately identifies the two factors:
\begin{align}
    j_{12}j_{34}&= (s^+_h\epsilon_\mathrm{crit,h} A_h) (s^+_n\epsilon_\mathrm{crit,n} A_n) \quad,\\
    j_{21}j_{43}-j_{23}j_{41}&= (\epsilon'_h + s^+_h \epsilon_\mathrm{crit,h})/A_h \, (\epsilon'_n + s^+_n \epsilon_\mathrm{crit,n})/A_n - \\
    &-( 2\omega_h N_{hn}A_n ) (2\omega_n N_{nh} A_h ) \quad.
\end{align} 
We now observe that the sign of the former factor is controlled by the product $s^+_h s^+_n$ (the other factors are nonnegative). Moreover, by using eqs.\eqref{eq:saturation amplitude mode h type 1}-\eqref{eq:saturation amplitude mode n type 1} in the latter equation to express $\epsilon'_h + s^+_h \epsilon_\mathrm{crit,h}=2\omega_h N_{hh} A_h^2$ and $\epsilon'_n + s^+_n \epsilon_\mathrm{crit,n}=2\omega_n N_{nn} A_n^2$, one obtains that the sign of the second factor is controlled by the quantity $N_{hh}N_{nn}-N_{hn}^2$. By combining these two results, one can derive the instability condition:
\begin{equation}
    s^+_h s^+_n (N_{hh}N_{nn}-N_{hn}^2) <0 \quad, \label{eq:RH stability close to saddle node}
\end{equation}
that is remarkably simple and effective to show that, for any coupled regime $c_{++}$ sufficiently close to the boundaries of its existence region $T_{++}$, the instability is controlled by the values of the NFS coefficients.

Thus, one can conclude that P-modes $c_{++}$ are unstable in the following conditions:
\begin{quote}
    $c_{++}$  is unstable close to $c_{++}\equiv u_h$ or $c_{++}\equiv u_n$ in $T_{++}$  if  $s^+_h s^+_n (N_{hh}N_{nn}-N_{hn}^2) <0$.
\end{quote}

Following the same line of reasoning considering $c_{+-},c_{-+},c_{--}$ P-modes lying close enough to the related pairs of lines expressing continuity  with $u_h,u_n$, one can see that, in their regions of existence, the aforementioned coupled regimes behave as follows: 
\begin{quote}
    $c_{+-}$ are unstable close to line $c_{+-}\equiv u_h$ for excitations in $T_{+-}$
     if  $s^+_h s^-_n (N_{hh}N_{nn}-N_{hn}^2) <0$. \\
    $c_{-+}$ are unstable close to line $c_{-+}\equiv u_n$ for excitations in 
    $T_{-+}$ if  $s^-_h s^+_n (N_{hh}N_{nn}-N_{hn}^2) <0$.   \\
    $c_{--}$ are never stable. 
\end{quote}
The above statements deserve a comment. In fact, they arise from the fact that the continuity conditions corresponding to stable uncoupled regimes are only $c_{+-}\equiv u_h$ and $c_{-+}\equiv u_n$, respectively. These lines correspond to a single side of the associated triangular region $T_{+-}$ (resp $T_{-+}$) where coupled regimes $c_{+-}$ (resp. $c_{-+}$) exist. As a consequence, considering the coupled regime $c_{--}$, the critical lines $c_{--}\equiv u_h, c_{--}\equiv u_n$ intersecting at the vertex $O_{--}$ would imply the continuity condition between the coupled mode $c_{--}$ and unstable uncoupled modes $A_\mathrm{h-,unc}^2$ and $A_\mathrm{n-,unc}^2$ that are never stable under any excitation condition. Finally, we notice that instability conditions 
for coupled regimes $c_{+-},c_{-+}$ are coincident since $s^+_h s^-_n=s^-_h s^+_n$.

It is also important to remark that the fulfillment of the above conditions guarantees the instability of the considered P-modes, but the converse is not true since they have been derived from a necessary condition for stability. 

A sufficient condition for proving the stability must occur through appropriate eigenvalue analysis. 

Nevertheless, for P-modes sufficiently close to the critical saddle-node bifurcation lines, which means that one eigenvalue is real and has amplitude sufficiently small $|\zeta|\ll 1$, one can neglect terms of order higher than one in $P(\zeta)$, thus also checking the sign of the coefficient $-S_3$ related with that of $S_4$. The coefficient $-S_3$ can be written as:
\begin{widetext}
\begin{align}
    -S_3=j_{44}j_{12}j_{21}+j_{22}j_{34}j_{43} &= 2\lambda_n s^\pm_h\epsilon_\mathrm{crit,h}A_h (2\omega_h N_{hh} A_h) + 2\lambda_h s^\pm_n\epsilon_\mathrm{crit,n}A_n (2\omega_n N_{nn} A_n) \nonumber \\
    &= 4\lambda_n s^\pm_h\epsilon_\mathrm{crit,h} \omega_h N_{hh} A_h^2 + 4\lambda_h s^\pm_n\epsilon_\mathrm{crit,n} \omega_n N_{nn} A_n^2 \nonumber \\
    &= \pm4\lambda_n \epsilon_\mathrm{crit,h} \omega_h |N_{hh}| A_h^2 \pm 4\lambda_h \epsilon_\mathrm{crit,n} \omega_n |N_{nn}| A_n^2\quad,
\end{align}
\end{widetext}
where the equality $s^\pm_i N_{ii}=\pm |N_{ii}|$ has been used.
From the latter equation, it is easy to infer that the sign of this coefficient $-S_3$ is positive close to the line $c_{+\pm}\equiv u_h$ (where $A_h>0$ and $A_n\approx 0$) and $c_{\pm+}\equiv u_n$ (where $A_n>0$ and $A_h\approx 0$). 

By combining the latter result with the former analysis for $S_4$, one obtains the following stability conditions for coupled regimes $c_{\pm\pm}$:
\begin{align}
    c_{++}& \text{ is stable close to $c_{++}\equiv u_h$ or $c_{++}\equiv u_n$ in $T_{++}$}\\ &\text{ if } s^+_h s^+_n (N_{hh}N_{nn}-N_{hn}^2) >0 \quad, \label{eq:c++ stable condition} 
\end{align}
\begin{align}
    c_{+-}& \text{ are stable close to  $c_{+-}\equiv u_h$ for excitations in $T_{+-}$}\\ &\text{ if } s^+_h s^-_n (N_{hh}N_{nn}-N_{hn}^2) >0 \quad, \label{eq:c+- stable condition} 
\end{align}
\begin{align}
    c_{-+}& \text{ are stable close to  $c_{-+}\equiv u_n$ for excitations in 
    $T_{-+}$}\\\ &\text{ if } s^-_h s^+_n (N_{hh}N_{nn}-N_{hn}^2) >0 \quad. \label{eq:c-+ stable condition}  
\end{align}
    
Up to this point, we have considered excitation conditions for coupled modes $c_{\pm\pm}$ (points within the blue triangles with vertices $O_{\pm\pm}$ in fig.\ref{fig:type1 phase diagram and hysteresis})  in the vicinity of the critical points $O_{\pm\pm}$ and of the critical lines $c_{\pm\pm}\equiv u_h$  and $c_{\pm\pm}\equiv u_n$. 

Now we relax this assumption and search for other types of co-dimension one bifurcations\cite{Perko_book} that may occur for generic coupled modes $c_{\pm\pm}$ everywhere in the corresponding blue triangular regions $T_{\pm\pm}$ with vertices $O_{\pm\pm}$.

A separate treatment is deserved for Hopf bifurcations that are only possible for coupled P-modes $c_{\pm\pm}$ given by \eqref{eq:Pmode coup} that have both $2\times 2$ off-diagonal blocks being nonzero in the Jacobian matrix $J$. In this respect, a supercritical Hopf bifurcation\cite{Perko_book} produces the change of stability of a stable coupled P-mode that becomes unstable while a stable limit cycle (termed Q-mode) appears in the (four-dimensional) state space. We observe that a limit cycle for the averaged dynamical eqs.\eqref{eq:averaged two nonlinear parametric Ah}-\eqref{eq:averaged two nonlinear parametric Phi_n} corresponds to a quasiperiodic signal that, for each mode, combines two frequencies. Specifically, for mode $h$ the Q-mode combines the applied tone frequency $\omega_\mathrm{rf1}/2$ with the (typically lower) frequency of the limit cycle, giving rise to low-frequency modulation. Analogous reasoning holds for mode $n$ that combines the applied tone frequency $\omega_\mathrm{rf2}/2$ with the (typically lower) frequency of the limit cycle.

We now derive the critical condition that leads to this bifurcation.

By using the notations introduced with eq.\eqref{eq:characteristic polynomial J coup}, the Hopf bifurcation condition, namely that the real part of two complex conjugate eigenvalues with nonzero imaginary part crosses zero, reads as\cite{kuznetsov2012,bosi2019local}:
\begin{equation}
 \left\{\begin{array}{c}
     S_2=\dfrac{S_3}{S_1}+\dfrac{S_4S_1}{S_3} \quad,  \\
       S_1 S_3>0 \quad.
 \end{array}   \right. \label{eq:Hopf bifurcation condition coup}
\end{equation}
The above condition can be imposed as function of the control parameters and leads to an additional critical curve in the phase diagram that can only lie within the (blue) region where coupled P-modes exist.

In generic conditions, after the above Hopf bifurcation occurs giving rise to a stable Q-mode, further change of the control parameters may lead this Q-mode to enlarge up to the point where it intersects a saddle P-mode (more precisely, one of the invariant manifolds departing from the saddle). This gives rise to a (global) homoclinic (saddle-connection) bifurcation\cite{Kuznetsov2023} that annihilates the Q-mode letting the system relax towards certain stable P-mode. This process may produce additional hysteretic behavior in the amplitude response.  
Finding analytical formulas for homoclinic bifurcations is generally very hard owing to their global character that prevents using linearization methods as is done for local bifurcations. This difficulty is even more pronounced when the dimensionality of the state space is larger than two. Thus, the critical condition for its occurrence is generally found using numerical techniques. In section \ref{sec:numerical results}, we quantitatively address examples of mode pairs interactions involving both P- and Q-modes. 

An important last consideration concerns the difference between the qualitative behavior of P-modes of type $c_{++}$ and P-modes of type $c_{+-},c_{-+}$ close to the vertices $O_{++}$ and $O_{+-},O_{-+}$, respectively. We have shown that, when $c_{++}$ P-modes are stable (condition \eqref{eq:c++ stable condition} is met), the point $O_{++}$ corresponds to the intersection of multiple critical lines $z\equiv u_h\equiv u_n \equiv c_{++}$ where all P-modes coincide with the zero P-mode $z$, which determines a degenerate situation where a unique stable P-mode exists. 

Let us now consider the critical points $O_{+-},O_{-+}$ and the associated triangular regions $T_{+-},T_{-+}$ associated with stable $c_{+-},c_{-+}$ regimes (condition \eqref{eq:c+- stable condition} or \eqref{eq:c-+ stable condition} is met). 
The first vertex $O_{+-}$ corresponds to conditions $z\equiv u_h,c_{+-}\equiv u_h$ but also to a nonzero stable P-mode $u_n$, while the second $O_{-+}$ corresponds to conditions $z\equiv u_n,c_{-+}\equiv u_n$ but also to a nonzero stable P-mode $u_h$. This means that the latter situations are not degenerate because in the neighborhood of $O_{+-},O_{-+}$ there can exist two different stable P-modes. In this situation, it can be shown that the codimension-two Bogdanov-Takens bifurcation occurs\cite{Kuznetsov2023}, namely one has two zero eigenvalues of the Jacobian matrix $J$ at points $O_{+-},O_{-+}$ in the presence of non-degenerate P-modes. 
This bifurcation implies that a Hopf bifurcation curve and a homoclinic bifurcation curve must originate from points $O_{+-},O_{-+}$ and must lie within the associated (blue) triangular regions $T_{+-},T_{-+}$. 
By using the notations for the principal minors of the Jacobian matrix $J$ introduced above, the condition for the realization of Bogdanov-Takens bifurcation is:
\begin{equation}
    S_3=S_4=0 \quad.
\end{equation}
The latter condition corresponds to a double zero eigenvalue of the matrix $J$, which can only happen in the vertices $O_{++},O_{+-},O_{-+}$. Excluding the degeneracy associated with $O_{++}$, this bifurcation will produce effects only when stable regimes of type $c_{+-},c_{-+}$ exist.

The results of the stability analysis carried out here are summarized in table \ref{tab:type1 interactions}.
The above considerations have general validity regardless of the particular sign and value of the NFS coefficients, which only determine the slopes of the critical lines as outlined before. Thus, the partitioning of the phase diagram will be different depending on the combination of signs and values.

\clearpage
\widetext
\begin{center}
\textbf{\large Supplementary Material \vspace{0.5cm}\\
Nonlinear interaction theory for parametrically-excited spin-wave modes in confined micromagnetic systems} \\
Massimiliano d'Aquino, Salvatore Perna, Hugo Merbouche, Grégoire de Loubens
\end{center}
\setcounter{equation}{0}
\setcounter{figure}{0}
\setcounter{table}{0}
\setcounter{page}{1}
\setcounter{section}{0}
\makeatletter
\@removefromreset{equation}{section}
\renewcommand\theHsection{S\arabic{section}}
\renewcommand\theHtable{S\arabic{table}}

\makeatother
\renewcommand{\theequation}{S\arabic{equation}}
\renewcommand{\thesection}{S\arabic{section}}
\renewcommand{\thesubsection}{\Alph{subsection}}
\renewcommand{\thefigure}{S\arabic{figure}}
\renewcommand{\thetable}{S\arabic{table}}
\renewcommand{\bibnumfmt}[1]{[S#1]}
\renewcommand{\citenumfont}[1]{S#1}

\widetext
\section{Some useful general results involving eigenmodes}

Let us consider two complex vector fields $\bm v(\bm r),\bm w(\bm r)\in\mathbb{C}^3$ lying in the plane perpendicular to $\bm m_0(\bm r)$. Then, the following triple product can be expressed as:
\begin{equation}\label{eq:triple product m0 v w}
    \bm v\cdot\bm m_0\times\bm w = (v_1\bm e_1+v_2\bm e_2)\cdot \bm m_0\times (w_1\bm e_1 + w_2\bm e_2) = (v_1\bm e_1+v_2\bm e_2)\cdot(w_1\bm e_2-w_2\bm e_1) = v_2 w_1-v_1 w_2 = \bm m_0\cdot(\bm w\times\bm v)\,.
\end{equation}
Let us now consider the eigenmodes $\bm\varphi_h^*,\bm\varphi_n$. Equation \eqref{eq:triple product m0 v w} yields:
\begin{align}
    \bm \varphi_h^*\cdot\bm m_0\times\bm \varphi_n = \bm m_0\cdot\bm \varphi_n\times\bm \varphi_h^* = \varphi_{h2}^* \varphi_{n1}-\varphi_{h1}^* \varphi_{n2} \quad. \label{eq:triple product phi_h* phi_n}
\end{align}
When considering the same eigenmode $h=n$, the latter equation ends up in:
\begin{equation}
    \bm \varphi_h^*\cdot\bm m_0\times\bm \varphi_h = \bm m_0\cdot\bm \varphi_h\times\bm \varphi_h^* = \varphi_{h2}^* \varphi_{h1}-\varphi_{h1}^* \varphi_{h2}= \varphi_{h2}^* \varphi_{h1}-(\varphi_{h2}^* \varphi_{h1})^*=2 j\imag{\varphi_{h2}^* \varphi_{h1}}\quad. \label{eq:triple product phi_h phi_h}
\end{equation}
When $n=-h$, that means $\bm\varphi_n=\bm\varphi_h^*$ in eq.\eqref{eq:triple product phi_h* phi_n}, one has:
\begin{align}
    \bm \varphi_h^*\cdot\bm m_0\times\bm \varphi_h^* = \bm m_0\cdot\bm \varphi_h^*\times\bm \varphi_h^* = \varphi_{h2}^* \varphi_{h1}^*-\varphi_{h1}^* \varphi_{h2}^*=0 \quad. \label{eq:triple product phi_h phi_h*}
\end{align}
Let us now consider the following scalar product:
\begin{equation}
    (\bm v, f \mathcal{C}\bm w) \quad, \label{eq:inner product C}
\end{equation}
where $\bm v(\bm r), \bm w(\bm r)$ are complex vector fields point-wise perpendicular to $\bm m_0(\bm r)$ and $f(\bm r)$ is a generic complex scalar field. We recall that any complex vector field transverse to $\bm m_0(\bm r)$ can be represented in the complete basis of eigenmodes $\bm\varphi_h$. 

It can be readily proved that the above inner product \eqref{eq:inner product C} depends only on the component of the effective field that is transverse to $\bm m_0(\bm r)$. In fact, expressing $\bm v=v_1\bm e_1+v_2\bm e_2$ and $\bm w=w_1\bm e_1+w_2\bm e_2$, one has:
\begin{align}
    \mathcal{C}\bm w&=(\mathcal{C}\bm w\cdot \bm e_1)\bm e_1 + (\mathcal{C}\bm w\cdot \bm e_2)\bm e_2 + (\mathcal{C}\bm w\cdot \bm m_0)\bm m_0 \quad,  \\
     (\bm v, f \mathcal{C}\bm w)&=(v_1\bm e_1+v_2\bm e_2, f \left[(\mathcal{C}\bm w\cdot \bm e_1)\bm e_1 + (\mathcal{C}\bm w\cdot \bm e_2)\bm e_2 + (\mathcal{C}\bm w\cdot \bm m_0)\bm m_0\right] )= \nonumber \\
      (\bm v, f \mathcal{C}\bm w)&=(v_1\bm e_1+v_2\bm e_2, f \left[(\mathcal{C}\bm w\cdot \bm e_1)\bm e_1 + (\mathcal{C}\bm w\cdot \bm e_2)\bm e_2 \right] ).    
\end{align}
which shows that the term parallel to $\bm m_0$ gives no contribution to the scalar product due to the geometric orthogonality between unit-vectors of the local reference triple $(\bm e_1,\bm e_2,\bm m_0)$.
As a consequence, in inner product terms of the form \eqref{eq:inner product C}, one is allowed to use the substitution
\begin{equation}
    \mathcal{C}\rightarrow \mathcal{A}_{0\perp}-h_0\mathcal{I} \quad,
\end{equation}
which allows to leverage all the properties of eigenmodes and, in particular, that of transforming the non-local operator $\mathcal{A}_{0\perp}$ into a local operator $-j\omega_h\bm m_0\times$ when acting on the eigenmode $\bm\varphi_h$. 

\section{Proof of reality of self-NFS for mode $h$} \label{sec:appendix NFS real}

Now we prove that $\tilde{N}_h$ is purely real. To this end, given the expression \eqref{eq:NMM coef d} 
\begin{equation*}
    d_{hijk} =-\left(\bm\varphi_h, \psi_{jk}\,\mathcal{C}\bm\varphi_i\right) + \left(\bm\varphi_h,\bm m_0\cdot\mathcal{C}\psi_{jk}\bm m_0\,\bm \varphi_i \right) \quad,\quad \psi_{jk} = \bm\varphi_j\cdot\bm\varphi_k \quad,
\end{equation*}
and the symmetry of the tensor $d_{hijk}$ with respect to the exchange of the indices $j,k$, the NFS coefficient \eqref{eq:NFS mode h} is reduced to:
\begin{equation}
    \tilde{N}_h= d_{h,h,-h,h} + \frac{d_{h,-h,h,h}}{2} \quad.   \label{eq:NSFS appendix}
\end{equation}
The first term of the sum in eq.\eqref{eq:NSFS appendix} is:
\begin{align}
    d_{h,h,-h,h}&= -\left(\bm\varphi_h, |\bm\varphi_h|^2\,\mathcal{C}\bm\varphi_h\right) + \left(\bm\varphi_h,\bm m_0\cdot\mathcal{C}|\bm\varphi_h|^2\bm m_0\,\bm \varphi_h \right) = \nonumber \\
    &=-\left(\bm\varphi_h, |\bm\varphi_h|^2\,(\mathcal{A}_{0\perp}-h_0\mathcal{I})\bm\varphi_h\right) + \left(\bm\varphi_h,\bm m_0\cdot\mathcal{C}|\bm\varphi_h|^2\bm m_0\,\bm \varphi_h \right) \nonumber \\
    &=-\frac{1}{V}\int_\Omega \left[|\bm\varphi_h|^2 \bm\varphi_h^*\cdot \mathcal{A}_{0\perp}\bm\varphi_h -h_0|\bm\varphi_h^*\cdot\bm\varphi_h|^2 - |\bm\varphi_h|^2 \bm m_0\cdot\mathcal{C}|\bm\varphi_h|^2\bm m_0 \right]\, dV = \nonumber \\
    &=-\frac{1}{V}\int_\Omega \left[|\bm\varphi_h|^2 \bm\varphi_h^*\cdot \mathcal{A}_{0\perp}\bm\varphi_h -h_0|\bm\varphi_h^*\cdot\bm\varphi_h|^2 - |\bm\varphi_h|^2 \bm m_0\cdot\mathcal{C}|\bm\varphi_h|^2\bm m_0 \right]\, dV = \nonumber\\
&=-\frac{1}{V}\int_\Omega \left[|\bm\varphi_h|^2 (-j\omega_h  \bm m_0)\cdot\bm\varphi_h\times \bm\varphi_h^* -h_0|\bm\varphi_h^*\cdot\bm\varphi_h|^2 - |\bm\varphi_h|^2 \bm m_0\cdot\mathcal{C}|\bm\varphi_h|^2\bm m_0 \right]\, dV \quad,
\end{align}
where the integrand is apparently a sum of real scalar fields, due to eq.\eqref{eq:triple product phi_h phi_h} and to the reality of the effective field operator $\mathcal{C}$.
The second coefficient in the sum in eq.\eqref{eq:NSFS appendix} is:
\begin{align}
    d_{h,-h,h,h}&= -\left(\bm\varphi_h, \bm\varphi_h^2\,\mathcal{C}\bm\varphi_h^*\right) + \left(\bm\varphi_h,\bm m_0\cdot\mathcal{C}\bm\varphi_h^2\bm m_0\,\bm \varphi_h^* \right) = \nonumber \\
&=-\frac{1}{V}\int_\Omega \left[\bm\varphi_h^*\cdot \bm\varphi_h^2 \mathcal{A}_{0\perp}\bm\varphi_h^*  -h_0 \bm\varphi_h^2 {\bm\varphi_h^*}^2 - \bm\varphi_h^*\cdot \bm\varphi_h^* \bm m_0\cdot\mathcal{C}\bm\varphi_h^2\bm m_0 \right]\, dV = \nonumber \\
&=-\frac{1}{V}\int_\Omega \left[\bm\varphi_h^*\cdot j\omega_h  \bm\varphi_h^2 \bm m_0\times \bm\varphi_h^*  -h_0 \bm\varphi_h^2 {\bm\varphi_h^*}^2 - \bm\varphi_h^*\cdot \bm\varphi_h^* \bm m_0\cdot\mathcal{C}\bm\varphi_h^2\bm m_0 \right]\, dV = \nonumber\\
&=-\frac{1}{V}\int_\Omega \left[j\omega_h  \bm\varphi_h^2\bm m_0\cdot \bm\varphi_h^* \times \bm\varphi_h^*  -h_0 \bm\varphi_h^2 {\bm\varphi_h^*}^2 - (\bm\varphi_h^2 \bm m_0)^*\cdot\mathcal{C}\bm\varphi_h^2\bm m_0 \right]\, dV \quad,
\end{align}
where the first term is zero due to eq.\eqref{eq:triple product phi_h phi_h*}, the second term is clearly real, and the third term is real when integrated over the magnetic body owing to the symmetry of the effective field operator $\mathcal{C}$. $ \blacksquare $

Finally, we report the complete analytical expression for the self-NFS coefficient obtained combining the above expressions:
\begin{equation}
    N_h= -\frac{1}{V}\!\!\!\int_\Omega \!\!\!\left\{\left[|\bm\varphi_h|^2 (-j\omega_h  \bm m_0)\cdot\bm\varphi_h\times \bm\varphi_h^* -h_0|\bm\varphi_h^*\cdot\bm\varphi_h|^2 - |\bm\varphi_h|^2 \bm m_0\cdot\mathcal{C}|\bm\varphi_h|^2\bm m_0 \right] -\frac{1}{2} \left[h_0 \bm\varphi_h^2 {\bm\varphi_h^*}^2 + (\bm\varphi_h^2 \bm m_0)^*\cdot\mathcal{C}\bm\varphi_h^2\bm m_0 \right]\right\} dV \,. \label{eq:self-NFS analytical}
\end{equation}

\section{Proof of reality for mutual NFS coefficients}
\label{sec:appendix mutual NFS real}

We prove here that the (complex) mutual NFS $\tilde{N}_{hn}$ are purely real numbers. To this end, let us recall the definition through eq.\eqref{eq:complex NMFS matrix} (which appropriately exploits the symmetry of $d_{hijk}$ w.r.t exchange of $j,k$):
\begin{equation}
    \tilde{N}_{hn}=d_{h,n,h,-n} + d_{h,h,n,-n}+ d_{h,-n,h,n} \quad, \label{eq:NMFS appendix}
\end{equation}
as well as the expression \eqref{eq:NMM coef d} 
\begin{equation*}
    d_{hijk} =-\left(\bm\varphi_h, \psi_{jk}\,\mathcal{C}\bm\varphi_i\right) + \left(\bm\varphi_h,\bm m_0\cdot\mathcal{C}\psi_{jk}\bm m_0\,\bm \varphi_i \right) \quad,\quad \psi_{jk} = \bm\varphi_j\cdot\bm\varphi_k \quad.
\end{equation*}
The first term of the sum in eq.\eqref{eq:NMFS appendix} is:
\begin{align}
    d_{h,n,h,-n}&= -\left(\bm\varphi_h, \bm\varphi_n^*\cdot\bm\varphi_h\,\mathcal{C}\bm\varphi_n\right) + \left(\bm\varphi_h,\bm m_0\cdot\mathcal{C}\bm\varphi_n^*\cdot\bm\varphi_h\bm m_0\,\bm \varphi_n \right) = \nonumber \\
    &=-\left(\bm\varphi_h, \bm\varphi_n^*\cdot\bm\varphi_h\,(\mathcal{A}_{0\perp}-h_0\mathcal{I})\bm\varphi_n\right) + \left(\bm\varphi_h,\bm m_0\cdot\mathcal{C}\bm\varphi_n^*\cdot\bm\varphi_h\bm m_0\,\bm \varphi_n \right) \nonumber \\
    &=-\frac{1}{V}\int_\Omega \left[\bm\varphi_n^*\cdot\bm\varphi_h \bm\varphi_h^*\cdot \mathcal{A}_{0\perp}\bm\varphi_n -h_0\bm\varphi_n^*\cdot\bm\varphi_h\,\bm\varphi_h^*\cdot\bm\varphi_n - \bm\varphi_h^*\cdot\bm\varphi_n \bm m_0\cdot\mathcal{C}\bm\varphi_n^*\cdot\bm\varphi_h\bm m_0 \right]\, dV = \nonumber \\
    &=-\frac{1}{V}\int_\Omega \left[\bm\varphi_n^*\cdot\bm\varphi_h \bm\varphi_h^*\cdot \mathcal{A}_{0\perp}\bm\varphi_n -h_0\bm\varphi_n^*\cdot\bm\varphi_h\,(\bm\varphi_h\cdot\bm\varphi_n)^* - \bm\varphi_h^*\cdot\bm\varphi_n \bm m_0\cdot\mathcal{C}\bm\varphi_n^*\cdot\bm\varphi_h\bm m_0 \right]\, dV = \nonumber\\
&=-\frac{1}{V}\int_\Omega \left[\bm\varphi_n^*\cdot\bm\varphi_h (-j\omega_n  \bm m_0)\cdot\bm\varphi_n\times \bm\varphi_h^* -h_0|\bm\varphi_n^*\cdot\bm\varphi_h|^2 - (\bm\varphi_h\cdot\bm\varphi_n^* \bm m_0)^*\cdot\mathcal{C}\bm\varphi_n^*\cdot\bm\varphi_h\bm m_0 \right]\, dV \quad, \label{eq:NMFS first d term}
\end{align}
where the second term is evidently real and the third term is also such (when integrated over the volume of $\Omega$) due to the symmetry of the real effective field operator. 
Let us now analyze the first term in the integrand of eq.\eqref{eq:NMFS first d term}:
\begin{align}
    &\bm\varphi_n^*\cdot\bm\varphi_h (-j\omega_n  \bm m_0)\cdot\bm\varphi_n\times \bm\varphi_h^*   \quad.
\end{align}
By expressing the eigenmodes $\bm\varphi_n,\bm\varphi_h$ using the local reference ($\bm e_1,\bm e_2,\bm m_0)$ defined $\forall\bm r\in\Omega$, one has:
\begin{align}
    \bm\varphi_n^*\cdot\bm\varphi_h&=(\varphi_{n1}^*\bm e_1 + \varphi_{n2}^*\bm e_2) \cdot (\varphi_{h1}\bm e_1 + \varphi_{h2}\bm e_2) = \varphi_{n1}^* \varphi_{h1} + \varphi_{n2}^* \varphi_{h2} \quad, \nonumber \\
    \bm\varphi_n\times \bm\varphi_h^*&= (\varphi_{n1}\bm e_1 + \varphi_{n2}\bm e_2) \times (\varphi_{h1}^*\bm e_1 + \varphi_{h2}^*\bm e_2)= (\varphi_{n1}\varphi_{h2}^* -\varphi_{n2} \varphi_{h1}^*)\bm m_0 \quad, \nonumber\\
    \bm\varphi_n^*\cdot\bm\varphi_h \bm m_0\cdot\bm\varphi_n\times \bm\varphi_h^*&=(\varphi_{n1}^* \varphi_{h1} + \varphi_{n2}^* \varphi_{h2}) (\varphi_{n1}\varphi_{h2}^* -\varphi_{n2} \varphi_{h1}^*) = \quad, \nonumber \\
    &=\varphi_{n1}^* \varphi_{h1} \varphi_{n1}\varphi_{h2}^* - \varphi_{n1}^* \varphi_{h1} \varphi_{n2} \varphi_{h1}^* + \varphi_{n2}^* \varphi_{h2} \varphi_{n1}\varphi_{h2}^* - \varphi_{n2}^* \varphi_{h2} \varphi_{n2} \varphi_{h1}^* =  \nonumber\\
      &=|\varphi_{n1}|^2 \varphi_{h1} \varphi_{h2}^* - \varphi_{n1}^* |\varphi_{h1}|^2 \varphi_{n2} + \varphi_{n2}^* |\varphi_{h2}|^2 \varphi_{n1} - |\varphi_{n2}|^2 \varphi_{h2} \varphi_{h1}^*   \quad.  \label{eq:first term first d}
\end{align}
The second term of the sum in eq.\eqref{eq:NMFS appendix} is:
\begin{align}
    d_{h,h,n,-n}&= -\left(\bm\varphi_h, \bm\varphi_n^*\cdot\bm\varphi_n\,\mathcal{C}\bm\varphi_h\right) + \left(\bm\varphi_h,\bm m_0\cdot\mathcal{C}\bm\varphi_n^*\cdot\bm\varphi_n\bm m_0\,\bm \varphi_h \right) = \nonumber \\
    &=-\left(\bm\varphi_h, |\bm\varphi_n|^2\,(\mathcal{A}_{0\perp}-h_0\mathcal{I})\bm\varphi_h\right) + \left(\bm\varphi_h,\bm m_0\cdot\mathcal{C}|\bm\varphi_n|^2\bm m_0\,\bm \varphi_h \right) \nonumber \\
    &=-\frac{1}{V}\int_\Omega \left[|\bm\varphi_n|^2 \bm\varphi_h^*\cdot \mathcal{A}_{0\perp}\bm\varphi_h -h_0|\bm\varphi_n|^2\,\bm\varphi_h^*\cdot\bm\varphi_h - |\bm\varphi_h|^2 \bm m_0\cdot\mathcal{C}|\bm\varphi_n|^2\bm m_0 \right]\, dV = \nonumber \\
&=-\frac{1}{V}\int_\Omega \left[|\bm\varphi_n|^2 (-j\omega_h  \bm m_0)\cdot\bm\varphi_h\times \bm\varphi_h^* -h_0|\bm\varphi_h|^2|\bm\varphi_n|^2 - |\bm\varphi_h|^2 \bm m_0 \cdot\mathcal{C}|\bm\varphi_n|^2\bm m_0 \right]\, dV \quad, \label{eq:NMFS second d term}
\end{align}
where all terms are real due to the same arguments used before. 
Finally, the third term of the sum in eq.\eqref{eq:NMFS appendix} is:
\begin{align}
    d_{h,-n,h,n}&= -\left(\bm\varphi_h, \bm\varphi_n\cdot\bm\varphi_h\,\mathcal{C}\bm\varphi_n^*\right) + \left(\bm\varphi_h,\bm m_0\cdot\mathcal{C}\bm\varphi_n\cdot\bm\varphi_h\bm m_0\,\bm \varphi_n^* \right) = \nonumber \\
    &=-\left(\bm\varphi_h, \bm\varphi_n\cdot\bm\varphi_h\,(\mathcal{A}_{0\perp}-h_0\mathcal{I})\bm\varphi_n^*\right) + \left(\bm\varphi_h,\bm m_0\cdot\mathcal{C}\bm\varphi_n\cdot\bm\varphi_h\bm m_0\,\bm \varphi_n^* \right) \nonumber \\
    &=-\frac{1}{V}\int_\Omega \left[\bm\varphi_n\cdot\bm\varphi_h \bm\varphi_h^*\cdot \mathcal{A}_{0\perp}\bm\varphi_n^* -h_0\bm\varphi_n\cdot\bm\varphi_h\,\bm\varphi_h^*\cdot\bm\varphi_n^* - \bm\varphi_h^*\cdot\bm\varphi_n^* \bm m_0\cdot\mathcal{C}\bm\varphi_n\cdot\bm\varphi_h\bm m_0 \right]\, dV = \nonumber \\
    &=-\frac{1}{V}\int_\Omega \left[\bm\varphi_n\cdot\bm\varphi_h \bm\varphi_h^*\cdot \mathcal{A}_{0\perp}\bm\varphi_n^* -h_0\bm\varphi_n\cdot\bm\varphi_h\,(\bm\varphi_h\cdot\bm\varphi_n)^* - \bm\varphi_h^*\cdot\bm\varphi_n^* \bm m_0\cdot\mathcal{C}\bm\varphi_n\cdot\bm\varphi_h\bm m_0 \right]\, dV = \nonumber\\
&=-\frac{1}{V}\int_\Omega \left[\bm\varphi_n\cdot\bm\varphi_h (j\omega_n  \bm m_0)\cdot\bm\varphi_n^*\times \bm\varphi_h^* -h_0|\bm\varphi_n\cdot\bm\varphi_h|^2 - (\bm\varphi_h\cdot\bm\varphi_n \bm m_0)^*\cdot\mathcal{C}\bm\varphi_n\cdot\bm\varphi_h\bm m_0 \right]\, dV \quad, \label{eq:NMFS third d term}
\end{align}
where the second and third term are real.
We express the first term in the integrand as:
\begin{align}
         \bm\varphi_n\cdot\bm\varphi_h \bm m_0 \cdot\bm\varphi_n^*\times \bm\varphi_h^* &= (\varphi_{n1} \varphi_{h1} + \varphi_{n2} \varphi_{h2}) (\varphi_{n1}^*\varphi_{h2}^* -\varphi_{n2}^* \varphi_{h1}^*) = \quad, \nonumber \\
         &=|\varphi_{n1}|^2 \varphi_{h1} \varphi_{h2}^* - \varphi_{n1} |\varphi_{h1}|^2 \varphi_{n2}^* + \varphi_{n2} |\varphi_{h2}|^2 \varphi_{n1}^* - |\varphi_{n2}|^2 \varphi_{h2} \varphi_{h1}^*  \quad. \label{eq:first term third d} 
\end{align}
Then, the difference between analogous leading terms in the sum of eqs.\eqref{eq:NMFS first d term},\eqref{eq:NMFS third d term} is from eqs.\eqref{eq:first term first d},\eqref{eq:first term third d}:
\begin{align}
     \bm\varphi_n^*\cdot\bm\varphi_h \bm m_0\cdot\bm\varphi_n\times \bm\varphi_h^* &- (\bm\varphi_n\cdot\bm\varphi_h \bm m_0 \cdot\bm\varphi_n^*\times \bm\varphi_h^*) =  \nonumber \\
      &=|\varphi_{n1}|^2 \varphi_{h1} \varphi_{h2}^* - \varphi_{n1}^* |\varphi_{h1}|^2 \varphi_{n2} + \varphi_{n2}^* |\varphi_{h2}|^2 \varphi_{n1} - |\varphi_{n2}|^2 \varphi_{h2} \varphi_{h1}^* - \nonumber \\
      &-(|\varphi_{n1}|^2 \varphi_{h1} \varphi_{h2}^* - \varphi_{n1} |\varphi_{h1}|^2 \varphi_{n2}^* + \varphi_{n2} |\varphi_{h2}|^2 \varphi_{n1}^* - |\varphi_{n2}|^2 \varphi_{h2} \varphi_{h1}^*) \nonumber = \\
      &= |\varphi_{h1}|^2 [\varphi_{n1}  \varphi_{n2}^*- (\varphi_{n1}  \varphi_{n2}^*)^*] +  |\varphi_{h2}|^2[\varphi_{n2}^* \varphi_{n1} - (\varphi_{n2}^* \varphi_{n1})^*] \nonumber \\
      &=2j \left[|\varphi_{h1}|^2 \imag{\varphi_{n1}  \varphi_{n2}^*} + |\varphi_{h2}|^2 \imag{\varphi_{n2}^* \varphi_{n1}}\right]=2j|\bm\varphi_h|^2\imag{\varphi_{n2}^* \varphi_{n1}} \quad,
\end{align}
which gives a purely imaginary number that, multiplied by $-j\omega_n$,  unconditionally produces a real result.  $ \blacksquare $

\section{Proof of symmetry of mutual NFS coefficients}
\label{sec:appendix symmetry NFS real}

Here we prove that the real coefficients impose reciprocal interactions between mode pairs $h,n$, namely:
\begin{equation}
    N_{hn}=N_{nh} \quad.
\end{equation}
To this end, by using the above results, let us explicitly express the coefficient $N_{hn}$:
\begin{align}
    N_{hn}&=d_{h,n,h,-n} + d_{h,h,n,-n}+ d_{h,-n,h,n} = \nonumber \\
    & -\left(\bm\varphi_h, \bm\varphi_n^*\cdot\bm\varphi_h\,\mathcal{C}\bm\varphi_n\right) + \left(\bm\varphi_h,\bm m_0\cdot\mathcal{C}\bm\varphi_n^*\cdot\bm\varphi_h\bm m_0\,\bm \varphi_n \right) + \nonumber \\
    & -\left(\bm\varphi_h, \bm\varphi_n^*\cdot\bm\varphi_n\,\mathcal{C}\bm\varphi_h\right) + \left(\bm\varphi_h,\bm m_0\cdot\mathcal{C}\bm\varphi_n^*\cdot\bm\varphi_n\bm m_0\,\bm \varphi_h \right) + \nonumber \\
    & -\left(\bm\varphi_h, \bm\varphi_n\cdot\bm\varphi_h\,\mathcal{C}\bm\varphi_n^*\right) + \left(\bm\varphi_h,\bm m_0\cdot\mathcal{C}\bm\varphi_n\cdot\bm\varphi_h\bm m_0\,\bm \varphi_n^* \right) \quad. \label{eq:Nhn}
\end{align}
The coefficient $N_{nh}$ is given by:
\begin{align}
    N_{nh}&=d_{n,h,n,-h} + d_{n,n,h,-h}+ d_{n,-h,n,h} = \nonumber \\
    & -\left(\bm\varphi_n, \bm\varphi_n\cdot\bm\varphi_h^*\,\mathcal{C}\bm\varphi_h\right) + \left(\bm\varphi_n,\bm m_0\cdot\mathcal{C}\bm\varphi_n\cdot\bm\varphi_h^*\bm m_0\,\bm \varphi_h \right) + \nonumber \\
    & -\left(\bm\varphi_n, \bm\varphi_h^*\cdot\bm\varphi_h\,\mathcal{C}\bm\varphi_n\right) + \left(\bm\varphi_n,\bm m_0\cdot\mathcal{C}\bm\varphi_h^*\cdot\bm\varphi_h\bm m_0\,\bm \varphi_n \right) + \nonumber \\
    & -\left(\bm\varphi_n, \bm\varphi_n\cdot\bm\varphi_h\,\mathcal{C}\bm\varphi_h^*\right) + \left(\bm\varphi_n,\bm m_0\cdot\mathcal{C}\bm\varphi_n\cdot\bm\varphi_h\bm m_0\,\bm \varphi_h^* \right) \quad.
\end{align}
Let us write the integrands of the above sums. For the former coefficient $N_{hn}$, one has:
\begin{align}
   &\bm\varphi_n^*\cdot\bm\varphi_h (-j\omega_n  \bm m_0)\cdot\bm\varphi_n\times \bm\varphi_h^* \underline{
   -h_0|\bm\varphi_n^*\cdot\bm\varphi_h|^2 - (\bm\varphi_h\cdot\bm\varphi_n^* \bm m_0)^*\cdot\mathcal{C}\bm\varphi_n^*\cdot\bm\varphi_h\bm m_0} + \nonumber \\
   &|\bm\varphi_n|^2 (-j\omega_h  \bm m_0)\cdot\bm\varphi_h\times \bm\varphi_h^* \underline{
   -h_0|\bm\varphi_h|^2|\bm\varphi_n|^2} - |\bm\varphi_h|^2 \bm m_0 \cdot\mathcal{C}|\bm\varphi_n|^2\bm m_0 + \nonumber \\
   &\bm\varphi_n\cdot\bm\varphi_h (j\omega_n  \bm m_0)\cdot\bm\varphi_n^*\times \bm\varphi_h^* \underline{
   -h_0|\bm\varphi_n\cdot\bm\varphi_h|^2 - (\bm\varphi_h\cdot\bm\varphi_n \bm m_0)^*\cdot\mathcal{C}\bm\varphi_n\cdot\bm\varphi_h\bm m_0} \quad,
\end{align}
whereas for the latter $N_{nh}$ one has:
\begin{align}
  &\bm\varphi_h^*\cdot\bm\varphi_n (-j\omega_h  \bm m_0)\cdot\bm\varphi_h\times \bm\varphi_n^* \underline{
  -h_0|\bm\varphi_h^*\cdot\bm\varphi_n|^2 - (\bm\varphi_n\cdot\bm\varphi_h^* \bm m_0)^*\cdot\mathcal{C}\bm\varphi_h^*\cdot\bm\varphi_n\bm m_0 } + \nonumber \\
   &|\bm\varphi_h|^2 (-j\omega_n  \bm m_0)\cdot\bm\varphi_n\times \bm\varphi_n^* \underline{
   -h_0|\bm\varphi_n|^2|\bm\varphi_h|^2 } - |\bm\varphi_n|^2 \bm m_0 \cdot\mathcal{C}|\bm\varphi_h|^2\bm m_0 + \nonumber \\
   &\bm\varphi_h\cdot\bm\varphi_n (j\omega_h  \bm m_0)\cdot\bm\varphi_h^*\times \bm\varphi_n^* \underline{
   -h_0|\bm\varphi_h\cdot\bm\varphi_n|^2 - (\bm\varphi_n\cdot\bm\varphi_h \bm m_0)^*\cdot\mathcal{C}\bm\varphi_h\cdot\bm\varphi_n\bm m_0 } \quad,
\end{align}
where terms underlined
will cancel out (when integrated over $\Omega$) in the difference $N_{hn}-N_{nh}$. 

By using eq.\eqref{eq:triple product phi_h phi_h} for terms multiplying $\bm\varphi_h\times\bm\varphi_h^*$ and $\bm\varphi_n\times\bm\varphi_n^*$, the rest of the difference can be written as:
\begin{align}
    N_{hn}-N_{nh}&= \frac{1}{V}\int_\Omega \Big(-2\omega_n|\bm\varphi_h|^2\imag{\varphi_{n2}^* \varphi_{n1}} +2\omega_h|\bm\varphi_n|^2\imag{\varphi_{h2}^* \varphi_{h1}} + \nonumber \\
    & -|\bm\varphi_n|^2 2\omega_h \imag{\varphi_{h2}^* \varphi_{h1}} +|\bm\varphi_h|^2 \bm m_0 \cdot\mathcal{C}|\bm\varphi_n|^2\bm m_0 + \nonumber \\
    & +|\bm\varphi_h|^2 2\omega_n \imag{\varphi_{n2}^* \varphi_{n1}} - |\bm\varphi_n|^2 \bm m_0 \cdot\mathcal{C}|\bm\varphi_h|^2\bm m_0\Big)\, dV=0 \quad,
\end{align}
where the self-adjointness of the effective field operator $\mathcal{C}$ (with natural boundary conditions for $\bm m_0$ and $\bm\varphi_h,\bm\varphi_n$) has been used. $ \blacksquare $

\clearpage 

\section{Validation of theory with numerical simulations}

Here we report some additional results of the simulations.

\subsection{Macrospin simulations}\label{sec:macrospin simulations}

For the macrospin description, the demagnetizing factors $N_x=N_y=6.341\times 10^{-2}$, $N_z=1-2N_x=0.8732$ and $h_{ax}=0.1697$ (corresponding to 30 mT). For testing and validation, we compare the developed theory with numerical simulations of the full NMM eq.\eqref{eq:NMM} including all coefficients, NMM with only NFS, and full LLG equation. 
We have determined the eigenfrequency of the Kittel mode(s) $\omega_{1,2}=\pm\sqrt{(h_{ax}+N_z-N_x)(h_{ax}+N_y-N_x)}=\pm 0.4077 \rightarrow f_{1,2}=\pm 2.0384\,$GHz.

The computation of the NFS coefficients $N_h$ with $h=1,2$ from eq.\eqref{eq:NFS mode h} and from the compact formula \eqref{eq:macrospin NFS} yields $N_h= -0.9256$. Thus, the Kittel mode has decreasing frequency for increasing amplitude. The ac field is applied with amplitude $\delta h=0.013$ (corresponding to 2.3 mT), which is slightly above the threshold $\delta h_\mathrm{thres,1}=0.0116$ computed from eq.\eqref{eq:threshold field parametric resonance}. 

The comparison between analytical and numerical calculations is reported in fig.\ref{fig:macrospin parametric resonance vs time}, where one can clearly see the excellent accuracy of formula \eqref{eq:saturation amplitude mode h} for the mode steady-state amplitude $|a_h|^2$ (cyan solid line) compared with the results of full NMM (blue), NMM with only NFS (green), macrospin LLG equation (red). Figure \ref{fig:macrospin parametric resonance vs time} also reports in the right panel the steady-state time-evolution of magnetization associated with the transverse component oscillation, that emphasizes how well the amplitude saturation behavior is effectively captured by the NMM using the only NFS coefficient.

\subsection{Micromagnetic simulations of single mode excitation}\label{sec:micromagnetic simulation single tone}

\begin{table}[t]
    \centering
    \begin{tabular}{c|c|c|c|c|c}
     frequency & value [GHz]& NFS  &  value & threshold & value\\
     \hline
     $f_1$ & 2.0263  &$N_1$   &  -0.8159  & $\delta h_\mathrm{thres,1}$ & 0.0115\\
     $f_2$ & 4.0673  &$N_2$   &  0.2853  & $\delta h_\mathrm{thres,2}$ & 0.0388\\
     $f_3$ & 4.5613  &$N_3$   &  -0.3390 & $\delta h_\mathrm{thres,3}$ & 0.0533\\
     $f_4$ & 7.5450  &$N_4$   &  0.1582 & $\delta h_\mathrm{thres,4}$ & 0.1457\\
     $f_5$ & 7.7726  &$N_5$   &  0.1332 &  $\delta h_\mathrm{thres,5}$ &  0.1575
    \end{tabular}
    \caption{Parametric resonance for the 100 nm YIG disk. Values of frequency, NFS coefficients and threshold field amplitudes (normalized by $M_s$, corresponding to 0.1768$\,$T) for the excitation of parametric resonance for modes $h=1,\ldots,5$. For reference, the threshold values refer to a damping parameter $\alpha=0.01$.}
    \label{tab:YIG disk 100nm}
\end{table}

\begin{figure}[t]
    \centering
    \includegraphics[width=5.5cm]{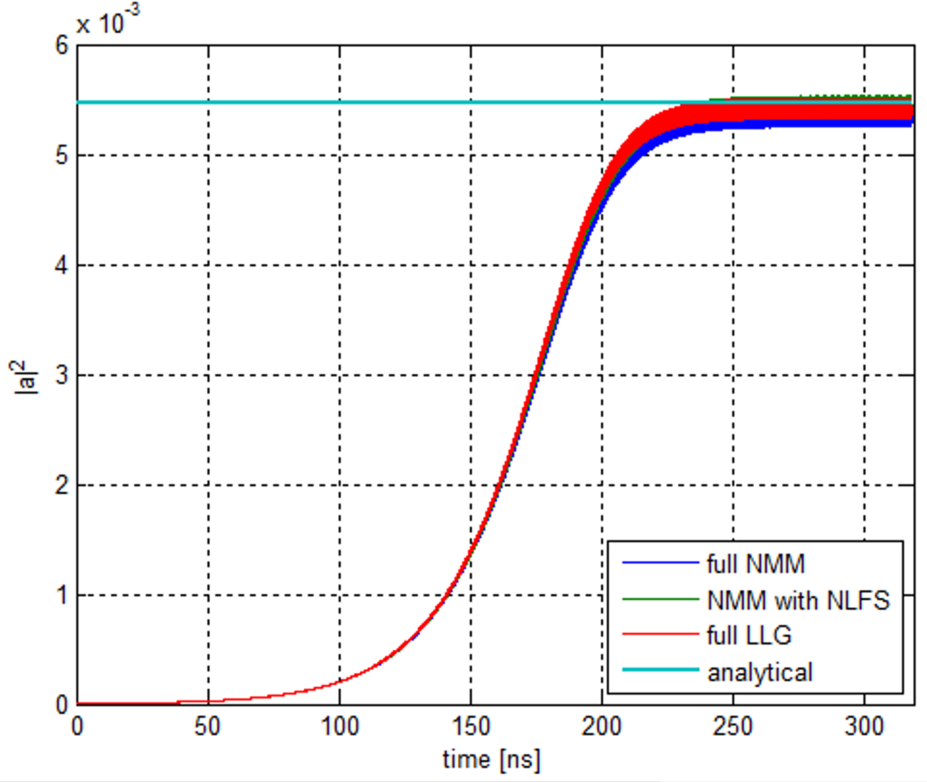} \includegraphics[width=6cm]{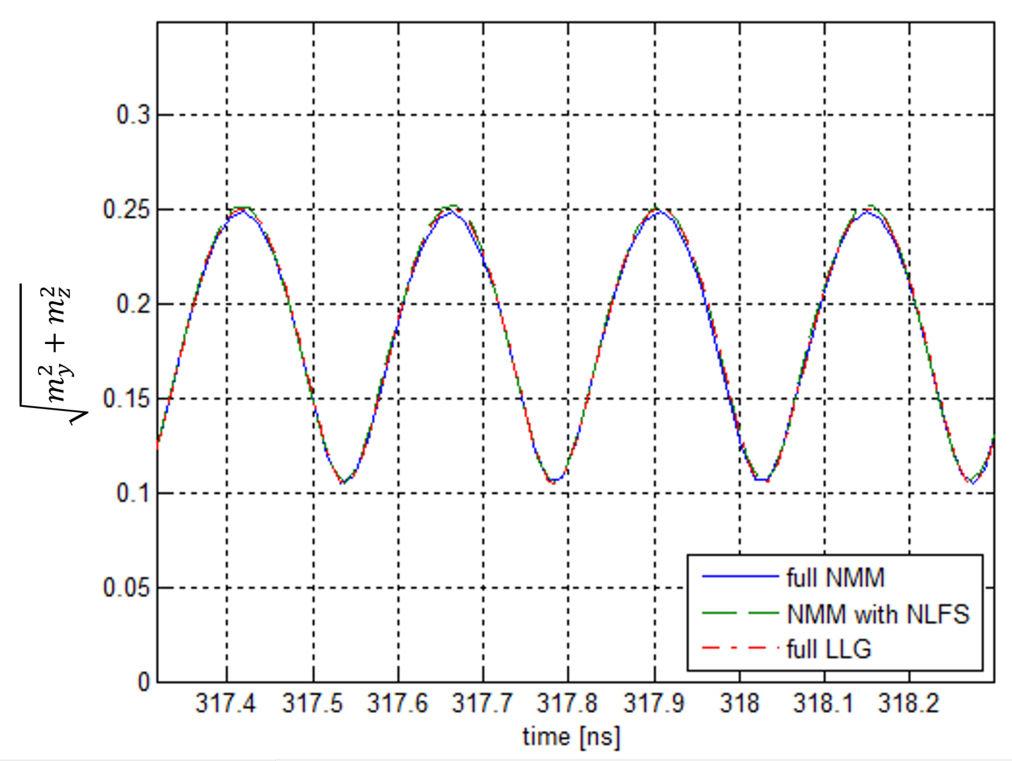}
    \caption{Parametric excitation of magnetization in a macrospin YIG disk. Left panel reports the time-evolution of mode 1 amplitude according to full NMM (solid blue) eq.\eqref{eq:ah NMM}, NMM eq.\eqref{eq:nonlinear parametric ah} with only NFS term (solid green), full macrospin LLG equation (solid red) and analytical formula \eqref{eq:saturation amplitude mode h} (solid cyan). Right panel reports the time evolution of transverse magnetization amplitude $\sqrt{m_y^2+m_z^2}$ for full NMM (solid blue), NMM with NFS (dashed green) and full LLG (dash-dotted red).}
    \label{fig:macrospin parametric resonance vs time}
\end{figure}

The eigenmodes are calculated by using the large-scale formulation described in ref.\cite{dAquino_JAP_2023} implemented in the numerical code MaGICo\cite{MaGICo}. The values of eigenfrequencies, self-NFS coefficients and threshold field amplitudes computed according to eq.\eqref{eq:threshold field parametric resonance} are reported in table \ref{tab:YIG disk 100nm}.

We have performed full micromagnetic simulations of parametric excitation of single modes using the time-domain LLG solver implemented in MaGICo\cite{MaGICo}. We have explored the parametric excitation for mode $h=1$ starting from a configuration $\mvec(\rvec,t=0)=(\bm e_x +\varepsilon\bm e_y)/|\bm e_x +\varepsilon\bm e_y|$, which is mostly aligned with the equilibrium $\mvec_0(\rvec)$ (we used $\varepsilon=0.1$). We recall that an initial deviation from the equilibrium $\mvec_0$ is required to trigger the exponential growth of mode amplitude due to parametric resonance. The ac applied field has frequency $2\times 2.0263\,$GHz and amplitude $\delta h=0.0136$ that corresponds to 2.4 mT.

\begin{figure}[t]
    \centering
    \includegraphics[width=8cm]{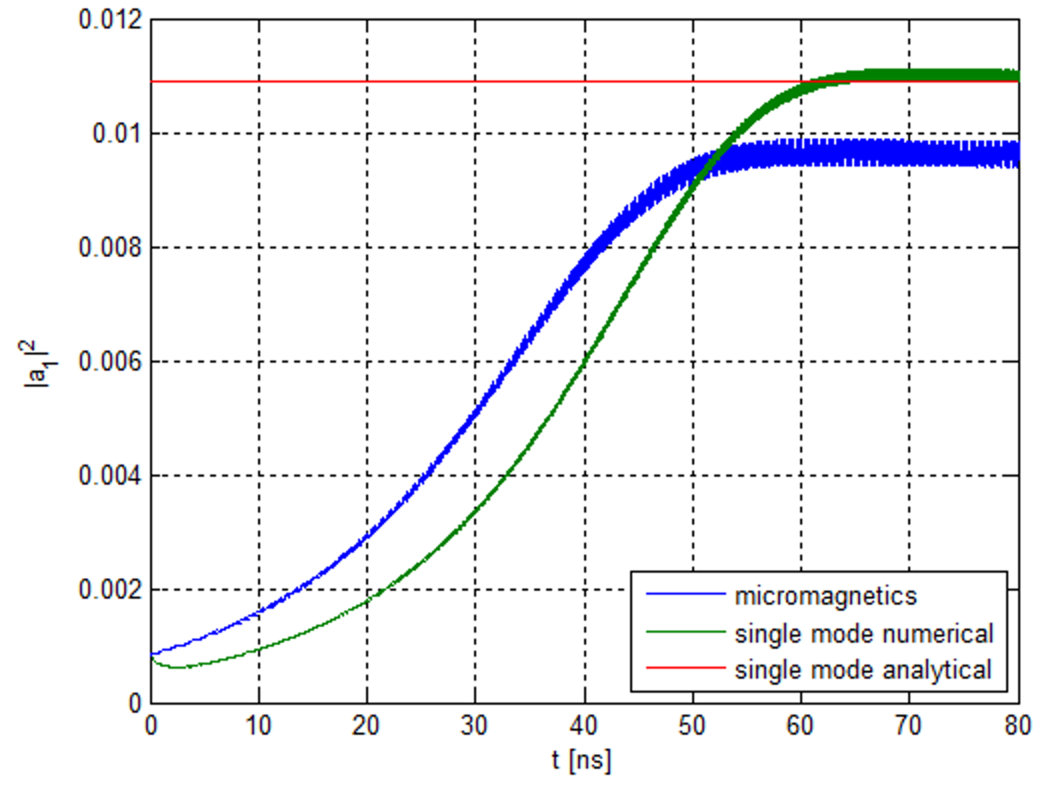} \includegraphics[width=8cm]{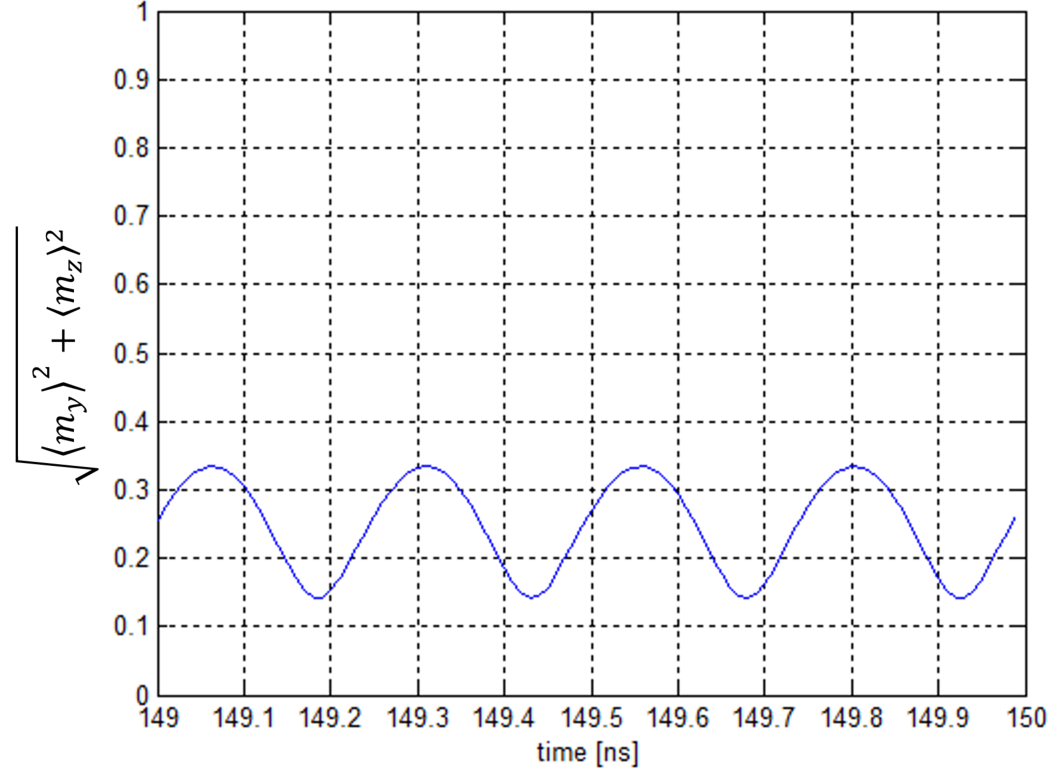}
    \caption{Parametric excitation of magnetization in a $100\times100\times 5$nm YIG disk. Left panel reports the time-evolution of mode 1 amplitude according to full micromagnetics (solid blue) through eq.\eqref{eq:A0 projection mumag}, single NMM eq.\eqref{eq:nonlinear parametric ah} (solid green) and analytical formula \eqref{eq:saturation amplitude mode h} (solid red). Right panel reports the time evolution of transverse magnetization amplitude $\sqrt{\langle m_y \rangle^2+\langle m_z \rangle^2}$ for full micromagnetic simulation (solid blue).}
    \label{fig:YIG disk 100 parametric resonance vs time}
\end{figure}

A first result is reported in fig.\ref{fig:YIG disk 100 parametric resonance vs time}, where the normal mode amplitude $|a_1|^2$ is computed using the following projection formula:
\begin{align}\label{eq:A0 projection mumag}
    a_h&=(\varphi_h,\mvec(\rvec,t)-\mvec_0)_{\mathcal{A}_{0\perp}}= \\
    &=j\omega_h(\mvec_0(\rvec)\times\bm\varphi_h^*,\mvec(\rvec,t)-\mvec_0(\rvec)) \quad,
\end{align}
which comes from the orthonormality property \eqref{eq:orthonormality} and from the definition of the operator $\mathcal{B}_0$ involved in the eigenproblem \eqref{eq:generlized eigenproblem}. In the left panel, one can see the time evolution of the mode amplitude $|a_1|^2$ computed using the real-time computation of the projection \eqref{eq:A0 projection mumag} implemented in the code MaGICo (solid blue line) which we compare with the numerical solution of the single mode dynamics described by eq.\eqref{eq:nonlinear parametric ah} (solid green line). The analytical expression of the steady-state amplitude $|a_1|_-$ given by eq.\eqref{eq:saturation amplitude mode h} (solid red line) shows a good agreement with the full micromagnetic simulation. For the sake of comparison with macrospin computations, the right panel reports the time-evolution of the amplitude of the (spatially-averaged) transverse magnetization component $\sqrt{\langle m_y\rangle^2+\langle m_z\rangle^2}$ (the notation $\langle\cdot\rangle$ means spatial average over the disk volume). By comparing this with the right panel of fig.\ref{fig:macrospin parametric resonance vs time} (corresponding to a slightly lower ac amplitude 2.3mT), one can see that there is reasonable agreement between full micromagnetics and macrospin description in the case of fundamental mode parametric excitation.

Next, we have investigated the dependence of the steady-state parametrically-excited amplitude of modes as function of frequency of the applied field.

\begin{figure}[h!]
    \centering
    \includegraphics[width=0.6\textwidth]{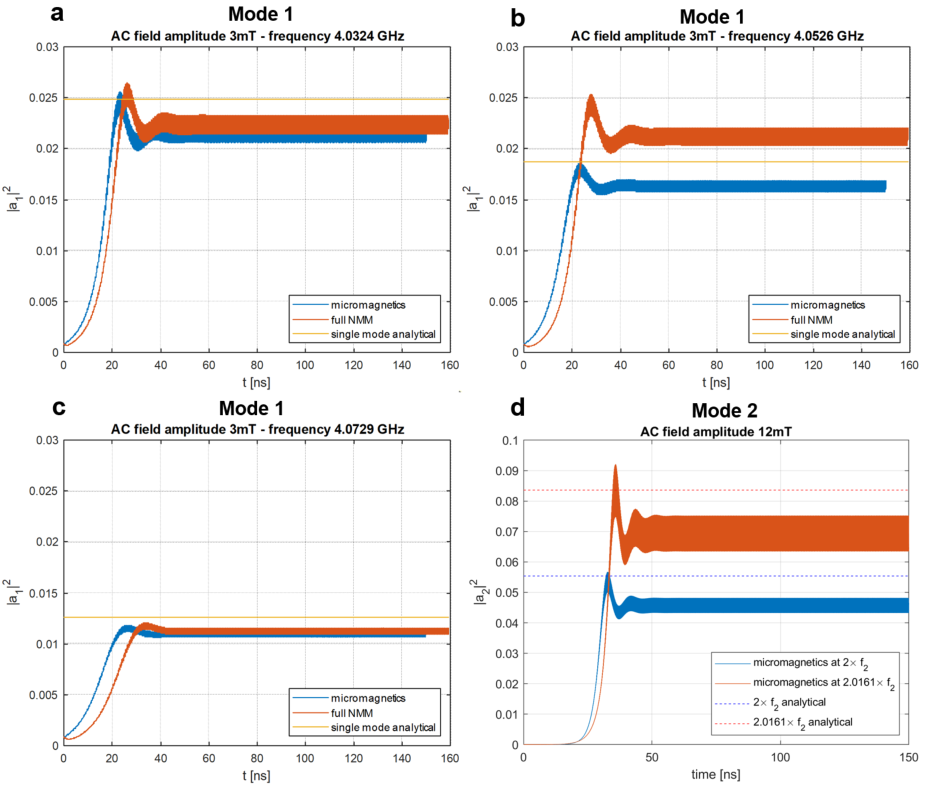} 
    \caption{Parametric excitation of magnetization in a $100\times100\times 5$nm YIG disk. Panels (a)-(c): comparison between micromagnetic simulations (solid blue), full NMM (solid brown) and analytical theory (solid yellow) for parametric excitation of mode 1. Panel (d): comparison between micromagnetic simulations (solid lines) and analytical theory (dashed lines) for parametric excitation of mode 2.}
    \label{fig:YIG disk 100 NFS test}
\end{figure}

Thus, we have simulated the excitation of the fundamental mode choosing the applied field frequency slightly detuned (both negatively and positively) with respect to the Arnold's tongue center frequency $2\times f_1$ while keeping the amplitude constant at 3 mT. Depending on the negative/positive sign of the detuning, according to eq.\eqref{eq:saturation amplitude mode h}, the steady-state amplitude of mode $h=1$ is expected to be larger/smaller than that associated with zero detuning (we remark that the NFS $N_1<0$ in this case). One can clearly see from panels \ref{fig:YIG disk 100 NFS test}(a)-(c) that this is exactly the case (panel (b) refers to the cone center frequency), confirming the significant predictive power of the analytical theory. The panels report comparison between full micromagnetic simulation and projection through eq.\eqref{eq:A0 projection mumag} (solid blue line), full single NMM eq.\eqref{eq:NMM} (solid brown line) and analytical formula \eqref{eq:saturation amplitude mode h} for the steady-state amplitude $|a_h|^2$ (solid yellow line). 

We have also tested the parametric excitation of mode $h=2$ (see panel (d) in fig.\ref{fig:YIG disk 100 NFS test}) using the center frequency $2\times f_2=8.1346\,$GHz (blue lines) and positively detuned frequency $2.0161\times f_2=8.2\,$GHz (brown lines) at constant ac field amplitude 12 mT. One can clearly observe that positively detuned frequencies yield larger amplitudes as it is expected since mode $h=2$ has positive NFS coefficient (see table \ref{tab:YIG disk 100nm}). 

\begin{figure}[t]
    \centering
    \includegraphics[width=\textwidth]{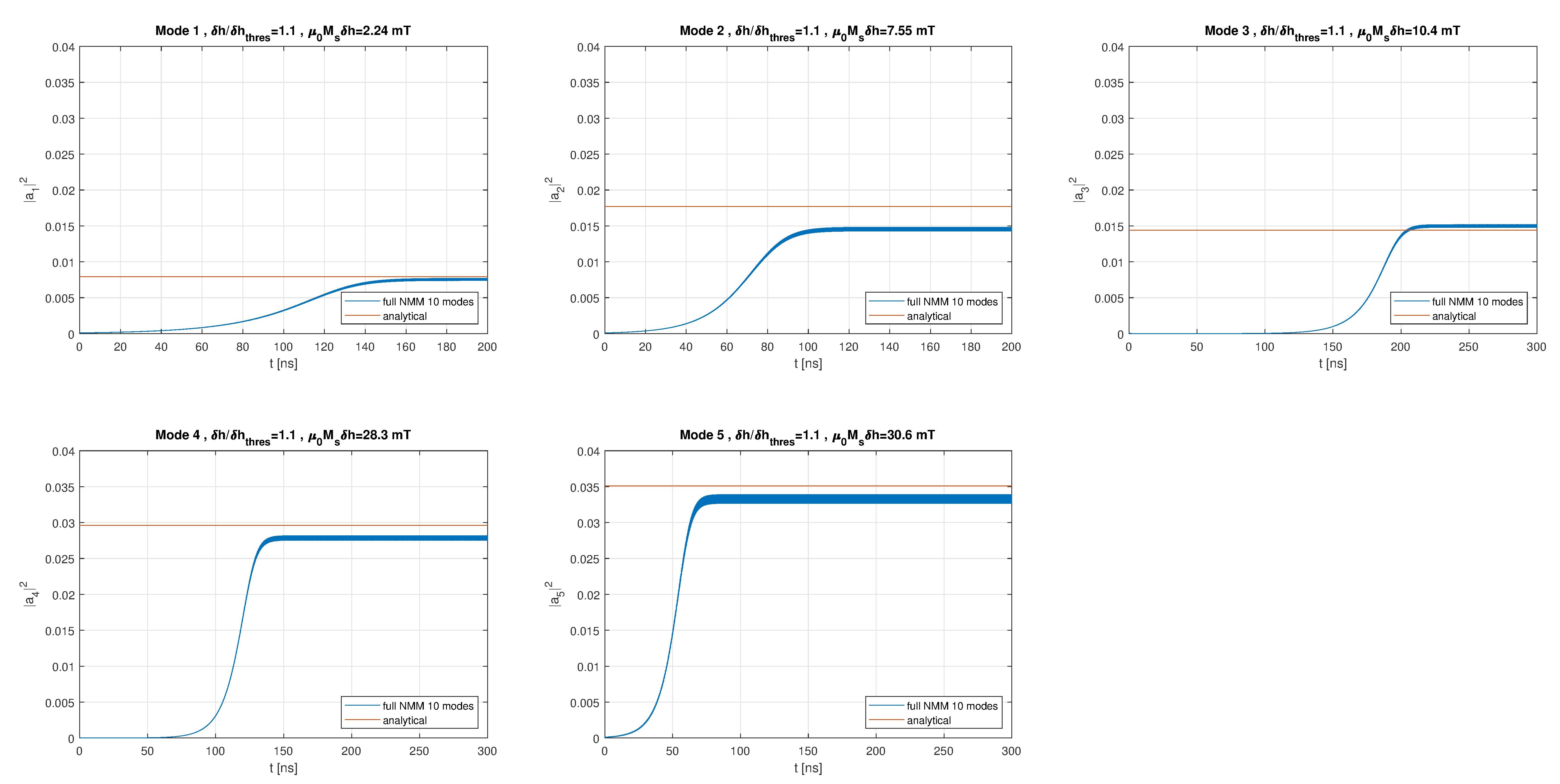} 
    \caption{Parametric excitation of magnetization modes 1-5 in a $100\times100\times 5$nm YIG disk: comparison between micromagnetic simulations (solid blue) and analytical theory (solid brown) for saturation amplitude given by eq.\eqref{eq:saturation amplitude mode h}.}
    \label{fig:YIG disk 100 saturation amplitude mumag}
\end{figure}

The results of further simulations reported in fig.\ref{fig:YIG disk 100 saturation amplitude mumag} exhibit very good agreement with the analytical formula \eqref{eq:saturation amplitude mode h} for amplitude saturation of parametrically-excited modes. As a general observation, we see that the analytical formula slightly overestimates the steady-state amplitude, but again we stress that it is purely analytical once the normal modes have been preliminary computed.

\clearpage

\section{Further results of simulations}\label{sec:further numerical results}

\subsection{Modes $h=1$, $n=5$}\label{sec:further numerical results h1n5}

\begin{figure*}[t]
    \centering
    \hspace{0.1\textwidth} NMM, sequence $51$ \hspace{0.1\textwidth} two-modes, sequence $51$  
    \\
    \includegraphics[width=0.3\textwidth]{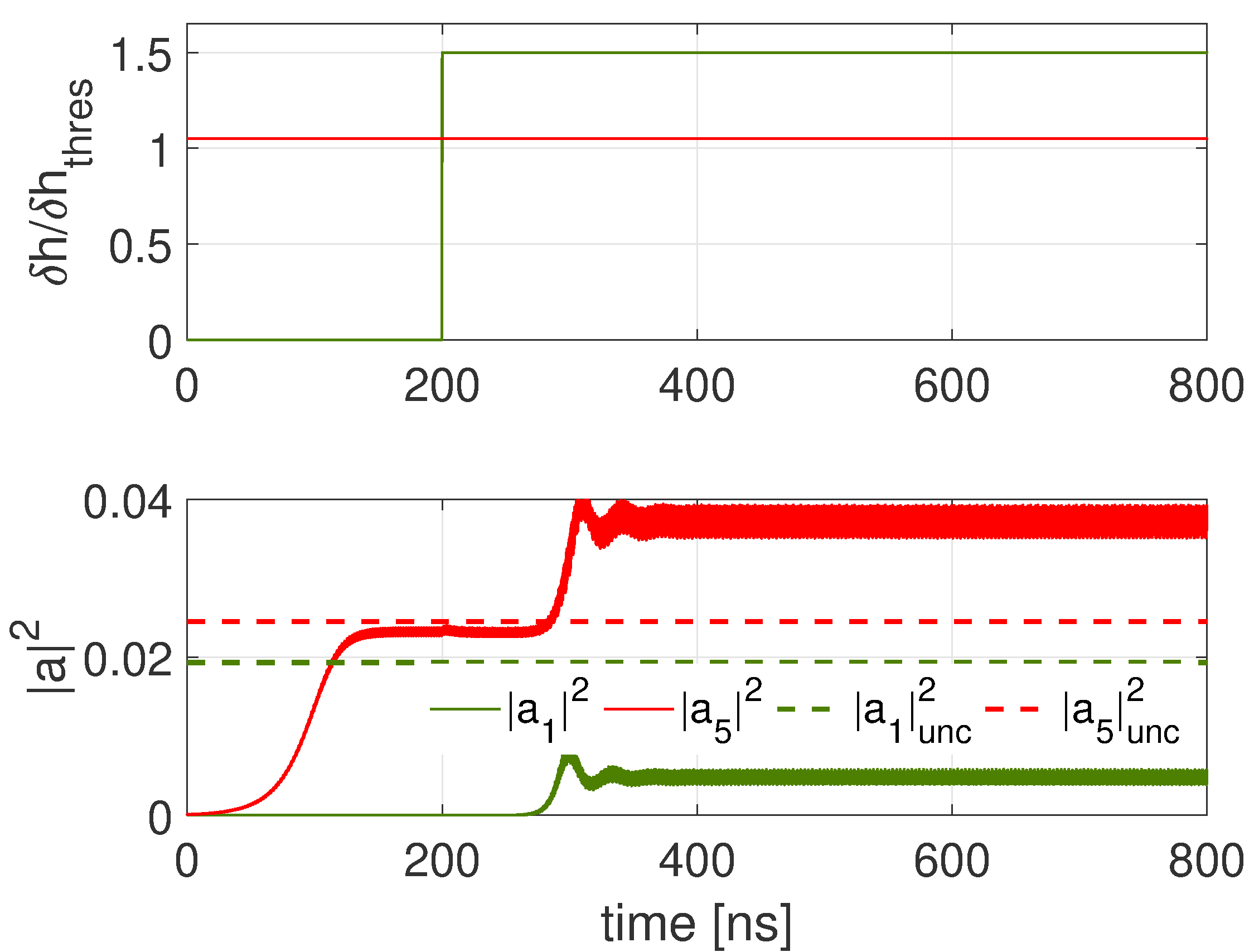} 
    \includegraphics[width=0.3\textwidth]{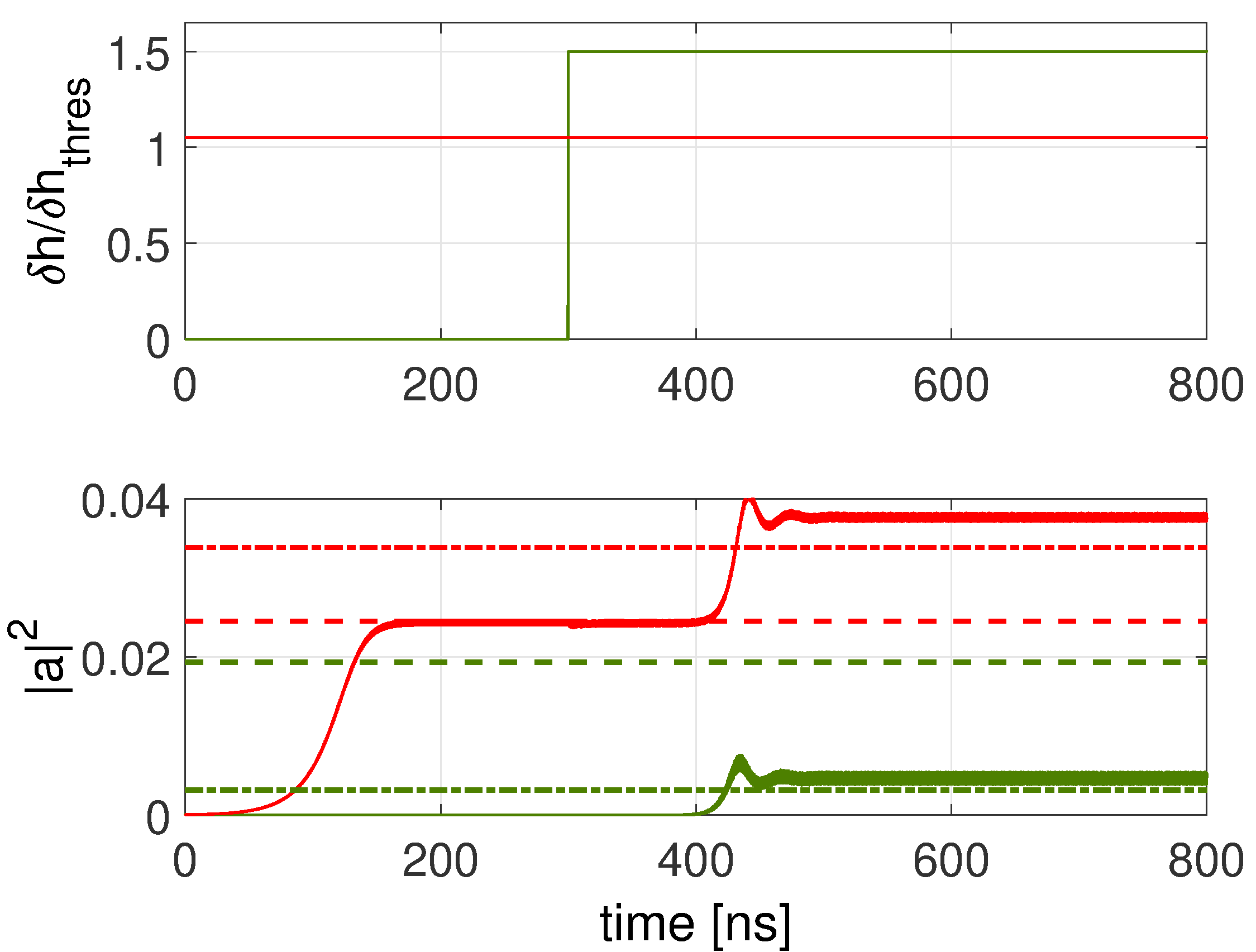}    
    \caption{Coupled interaction of parametrically-excited modes 1 and 5.  The ac field frequencies are $\omega_\mathrm{rf1}=2\omega_1, \omega_\mathrm{rf2}=2\omega_5$ while the field amplitudes are $\delta h_1=1.5\times\delta h_\mathrm{thres,1}, \delta h_2=1.05\times\delta h_\mathrm{thres,5}$. 
    The left panel reports the result of numerical simulation of full NMM \eqref{eq:NMM} with 10 modes.
    The right panel reports the result of the time-domain simulation of the two-modes model \eqref{eq:nonlinear coupled parametric ah}-\eqref{eq:nonlinear coupled parametric an} and analytical formulas for steady-state amplitudes of coupled modes \eqref{eq:saturation amplitude mode h type 1 linear}-\eqref{eq:saturation amplitude mode n type 1 linear} (dash-dotted horizontal lines) and uncoupled modes \eqref{eq:saturation amplitude mode h} (dashed horizontal lines). 
    }
    \label{fig:YIG_disk_100_mode5_1p05_mode1_1p5}
\end{figure*}

\begin{figure}[t]
    \centering
     \hspace{0.28\textwidth} two-modes, sequence $15$ \hfill \mbox{ } \\
    \includegraphics[width=0.3\textwidth]{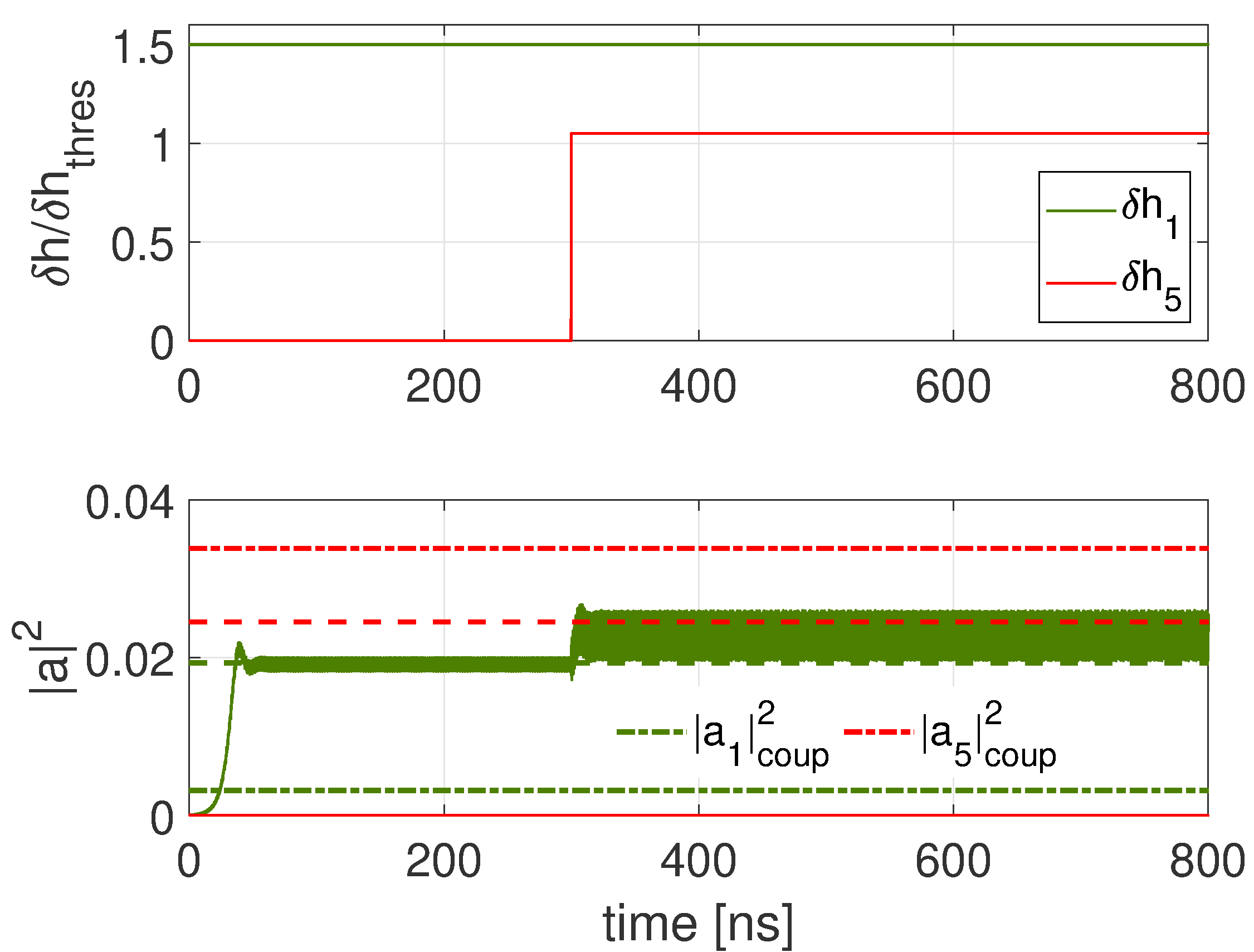} \includegraphics[width=0.31\textwidth]{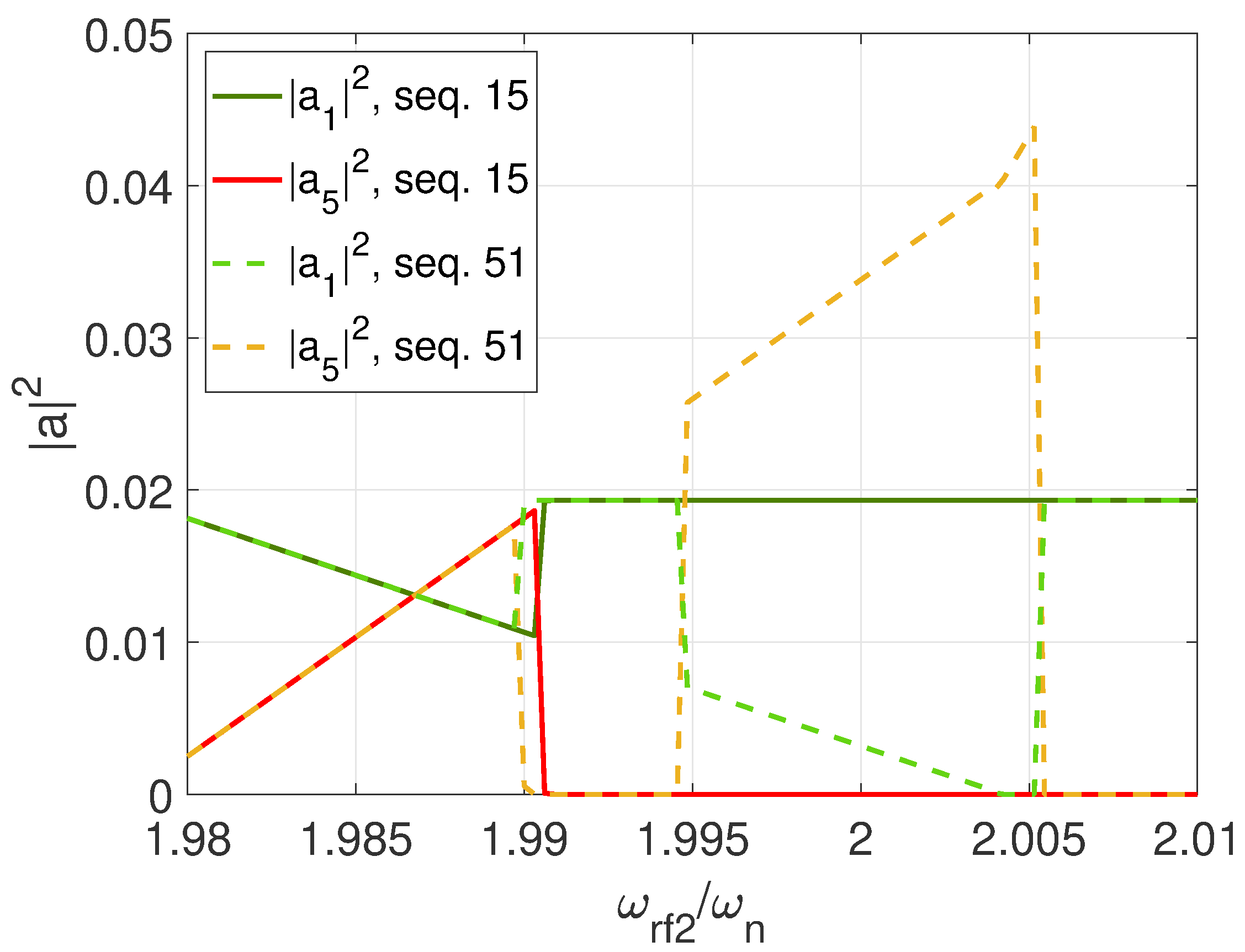}  
    \caption{Coupled interaction of parametrically-excited modes $h=1$ and $n=5$.  The  left panel reports time-domain simulation of two-modes model with excitation sequence reversed compared to fig.\ref{fig:YIG_disk_100_mode5_1p05_mode1_1p5}. The ac field frequency is fixed $\omega_\mathrm{rf2}=2\omega_5$, the field amplitudes are $\delta h_1=1.5\times\delta h_\mathrm{thres,1}, \delta h_2=1.05\times\delta h_\mathrm{thres,5}$. The right panel reports the result of numerical integration of averaged dynamical eqs.\eqref{eq:averaged two nonlinear parametric Ah}-\eqref{eq:averaged two nonlinear parametric Phi_n}. It is apparent that, when $\omega_\mathrm{rf1}=2\omega_1$ two possible regimes exist (coupled or uncoupled mode 1) depending on the sequence of excitation of tones. Solid  lines refer to $|a_h|^2,|a_n|^2$ when mode $h$ excited before mode $n$, dashed lines refer to the case where mode $n$ is excited before mode $h$.}
    \label{fig:YIG_disk_100_mode5_1p05_mode1_1p5_coupled_non_comm}
\end{figure}

Here we report additional results concerning comparison between time-domain simulations performed using the full NMM model \eqref{eq:NMM}, the two-modes model \eqref{eq:nonlinear coupled parametric ah}-\eqref{eq:nonlinear coupled parametric an} and analytical formulas for steady-state amplitudes of uncoupled modes \eqref{eq:saturation amplitude mode h} in figs.\ref{fig:YIG_disk_100_mode5_1p05_mode1_1p5} and \ref{fig:YIG_disk_100_mode5_1p05_mode1_1p5_coupled_non_comm}.

The left panel of fig.\ref{fig:YIG_disk_100_mode5_1p05_mode1_1p5} refers 
to the result of numerical simulation of the full NMM \eqref{eq:NMM} with 10 modes.
The right panel reports 
the time-domain simulation of the two-modes model \eqref{eq:nonlinear coupled parametric ah}-\eqref{eq:nonlinear coupled parametric an} for the same excitation parameters compared with the analytical formulas for coupled modes amplitudes \eqref{eq:saturation amplitude mode h type 1 linear}-\eqref{eq:saturation amplitude mode n type 1 linear}, showing very good agreement in the prediction of mode 1 suppression and enhancement of mode 5. 
We evidence the good agreement between NMM, two-modes model and the analytical theory. 

If one reverses the sequence of excitation tones, the result is different (see the left panel of fig.\ref{fig:YIG_disk_100_mode5_1p05_mode1_1p5_coupled_non_comm}) since mode 1 remains in the uncoupled steady-state that is coexistent with the coupled solution, giving rise to non-commutativity even in the coupled regime. Numerical simulation of averaged dynamical eqs.\eqref{eq:averaged two nonlinear parametric Ah}-\eqref{eq:averaged two nonlinear parametric Phi_n} confirms this occurrence (see the left panel of fig.\ref{fig:YIG_disk_100_mode5_1p05_mode1_1p5_coupled_non_comm}).

\begin{figure*}[t]
    \centering
     CW vertical sweeps\\ 
     \includegraphics[width=0.32\linewidth]{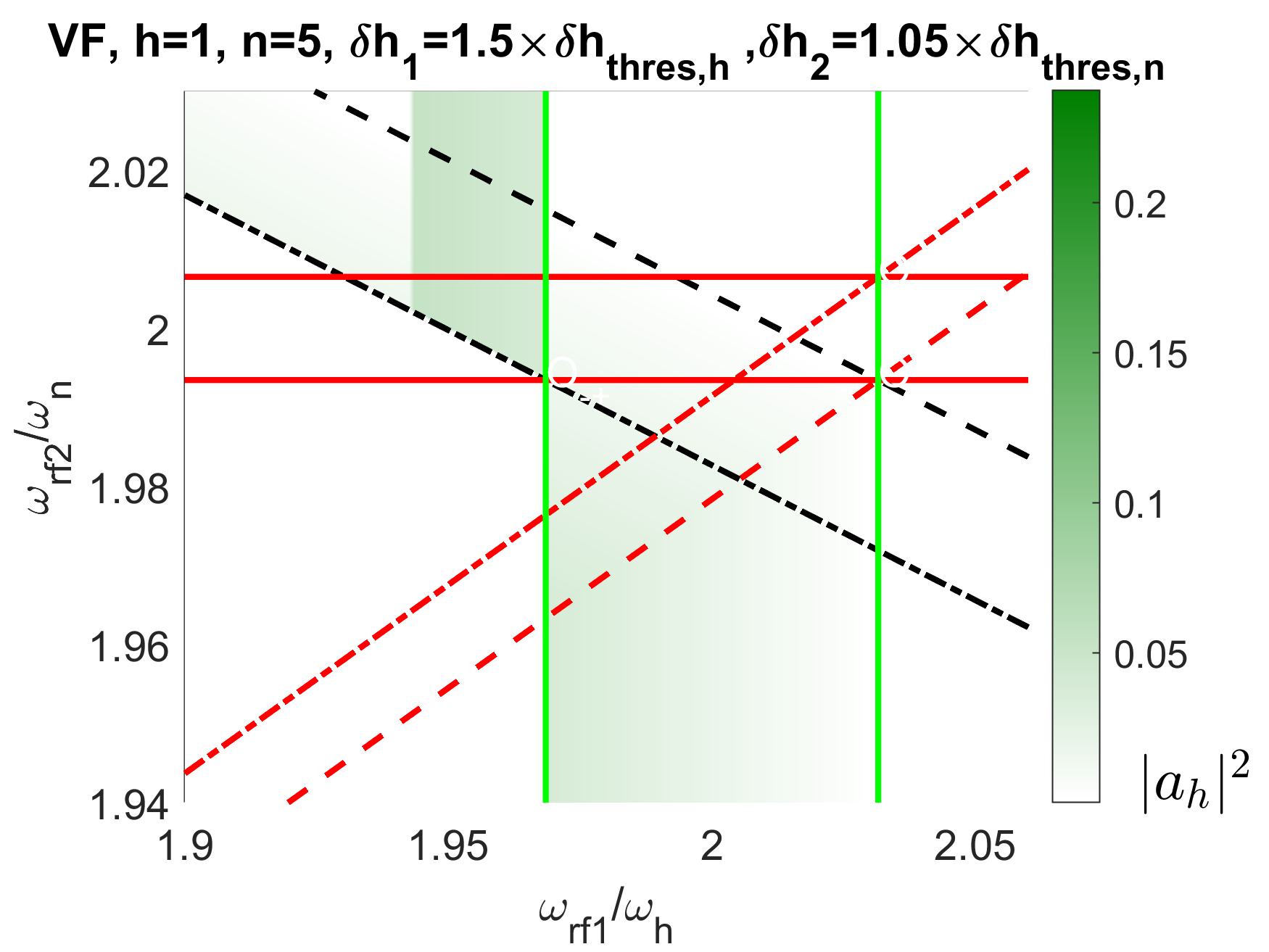}     \includegraphics[width=0.32\linewidth]{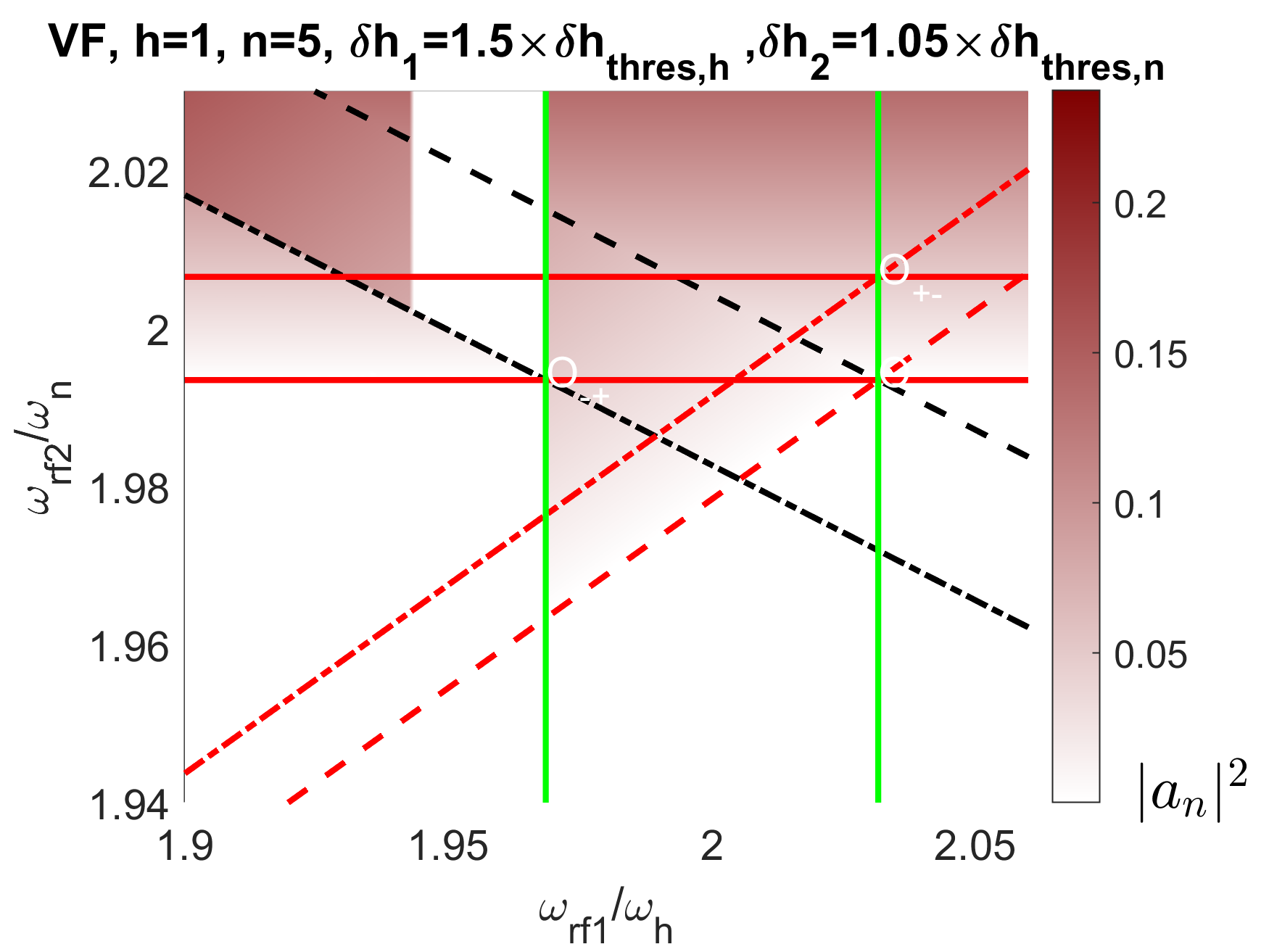}    \includegraphics[width=0.32\linewidth]{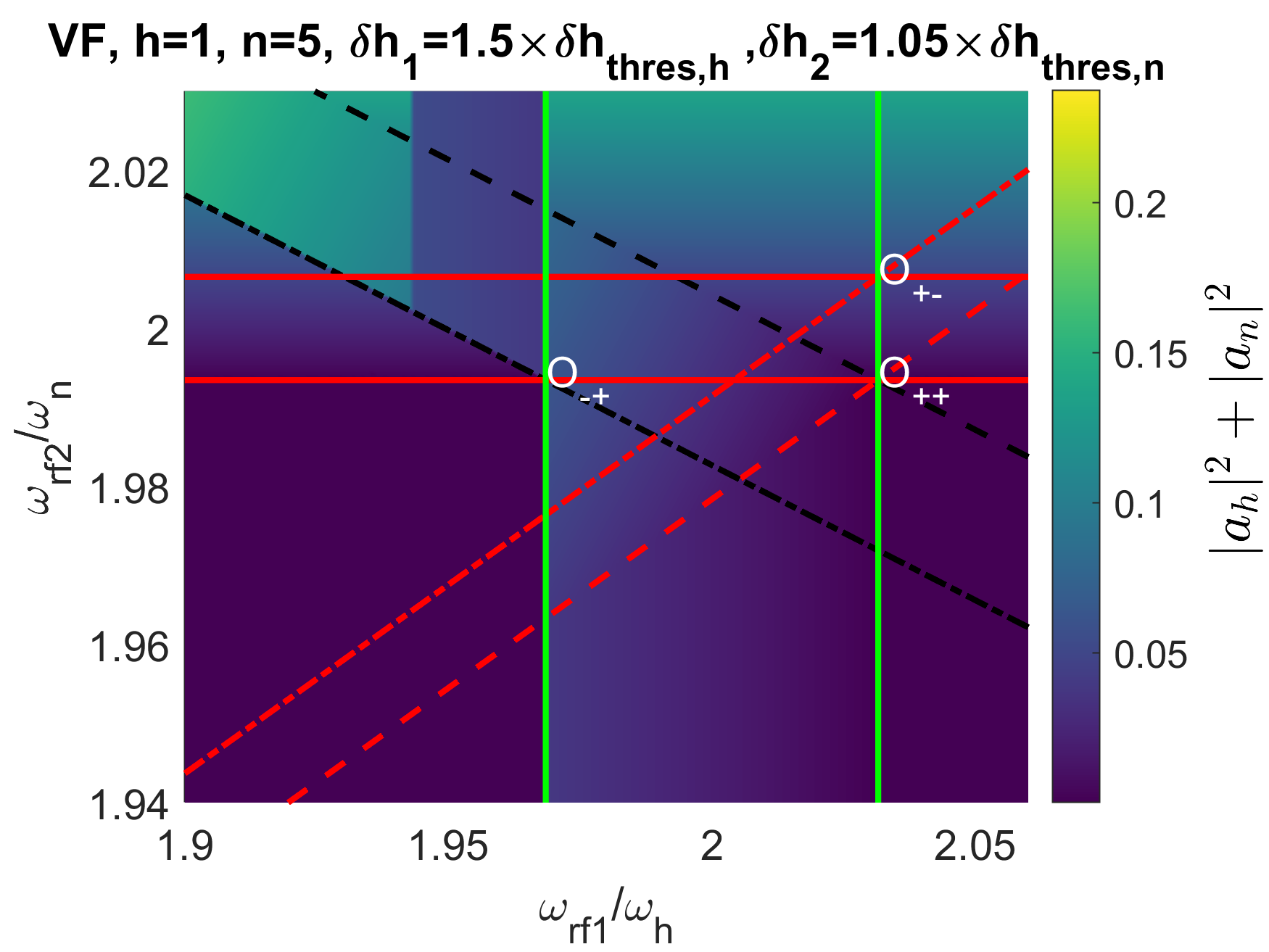} \\
    \includegraphics[width=0.32\linewidth]{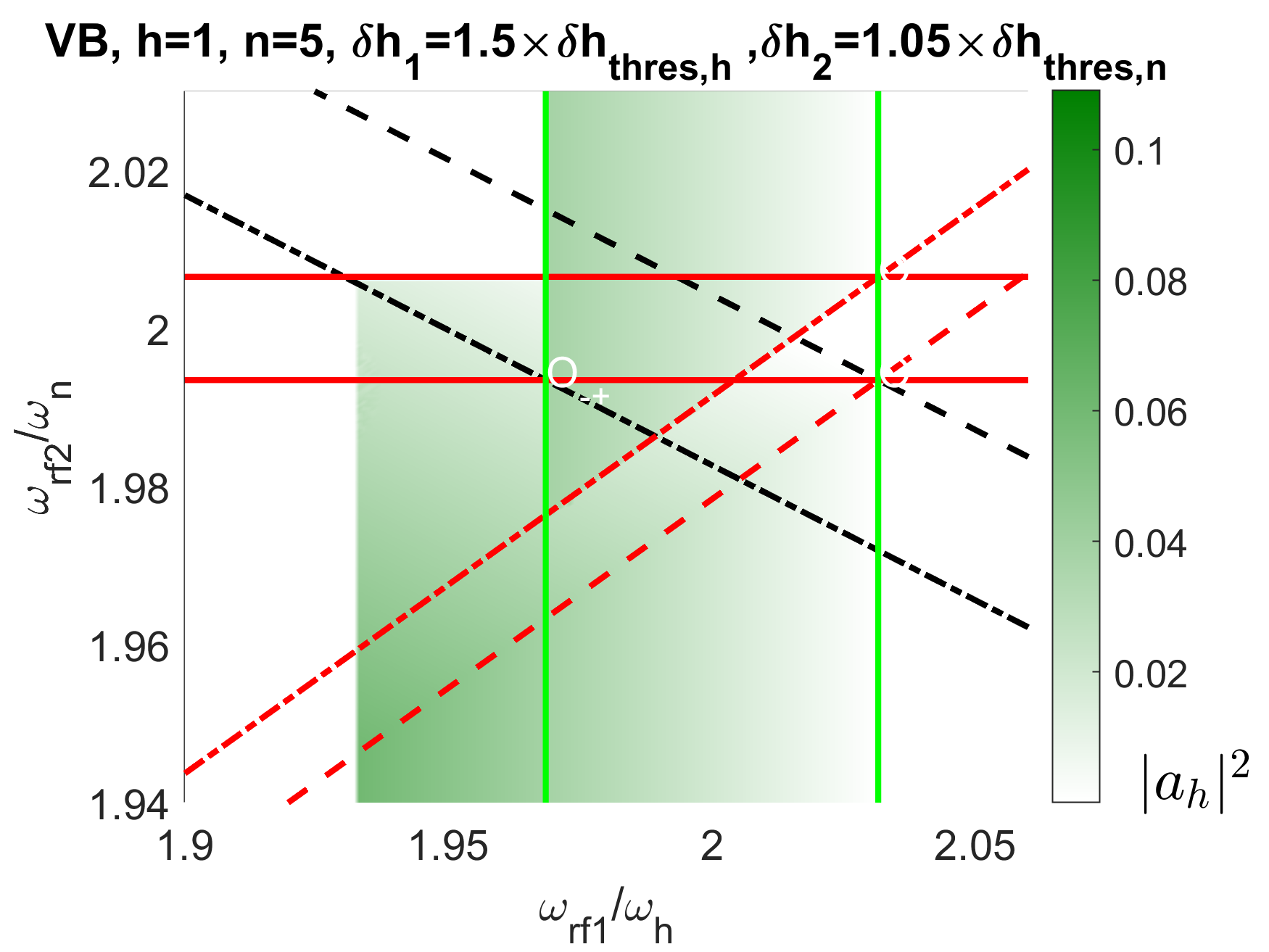}     \includegraphics[width=0.32\linewidth]{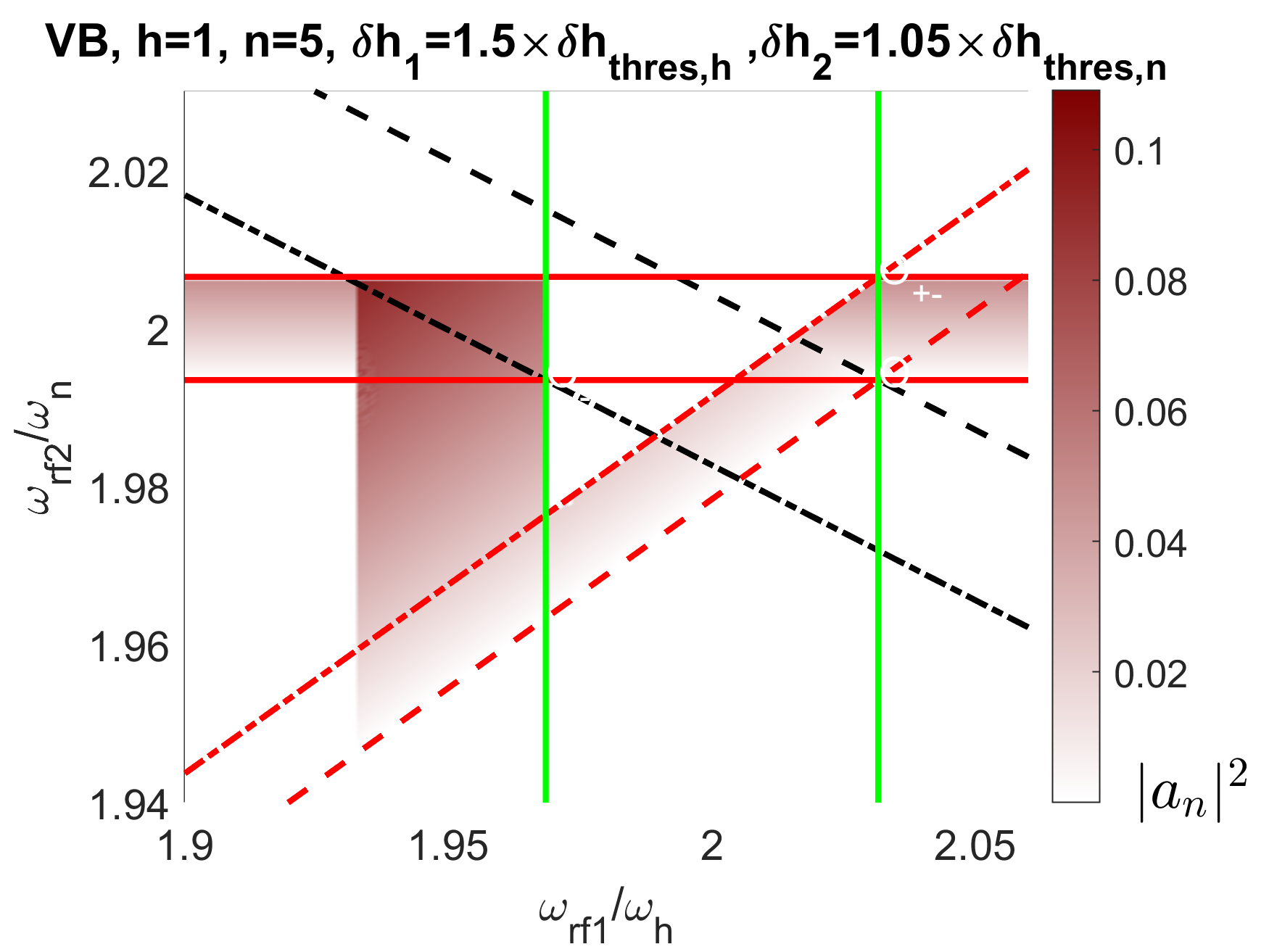}    \includegraphics[width=0.32\linewidth]{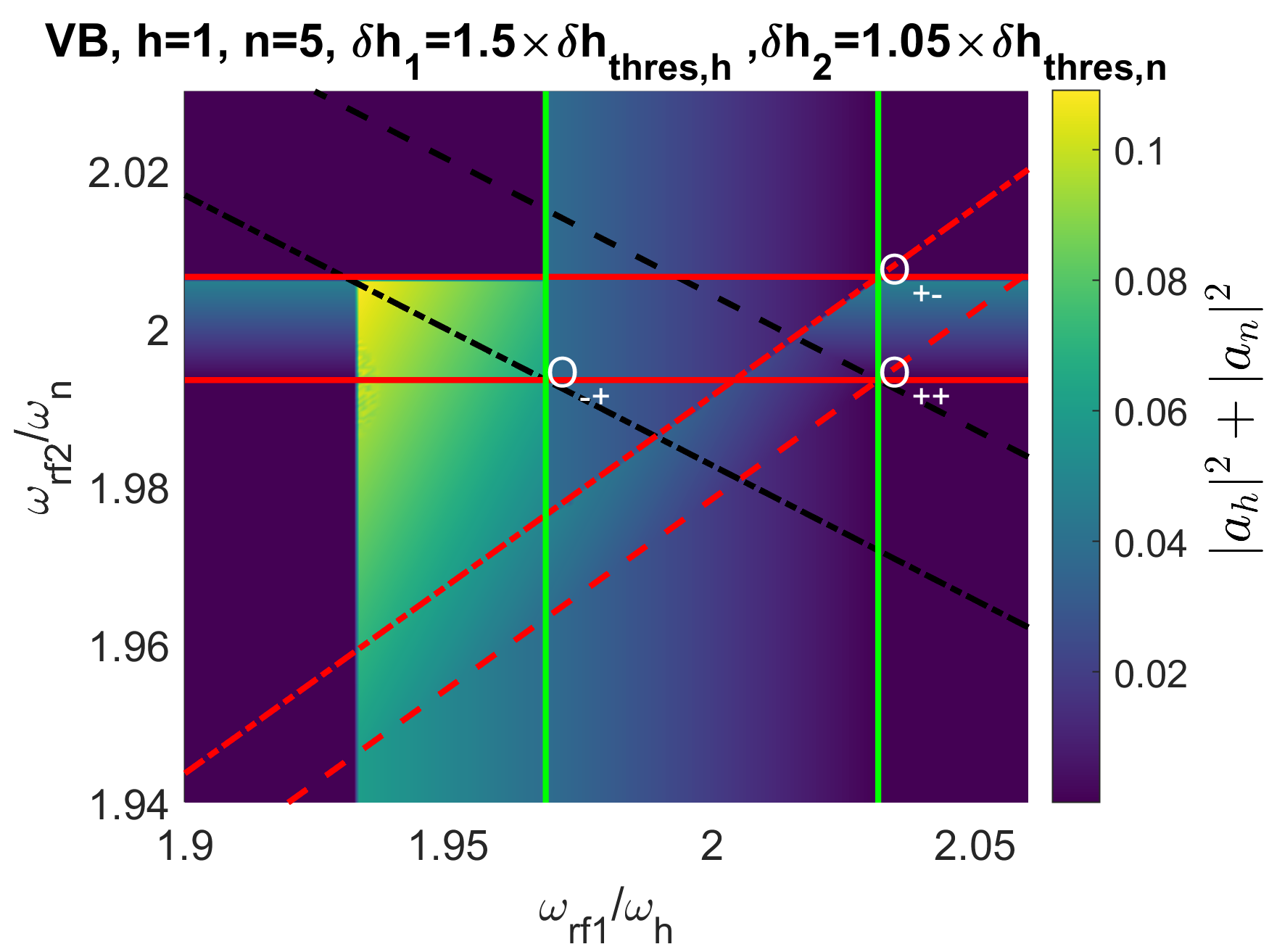}   
    \caption{Continuous wave (CW) dynamics for parametrically-excited modes $h=1$ and $n=5$ via parallel pumping with two-tone signal at fixed above-threshold amplitudes $\delta h_1=1.5\times\delta h_\mathrm{thres,h},\delta h_2=1.05\times\delta h_\mathrm{thres,n}$. Each point in the colormap is obtained integrating the averaged two-modes model \eqref{eq:averaged two nonlinear parametric Ah}-\eqref{eq:averaged two nonlinear parametric Phi_n} for long enough time to approach the steady-state. The panels report the amplitude diagrams associated to steady-state for: (left) mode $h$, amplitude  $|a_h|^2$; (middle) mode $n$, amplitude $|a_n|^2$; (right) summed amplitudes $|a_h|^2+|a_n|^2$. Notation VF/VB in the panel title bar means Vertical Forward/Backward frequency sweep.  Black dashed line corresponds to the condition $c_{++}\equiv u_n$ while red dashed line refers to $c_{++}\equiv u_h$. The  boundaries of regions $C_h$ and $C_n$ are marked with green and red solid lines, respectively.
    Black dash-dotted line refers to the condition $c_{-+}\equiv u_n$ ('no $u_n$'), while red dash-dotted line refers to the condition $c_{+-}\equiv u_h$ ('no $u_h$'). Solid yellow (Hopf bifurcation) line $c\leftrightarrow q$ refer to the boundary of region $T_{++}$ due to the onset of Q-modes, solid blue (homoclinic bifurcation) line $q\rightarrow u$ denotes the boundary of the region where Q-modes can exist.}
    \label{fig:CW array h1 n5 vertical}
\end{figure*}

Now we briefly focus on forward (resp. backward) vertical (denoted as VF/VB in figure panels) sweeps, which have been performed by exciting both modes at constant $\omega_\mathrm{rf1}$ starting from sufficiently small (resp. large) frequency $\omega_\mathrm{rf2}$ and then linearly increasing (resp. decreasing) it in a CW fashion. First of all, we point out that Q-modes are not involved since they can only originate from $c_{++}$ P-modes, which in turn cannot be excited with vertical excitation outside the region $C_h$ (enclosed within green solid lines). Thus, the critical lines $c\leftrightarrow q$ and $q\rightarrow u$ (yellow and blue) associated with Hopf and homoclinic bifurcation do not affect the response to vertical frequency sweeps and are not reported in the related panels of fig.\ref{fig:CW array h1 n5 vertical}.  

One can see that forward sweeps mainly involve mode $h$ in the region below the (dashed red) line $c_{++}\equiv u_h$, which indicates the continuity with the coupled P-mode $c_{++}$ that gives rise to a smooth transition of the total amplitude, except for a vertical strip between $\omega_\mathrm{rf1}\approx 1.94\omega_h$ and $\omega_\mathrm{rf1}\approx 1.96\omega_h$ (left solid green line) and above the (black dash-dotted) 'no $u_n$' line, where the $u_n$ P-mode becomes unstable and the system ends up in the $u_h$ regime. This occurs since $u_h$ and $c_{++}$ coexist in that region and the determination of the actual steady-state is governed by the transient system dynamics starting from $u_n$ regime. This is similar to what has been outlined before concerning horizontal forward frequency sweeps. 

Finally, we analyze vertical backward sweeps. We start from those occurring for small frequency $\omega_\mathrm{rf1}$, where only $u_n$ is excited within region $C_n$. The limit situation occurs at the intersection of the (dash-dotted black) line 'no $u_n$' with the upper solid red line. In fact, crossing the red line before the black dash-dotted excites $c_{++}$ regimes since $u_n$ is forbidden within black lines. Then, by further decreasing $\omega_\mathrm{rf2}$, the system persists in the coupled regime $c_{++}$. For VB sweeps inside region $C_h$ (enclosed within green lines), $u_h$ is excited until the (red dash-dotted) line 'no $u_h$' is crossed, giving rise to $c_{++}$ regimes within the red lines where $u_h$ is forbidden. After crossing the dashed red line $c_{++}\equiv u_h$, the system smoothly transitions again to $u_h$ P-mode. In the region at the right of $C_h$, only $u_n$ P-mode is excited within the region $C_n$.

\begin{figure*}[t]
    \centering
        \includegraphics[width=0.35\textwidth]{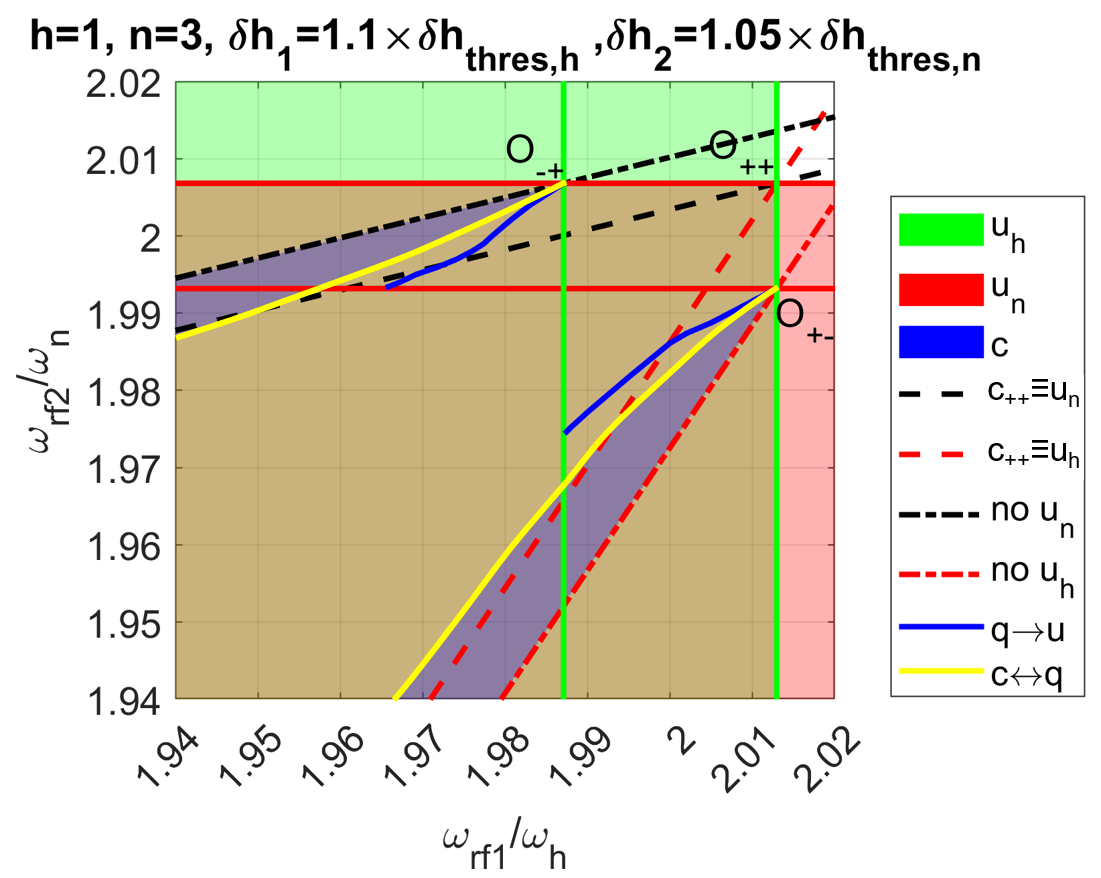} \includegraphics[width=0.3\textwidth]{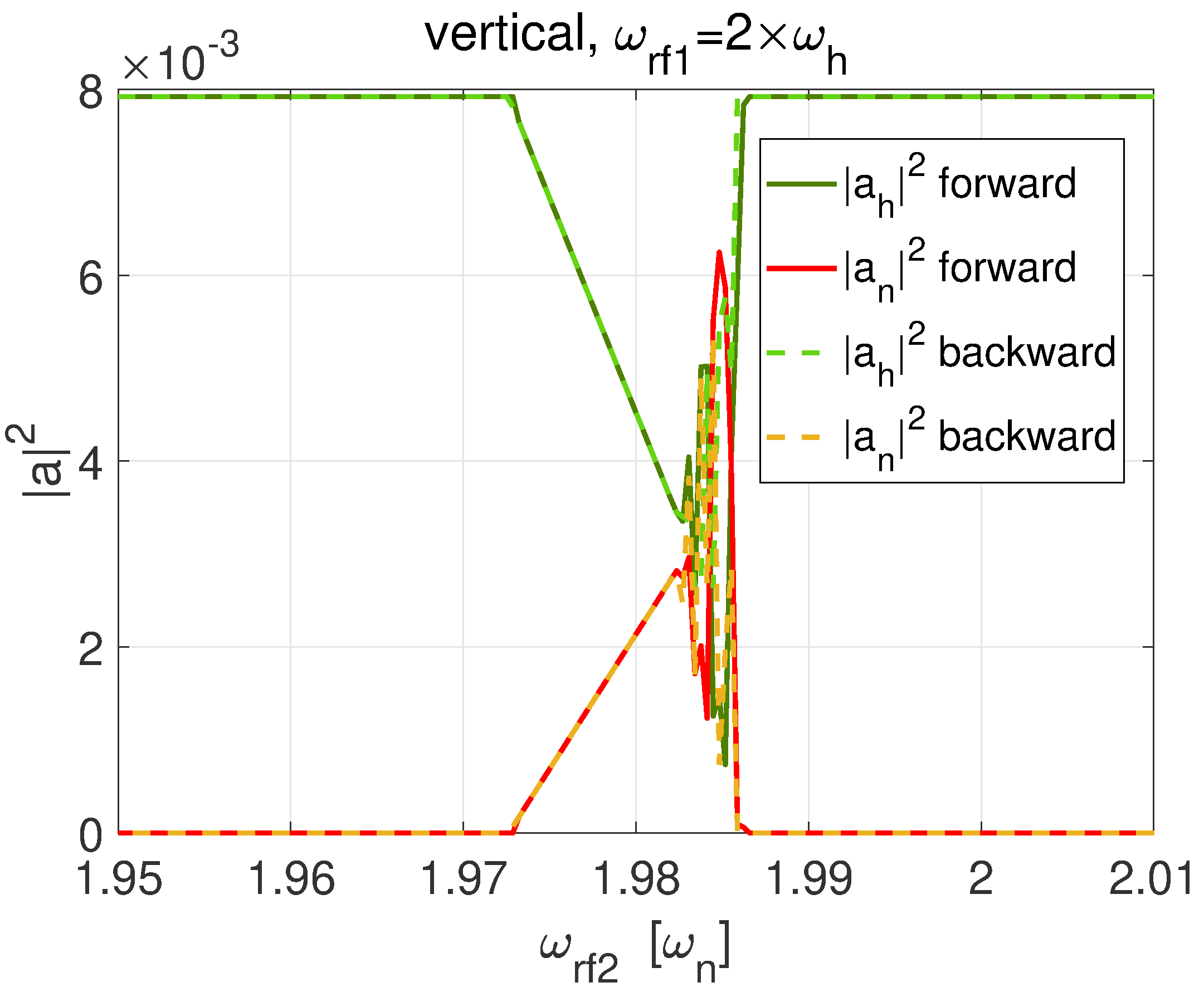}\includegraphics[width=0.31\textwidth]{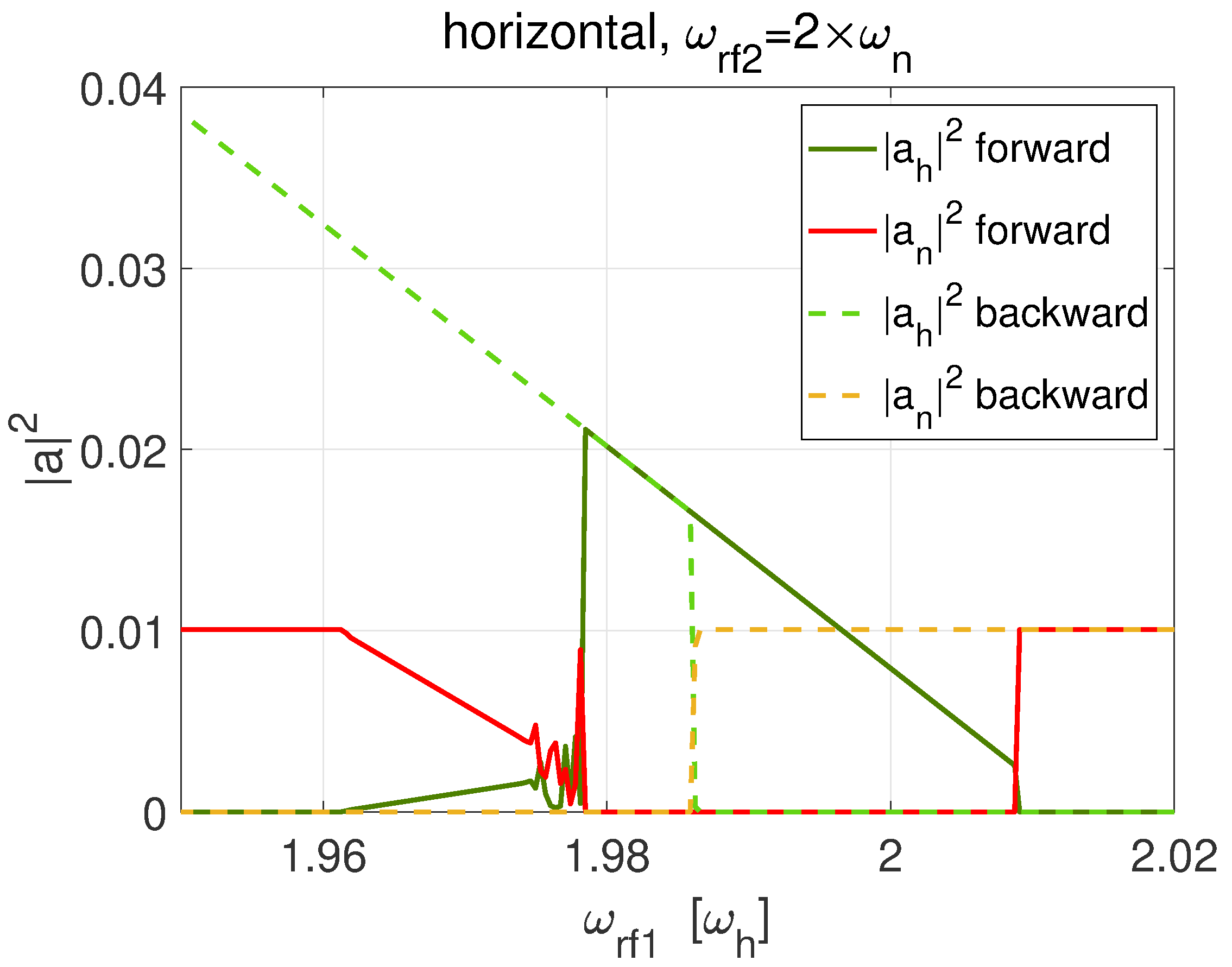}
    \caption{Phase diagram of interaction in $(\omega_\mathrm{rf1},\omega_\mathrm{rf2})$ plane for parametrically-excited modes $h=1$ and $n=3$ via parallel pumping with two-tone signal at fixed above-threshold amplitudes $\delta h_1=1.1\times\delta h_\mathrm{thres,h},\delta h_2=1.05\times\delta h_\mathrm{thres,n}$. NFS coefficients are $N_{hh}=-0.8159, N_{hn}=-1.2972=N_{nh}, N_{nn}=-0.3390$. Left panel depicts different regions (green, red, blue) corresponding to the conditions \eqref{eq:constraint real amplitudes} 
    for which modes $u_h,u_n,c_{+-},c_{-+}$ exist and are stable (the boundaries of green and red regions are marked with green and red solid lines, respectively). Stable Q-modes exist in the region enclosed between yellow $c\leftrightarrow q$ and blue $q\rightarrow u$ lines associated with Hopf and homoclinic bifurcation, respectively (notice that homoclinic bifurcation lines have been numerically computed only within the red and green regions). 
    Middle and right panels report steady-state amplitudes $|a_h|^2,|a_n|^2$ as function of frequency corresponding to 'vertical' ('horizontal') slow variation of CW excitation frequency $\omega_\mathrm{rf2}$ (resp. $\omega_\mathrm{rf1}$) while $\omega_\mathrm{rf1}=2\omega_h$ (resp. $\omega_\mathrm{rf2}=2\omega_n$). Both amplitude responses exhibit hysteretic behavior (see solid/dashed lines referring to forward/backward frequency sweep).}
    \label{fig:type1 phase diagram and hysteresis modes 1 and 3}
\end{figure*}

\subsection{Modes $h=1,n=3$}\label{sec:further numerical results h1n3}

In order to explore the range of possibilities offered by the choice of different parameters, we have also analyzed the case where the two modes have NFS coefficients $N_{hh},N_{nn}$ of the same sign. In this respect, we choose $h=1$ and $n=3$, which have $N_{hh}=-0.8159<0, N_{hn}=-1.2972=N_{nh}<0, N_{nn}=-0.3390<0$ (see the matrix \eqref{eq:NMFS YIG100 correct}). The result is reported in figure \ref{fig:type1 phase diagram and hysteresis modes 1 and 3}. 

Compared with the previous modes' pair, we can infer from the theory that now the (dashed black) line $c_{++}\equiv u_n$ corresponding to the condition $|a_h|^2_\mathrm{coup}=0,|a_n|^2=|a_n|^2_\mathrm{unc}$ (see eq. \eqref{eq:saturation amplitude mode h type 1 linear}) and the (red dashed) line $c_{++}\equiv u_h$  referring to $|a_n|^2_\mathrm{coup}=0,|a_h|^2=|a_h|^2_\mathrm{unc}$ (see eq. \eqref{eq:saturation amplitude mode n type 1 linear}) have both positive slopes and intersect in the upper right corner $O_{++}$ of the brown region in the top panel of fig.\ref{fig:type1 phase diagram and hysteresis modes 1 and 3}.
In fact, the black dashed line corresponding to eq.\eqref{eq:saturation amplitude mode h type 1 linear} has slope sign determined by $N_{hn}/N_{nn}>0$, whereas the red dashed line associated with eq.\eqref{eq:saturation amplitude mode n type 1 linear} has slope sign determined by $N_{nh}/N_{hh}>0$.

One can immediately see that, as happened for the previous modes' pair, this case produces hysteresis in both 'horizontal' and 'vertical' excitations. The phase diagram reported in the upper panel of fig.\ref{fig:type1 phase diagram and hysteresis modes 1 and 3} exhibits different features compared to the analogous diagrams of figs.\ref{fig:type1 phase diagram and hysteresis modes 1 and 5} and \ref{fig:type1 phase diagram and hysteresis modes 1 and 5 large power} that can be explained using the theory developed in section \ref{sec:mutual modes interaction}.

The coefficient $L_{hn}=N_{hh}N_{nn}/(N_{hh}N_{nn}-N_{hn}^2)$ is negative, meaning that triangular regions $T_{\pm\pm}$ are located to the right of the lines $c_{\pm\pm}\equiv u_n$ and above the lines $c_{\pm\pm}\equiv u_h$ that are grouped in pairs with the same slopes $\omega_h N_{hn}/(\omega_n N_{nn})>0$ and $\omega_nN_{nh}/(\omega_h N_{hh})>0$, respectively. In the figure, for the sake of simplicity, we have only reported the lines $c_{++}\equiv u_n$, $c_{++}\equiv u_h$, $c_{-+}\equiv u_n$ ('no $u_n$'), $c_{+-}\equiv u_h$ ('no $u_h$'),

Moreover, the stability condition $\text{sign}(N_{hh}) \text{sign}(N_{nn})(N_{hh}N_{nn}-N_{hn}^2)<0$ means that the coupled P-modes of type $c_{++}$ are unstable, whereas coupled modes of type $c_{+-}, c_{-+}$ are stable close to the vertices $O_{+-},O_{-+}$ and lines $c_{+-}\equiv u_h, c_{-+}\equiv u_n$, respectively. As discussed in appendix \ref{sec:stability multi P modes}, these vertices are critical points associated with the Bogdanov-Takens bifurcation, which in its neighborhood involves multiple stable P-modes along with a Q-mode. For illustrative purpose, let us consider the vertex $O_{+-}$. In the vicinity of $O_{+-}$, uncoupled P-modes $u_h,u_n$ as well as $c_{+-}$ are stable, as one can see inspecting the 
left panel of fig.\ref{fig:type1 phase diagram and hysteresis modes 1 and 3}. We remark that, according to the rule \ref{item:S6} reported in table \ref{tab:type1 interactions} (also discussed in appendix \ref{sec:stability multi P modes}), the uncoupled $u_h$ regime is unstable between the critical red lines $c_{++}\equiv u_h,c_{+-}\equiv u_h$ ('no $u_h$').

Then, we consider a vertical CW excitation with forward frequency sweep at constant $\omega_\mathrm{rf1}=2\omega_h$. One can clearly see
(middle panel of Fig.\ref{fig:type1 phase diagram and hysteresis modes 1 and 3}) that the system, initially excited in the uncoupled $u_h$ regime, undergoes a continuous transition to a $c_{+-}$ regime crossing the '$\text{no }u_h$' (red dash-dotted) line and exhibits  linear behavior of the squared amplitudes until the frequency matches that of the yellow $c\leftrightarrow q$ Hopf bifurcation line. At that point, the $c_{+-}$ P-mode becomes unstable and a Q-mode is created in around it. This results in the oscillatory behavior of the squared amplitudes $|a_h|^2,|a_n|^2$ around $\omega_\mathrm{rf2}\sim 1.985\omega_n$. This behavior is observed in the same range of frequencies for both the forward and backward frequency sweeps. This occurs since the homoclinic bifurcation (blue) line is almost overlapped with the $c_{++}\equiv u_h$ line at $\omega_\mathrm{rf1}=2\omega_h$. If one performs the same calculation choosing slightly lower frequency $\omega_\mathrm{rf1}$, the qualitative behavior of the dynamics is different, as one can clearly see in fig.\ref{fig:type1 hysteresis Qmodes 1 and 3} in the supplementary material. In the left panel the frequency is smaller $\omega_\mathrm{rf1}=1.998\omega_h$, but nevertheless this produces a behavior qualitatively similar to that appearing in the right panel of fig.\ref{fig:type1 phase diagram and hysteresis modes 1 and 3}, only exhibiting hysteresis in the activation of the Q-mode. Conversely, for slightly larger frequency $\omega_\mathrm{rf1}=2.002\omega_h$, the red dashed line is crossed before the homoclinic (blue) bifurcation line, resulting in the jump of the system into the uncoupled regime $u_n$, completely avoiding the activation of the Q-mode for decreasing frequency.

\begin{figure*}
    \centering
     CW horizontal sweeps\\ 
    \includegraphics[width=0.32\linewidth] {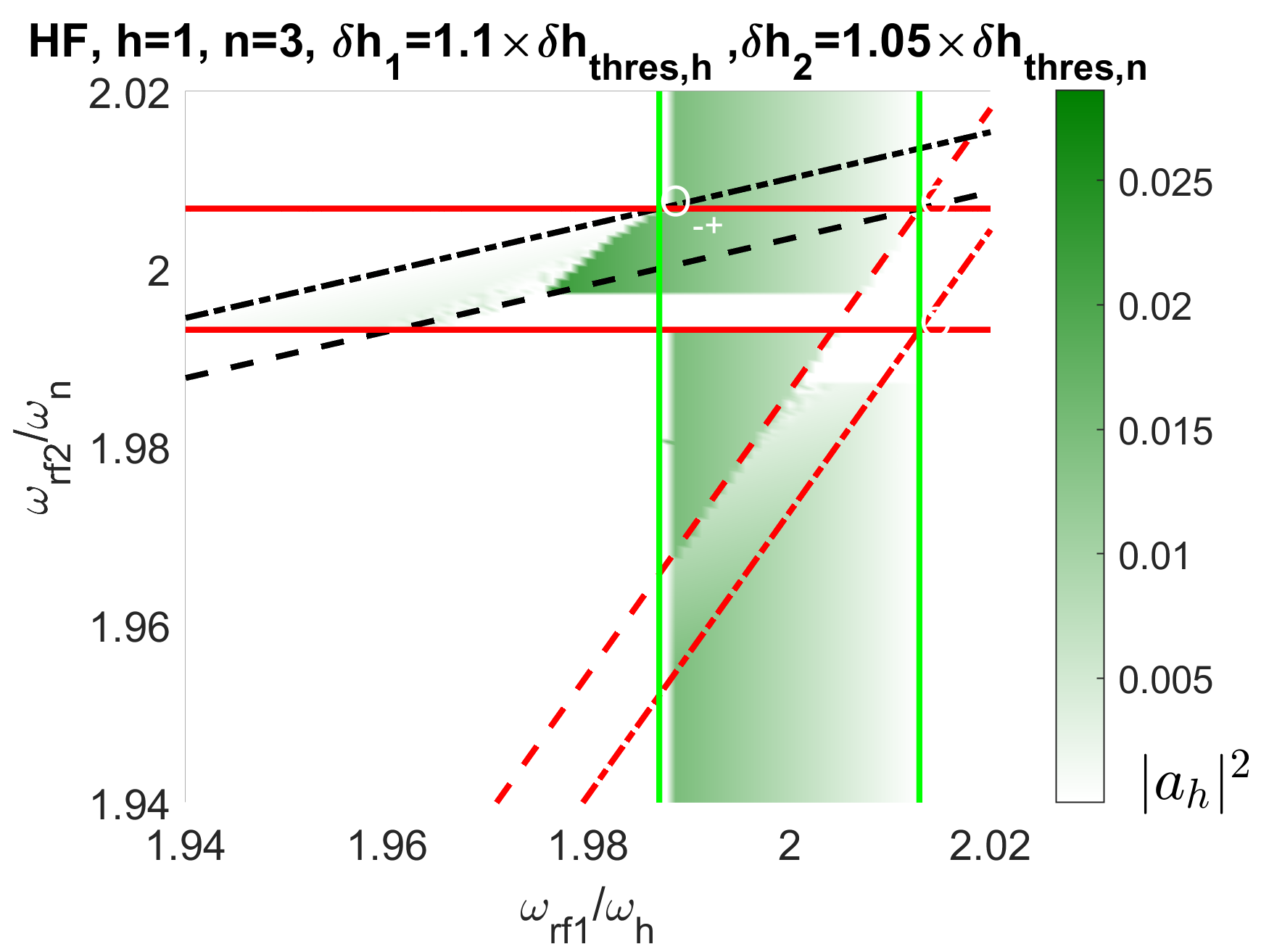}     \includegraphics[width=0.32\linewidth]{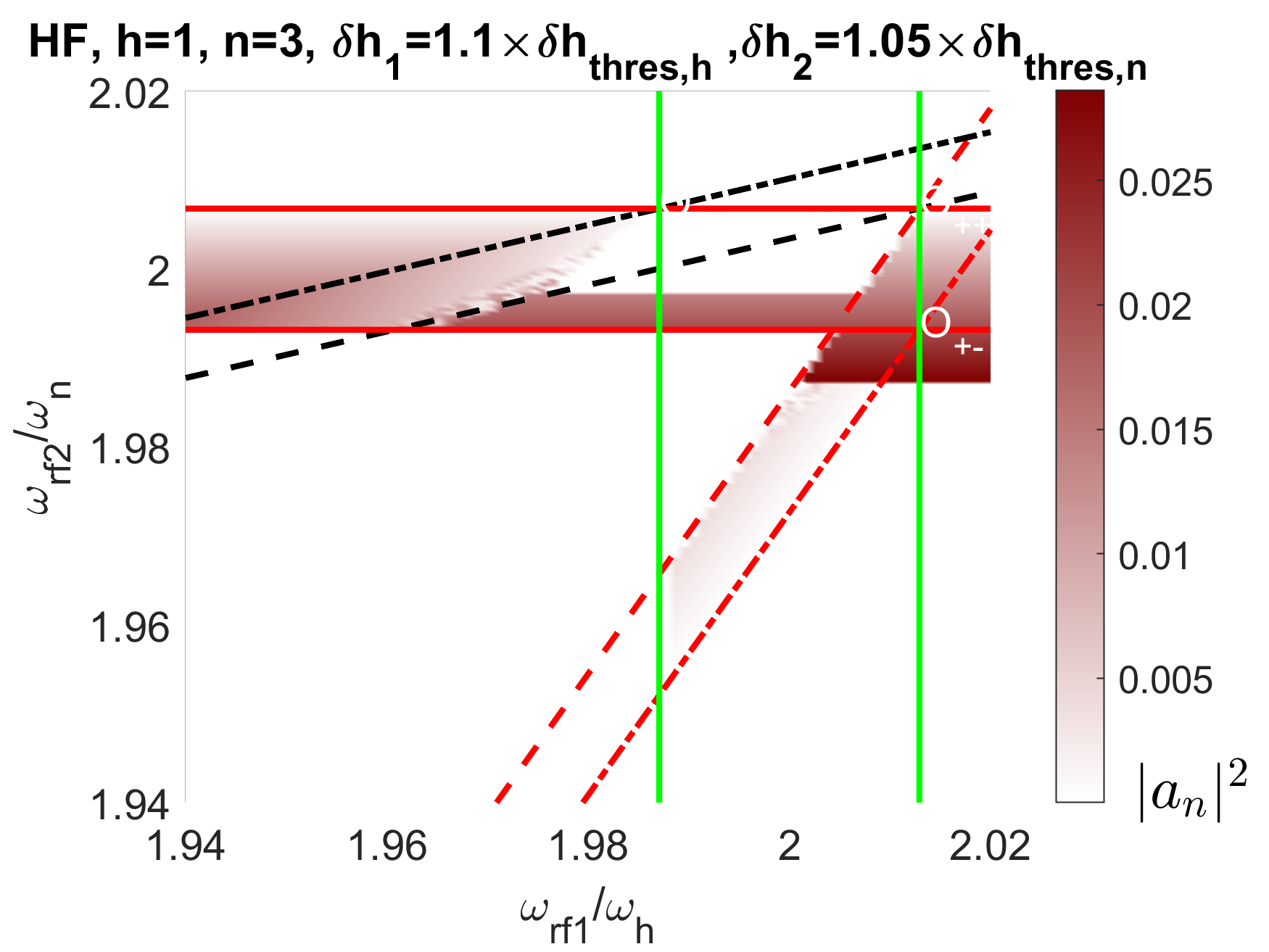}    \includegraphics[width=0.32\linewidth]{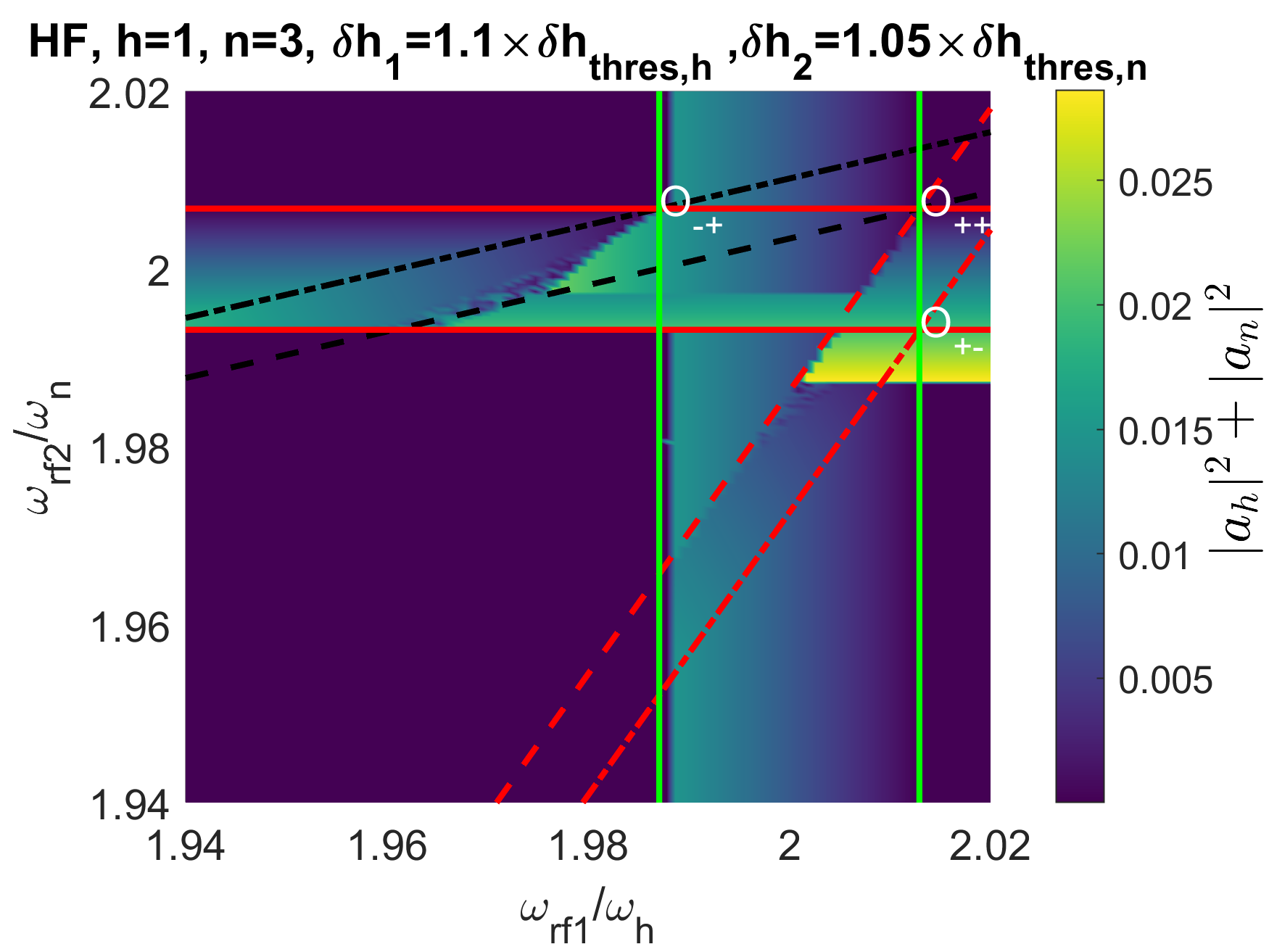} \\
    \includegraphics[width=0.32\linewidth]{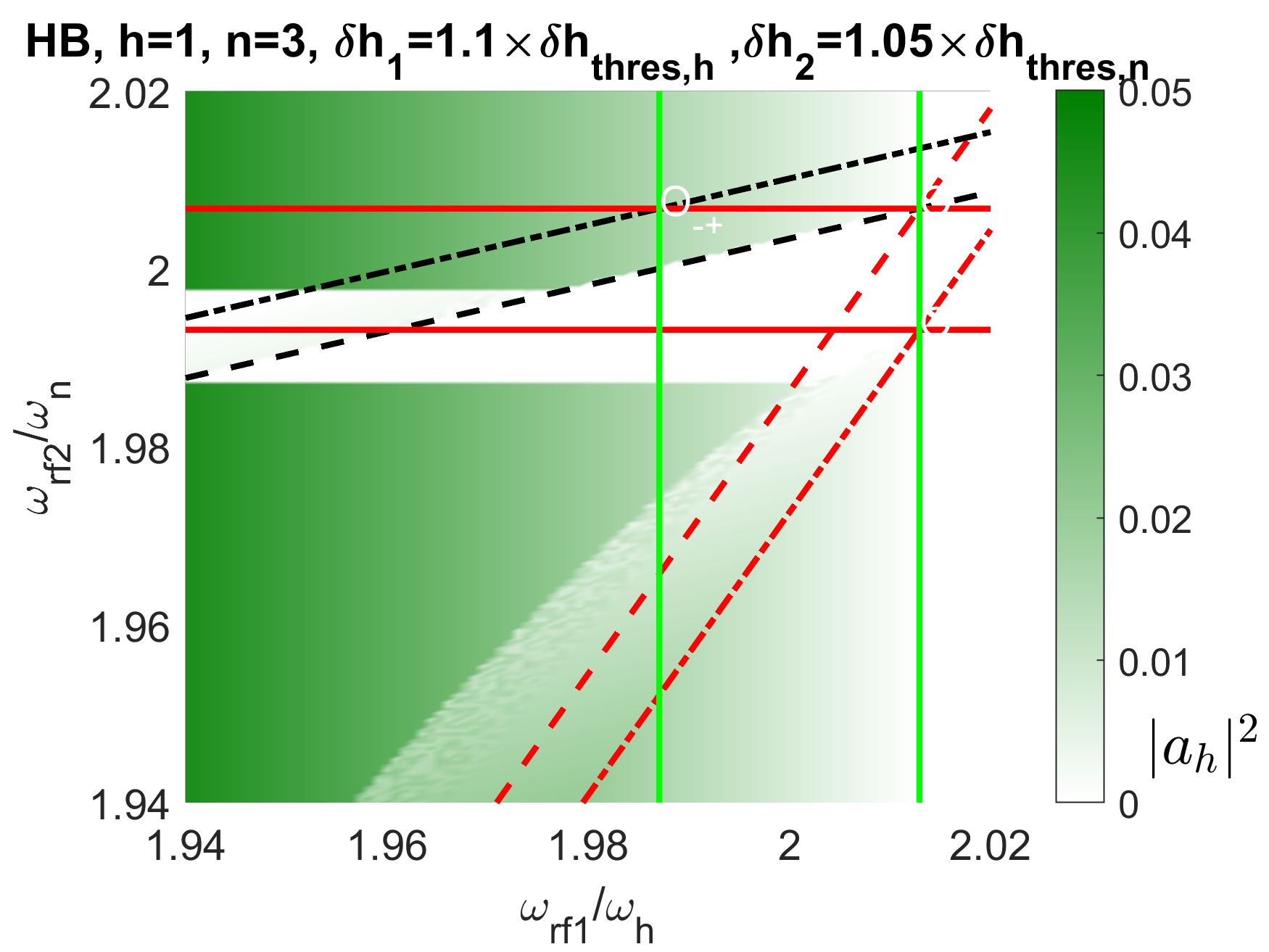}     \includegraphics[width=0.32\linewidth]{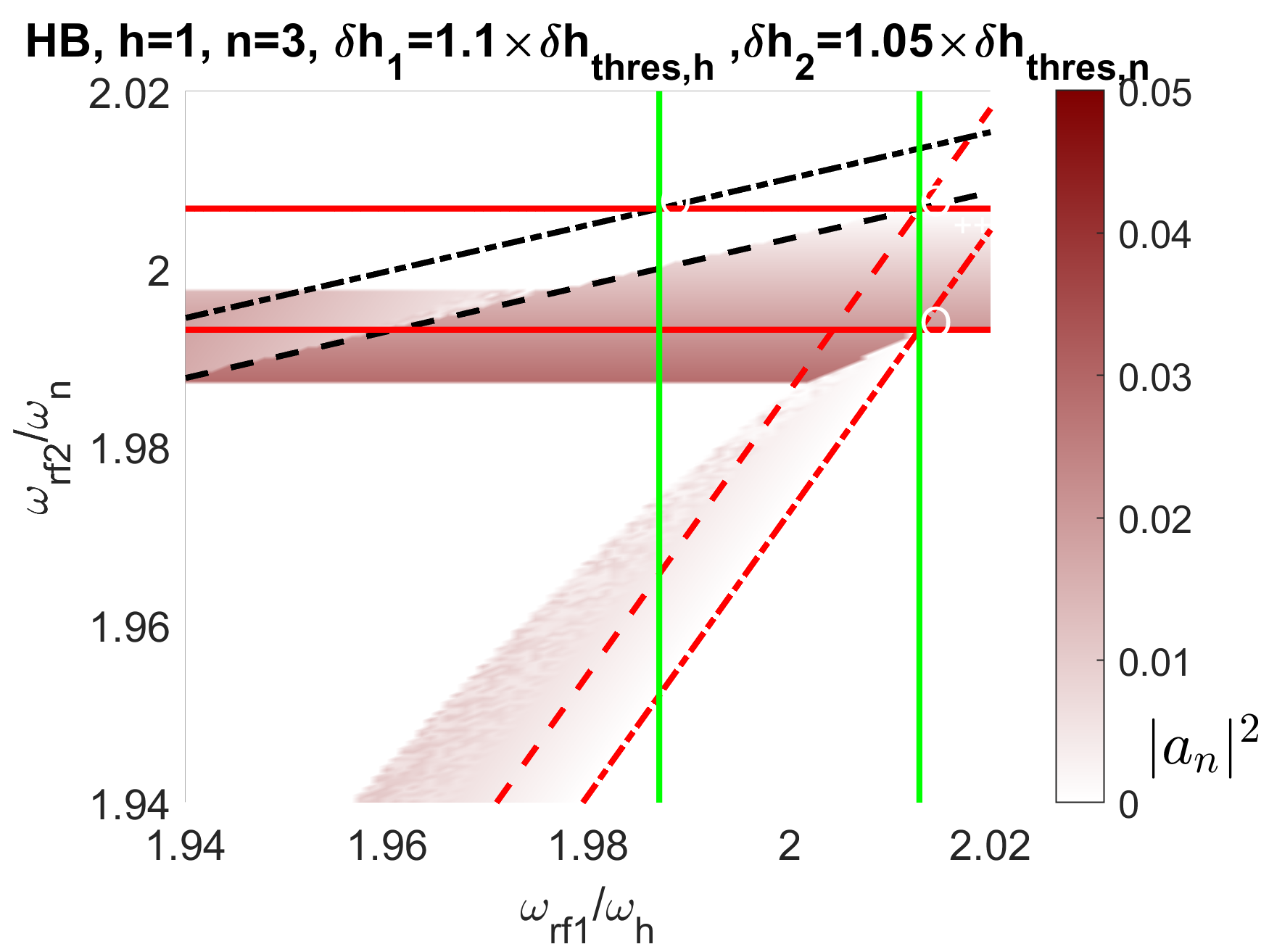}    \includegraphics[width=0.32\linewidth]{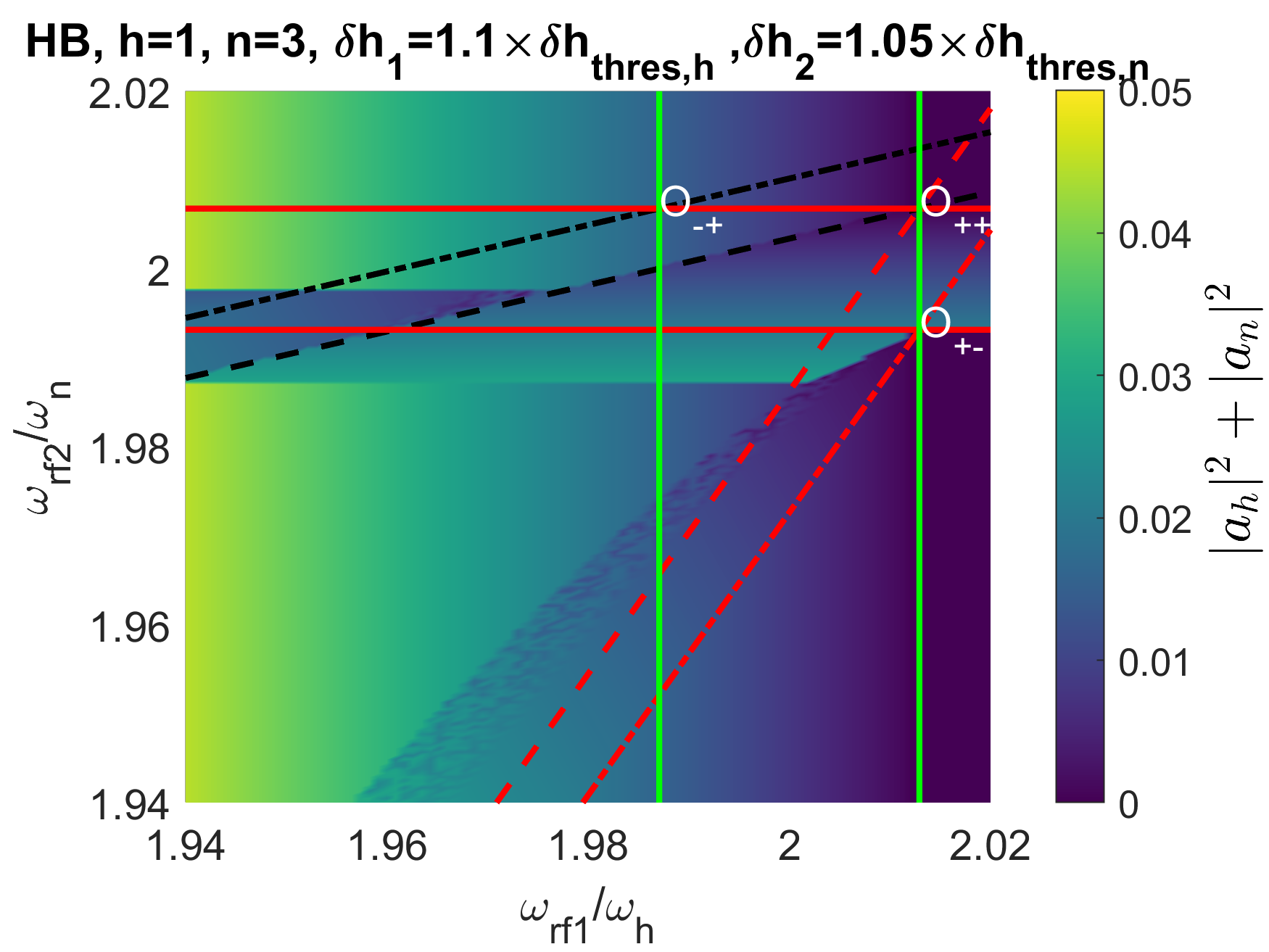} \\

     CW vertical sweeps\\ 
     \includegraphics[width=0.32\linewidth]{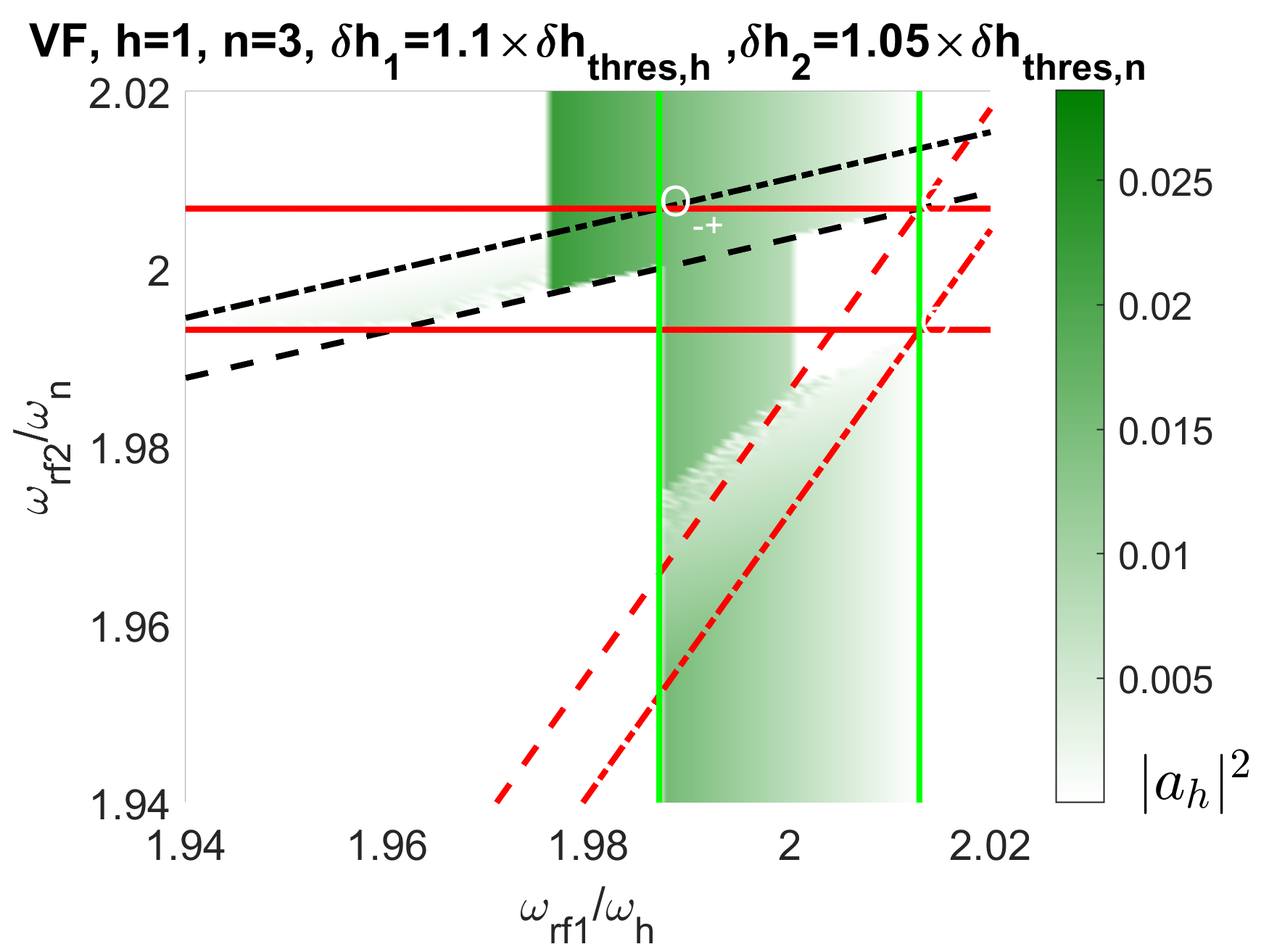}     \includegraphics[width=0.32\linewidth]{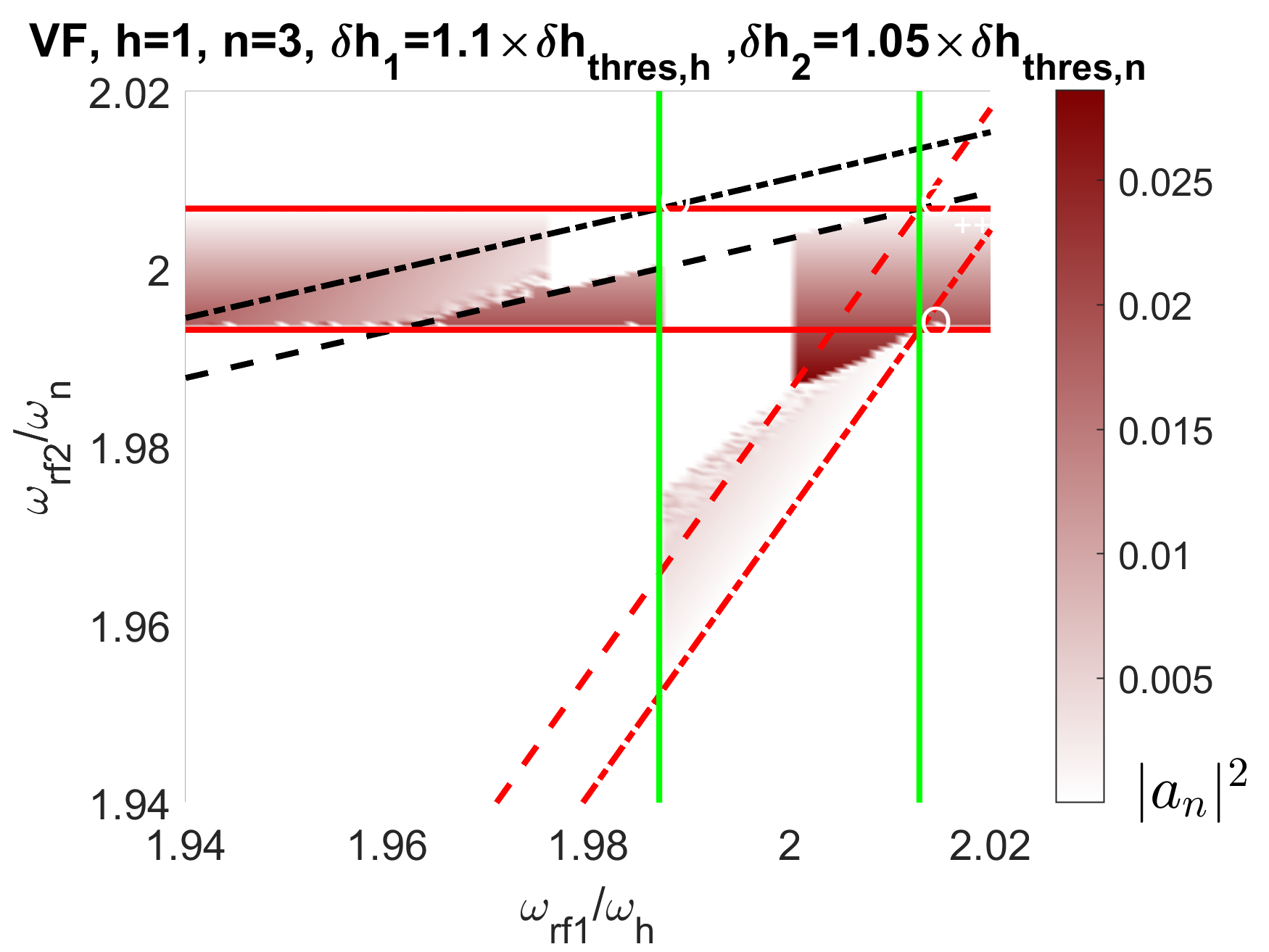}    \includegraphics[width=0.32\linewidth]{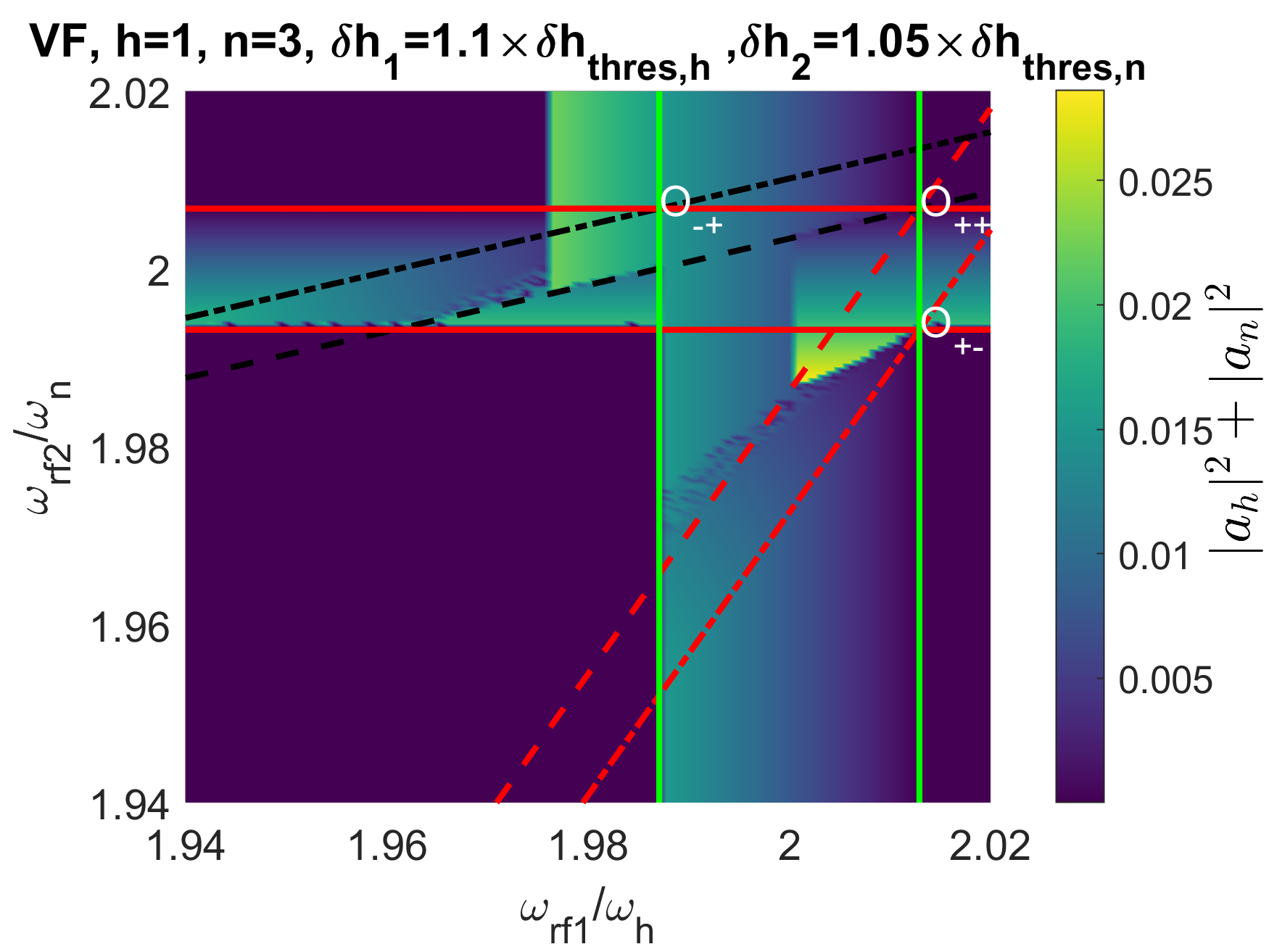} \\
    \includegraphics[width=0.32\linewidth]{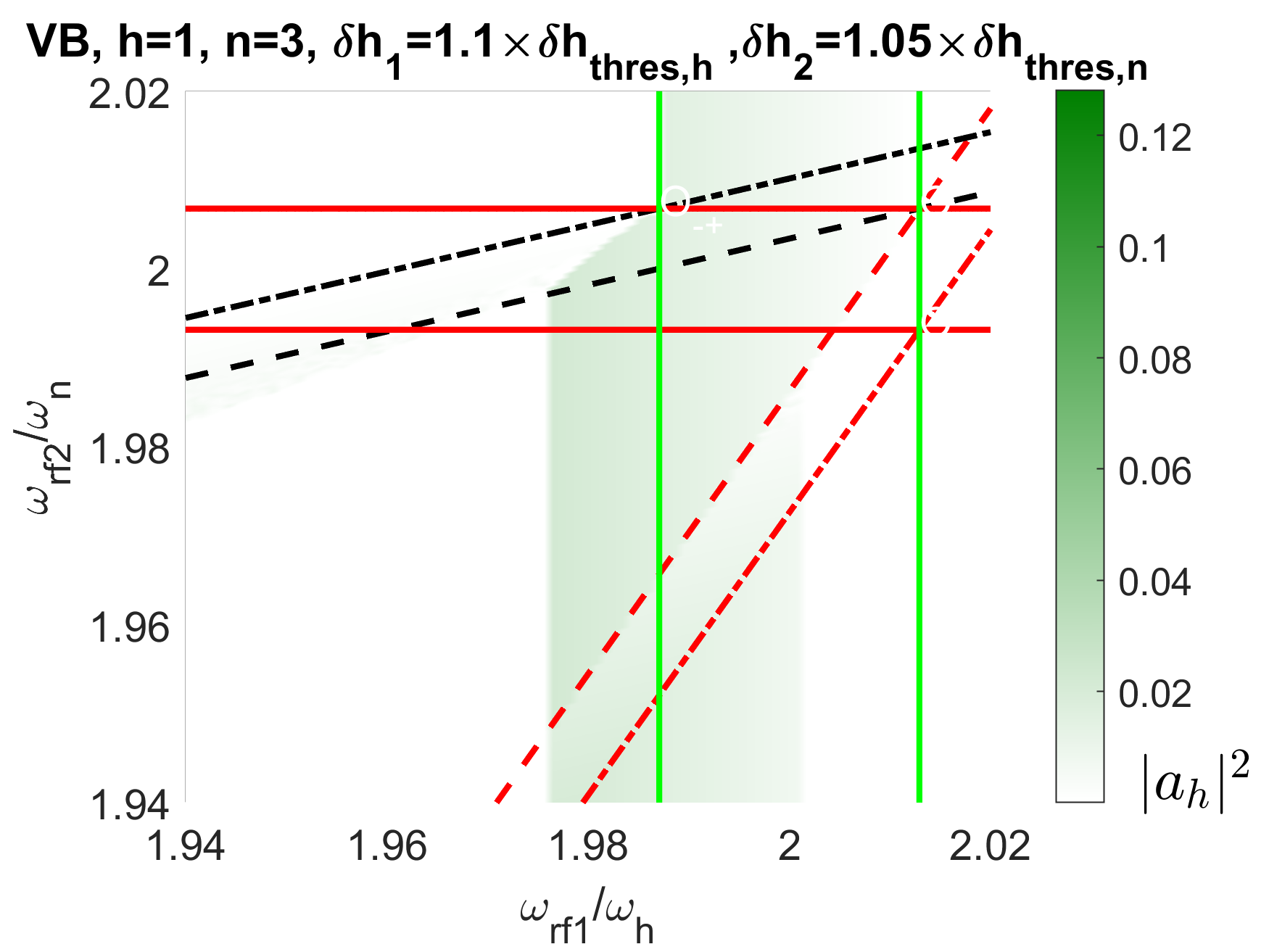}     \includegraphics[width=0.32\linewidth]{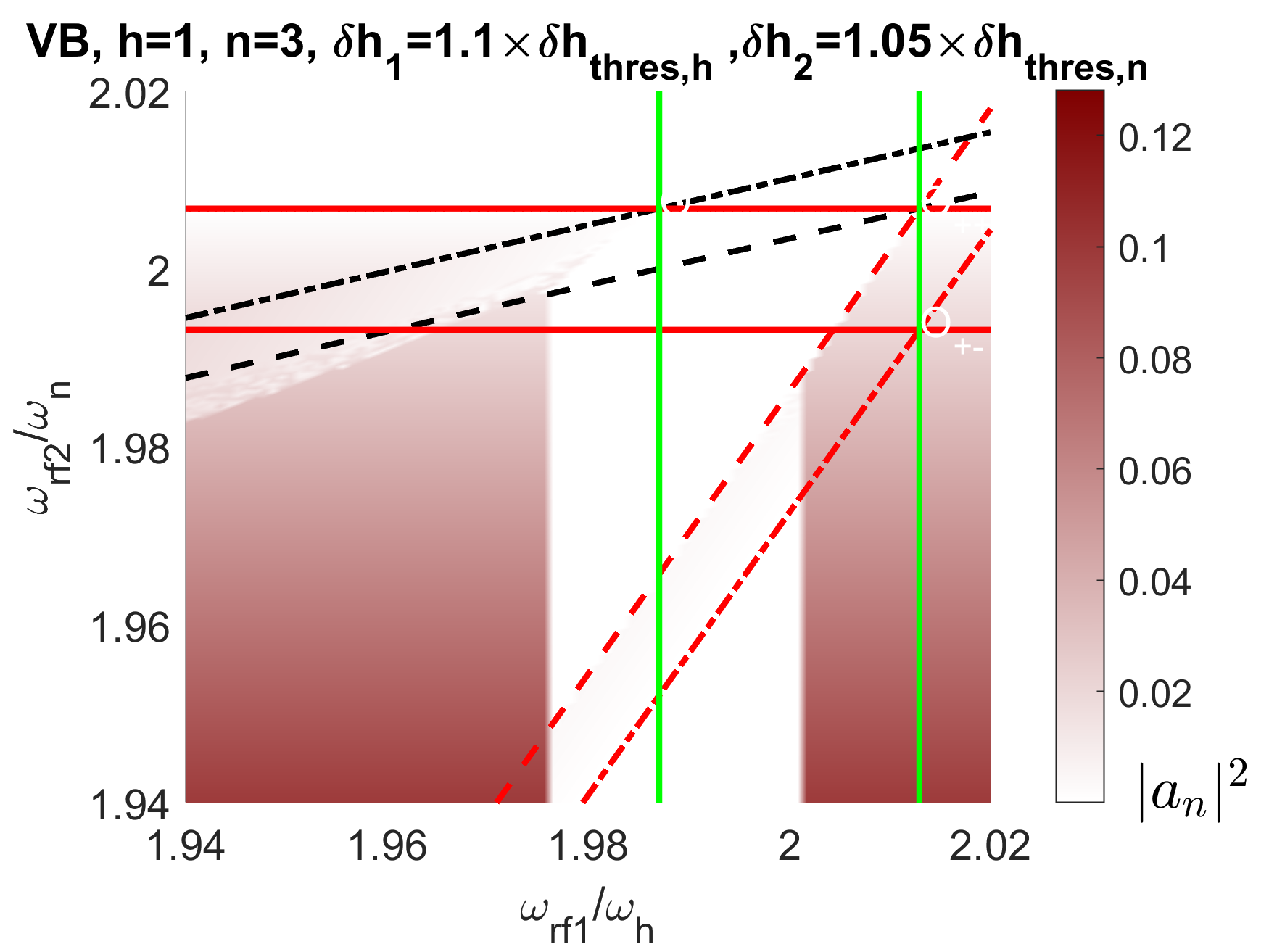}    \includegraphics[width=0.32\linewidth]{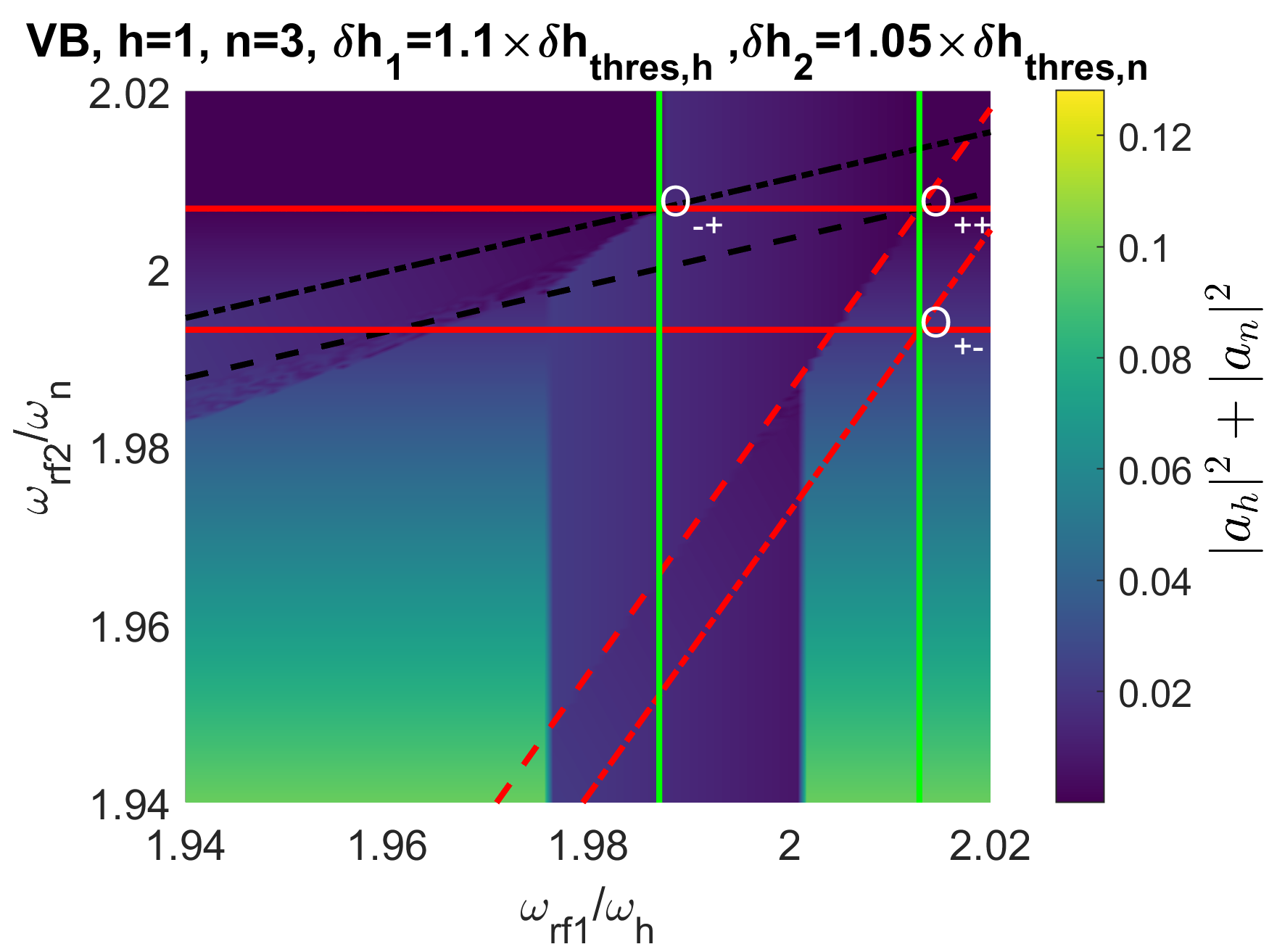}   
    \caption{Continuous wave (CW) dynamics for parametrically-excited modes $h=1$ and $n=3$ via parallel pumping with two-tone signal at fixed above-threshold amplitudes $\delta h_1=1.1\times\delta h_\mathrm{thres,h},\delta h_2=1.05\times\delta h_\mathrm{thres,n}$. Each point in the colormap is obtained integrating the averaged two-modes model \eqref{eq:averaged two nonlinear parametric Ah}-\eqref{eq:averaged two nonlinear parametric Phi_n} for long enough time to approach the steady-state. The panels report the amplitude diagrams associated to steady-state for: (left) mode $h$, amplitude  $|a_h|^2$; (middle) mode $n$, amplitude $|a_n|^2$; (right) summed amplitudes $|a_h|^2+|a_n|^2$. Notation HF/HB or VF/VB in the panel title bar means Horizontal Forward/Backward and Vertical Forward/Backward frequency sweep, respectively.  Black dashed line corresponds to the condition $c_{++}\equiv u_n$ while red dashed line refers to $c_{++}\equiv u_h$. The  boundaries of regions $C_h$ and $C_n$ are marked with green and red solid lines, respectively.
    Black dash-dotted line refers to the condition $c_{-+}\equiv u_n$ ('no $u_n$'), while red dash-dotted line refers to the condition $c_{+-}\equiv u_h$ ('no $u_h$'). Solid yellow (Hopf bifurcation) lines $c\leftrightarrow q$ refer to the boundary of regions $T_{+-},T_{-+}$ due to the onset of Q-modes, solid blue (homoclinic bifurcation) lines $q\rightarrow u$ denote the boundary of the regions where Q-modes can exist. The phase diagram in fig.\ref{fig:type1 phase diagram and hysteresis modes 1 and 3} is instrumental for the interpretation of the amplitude patterns.}
    \label{fig:CW array h1 n3}
\end{figure*}

We have also performed numerical simulations of forward/backward horizontal and vertical frequency sweeps by time-integrating the averaged two-modes model \eqref{eq:averaged two nonlinear parametric Ah}-\eqref{eq:averaged two nonlinear parametric Phi_n}. The integration has been carried out for a time long enough to let the system go to the steady state. The final amplitudes $|a_h|^2,|a_n|^2$ have been used to compute the amplitude maps in the control plane $(\omega_\mathrm{rf1},\omega_\mathrm{rf2})$. The results are reported in figure \ref{fig:CW array h1 n3} and can be interpreted with the help of the phase diagram reported in figure \ref{fig:type1 phase diagram and hysteresis modes 1 and 3}. 

\begin{figure*}
    \centering
     PWM sequence $hn$\\ 
    \includegraphics[width=0.32\linewidth] {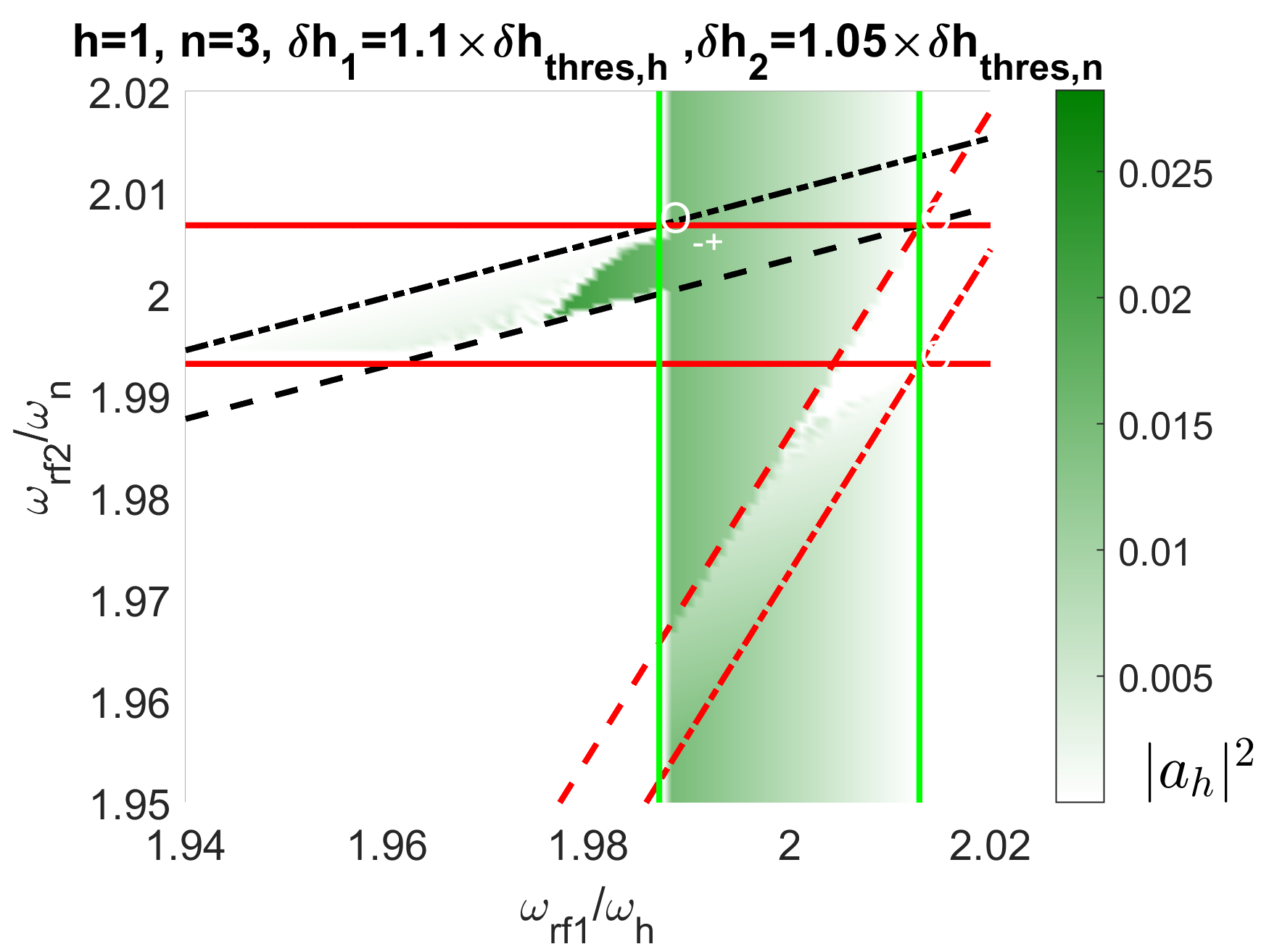}     \includegraphics[width=0.32\linewidth]{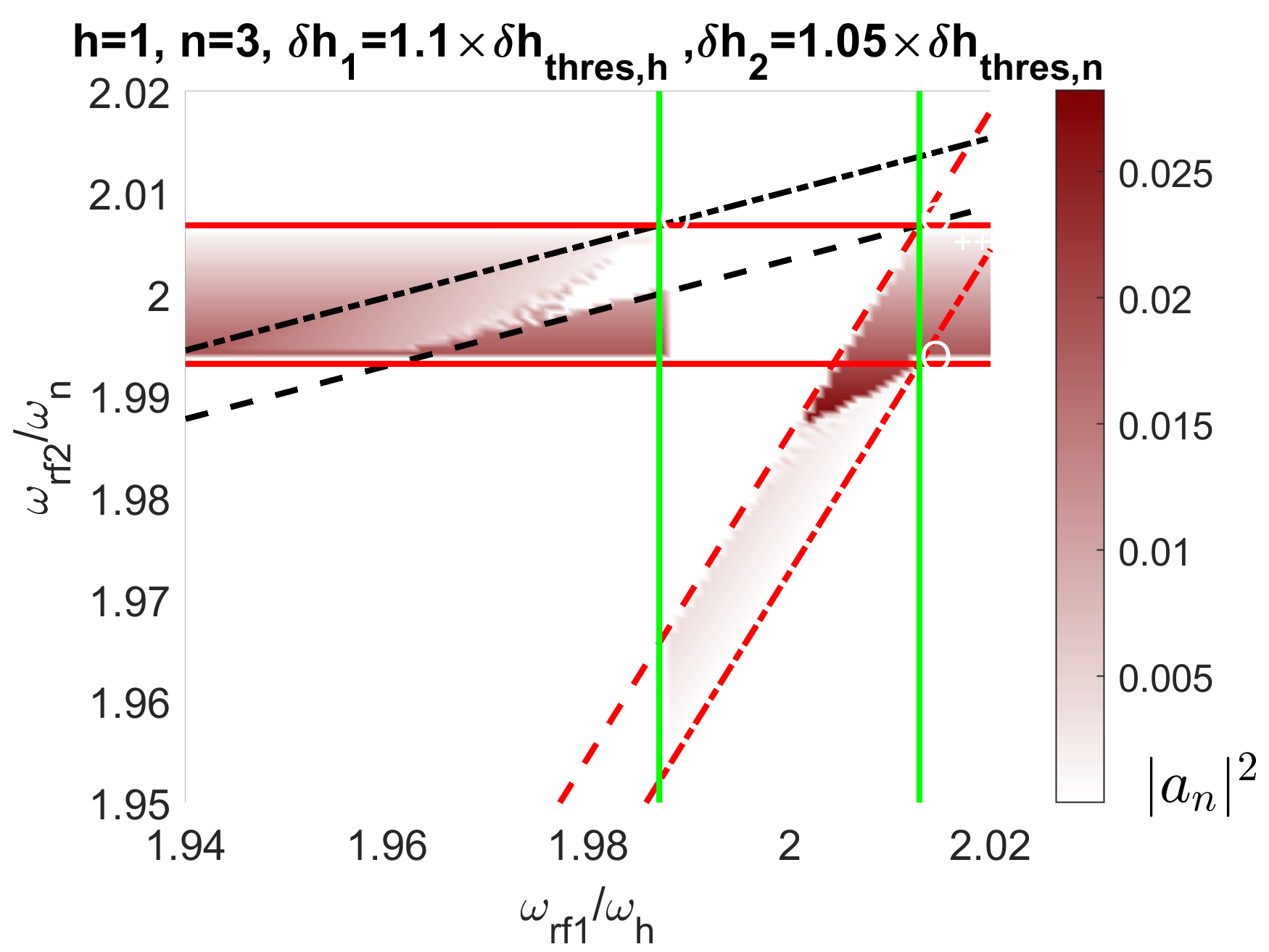}    \includegraphics[width=0.32\linewidth]{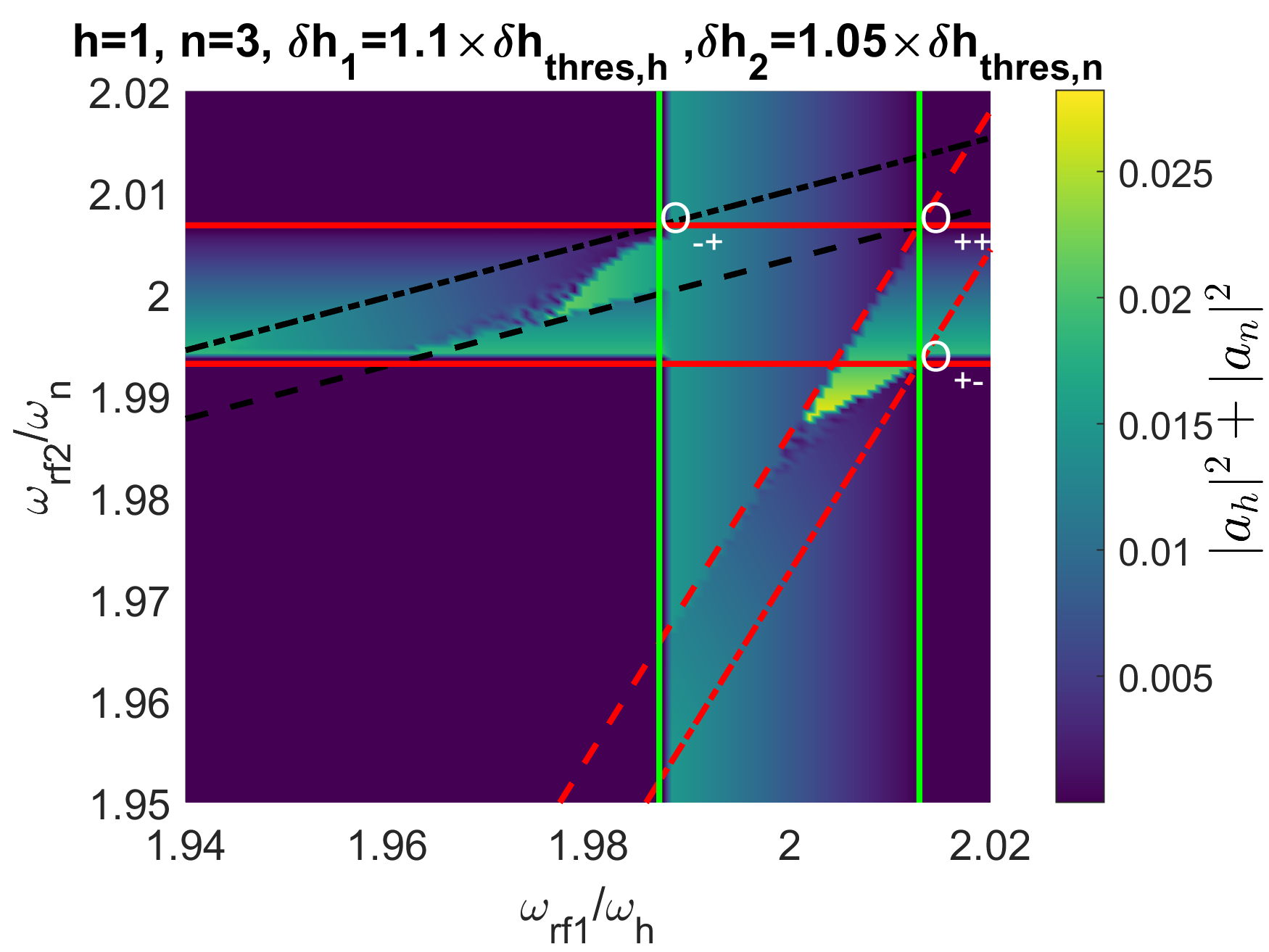} \\
   PWM sequence $nh$\\ 
    \includegraphics[width=0.32\linewidth] {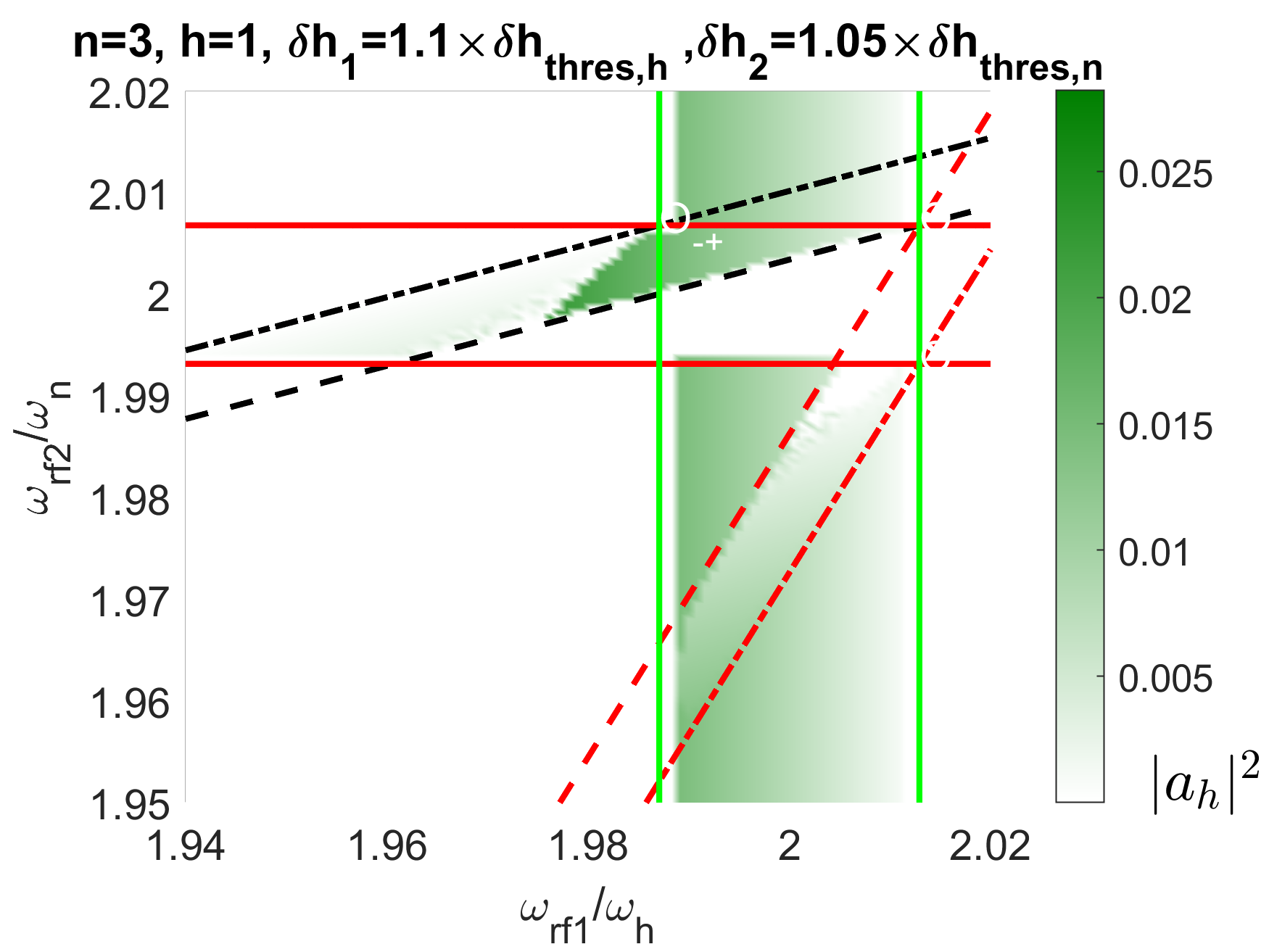}     \includegraphics[width=0.32\linewidth]{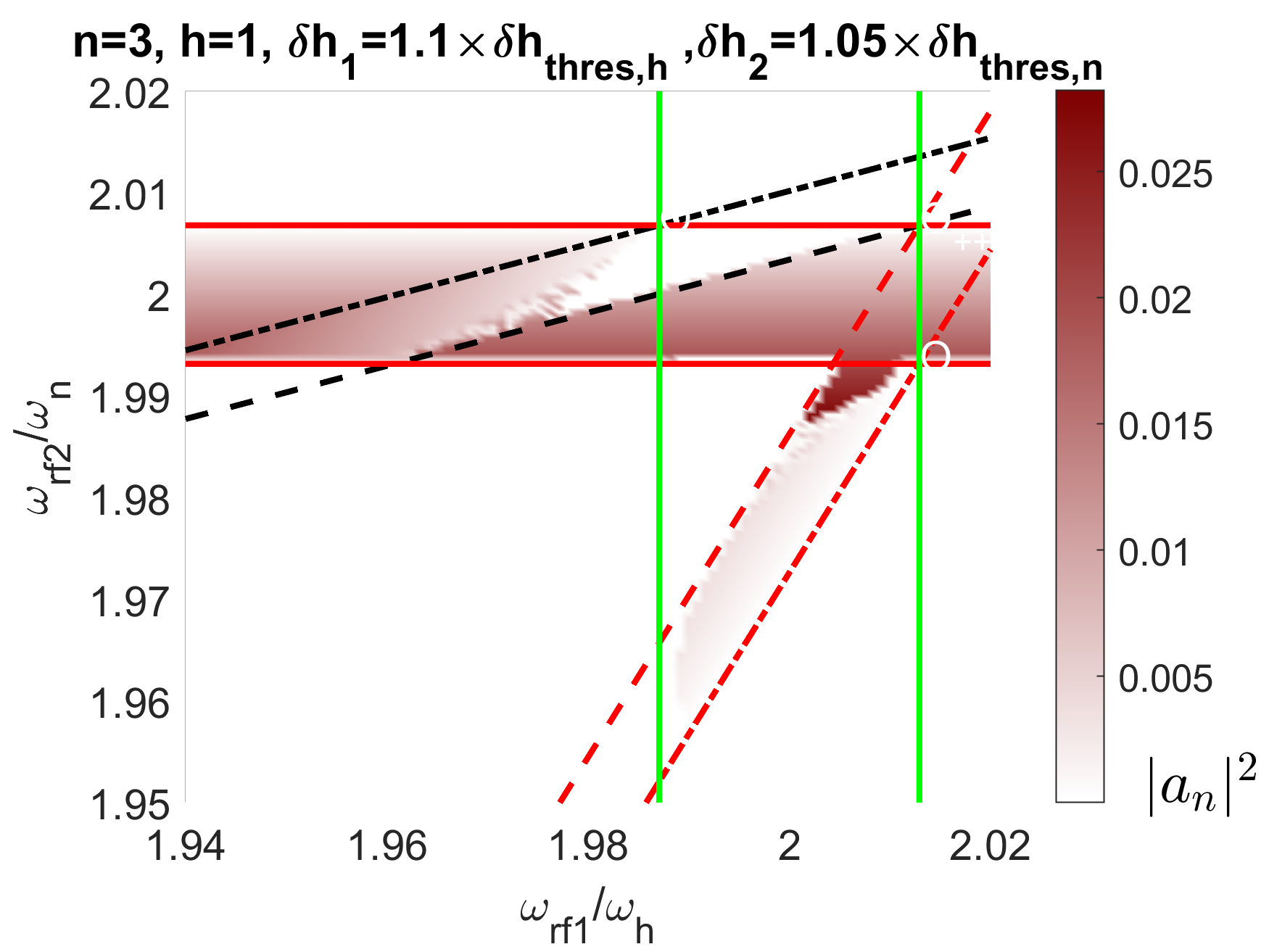}    \includegraphics[width=0.32\linewidth]{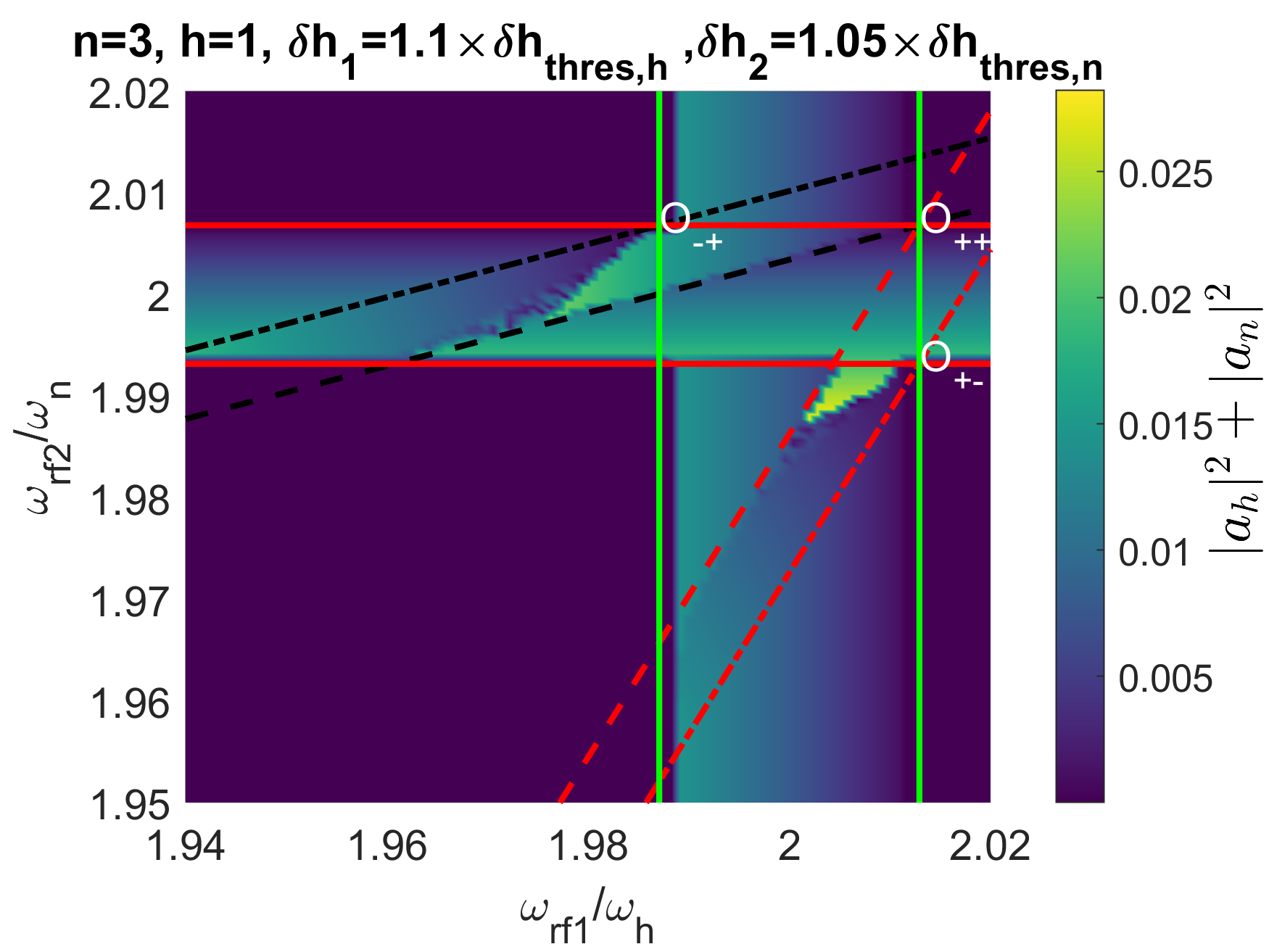} 

    \caption{Pulse-width Modulated (PWM) dynamics for parametrically-excited modes $h=1$ and $n=3$ via parallel pumping with two-tone signal at fixed above-threshold amplitudes $\delta h_1=1.1\times\delta h_\mathrm{thres,h},\delta h_2=1.05\times\delta h_\mathrm{thres,n}$. Each point in the colormap is obtained integrating the averaged two-modes model \eqref{eq:averaged two nonlinear parametric Ah}-\eqref{eq:averaged two nonlinear parametric Phi_n} for long enough time to approach the steady-state.  The panels report the amplitude diagrams associated to steady-state for: (left) mode $h$, amplitude  $|a_h|^2$; (middle) mode $n$, amplitude $|a_n|^2$; (right) summed amplitudes $|a_h|^2+|a_n|^2$. Notation $hn$ or $nh$ in the panel title bar refers to the tone sequence in the PWM excitation.  Black dashed line corresponds to the condition $c_{++}\equiv u_n$ while red dashed line refers to $c_{++}\equiv u_h$. The  boundaries of regions $C_h$ and $C_n$ are marked with green and red solid lines, respectively.
    Black dash-dotted line refers to the condition $c_{-+}\equiv u_n$ ('no $u_n$'), while red dash-dotted line refers to the condition $c_{+-}\equiv u_h$ ('no $u_h$'). The phase diagram in fig.\ref{fig:type1 phase diagram and hysteresis modes 1 and 3} is instrumental for the interpretation  of the amplitude patterns.}
    \label{fig:PWM array h1 n3}
\end{figure*}

We have performed also numerical simulations of mode pairs dynamics driven by PWM excitation. The system has been excited with two sequences, denoted as $hn$ and $nh$, which refer to the order of activation of each tone of the driving signal. The results are reported in figure \ref{fig:PWM array h1 n3}  and can be interpreted with the help of figure  \ref{fig:type1 phase diagram and hysteresis modes 1 and 3}. 

In addition, CW simulations involving Q-modes, CW and PWM simulations of the averaged two modes model for modes $h=1,n=3$ are reported in figs.\ref{fig:type1 hysteresis Qmodes 1 and 3},\ref{fig:CW array h1 n3},\ref{fig:PWM array h1 n3}.

\begin{figure*}[t]
    \centering    \includegraphics[width=0.3\textwidth]{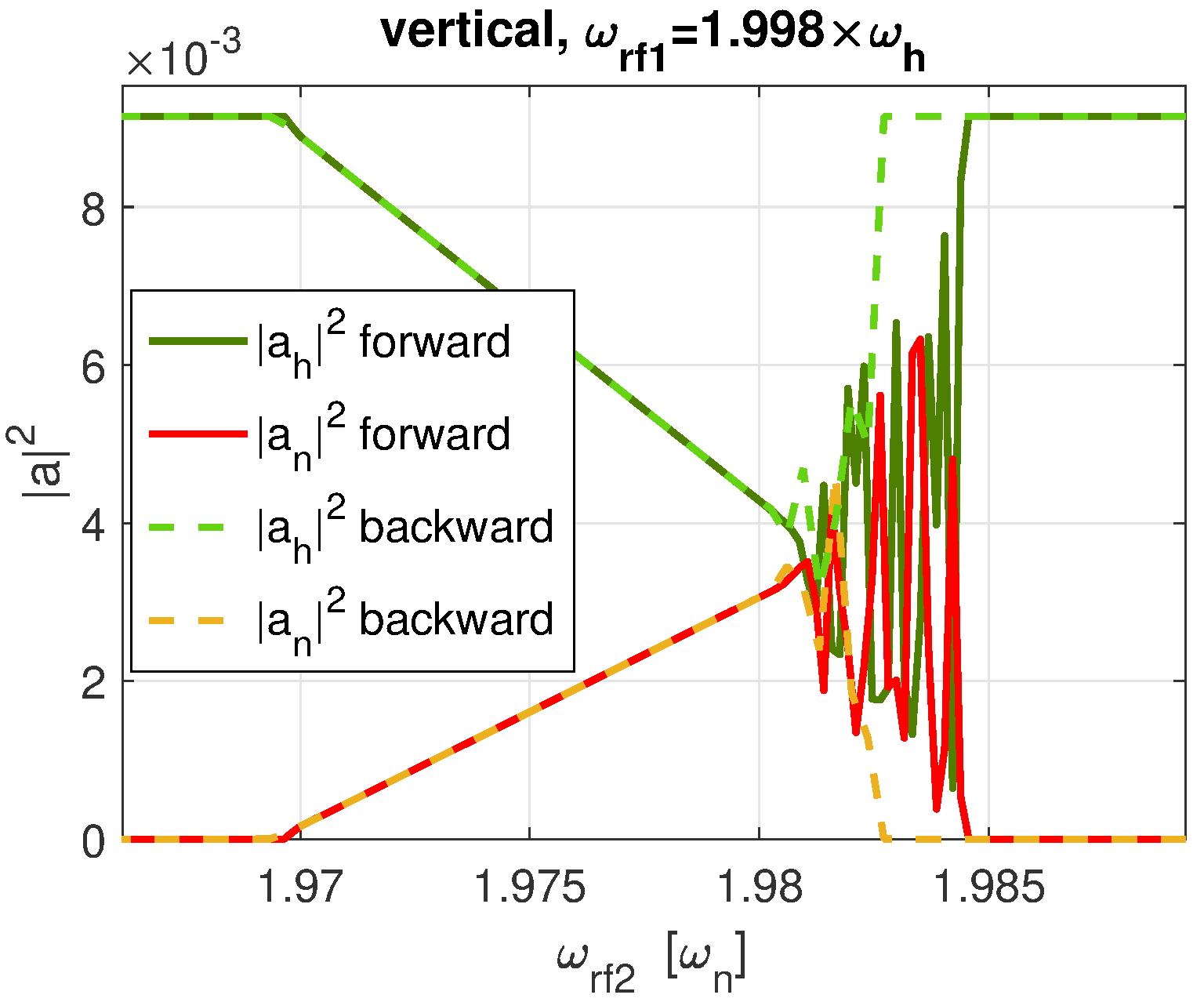}\includegraphics[width=0.31\textwidth]{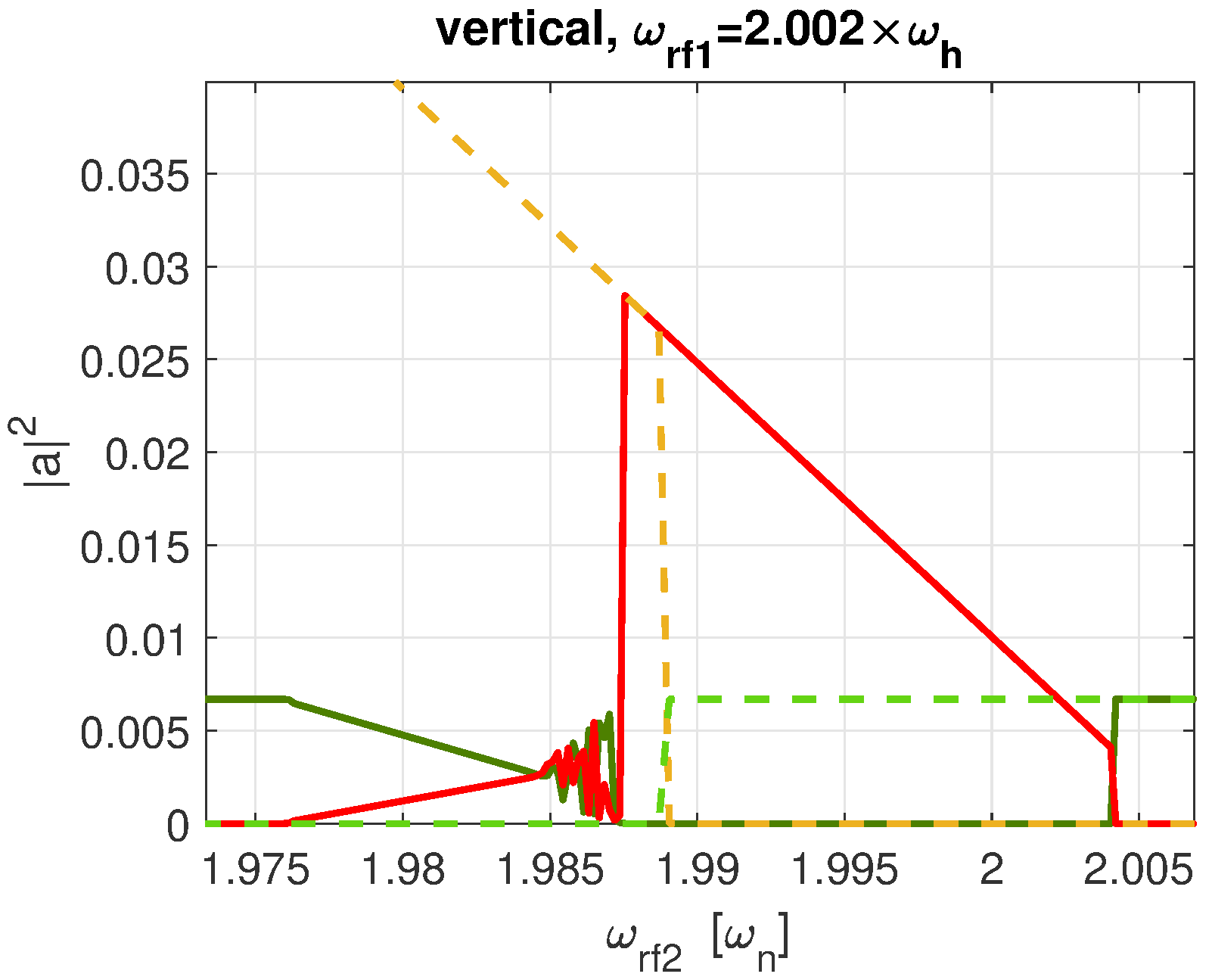}
    \caption{Parametrically-excited modes $h=1$ and $n=3$ via parallel pumping with two-tone signal at fixed above-threshold amplitudes $\delta h_1=1.1\times\delta h_\mathrm{thres,h},\delta h_2=1.05\times\delta h_\mathrm{thres,n}$. NFS coefficients are $N_{hh}=-0.8159, N_{hn}=-1.2972=N_{nh}, N_{nn}=-0.3390$. The panels report steady-state amplitudes $|a_h|^2,|a_n|^2$ as function of frequency corresponding to 'vertical' slow variation of excitation frequency $\omega_\mathrm{rf2}$ while  $\omega_\mathrm{rf1}=1.998\omega_h$ (left) and $\omega_\mathrm{rf1}=2.002\omega_h$ (right). Both amplitude responses exhibit hysteretic behavior (see solid/dashed lines referring to forward/backward frequency sweep), but with different qualitative behavior.}
    \label{fig:type1 hysteresis Qmodes 1 and 3}
\end{figure*}

\end{document}